\documentclass[useAMS,usenatbib]{mn2e-phd}
\usepackage{amssymb}
\usepackage{color}
\usepackage{graphics}
\usepackage{tabularx}
\usepackage{natbib}
\usepackage{fancyhdr}
\usepackage[english]{babel}

\def\phdchapter#1{\section{#1}}
\def\phdsection#1{\subsection{#1}}
\def\phdsubsection#1{\subsubsection{#1}}
\def\phdsubsubsection#1{\paragraph{#1}}
\def\xitem{\item}

\def\phdtitle{Tools for discovering and characterizing extrasolar planets}
\def\phdshorttitle{Tools for discovering and characterizing extrasolar planets}

\def\phdauthor{Andr\'as P\'al}

\def\phdshortauthor{A. P\'al}
\def\phdauthoremail{apal@szofi.net}

\newcommand{\fihat}{\pack{FI/fihat}}

\def\eqref#1{equation~(\ref{#1})}
\def\eqrefs#1#2{equations~(\ref{#1})-(\ref{#2})}

\def\binom#1#2{{#1 \choose #2}}

\def\eoq{\mathrm{q}}
\def\eop{\mathrm{p}}
\def\ordo{\mathcal{O}}

\def\aref{\ensuremath{\mathcal{R}}}				
\def\areft{\ensuremath{\mathcal{R^{\prime}}}}			
\def\aimg{\ensuremath{\mathcal{I}}}				
\def\atran{\ensuremath{\mathcal{F}_{R\rightarrow I}}}		
\def\atranone{\ensuremath{\mathcal{F}^{(1)}_{R\rightarrow I}}}	
\def\atrani{\ensuremath{\mathcal{F}^{(i)}_{R\rightarrow I}}}	
\def\tmixed{\ensuremath{T^{\rm (mix)}}}				
\def\tchir{\ensuremath{T^{\rm (chir)}}}
\def\tcont{\ensuremath{T^{\rm (cont)}}}

\newcommand{\pack}[1]{\textsc{\lowercase{#1}}}
\newcommand{\prog}[1]{\texttt{\lowercase{#1}}}

\newcommand{\arcdeg}{\ensuremath{^{\circ}}}

\newcommand{\lc}{light curve}
\newcommand{\lcs}{light curves}

\newcommand{\kms}{\ensuremath{\rm km\,s^{-1}}}
\newcommand{\ms}{\ensuremath{\rm m\,s^{-1}}}

\newcommand{\gcmc}{\ensuremath{\rm g\,cm^{-3}}}

\newcommand{\logg}{\ensuremath{\log{g}}}
\newcommand{\vsini}{\ensuremath{v \sin{i}}}

\newcommand{\teffstar}{\ensuremath{T_{\rm eff}}}

\newcommand{\rjup}{\ensuremath{R_{\rm J}}}
\newcommand{\mjup}{\ensuremath{M_{\rm J}}}

\newcommand{\flwof}{\mbox{FLWO 1.2 m}}		
\newcommand{\flwos}{\mbox{FLWO 1.5 m}}		

\newcommand{\sophie}{OHP}

\newcommand{\band}[1]{\ensuremath{#1}-band}	

\newsavebox{\boxcmdbox}
\newenvironment{lcmd}{\setlength{\fboxsep}{0pt}\begin{lrbox}{\boxcmdbox}\hspace*{6mm}\begin{minipage}{160mm}\vspace*{2mm}\tt}{\vspace*{2mm}\end{minipage}\hspace*{5mm}\end{lrbox}\begin{center}\framebox{\colorbox[gray]{0.9}{\usebox{\boxcmdbox}}}\end{center}}
\newenvironment{scmd}{\setlength{\fboxsep}{0pt}\begin{lrbox}{\boxcmdbox}\hspace*{4mm}\begin{minipage}{74mm}\vspace*{2mm}\tt}{\vspace*{2mm}\end{minipage}\hspace*{4mm}\end{lrbox}\begin{center}\framebox{\colorbox[gray]{0.9}{\usebox{\boxcmdbox}}}\end{center}}
\newenvironment{ccmd}{\setlength{\fboxsep}{0pt}\begin{lrbox}{\boxcmdbox}\hspace*{4mm}\begin{minipage}{60mm}\vspace*{2mm}\tt}{\vspace*{2mm}\end{minipage}\hspace*{4mm}\end{lrbox}\framebox{\colorbox[gray]{0.9}{\usebox{\boxcmdbox}}}}
\def\ptab{\hspace*{10ex}}

\def\dgmatrix#1#2#3#4{\begin{tabular}{cc}\ensuremath{#1}&\ensuremath{#2}\\\ensuremath{#3}&\ensuremath{#4}\end{tabular}}
\def\dpmatrix#1#2#3#4{\left(\dgmatrix{#1}{#2}{#3}{#4}\right)}

\title[\phdshorttitle]{\phdtitle}
\author[\phdshortauthor]{\phdauthor\ensuremath{^{1,2}}\thanks{E-mail: \phdauthoremail}\\
$^{1}$Department of Astronomy, Lor\'and E\"otv\"os University,
	P\'azm\'any P. st. 1/A,
	Budapest H-1117, Hungary \\
$^{2}$Harvard-Smithsonian Center for Astrophysics,
	60 Garden street,
	Cambridge, MA, 02138, USA}

\begin{document}

\sloppy

\selectlanguage{english}

\maketitle

\begin{abstract}	
Among the group of extrasolar planets, transiting planets provide 
a great opportunity to obtain direct measurements for the basic 
physical properties, such as mass and radius of these objects. 
These planets are therefore highly important in the understanding of 
the evolution and formation of planetary systems: from
the observations of photometric transits, the interior structure of the planet
and atmospheric properties can also be constrained. 
The most efficient way to search for transiting 
extrasolar planets is based on wide-field surveys by hunting
for short and shallow periodic dips in light curves covering 
quite long time intervals. These surveys
monitor fields with several degrees in diameter and tens or hundreds
of thousands of objects simultaneously. In the practice of astronomical
observations, surveys of large field-of-view are rather new and therefore
require special methods for photometric data reduction that have not been used 
before. Since 2004, I participate in the HATNet project, one of 
the leading initiatives in the competitive search for transiting
planets. Due to the lack of software solution which is capable
to handle and properly reduce the yield of such a wide-field survey,
I have started to develop a new package designed to perform the
related data processing and analysis. After several years of improvement,
the software package became sufficiently robust and played a key
role in the discovery of several transiting planets. In addition,
various new algorithms for data reduction had to be developed, 
implemented and tested which were relevant during the reduction
and the interpretation of data. 

In this PhD thesis, I summarize my efforts related to the 
development of a complete software solution for high precision photometric 
reduction of astronomical images. I also demonstrate the role of
this newly developed package and the related algorithms
in the case of particular discoveries of the HATNet project. 
\end{abstract}

%
%
%




\phdchapter{Introduction}
\label{chapter:introduction}

In the last two decades, the discovery and characterization of 
extrasolar planets became an exciting field of astronomy. The 
first companion that was thought to be an object roughly 10 
times more massive than Earth,
had been detected around the pulsar PSR1829-10 
\citep{bailes1991}. Although this detection 
turned out to be a false one \citep{lyne1992}, shortly after the 
method of detecting planetary companions involving 
the analysis of pulsar timing variations led to the successful confirmation
of the multiple planetary system around PSR1257+12 
\citep{wolszczan1992}. The pioneering discovery of a planet 
orbiting a main sequence star was announced 
by \citet{mayor1995}. They reported the presence of a short-period planet 
orbiting the Sun-like star 51~Peg. This detection was based on precise
radial velocity measurements with uncertainties at the level of meter 
per second. Both 
discovery methods mentioned above are based on the fact that all components
in a single or multiple planetary system, including the host star itself,
revolve around the common barycenter, that is the point in the system having
inertial motion. Thus, companions with smaller masses offset the barycenter
only slightly from the host star whose motion is detected, either by
the analysis of pulsar timing variations or by radial velocity measurements.
Therefore, such methods -- which are otherwise fairly common among the 
investigation techniques of binary or multiple stellar systems -- yielded
success in the form of confirming planets only after the evolution
of instrumentation. Due to the physical constraints found in these
methods, the masses of the planets can only be constrained by a lower
limit, while we even do not have any information on the sizes of these
objects.

The discovery of 51~Peg~b was followed by numerous other detections,
mainly by the method of radial velocity analysis, yielding the discovery,
for instance, of the
first planetary system with two planets around 47~UMa 
\citep{butler1996b,fischer2002}, and the first multiple planetary system 
around $\upsilon$~And \citep{butler1999}. Until the first photometric 
detection of planetary transits in the system of HD~209458(b) 
\citep{henry2000,charbonneau2000}, 
no radius estimations could be given to the detected planets, and all 
of these had only lower limits for their masses. Transiting planets 
provide the opportunity to characterize the size of the planet, and 
by the known inclination of its orbit, one can derive the mass 
of the planet without any ambiguity by combining the 
results of transit photometry with the radial velocity measurements.
The planetary nature of HD~209458b was first confirmed by the
analysis of radial velocity variations alone. The first discovery
based on photometric detection of periodic dips
in light curves was the discovery of OGLE-TR-56b 
\citep{konacki2003}. Since several scenarios can mimic events that have 
similar light curves to transiting planets, confirmation spectroscopy
and subsequent analysis of radial velocity data is still necessary
to verify the planetary nature of objects found by transit searches
\citep{queloz2001,torres2005}.

Since the first identification of planetary objects transiting 
their parent stars, numerous additional systems have been discovered
either by the detection of transits after a confirmation based on 
radial velocity measurements or by searching transit-like signals in
photometric time series and confirming the planetary nature with follow-up
spectroscopic and radial velocity data. The former method led to
the discovery of transits for many well-studied systems, such 
as HD~189733 \citep[planet transiting a nearby K dwarf;][]{bouchy2005}, 
GJ~436 \citep{butler2004}, HD~17156 \citep{fischer2007,barbieri2007}
or HD~80606 \citep[the transiting planet with the longest known
orbital period of $\sim111$\,days;][]{naef2001,moutou2009}. These planets
with transits confirmed later on are found around brighter stars since
surveys for radial velocity variations mainly focus on these. 
However, the vast majority of
the currently known transiting extrasolar planets have been detected
by systematic photometric surveys, fully or partially dedicated for 
planet searches. Such projects monitor either faint targets
using telescopes with small field-of-view or bright targets involving
large field-of-view optical instrumentation. Some of the projects 
focused on the monitoring of smaller fields are 
the Monitor project \citep{irwin2006,aigrain2007}, 
Deep MMT Transit Survey \citep{hartman2008},
a survey for planetary transits in the field of NGC 7789 by \cite{bramich2005},
the ``Survey for Transiting Extrasolar Planets in Stellar Systems'' by \cite{burke2004},
``Planets in Stellar Clusters Extensive Search'' \citep{mochejska2002,mochejska2006},
``Single-field transit survey toward the Lupus'' of \cite{weldrake2008},
the SWEEPS project \citep{sahu2006}, and
the Optical Gravitational Lensing Experiment (OGLE)
\citep{udalski1993,udalski2002,konacki2003}. 
Projects monitoring wide fields are the
Wide Angle Search for Planets \citep[WASP, SuperWASP, see][]{street2003,pollacco2004,cameron2007},
the XO project \citep{mccullough2005,mccullough2006},
the Hungarian-made Automated Telescope project \citep[HATNet,][]{bakos2002,bakos2004},
the Transatlantic Exoplanet Survey \citep[TrES,][]{alonso2004},
the Kilodegree Extremely Little Telescope \citep[KELT,][]{pepper2004,pepper2007},
and the Berlin Extrasolar Transit Search project \citep[BEST,][]{rauer2004}.
One should mention here the existing space-borne project, the {\it CoRoT} mission
\citep{barge2008} and the {\it Kepler} mission, launched successfully
on 7 March 2009 \citep{borucki2007}. Both missions are dedicated (in part time)
to searching for transiting extrasolar planets. As of March 2009,
the above mentioned projects announced $57$ planets. 
$6$ planets were found by radial velocity surveys where transits were
confirmed after the detection of RV variations (GJ~436b,  HD~149026b,
HD~17156b, HD~80606b, HD~189733b and HD~209458b), while the other $51$ were
discovered and announced by one of the above mentioned surveys. The {\it CoRoT}
mission announced $7$ planets, for which $4$ had published orbital and planetary
data; the OGLE project reported data for $7$ planets and an additional
planet with existing photometry in the OGLE archive has also been confirmed 
by an independent group \citep{snellen2008}; 
the Transatlantic Exoplanet Survey reported the discovery
of $4$ planets; the XO project has detected and confirmed $5$ planets;
the SWEEPS project found $2$ planets; the SuperWASP project
announced $14+1$ planets, however, $2$ of them are known only from conference
announcements; and the HATNet project has $10+1$ confirmed planets. The
planet WASP-11/HAT-P-10b had a shared discovery, it was confirmed 
independently by the SuperWASP and HATNet groups (this
common discovery has been denoted earliet by the $+1$ term). The HATNet project
also confirmed independently the planetary nature of the object 
XO-5b \citep{pal2008xo5}.

All of the above mentioned wide-field surveys involve optical designs
that yield a field-of-view of several degrees, moreover, 
the KELT project monitors areas having a size of thousand square degrees
(hence the name, ``Kilodegree Extremely Little Telescope''). 
The calibration and data reduction for
such surveys revealed various problems that were not
present on the image processing of ``classical'' data (obtained by 
telescopes with fast focal ratios and therefore smaller field-of-view).
Some of the difficulties that occur are the following. 
Even the calibration frames themselves have to be filtered carefully, in order
to avoid any significant structures (such as patches of clouds in the
flat field images). Images taken by 
fast focal ratio optics have significant vignetting, therefore the 
calibration process should track its side effects, such as the 
variations in the signal-to-noise level across the image. 
Moreover, fast focal ratio yields comatic aberration
and therefore systematic spatial variations in the stellar profiles.
Such variations make the source extraction and star detection algorithms 
not only more sensitive but also are one of the major sources of 
the correlated noise (or red noise) presented in the final light curves
\footnote{The time variation of stellar profiles is what causes red noise}.
Due to the large field-of-view and the numerous individual objects
presented in the image, the source identification and the derivation
of the proper ``plate solution'' for these images is also a non-trivial
issue. The photometry itself is hardened by the very narrow and therefore
undersampled sources. Unless the effects of the undersampled profiles
and the spatial motions of the stellar profiles are handled with care, 
photometric time series are affected by strong systematics. 
Due to the short fractional duration 
and the shallow flux decrease of the planetary transits,
several thousands of individual frames with proper photometry are 
required for significant and reliable detection. Since hundreds
of thousands of stars are monitored simultaneously during the observation
of a single field, the image reduction process yields enormous amount 
of photometric data, i.e. billions of individual photometric measurements.
In fact, hundreds of gigabytes up to terabytes of processed images 
and tabulated data can be associated in a single monitored field. 
Even the most common operations on such a large amount of data 
require special methods. 

The Hungarian Automated Telescope (HAT) project was 
initiated by Bohdan Pacz\'yski and G\'asp\'ar Bakos \citep{bakos2002}.
Its successor, the Hungarian-made Automated Telescope Network 
\citep{bakos2004}
is a network of small class of telescopes with large field-of-view, dedicated
to an all-sky variability survey and search for planetary transits. In
the past years, the project has became one of the most successful 
projects in the discovery of almost one fifth of the known transiting
extrasolar planets. After joining the project in 2004, the author's goal
was to overcome the above mentioned issues and problems,
related to the image processing of the HATNet data. In this thesis, 
the efforts for the development of a software package and its related
applications in the HATNet project are summarized.

This PhD thesis has five chapters. Following the Introduction,
the second chapter, ``Algorithms and Software environment'' 
discusses the newly developed
and applied algorithms that form the basis of the photometry pipeline,
and gives a description on the primary concepts of the 
related software package in which these algorithms are implemented.
The third chapter, ``HATNet discoveries'' describes a particular example for 
the application of the software on the analysis of the HATNet data. This
application and the discussion is related to the discovery of 
the planet HAT-P-7b, transiting a late F star on a quite tight orbit. 
The fourth chapter, ``Follow-up observations'' focuses on the 
post-discovery measurements (including photometric and radial velocity 
data) of the eccentric transiting exoplanetary system of HAT-P-2b. 
The goals, methods and theses are summarized in the fifth chapter.


\phdchapter{Algorithms and Software environment}
\label{chapter:swenv}

In principle, \emph{data reduction} or simply \emph{reduction} 
is the process when the raw data obtained by the instrumentation
are transformed into a more useful form. In fact, raw data can 
be analyzed during acquisition in order do modify the
instrumentation parameters for the subsequent 
measurements\footnote{For instance, in the case of HATNet, real-time
astrometric guiding is used to tweak the mount coordinates 
in the cases when the telescope drifts away from the desired celestial
position. This guiding basically uses the same algorithms and routines
that are involved in the photometric reduction. Like so, simplified
forms of photometry can be used in the case of follow-up 
measurements of exoplanetary candidate host stars: if light curve
variations show unexpected signals, the observation schedule could be changed 
accordingly to save expensive telescope time.}.
However, in the practice of astronomical data analysis,
all raw data are treated to be known in advance of the reduction
process. Moreover, the term ``more useful form'' of data is highly 
specific and depends on our needs. Regarding to photometric exoplanetary studies,
this ``more useful form'' means two things. 
First -- as in the case of HATNet where
the discoveries are based on long-term photometric time series --, 
reduction ends at the stage of analyzed light curves, where 
transit candidates are recovered by the result of this analysis. Second, 
additional high-precision photometry\footnote{Combined with 
additional techniques, such as spectroscopy or stellar evolution modelling.
The confirmation the planetary nature by radial velocity measurements
is essiential.}
yields precise information directly about the planet itself. 
One should mention here that other types of measurements
involving advanced and/or space-borne techniques (for instance,
near-infrared photometry of secondary eclipses) have same 
principles of the reduction. The basics of the reductions are 
roughly the same and such observations yield even more types of 
planetary characteristics, such as brightness contrast or
surface temperature distribution.

The primary platform for data reduction is computers and the reduction
processes are performed by dedicated software systems. As it was mentioned
in the introduction, existing software solutions lack several relevant
components that are needed for a consistent analysis of the HATNet 
data flow. One of our goals was to develop a software package that
features all of the functionality required by the proper reduction of 
the HATNet and the related follow-up photometry. The package itself
is named \fihat{}, referring to both the HATNet project as well as the
invocation of the related individual programs. 

In the first major chapter of this PhD thesis, I summarize both the 
algorithms and their implementations that form the base of the
\fihat{} software package. Due to the difficulties of the 
undersampled and wide-field photometry, several new methods
and algorithms should have been developed, tested and implemented
that were missing from existing and available image reduction packages.
These difficulties are summarized in the next section 
(Sec.~\ref{subsec:difficulties}) while 
the capabilities and related problems of existing software solutions
are discussed in Sec.~\ref{subsec:availablesw}. 

The following sections describe the details of the algorithms and methods,
focusing primarily on those that do not have any known implementation in 
any publicly available and/or commercial software. 
Sec.~\ref{subsec:calibration} discusses the details of the calibration process, 
Sec.~\ref{subsec:stardetection} describes how the point-like sources 
(stars) are detected, extracted and characterized from the images, 
the details of the astrometry and the related problems -- such as automatic
source identification and obtaining the plate solution -- are explained in 
Sec.~\ref{subsec:astrometry},
the details of the image registration process is discussed 
in Sec.~\ref{subsec:imageregistering},
Sec.~\ref{subsec:photometry} summarizes the problems related to the 
instrumental photometry,
Sec.~\ref{subsec:convolution} describes the concepts of the 
``image subtraction'' process, that is mainly the derivation of a proper
convolution transformation between two registered images,
Sec.~\ref{subsec:subtractedphotometry} explains how can the 
photometry be optimally performed on convolved or subtracted images.
Sec.~\ref{subsec:trendfiltering} describes the major concepts of 
how the still remaining systematic light curve variations can be removed.

In Sec.~\ref{subsec:majorconcepts}, after the above listed description 
of the crucial steps of the 
whole image reduction and photometry process, I outline the major 
principles of the newly developed software package. 
This part is then followed by the
detailed description of the individual components of the software package.
And finally, the chapter ends with the practices about how this package
can be used in order to perform the complete image reduction
process.


\begin{figure}
\begin{center}
\noindent
\resizebox{80mm}{!}{\includegraphics{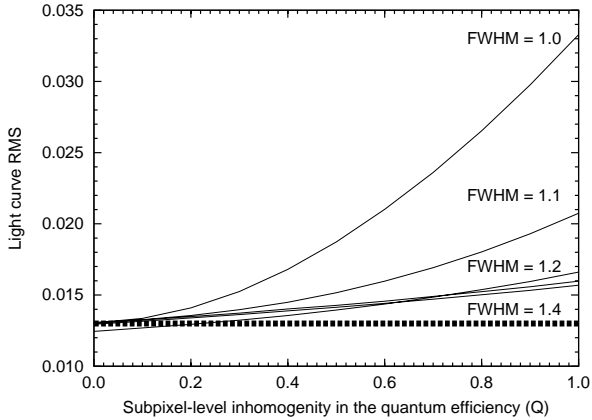}}
\end{center}
\caption{Plot showing the light curve scatter (rms, in magnitudes) 
of a mock star with various FWHMs and having a flux of $I=10000$ (in electrons).
The light curve rms is plotted as the function of the subpixel-level 
inhomogeneity, $Q$. Supposing a pixel structure where the pixel is 
divided to two rectangles of the same size, $Q$ is defined as 
the difference in the normalized quantum efficiencies of these 
two parts (i.e. $Q=0$ represents completely uniform sensitivity 
and at $Q=1$ one of the parts is completely insensitive (that is
typical for front-illuminated detector).}
\label{fig:subpixphot}
\end{figure}

\phdsection{Difficulties with undersampled, crowded and wide-field images}
\markright{2.1. UNDERSAMPLED, CROWDED AND WIDE-FIELD IMAGES}
\label{subsec:difficulties}

In this section we summarize effects that are prominent 
in the reduction of the HATNet frames when compared to the ``classical''
style of image reductions. The difficulties can be categorized into three
major groups. These groups do represent almost completely different 
kind of problems, however, all of these are the result of the type
of the survey. Namely, these problems are related to 
the \emph{undersampled} property, the \emph{crowding} of the 
sources that are the point of interest and the \emph{large field-of-view}
of the images. In this section we examine what particular problems 
arise due to these properties.

\phdsubsection{Undersampled images}
\label{sec:undersampledproblems}

At a first glance, an image can be considered to be undersampled if the source 
profiles are ``sharp''. The most prevalent quantity that characterizes
the sharpness of the (mostly stellar) profiles is the \emph{full width
at half magnitude} (FWHM). This parameter is the diameter of the contour
that connects the points having the half of the source's peak intensity.
Undersampled images therefore have (stellar) profiles with
small FWHM, basically comparable to the pixel size. In the following,
we list the most prominent effects of such a ``small'' FWHM and also
check what is the practical limit below which this ``small'' is really small.
In this short section we demonstrate the yields of various effects
that are prominent in the photometry for stellar profiles with small FWHMs.
All of these effects worsen the quality of the photometry unless special
attention is made for their reduction.

\begin{figure*}
\begin{center}
\noindent
\resizebox{160mm}{!}{\includegraphics{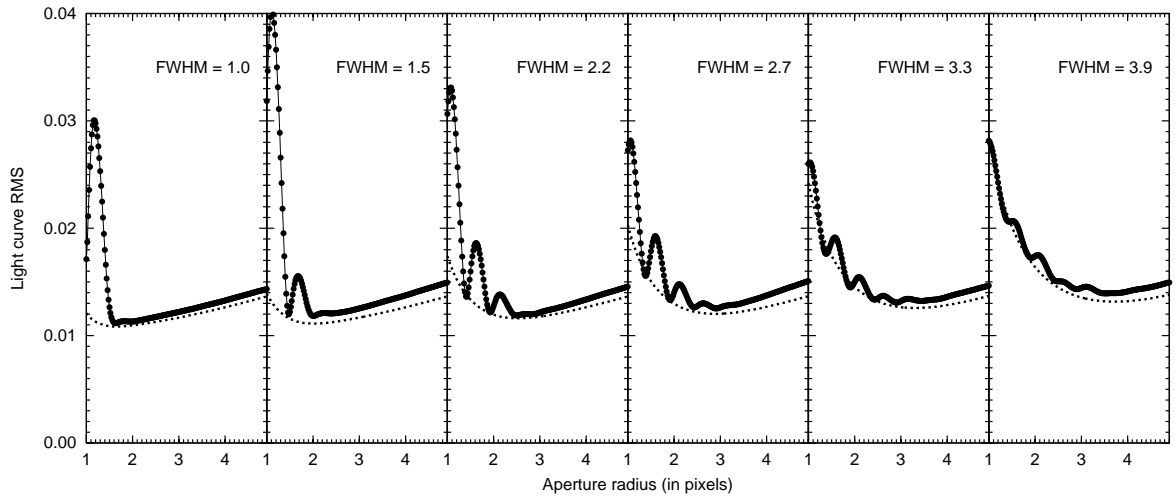}}
\end{center}
\caption{The graphs are showing the light curve scatters for mock
stars (with 1\% photon noise rms) 
when their flux is derived using aperture photometry. The subsequent
panels shows the scatter for increasing stellar profile FWHM, assuming
an aperture size between $1$ and $5$ pixels. The thick dots show the 
actual measured scatter while the dashed lines represent the lower 
limit of the light curve rms, derived from the photon noise and the
background noise.}
\label{fig:optaper}
\end{figure*}

\phdsubsubsection{Subpixel structure}
The effect of the subpixel structure is relevant when the 
characteristic length of the flux variations becomes
comparable to the scale length of the pixel-level sensitivity 
variations in the CCD detector. The latter is resulted mostly by
the presence of the gate electrodes on the surface of the detector, 
that block the photons at certain regions of a given pixel. Therefore,
this structure not only reduces the \emph{quantum efficiency} of the
chip but the signal depends on the centroid position of the incoming flux:
the sharper the profile, the larger the dependence on the centroid 
positions.
As regards to photometry, subpixel structure yields a non-negligible
correlation between the raw and/or instrumental magnitudes and the fractional
centroid positions. Advanced detectors such as \emph{back-illuminated}
CCD chips reduce the side effects of subpixel structure and also have
larger quantum efficiency.
Fig.~\ref{fig:subpixphot} shows that the effect of the subpixel structure
on the quality of the photometry highly dominates for sharp stars,
where ${\rm FWHM} \lesssim 1.2$\,pixels.

\begin{figure}
\begin{center}
\noindent
\resizebox{80mm}{!}{\includegraphics{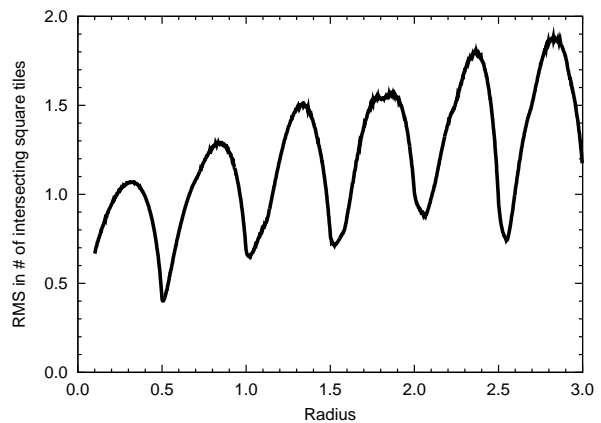}}
\end{center}
\caption{If circles with a fixed radius are drawn randomly and uniformly to 
a grid of squares, the number of intersecting squares has a 
well-defined scatter (since the number of squares intersecting 
the circle depends not only the radius of the circle but on the 
centroid position). The plot shows this scatter as the function of the 
radius.}
\label{fig:cintersec}
\end{figure}

\phdsubsubsection{Spatial quantization and the size of the aperture}
On CCD images, aperture photometry is the simplest technique to derive 
fluxes of individual point sources. Moreover, advanced 
methods such as photometry based on PSF fitting or image subtraction 
also involve aperture photometry on the fit residuals and the 
difference images, thus the properties of this basic 
method should be well understood. In principle, aperture 
is a certain region around a source. For nearly symmetric sources,
this aperture is generally a circular region with a pre-defined radius.
Since the image itself is quantized (i.e. the fluxes
are known only for each pixel) at the boundary of the aperture, the
per pixel flux must be properly weighted by the area of the intersection
between the aperture and the pixel. Aperture photometry is implemented
in almost all of the astronomical data reduction software packages
\citep[see e.g.][]{stetson1987}. 
As it is known from the literature \citep{howell1989}, both small
and large apertures yield small signal-to-noise ratio (SNR) or
relatively high light curve scatter (or \emph{root mean square}, rms). 
Small aperture contains small amount of flux therefore 
Poisson noise dominates. For large apertures,
the background noise reduces the SNR ratio. Of course, the size of 
the \emph{optimal aperture} depends on the total flux of the source as 
well as on the magnitude of the background noise. For fainter sources,
this optimal aperture is smaller, approximately its radius is in the range
of the profile FWHM, while for brighter stars it is few times larger
than the FWHM \citep[see also][]{howell1989}. However, for very narrow/sharp
sources, the above mentioned naive noise estimation becomes misleading.
As it is seen in the subsequent panels of Fig.~\ref{fig:optaper},
the actual light curve scatter is a non-trivial oscillating function
of the aperture size and this oscillation reduces and becomes
negligible only for stellar profiles wider than ${\rm FWHM}\gtrsim4.0$\,pixels.
Moreover, a ``bad'' aperture can yield a light curve rms about $3$ times
higher than the expected for very narrow profiles. 
The oscillation has a characteristic period of roughly $0.5$\,pixels. 
It is worth to mention that this dependence of the light curve scatter on
the aperture radius 
is a direct consequence of the topology of intersecting circles and squares.
Let us consider a bunch of circles with the same radius, drawn randomly 
to a grid of squares. The actual number of the squares that intersect
a given circle depends on the circle centroid position. Therefore,
if the circles are drawn uniformly, this number of intersecting squares
has a well defined scatter. In Fig.~\ref{fig:cintersec} this scatter is 
plotted as the function of the circle radius. As it can be seen, 
this scatter oscillates with a period of nearly $0.5$\,pixels. Albeit
this problem is much more simpler than the problem of light curve scatter 
discussed above, the function that describes the dependence of the 
scatter in the number of intersecting squares on the circle radius
has the same qualitative behavior (with the same period and positions
of local minima). This is an indication of a non-trivial source of 
noise presented in the light curves if the data reduction 
is performed (at least partially) using the method of aperture photometry.
In the case of HATNet, the typical FWHM is between $\sim 2 - 3$\,pixels.
Thus the selection of a proper aperture in the case of simple and image 
subtraction based photometry is essential. The methods intended to 
reduce the effects of this \emph{quantization noise} are going to be 
discussed later on, see Sec.~\ref{subsec:trendfiltering}.

\begin{figure*}
\begin{center}
\noindent
\resizebox{160mm}{!}{\includegraphics{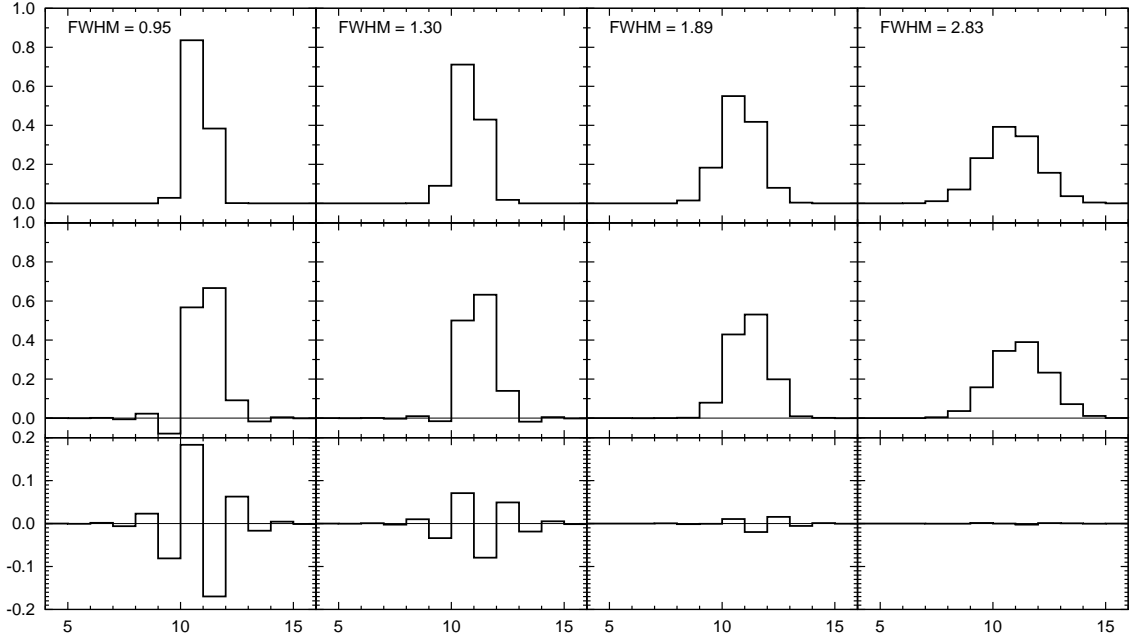}}
\end{center}
\caption{One-dimensional stellar profiles for various FWHMs,
shifted using spline interpolation. The profiles on the upper stripe
show the original profile while the plots in the middle stripe show
the shifted ones. All of the profiles are Gaussian profiles (with the 
same total flux) and centered at $x_0=10.7$. The shift is done rightwards
with an amplitude of $\Delta x=0.4$. The plots in the lower stripe
show the difference between the shifted profiles and a fiducial sampled profile
centered at $x=x_0+\Delta x=10.7+0.4=11.1$.}
\label{fig:samplespline}
\end{figure*}

\phdsubsubsection{Spline interpolation}
As it is discussed later on, one of the relevant steps in the
photometry based on image subtraction is the \emph{registration} process,
when the images to be analyzed are spatially shifted to the 
same reference system.
As it is known, the most efficient way to perform such a registration 
is based on quadratic or cubic spline interpolations. Let us suppose 
a sharp structure (such as a narrow, undersampled stellar profile)
that is shifted using a transformation aided by cubic spline interpolation.
In Fig.~\ref{fig:samplespline} a series of one-dimensional sharp profiles
are shown for various FWHMs between $\sim 1$ and $\sim 3$\,pixels, before and
after the transformation. As it can be seen well, for very narrow
stars, the resulted structure has values smaller than the 
baseline of the original profile. For extremely sharp (${\rm FWHM}\approx 1$)
profiles, the magnitude of these undershoots can be as high as $10-15\%$
of the peak intensity. Moreover, the difference between the shifted
structure and a fiducial profile centered on the shifted position
also has a specific oscillating structure. The magnitude of such
oscillations decreases dramatically if the FWHM is increased. For
profiles with ${\rm FWHM}\approx 3$, the amplitude of such oscillation
is about a few thousandths of the peak intensity (of the original profile).
If the photometry is performed by the technique of image subtraction,
such effects yield systematics in the photometry. Attempts to reduce these 
effects are discussed later on (see Sec.~\ref{subsec:subtractedphotometry}).

\begin{figure}
\begin{center}
\noindent
\resizebox{80mm}{!}{\includegraphics{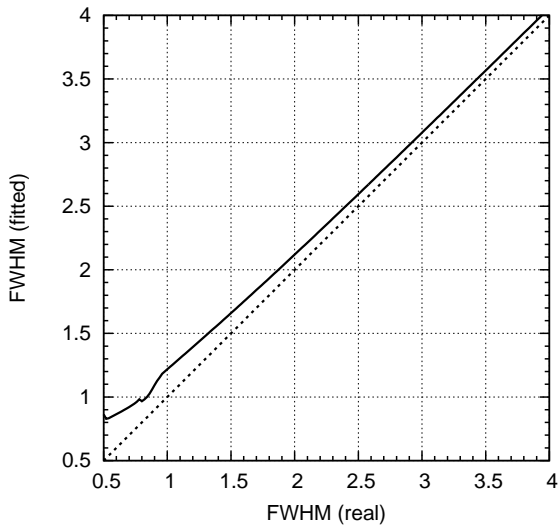}}
\end{center}
\caption{This plot shows how the profile FWHM is overestimated by 
the simplification of the fit. The continuous line shows the fitted FWHM
if the model function is sampled at the pixel centers (instead of integrated 
properly on the pixels). The dashed line shows the identity function, for
comparison purposes.}
\label{fig:modelfwhm}
\end{figure}

\phdsubsubsection{Profile modelling}
Regarding to undersampled images, one should mention some relevant details
of profile modelling. In most of the data reduction processes, stellar 
profiles detected on CCD images are characterized by simple 
analytic model functions. These functions (such as Gaussian or 
Moffat function) have a few parameters that are related to the centroid
position, peak intensity and the profile shape parameters. During the
extraction of stellar sources the parameters of such model functions
are adjusted to have a best fit solution for the profile. In order 
to perform a self-consistent modelling, one should derive the integrated 
value of the model function to adjacent pixels and fit these integrals
to the pixel values instead of sampling the model function on a 
square grid and fit these samples to the pixel values. Although the
calculations of such integrals and its parametric 
derivatives\footnote{Parametric derivatives of the model functions
are required by most of the fitting methods.} are computationally
expensive, neglecting this effect yields systematic offsets in
the centroid positions and a systematic overestimation of the
profile size (FWHM). Since the plate solution is based on 
the individual profile centroid coordinates, such simplification in the profile
modelling yields additional systematics in the final light 
curves\footnote{For photometry, the final centroid positions 
are derived from the plate solution and a catalogue. Therefore,
systematic variations in the plate solution indirectly yield systematic
variations in the photometry and in the light curves.}
Moreover, precise profile modelling is essential in the reduction
of the previously discussed spline interpolation side effect.
As an example, in Fig.~\ref{fig:modelfwhm} we show how the fitted 
FWHM is overestimated by the ignorance of the proper profile modelling,
if the profile model function is Gaussian.

\phdsubsection{Crowded images}

Since the ``CCD era'', various dense stellar fields, such as globular
clusters or open clusters are monitored for generic photometric 
analysis and variability search. The main problems of such crowded images
are well known and several attempts have been done in order to 
reduce the side effects resulted from the merging profiles. In this
section some of the problems are discussed briefly.

\phdsubsubsection{Merging and sensitivity to profile sharpness}
Merging of the adjacent stellar profiles have basically two 
consequences in the point of photometry. First, it is hard 
to derive the background intensity
and noise level around a given target star.
Stars in the background area can be excluded by two ways: the pixels
belonging to such profiles can be ignored either by treating them
outliers or the proximity of other photometric centroids are removed
from the set of background pixels. The second consequence of the profile 
merging is the fact that flux from adjacent stars is likely to 
be ``flowed'' underneath to the target aperture. Moreover,
the magnitude of such additional flux depends extremely strongly 
on the profile FWHMs and therefore variations in the widths of the profiles
cause significant increase in the light curve scatter. 

\phdsubsubsection{Modelling}
The modelling of stellar profiles, both by analytical model
functions and empirical point-spread functions are definitely hardened 
in the case of merging sources. In this case the detected stars cannot
be modelled separately, thus a joint fit should be performed simultaneously
on all of the stars or at least on the ones that are relatively 
close to each other to have significant overlapping in the model functions.
In the case of extremely crowded fields, sophisticated grouping and/or
iterative procedures should be employed, otherwise the computation
of the inverse matrices (associated with the parameter fitting) is not
feasible \citep[see also][]{stetson1987}.

\paragraph*{}%
As we will see later on, the method of difference image photometry 
helps efficiently to reduce these side effects related to the crowdness
of the images. However, it is true only for \emph{differential} 
photometry, i.e. during the photometry of the reference 
frames\footnote{In principle, the method of differential photometry
derives the flux of objects on a target image by adding the flux
of the objects on a reference image to the flux of the residual
on the image calculacted as the difference between the target 
and the reference image.} these
problems still emerge.

\phdsubsection{Large field-of-view}
\label{sec:difficulties:largefov}

Additionally to the previously discussed issues, the large size
of the field-of-view also introduces various difficulties. 

\phdsubsubsection{Background variations}

Images covering large field-of-view on the sky are supposed to have
various background structures, such as  
thin cirrus clouds, or scattered light due to dusk, dawn or the 
proximity of the Moon or even interstellar clouds\footnote{Although
interstellar clouds are steady background structures, in the point
of the analysis of a \emph{single} image, these cause the same kind
of features on the image.}.
These background variations make impossible
the derivation of a generic background level. Moreover the background
level cannot be characterized by simple functions such as polynomials
or splines since it has no any specific scale length. 
Because the lack of a well-defined background level, the source extraction 
algorithm is required to be purely topological (see also
Sec.~\ref{subsec:stardetection}).

\phdsubsubsection{Vignetting, signal-to-noise level and effective gain}

The large field-of-view can only be achieved by fast focal ratio
optical designs. Such optical systems do not have negligible vignetting,
i.e. the effective sensitivity of the whole system decreases at the
corners of the image. In the case of HATNet optics, such vignetting
can be as strong as $1$ to $10$. Namely, the total incoming flux at 
the corners of the image can be as small as the tenth of the flux
at the center of the image. Although flat-field corrections 
eliminate this vignetting, the signal-to-noise ratio is unchanged. Since
the latter is determined by the \emph{electron count}, increasing the
flux level reduces the effective \emph{gain}\footnote{The 
gain is defined as the joint electron/ADU conversion ratio of the 
amplifier and the A/D converter. A certain CCD camera may have a variable gain
if the amplification level of the signal read from the detector can be varied
before digitization.} at the corner of the images. 
Since the expectations of the photometric
quality (light curve scatter and/or signal-to-noise) 
highly depends on this specific gain
value, the information about this yield of vignetting should be 
propagated through the whole photometric process. 

\begin{table}
\caption{Typical astrometric residuals in the function of 
polynomial transformation order, for absolute and relative 
transformations. For absolute transformations the 
reference is an external catalog while for relative
transformations, the reference is one of the frames. 
\label{tab:astromres}}
\begin{center}
\footnotesize
\begin{tabular}{lll}
\hline
Order	&	Absolute	& Relative		\\
\hline
1 	&	$0.841 - 0.859$	& $0.117 - 0.132$	\\
2 	&	$0.795 - 0.804$	& $0.049 - 0.061$	\\
3 	&	$0.255 - 0.260$	& $0.048 - 0.061$	\\
4 	&	$0.252 - 0.259$	& $0.038 - 0.053$	\\
5 	&	$0.086 - 0.096$	& $0.038 - 0.053$	\\
6 	&	$0.085 - 0.096$	& $0.038 - 0.053$	\\
7 	&	$0.085 - 0.095$	& $0.038 - 0.053$	\\
8	&	$0.085 - 0.095$	& $0.038 - 0.053$	\\
9	&	$0.085 - 0.095$	& $0.038 - 0.053$	\\
\hline
\end{tabular}
\end{center}
\end{table}

\phdsubsubsection{Astrometry}

Distortions due to the large field-of-view affects the astrometry
and the source identification. Such distortions can efficiently
be quantified with polynomial functions. After the sources
are identified, the optimal polynomial degree (the order of the fit) can easily
be obtained by calculating the unbiased fit residuals. For a sample series
of HATNet images we computed these fit residuals, as it is shown
in Table~\ref{tab:astromres}. It can easily be seen that the residuals
do not decrease significantly after the $5-6$th order if an external
catalogue is used as a reference, while the optimal polynomial degree
is around $\sim 3-4$ if one of the images is used as a reference. 
The complex problem of the astrometry is discussed in 
Sec.~\ref{subsec:astrometry} in more detail.

\phdsubsubsection{Variations in the profile shape parameters}

Fast focal ratio optical instruments have significant comatic aberrations.
The comatic aberration yields not only elongated stellar profiles 
but the elongation parameters (as well as the FWHMs themselves) vary
across the image. As it was demonstrated, many steps of a complete
photometric reduction depends on the profile sizes and shapes, the
proper derivation of the shape variations is also a relevant issue.

\subsection*{Summary}

In this section we have summarized various influences of image
undersampling, crowdness and large field-of-view 
that directly or indirectly affects the quality of the photometry.
Although each of the distinct effects can be well quantified, in 
practice all of these occur simultaneously. The lack of a complete
and consistent software solution that would be capable to overcome
these and further related problems lead us to start the development
of a program designed for these specific problems. 

In the next section we review the most wide-spread software
solutions in the field of astronomical photometric data reduction.


\begin{table*}
\caption{Comparison of some of the existing software solutions for astronomical image
processing and data reduction. All of these software systems are available for the 
general public, however it does not mean automatically that 
the particular software is free or open source. This table
focuses on the most wide-spread softwares, and we omit the 
``wrappers'', that otherwise allows the access of such programs from
different environments (for instance, processing of astronomical images in
IDL use IRAF as a back-end).
}
\label{tab:software}
\begin{center}
\begin{tabularx}{164mm}{p{63mm}p{93mm}}
\hline
Pros	& Cons \\
\hline
{\bf\small IRAF}$^{1}$

\footnotesize
\begin{itemize}
\xitem Image Reduction and Analysis Facility.
The most commonly recognized software for astronomical
data reduction, with large literature and numerous references.

\xitem IRAF supports the functionality of the package DAOPHOT$^{2}$,
one of the most frequently used software solution for aperture photometry
and PSF photometry with various fine-tune parameters. 

\xitem IRAF is a complete solution for image analysis, no additional
software is required if the general functionality and built-in algorithms 
of IRAF (up to instrumental photometry) are sufficient for our demands.
\end{itemize}

&

\footnotesize
\begin{itemize}
\xitem Not an open source software. Although the higher level modules
and tasks are implemented in the own programming language of IRAF,
the back-end programs have non-published source code. Therefore,
many of the tasks and jobs are done by a kind of ``black box'', with
no real assumption about its actual implementation.

\xitem Old-style user interface. The primary user interface of IRAF
follows the archaic designs and concepts from the eighties. Moreover, 
many options and parameters reflect the hardware conditions 
at that time (for instance, reading and writing data from/to tapes,
assuming very small memory size in which the images do not fit and so on).

\xitem Lack of functionality required by the proper processing 
of wide-field images. For instance, there is no particular effective
implementation for astrometry or for light curve processing (such as transposing
photometric data to light curves and doing some sort of manipulation
on the light curves, such as de-trending).
\end{itemize}
\\
\hline
{\bf\small ISIS}$^{3}$.

\footnotesize
\begin{itemize}
\xitem Image subtraction package. 
The first software solution employing image subtraction
based photometry. 

\xitem The program performs all of the necessary steps related to
the image subtraction algorithm itself and the photometry as well.

\xitem Fully open source software, comes with some shell scripts
(written in C shell), that demonstrate the usage of the program,
as well as these scripts intend to perform the whole process (including
image registration, a fit for convolution kernel and photometry).
\end{itemize}
&

\footnotesize
\begin{itemize}
\xitem Not a complete software solution in a wider context. Additional
software is required for image calibration, source detection and 
identification and also for the manipulation of the photometric results.

\xitem Although this piece of software has open source codebase, the
algorithmic details and some tricks related to the photometry on 
subtracted images are not documented (i.e. neither in the reference scientific
articles nor in the program itself).

\xitem The kernel basis used by ISIS is fixed. The built-in basis
involves a set of functions that can easily and successfully be applied on 
images with wider stellar profiles, but not efficient on images with
narrow and/or undersampled profiles.

\xitem Some intermediate data are stored in blobs. Such blobs may 
contain useful information for further processing (such as the kernel
solution itself), but the access to these blobs is highly inconvenient.
\end{itemize}
\\
\hline
{\bf\small SExtractor}$^{4}$.

\footnotesize
\begin{itemize}
\xitem Source-Extractor. Widely used software package for
extracting and classifying various kind of sources from astronomical
images.

\xitem Open source software.

\xitem Ability to perform photometry on the detected sources.
\end{itemize}
&

\footnotesize
\begin{itemize}
\xitem The primary goal of SExtractor was to be a package that 
focuses on source classification. Therefore, this package is not
a complete solution for the general problem, it can be used only
for certain steps of the whole data reduction.

\xitem Photometry is also designed for extended sources.
\end{itemize}
\\
%
%
%
%
%

\hline
\end{tabularx}
\end{center}
\begin{flushleft}
\scriptsize
\noindent $^{1}$ IRAF is distributed by the National Optical Astronomy
Observatories, which are operated by the Association of Universities
for Research in Astronomy, Inc., under cooperative agreement with the
National Science Foundation. See also \texttt{http://iraf.net/}.

\noindent $^{2}$ DAOPHOT is a standalone photometry package, written
by Peter Stetson at the Dominion Astrophysical Observatory \citep{stetson1987}.

\noindent $^{3}$ ISIS is available from \texttt{http://www2.iap.fr/users/alard/package.html}
with additional tutorials and documentation \citep{alard1998,alard2000}.

\noindent $^{4}$ SExtractor is available from \texttt{http://sextractor.sourceforge.net/},
see also \cite{bertin1996}.
\end{flushleft}
\end{table*}

\phdsection{Problems with available software solutions}
\label{subsec:availablesw}

In the past decades, several software packages became available
for the general public, intended to perform astronomical data reductions.
The most widely recognized package is the Image Reduction and Analysis
Facility (IRAF), distributed by the National Optical Astronomy 
Observatories (NOAO). With the exception of photometry methods
of image subtraction, many algorithms related to photometric
data reductions have been implemented in the framework of IRAF
\citep[for instance, DAOPHOT, see e.g.][]{stetson1987,stetson1989}.
The first public implementation of the image convolution and 
the related photometry was given by \cite{alard1998}, in the
form of the ISIS package. This package is focusing on
certain steps of the procedure but it is not a complete 
solution for data reduction (i.e. ISIS alone is not sufficient if 
one should derive light curves from raw CCD frames, in this case
other packages must be involved due to the lack of several functionalities
in the ISIS package). The SExtractor package of \cite{bertin1996}
intends to search, classify and characterize sources of various kind of shape
and brightness. This program was designed for extragalactic surveys,
however, it has several built-in methods for photometry as well. 
Of course, there are several other independent packages or 
wrappers for the previously mentioned ones\footnote{These wrappers allow
the user to access functionalities from external data processing
environments. For instance, the astronomical reduction package
of the IDL environment uses the IRAF as a back-end, or the
package PyRAF provides access to IRAF tasks within the Python language.}.
Table~\ref{tab:software} gives a general overview of the advantages
and disadvantages of the previously discussed software packages.
Currently, one can say that these packages alone \emph{do not} provide
sufficient functionality for the complete and consistent photometric
reduction of the HATNet frames. In the following, we are focusing
on those particular problems that arise during the photometric
reductions of images similar to the HATNet frames and as of this
writing, do not have any publicly available software solutions
to overcome.


\phdsection{Calibration and masking}
\label{subsec:calibration}

For astronomical images acquired by CCD detectors, the aim of the 
calibration process is twofold. The first goal is to reduce the effect of 
both the differential light sensitivity characteristics of the pixels and
the large-scale variations yielded by the telescope optics. The second
goal is to mark the pixels that must be excluded from the further 
data reduction since the previously mentioned corrections cannot be 
e performed because of various reasons. The most common sources
of such reasons are the saturation and blooming of bright pixels, 
cosmic ray events or malfunctioning pixels (such as pixels with highly nonlinear 
response or with extraordinary dark current). Of course, some of these
effects vary from image to image (e.g. saturation or cosmic ray events)
while other ones (such as nonlinear pixels) have constant structure. 

In this section the process of the calibration is described,
briefly discussing the sensitivity corrections, followed by a bit more detailed
explanation of the masking procedure (since it is a relevant improvement
comparing to the existing software solutions). Finally, we show how
these masks are realized and stored in practice. The actual software
implementation related to the calibration process are described later 
in Sec.~\ref{subsec:implementation}.

\phdsubsection{Steps of the calibration process}

Basically, the calibration of all of the image frames, almost 
independently from the instrumentation, has been done involving
bias, dark and flat images and overscan 
correction (where an appropriate overscan section is available 
on the detector). These calibration steps correct for the light sensitivity
inhomogeneities with the exception of nonlinear responses, effects
due to the dependence on the spatial and/or temporal variations in the 
telescope position or in the sky background\footnote{Such as scattered 
light, multiple reflections in the optics or fringing yielded 
by the variations in the sky background spectrum} and 
second-order sensitivity effects\footnote{such as the shutter effect}. 
In practice, the linear corrections provided by
the classic calibration procedure are acceptable, as in the case of HATNet
image calibrations.

Let us consider an image $I$ and denote its calibrated form by
$\mathrm{C}(I)$. If the basic arithmetic
operators between two images are defined as \emph{per pixel} operations,
$\mathrm{C}(I)$ can be derived as 
\begin{equation}
\mathrm{C}(I)=\frac{I-\mathrm{O}(I)-B_0-(\tau[I]/\tau[D_0])D_0}{F_0/\|F_0\|}, \label{eq:calibgeneral}
\end{equation}
where $\mathrm{O}(I)$ is the overscan level\footnote{Derived from the 
pixel values of the overscan area. The large scale structure of the
overscan level is modelled by a simple function (such as spline 
or polynomial) and this function is then extrapolated 
to the image area.} of the image $I$; $B_0$, $D_0$
and $F_0$ are the \emph{master} calibration images of bias, 
dark and flat, respectively. We denote the exposure time of the image $x$
by $\tau[x]$. $\|x\|$ denotes the norm of the image $x$, that 
is simply the mean or median of the pixel values. In practice, 
when any of the above master calibration images
does not exist in advance, one can substitute for these by zero, or in the 
case of flat images, by arbitrary positive constant value. The master 
calibration frames are the \emph{per pixel} mean or median averages 
(with optional $n$-$\sigma$ rejection) of individual frames:
\begin{eqnarray}
\mathrm{C}(B_i) & = &  B_i - \mathrm{O}(B_i) , \label{eq:calbias} \\
B_0 & = & \left< \mathrm{C}(B_i) \right>, \label{eq:avgbias} \\
\mathrm{C}(D_i) & = &  D_i - \mathrm{O}(D_i) - B_0 , \label{eq:caldark} \\
D_0 & = & \left< \mathrm{C}(D_i) \right>, \label{eq:avgdark} \\
\mathrm{C}(F_i) & = &  F_i - \mathrm{O}(F_i) - B_0 - \frac{\tau[F_i]}{\tau[D_0]}D_0, \label{eq:calflat}\\ 
F_0 & = & \left< \mathrm{C}(F_i) \right>. \label{eq:avgflat}
\end{eqnarray}
Equations (\ref{eq:calbias}), (\ref{eq:caldark}) and (\ref{eq:calflat})
clearly show that during the calibration of the individual bias, dark 
and flat frames, only overscan correction, overscan correction and a master 
bias frame, and overscan correction, a master bias and a master dark 
frame are used, respectively.

\phdsubsection{Masking}
\label{subsec:masking}

As it was mentioned earlier, pixels having some undesirable properties must be 
masked in order to exclude them from further processing.
The \fihat{} package and therefore the pipeline of the whole 
reduction supports various kind of masks. These masks are transparently 
stored in the image headers (using special keywords)
and preserved even if an independent software modifies the image. 
Technically, this mask is a bit-wise combination of Boolean flags, assigned
to various properties of the pixels. In this paragraph we briefly summarize
our masking method.

First, before any further processing and
right after the readout of the images, a mask is added to
mark the bad pixels of the image. Bad pixels are not only hot pixels
but pixels where the readout is highly nonlinear or the readout noise
is definitely larger than the average for the given detector. These 
\emph{bad mask}s are determined after a couple of sky flats were 
acquired. Using sky flats for the estimation of nonlinearity and readout noise 
deviances are fairly good, since during dusk or dawn, images are exposed 
with different exposure times yielding approximately the same flux and 
all of the pixels have a locally uniform incoming flux. See \cite{bakos2004phd}
for further details. 

Second, all saturated pixels are marked with a \emph{saturation mask}. 
In practice, there are two kind of effects related to the saturation: 
1) when the pixel itself
has an intensity that reaches the maximum expected ADU value or 2) if
there is no support for anti-blooming in the detector, charges 
from saturated pixels can overflow into the adjacent ones during readout. 
These two types of saturation are distinguished in the 
\emph{oversaturation mask} and \emph{blooming mask}. If any of these mask are 
set, the pixel itself is treated as saturated. We note that this saturation 
masking procedure is also done before any calibration.

Third, after the calibration is done, additional masks can be added to mark
the hot pixels (that were not corrected by subtracting the dark image), 
cosmic ray events and so on. 

Actually, the latest version of the package supports the following masks:
\begin{itemize}
\item \emph{Mask for faulty pixels.} These pixels show strong non-linearity. 
These masks are derived occasionally from the ratios of flat field
images with low and high intensities.
\item \emph{Mask for hot pixels.} The mean dark current for these pixels
is significantly higher than the dark current of normal pixels.
\item \emph{Mask for cosmic rays.} Cosmic rays cause sharp structures,
these structures mostly resemble hot or bad pixels, but these
does not have a fixed structure that is known in advance.
\item \emph{Mask for outer pixels.} After a geometric transformation (dilation,
rotation, registration between two images), certain pixels near the
edges of the frame have no corresponding pixels in the original 
frame. These pixels are masked as ``outer'' pixels. 
\item \emph{Mask for oversaturated pixels.} These pixels have an ADU value
that is above a certain limit defined near the maximum value of the A/D
conversion (or below if the detector shows a general nonlinear response
at higher signal levels).
\item \emph{Mask for blooming.} In the cases when the detector has 
no antiblooming feature or this feature is turned off, extremely 
saturated pixels causes ``blooming'' in certain directions (usually 
parallel to the readout direction). The A/D conversion value of the 
blooming pixels does not reach the maximum value of the A/D conversion,
but these pixels also should be treated as somehow saturated. 
The ``blooming'' and ``oversaturated'' pixels are commonly referred
as ``saturated'' pixels, i.e. the logical combination of these two respective
masks indicates pixels that are related to the saturation and its 
side effects.
\item \emph{Mask for interpolated pixels.} Since the cosmic rays and 
hot pixels can be easily detected, in some cases it is worth to 
replace these pixels with an interpolated value derived from the
neighboring pixels. However, these pixels should only be used with
caution, therefore these are indicated by such a mask for the further
processes.
\end{itemize}
We found that the above categories of 7 distinct masks are feasible
for all kind of applications appearing in the data processing. The fact
that there are 7 masks -- all of which can be stored in a single bit for
a given pixel -- makes the implementation quite easy. All bits of the mask
corresponding to a pixel fit in a byte and we still have an additional bit.
It is rather convenient during the implementation of certain steps (e.g.
the derivation of the blooming mask from the oversaturated mask), since
there is a temporary storage space for a bit that can be used for 
arbitrary purpose.

\begin{figure*}
\begin{center}
\noindent
\resizebox{80mm}{!}{\includegraphics{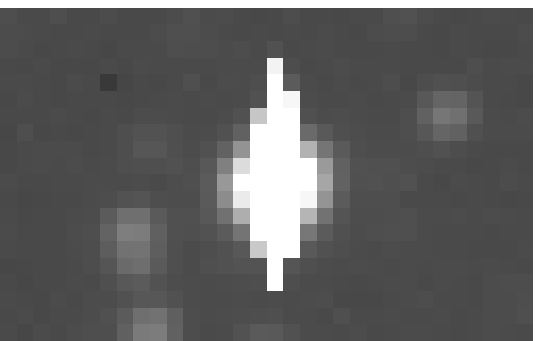}}\hspace*{4mm}%
\resizebox{80mm}{!}{\includegraphics{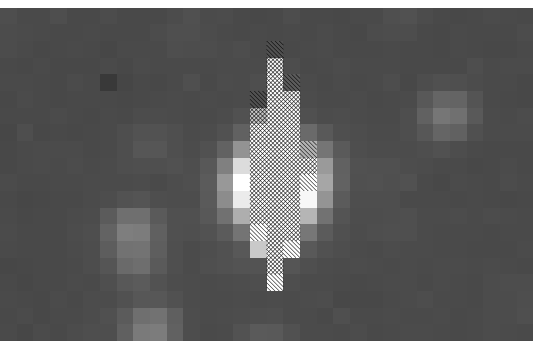}}
\begin{lcmd}
\small
MASKINFO= '1 -32 16,8 -16 0,1:-2 -32 1,1 -2,1 -16 1,0:2 -1,1:3,3 -32 3,2' \\
MASKINFO= '-16 -3,1:4 0,1:3,3 -32 3,0 -3,3 -16 1,0:2 0,1:-2 -32 1,0 -1,2'
\end{lcmd}
\end{center}
\caption{Stamp showing a typical saturated star. The images cover an 
approximately $8'\times 5'$ area ($32\times20$ pixels) of the sky, taken by one
of the HATNet telescopes. The blooming structure can be seen well. The left
panel shows the original image itself. In the right panel, 
oversaturated pixels (where the actual ADU values reach the maximum
of the A/D converter) are marked with right-diagonal stripes while 
pixels affected by blooming are marked with left-diagonal stripes. Note that
most  of the oversaturated pixels are also blooming ones, since their 
lower and/or upper neighboring pixels are also oversaturated. Such
pixels are therefore marked with both left- and right-diagonal stripes.
Since the readout direction in this particular detector was vertical,
the saturation/blooming structure is also vertical. The \texttt{``MASKINFO''}
blocks seen below the two stamps show how this particular masking information
is stored in the FITS headers in a form of special keywords. }
\label{fig:masksaturated}
\end{figure*}

\begin{figure*}
\begin{center}
\footnotesize
\begin{tabularx}{150mm}{lX}
\hline
Value & Interpretation \\
\hline
$T$ 	 	&
	Use type $T$ encoding. $T=0$ implies absolute cursor movements,
	$T=1$ implies relative cursor movements. Other values of $T$ are
	reserved for optional further improvements. \\
$-M$		&
	Set the current bitmask to $M$. $M$ must be between 1 and 127 and 
	it is a bit-wise combination of the numbers 1, 2, 4, 8, 16, 
	32 and 64, for faulty, hot, cosmic, outer, oversaturated,
	blooming and interpolated pixels, respectively. \\
$x,y$	&
	Move the cursor to the position $(x,y)$ (in the case of $T=0$)
	or shift the cursor position by $(x,y)$ (in the case of $T=1$)
	and mark the pixel with the mask value of $M$. \\
$x,y:h$ 	&
	Move/shift the cursor to/by $(x,y)$ and mark the horizontal line
	having the length of $h$ and left endpoint at the actual
	position. \\
$x,y:-v$	&
	Move/shift the cursor to/by $(x,y)$ and mark the vertical line 
	having the length of $v$ and lower endpoint at the actual
	position. \\
$x,y:h,w$ 	&
	Move/shift the cursor to/by $(x,y)$ and mark the rectangle
	having a size of $h\times w$ and lower-left corner at the actual
	cursor position. \\
\hline
\end{tabularx}
\end{center}
\caption{Interpretation of the tags found 
\texttt{MASKINFO} keywords in order to decode the respective mask. The
values of $M$, $h$, $v$ and $w$ must be always positive.}
\label{fig:maskalgorithm}
\end{figure*}

\phdsubsection{Implementation}

The basic \emph{per pixel} arithmetic operations required by the calibration
process are implemented in the program \texttt{fiarith} 
(see Sec.~\ref{sec:prog:fiarith}), while individual
operations on associated masks can be performed using the \texttt{fiign}
program (Sec.~\ref{sec:prog:fiign}). Although the distinct steps of 
the calibration can be performed by the appropriate subsequent invocation 
of the above two programs, a more efficient implementation is given 
by \texttt{ficalib} (Sec.~\ref{sec:prog:ficalib}), that allows
fast evaluation of \eqref{eq:calibgeneral} on a large set of images.
Moreover, \texttt{ficalib} also creates the appropriate masks upon request.
The master calibration frames (referred as $B_0$, $D_0$ and $F_0$
in equation~\ref{eq:calibgeneral})
are created by the combination of the individual calibration images
(see equations \ref{eq:avgbias}, \ref{eq:avgdark} and \ref{eq:avgflat}), 
involving the program \texttt{ficombine} (Sec.~\ref{sec:prog:ficombine}).
See also Sec.~\ref{sec:prog:ficalib} for more specific examples about
the application of these programs. 

As it was mentioned earlier, the masks are stored in the FITS header
using special keywords. Since pixels needed to be masked represent
a little fraction of the total CCD area, only information (i.e. mask
type and coordinates) about these masked pixels are written to the
header. By default, all other pixels are ``good''. A special
form of run-length encoding is used to compress the mask itself,
and the compressed mask is then represented by a series of integer numbers. 
This series of integers should be interpreted as follows. Depending on the values 
of these numbers, a virtual ``cursor'' is moved along the image.
After each movement, the pixel under the cursor or a rectangle
whose lower-left corner is at the current cursor position is masked
accordingly. In Fig.~\ref{fig:masksaturated} a certain example
is shown demonstrating the masks in the case of a saturated star 
(from one of the HATNet images). The respective encoded masks (as 
stored literally in the FITS header) can be seen below the image stamps.
The encoding scheme is summarized in Fig.~\ref{fig:maskalgorithm}. We found
that this type of encoding (and the related implementation) provides an
efficient way of storing such masks. Namely, the encoding and decoding
requires negligible amount of computing time and the total information
about the masking requires a few dozens from these ``\texttt{MASKINFO}''
keywords, i.e. the size of the FITS image files increases only 
by $3-5$\,kbytes (i.e. by less than 1\%).


\phdsection{Detection of stars}
\label{subsec:stardetection}

Calibration of the images is followed by detection of stars.
A successful detection of star-like objects is not only important because of
the reduction of the data but for the telescopes of HATNet it is 
used \emph{in situ} for guiding and slewing corrections. 

In the typical field-of-view of a HATNet telescope there are
$10^4$ -- $10^5$ stars with suitable signal-to-noise ratio (SNR), 
which are proper candidates for photometry. Additionally, there are
several hundreds of thousands, or millions of stars which 
are also easy to detect and characterize but not used for further photometry.
The HATNet telescopes acquire images that are highly crowded and 
undersampled, due to the fast focal ratio instrumentation ($f/1.8$ for
the lenses used by the HAT telescopes). Because of the large field-of-view,
the sky background does also vary rapidly on an ordinary image 
frame, due to the large-scale structure of the Milky Way, 
atmospheric clouds, differential extinction or another light scattering effects. 
Due to the fast focal ratio, the vignetting effects are
also strong, yielding stars of the same magnitude to have 
different SNR in the center of the images 
and the corners. This focal ratio also results in stars with different
shape parameters, i.e. systematically and heavily varying FWHM and elongation
even for focused images (this effect is known as \emph{comatic aberration} or
\emph{coma}). These parameters may also vary due to the
different sky conditions, e.g. airmass resulted also by the large 
FOV (12-15 degrees in the diameter).

Because of these, one should expect the following properties
of a star detection and characterization algorithm that are thereafter able to
overcome the above mentioned problems. 
\begin{description}
\item[A)] The method should
be local both in the sense of pixel positions and in the intensity. 
Namely, the result
of an object detection must not differ if one applies an affine transformation
in the intensity or after a spatial shift of the image.
\item[B)] The method should not contain
any characteristic scale length, due to the unpredictable scale length
of the background. It also implies that there
should not be any kind of "partitioning the image into blocks"
during the detection, i.e. one
should not expect that any of the above mentioned affects disappear
if the image is divided into certain blocks and some quantities are
treated as constants in such a block. Moreover, there is no "background"
for the image, even one cannot do any kind of interpolation to determine
a smooth background. 
\item[C)] The algorithm should be fast. Namely, it is expected
to be an $\ordo(N)$ algorithm, where $N=S_x\times S_y$, the total number of
the image pixels. In other words, the computing time is expected to be nearly 
independent from the number and/or the density of the detected objects. 
\item[D)] On highly crowded \emph{and} undersampled images,
stars should be distinguished even if they are very close to each other. 
Thus, the direct detection should not be preceded by a convolution with a 
kernel function, as it is done in the most common algorithms and 
software \citep[e.g. as it is used in DAOPHOT/FIND, see][]{stetson1987}.
Although this preliminary convolution increases the detectability
of low surface brightness object, in our case it would fuse nearby stars. 
\item[E)] The algorithm may have as few as possible
external fine-tune parameters. 
\item[F)] The algorithm should explicitly assign
the pixels to the appropriate detected objects.
\item[G)] Last but not least, the algorithm should
work properly not only for undersampled and crowded images but for
images acquired by ``classic'' types of telescopes where the average
FWHM of the stars are higher and/or the number density is lower. Additionally,
one may expect from such an algorithm to handle the cases of smeared or
defocused images, even when the star profiles have ``doughnut'' shape, 
as well as the proper characterization of digitized 
photographic data (e.g. POSS/DSS).
\end{description}

In this section we give an algorithm that is suitable for the
above criteria. Moreover, it is purely topological since
considers only "less than" or "greater than" relations between adjacent
pixel intensities. Obviously, an algorithm that relies only the topology,
automatically satisfies the conditions A and B above. The first part of this section
discusses how can the image be partitioned to smaller partitions that
are sets of cohesive pixels that belong to the same star (see 
also condition F above). The second part of the section describes how
these partitions/stars can be characterized by a couple of numbers,
such as centroid coordinates, flux, and shape parameters.

\phdsubsection{Image partitioning}

\phdsubsubsection{Pixel links and equivalence classes}
The first step of the detection algorithm is to define local pixel 
connections with the following properties. An ordinary pixel has 8 neighbors, 
and the number of neighbors is less only
if the pixel is a boundary pixel (in this case there can be 5 or 3 neighbors)
or if any of the neighboring pixels are excluded due to a mask of 
bad, hot or saturated pixel. Including the examined pixel with the
coordinates of $x$ and $y$, we select the one with the largest intensity 
from this set. Let us denote the coordinates of this pixel by $n_x(x,y)$ 
and $n_y(x,y)$. For a shorter notation, we introduce $\mathbf{x}=(x,y)$
and $\mathbf{n}(\mathbf{x})=[n_x(x,y),n_y(x,y)]$.
Obviously, $|n_x-x|\le1$ and $|n_y-y|\le1$, i.e. 
$\|\mathbf{n}(\mathbf{x})-\mathbf{x}\|_\infty\le1$, where
$\|\mathbf{x}\|_\infty=\max(|x|,|y|)$, the maximal norm. The derivation
of this set of $\mathbf{n}=(n_x,n_y)$ points requires $\ordo(N)$ time. 
Second, we define $\mathbf{m}(\mathbf{x})=[m_x(x,y),m_y(x,y)]$ 
for a given pixel by
\begin{equation}
\mathbf{m}(\mathbf{x})=\left\{
        \begin{tabular}{ll}
                 $\mathbf{x}$ & \hspace*{2mm} if $\mathbf{n}(\mathbf{x})=\mathbf{x}$, \\
                 $\mathbf{m}(\mathbf{n}(\mathbf{x}))$ & \hspace*{2mm} otherwise.
        \end{tabular}\right.
\end{equation}
Note that this definition of $\mathbf{m}(\mathbf{x})$ is only a functional
of the relation $\mathbf{x} \to \mathbf{n}(\mathbf{x})$: there is no need
for the knowledge of the underlying neighboring and the partial ordering
between pixels.
This definition results a set of finite pixel links 
$\mathbf{x}$,
$\mathbf{n}(\mathbf{x})$, $\mathbf{n}(\mathbf{n}(\mathbf{x}))\equiv
\mathbf{n}^2(\mathbf{x})$, \dots where the length $L$ of this link
is the smallest value where $\mathbf{n}^L(\mathbf{x})=\mathbf{n}^{L+1}(\mathbf{x})=
\mathbf{m}(\mathbf{x})$.
Third, we define two pixels, say, $\mathbf{x}_1=(x_1,y_1)$ and 
$\mathbf{x}_2=(x_2,y_2)$ to be \emph{equivalent} if 
$\mathbf{m}(\mathbf{x}_1)=\mathbf{m}(\mathbf{x}_2)$. This equivalence relation
partitions the image into disjoint sets, equivalence classes. 
In other words, each equivalence contains links with
the same endpoint. Let us
denote these classes by $C_i$.

\begin{figure*}
\begin{center}
\resizebox{50mm}{!}{\includegraphics{img/star/5-153695-trim3-img.eps}}\hspace*{5mm}%
\resizebox{50mm}{!}{\includegraphics{img/star/5-153695-trim2-img.eps}}\hspace*{5mm}%
\resizebox{50mm}{!}{\includegraphics{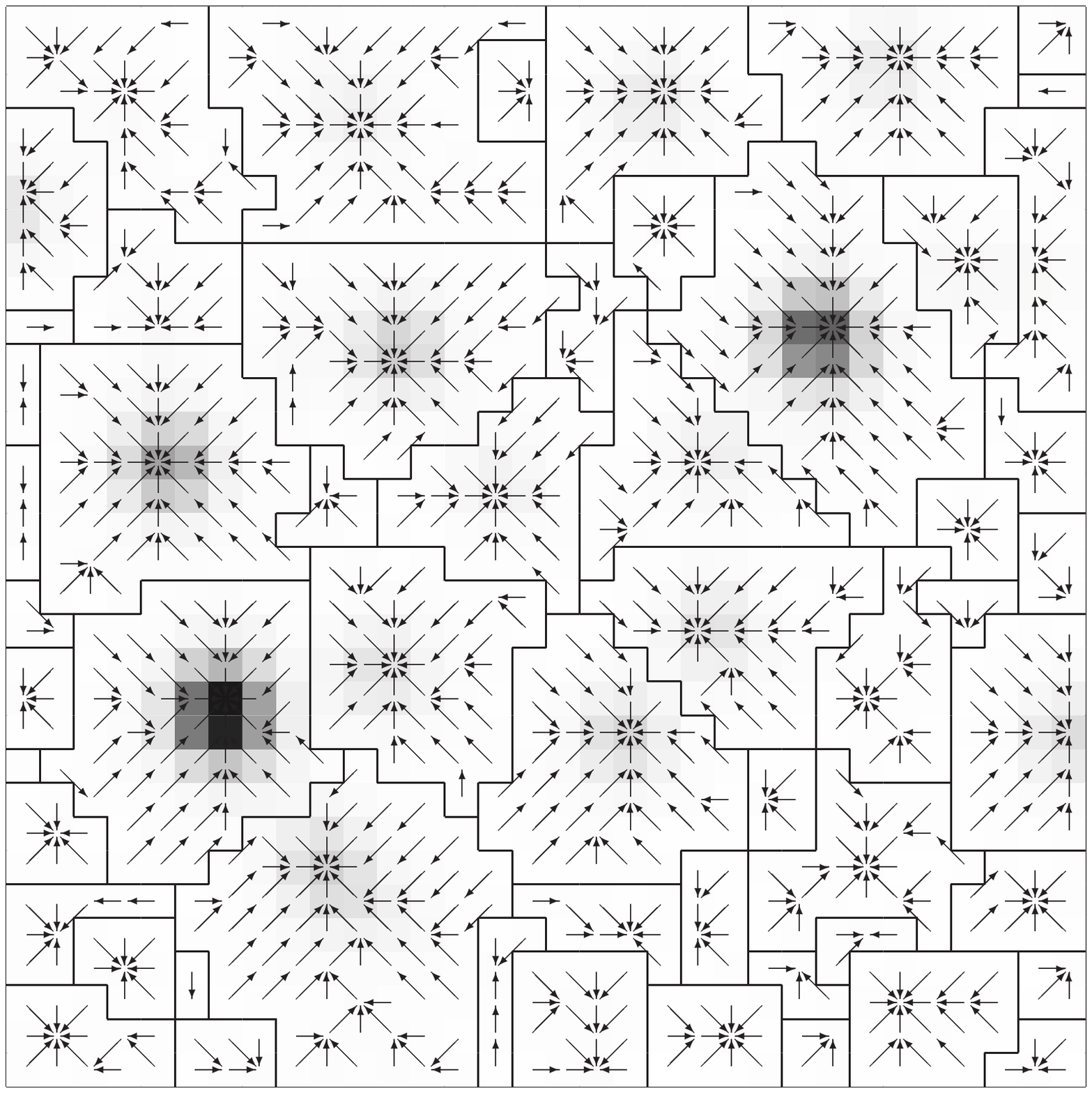}}
\end{center}
\caption{Left panel: a stamp of $128\times128$ pixels from a typical
crowded HAT image, covering approximately $0.5^\circ\times 0.5^\circ$ 
area on the sky. Middle panel: the central are of the stamp shown in the
left panel, covering approximately an area of $7'\times 7'$. This smaller
stamp has a size of $32\times 32$ pixels. Right panel: the links 
and equivalence classes generated from the smaller stamp. Note that
even the faintest stars are detected and the belonging pixels form
separated partitions (for an example see the stars encircled on the middle
panel). }
\label{fig:starlinkbasic}
\end{figure*}

Each class is represented by the appropriate 
$\mathbf{m}_i\equiv\mathbf{m}(C_i)$ pixel, that is, by definition, 
a local maximum. Each equivalence class
can be considered as a possible star, or a part of a star if the image 
was defocused or smeared. 
In Fig.~\ref{fig:starlinkbasic}, one can see stamps from a typical image 
obtained by one of the HATNet telescopes and the derived pixel links 
and the respective equivalence classes. In the figure, the mapping
$\mathbf{x} \to \mathbf{n}(\mathbf{x})$ is represented by the
$\mathbf{n}(\mathbf{x})-\mathbf{x}$ vectors, originating from the pixel 
$\mathbf{x}$.

\phdsubsubsection{Background}

Let us define the number of possible neighbors of a given pixel $\mathbf{x}$
by $K_0(\mathbf{x})$. As it was described above, in average it is 8, for 
boundary pixels it is 5 or 3, and it can be less if there are surrounding
masked ones. The quantity $R(\mathbf{x})$ is defined by the 
cardinality of the set 
$\{\mathbf{x}'\in\mathrm{Image}~:~\mathbf{n}(\mathbf{x}')=\mathbf{x}\}$.
Let us also define $K(\mathbf{x})=K_0(\mathbf{x})+1$ and $G(\mathbf{x})$
as the cardinality of 
\begin{equation}
\{\mathbf{x}'\in\mathrm{Image}~:~\|\mathbf{x}'-\mathbf{x}\|_\infty\le1 {\rm ~~and~~}
\mathbf{m}(\mathbf{x}')=\mathbf{m}(\mathbf{x})\},
\end{equation}
which is the number of surrounding pixels in the same class.
For a given equivalence class $C$ we can define its \emph{background pixels}
by 
\begin{equation}
\mathbf{B}(C)=\{\mathbf{x}'\in C~:~R(\mathbf{x}')=0~\mathrm{and}~G(\mathbf{x}')<K(\mathbf{x}')\}. 
\end{equation}
This set of pixels are the boundary starting points of pixel links in 
this equivalence class. Note that this definition may not reflect the 
true background if there are merging stars in the vicinity. 
In such case these pixels are saddle points between two or more stars. 
However, the median
of the pixel intensities in the set $\mathbf{B}(C)$ is a good assumption
of the local background for the star candidate $C$, even for highly crowded
images. For simplicity, let us denote the background of $C$ by
\begin{equation}
b(C)\equiv\left<\mathbf{B}(C)\right>. \label{eq:partitionbackground}
\end{equation}

\phdsubsubsection{Collection of subsets}

The above definitions of equivalence classes and background pixels 
are quite robust ones, but still there are some demands for certain cases.
First, in extremely crowded fields, the number of background pixels can be 
too small for a local background assumption. Second, the defocused
or smeared stars may consist of several separate local maxima that yield
distinct equivalence classes instead of one cohesive set of pixels.
To overcome these problems, we make some other definitions.
An equivalence class $C$ is \emph{degenerated} if
\begin{equation}
R(\mathbf{m}(C)) < K(\mathbf{m}(C)).
\end{equation}
In other words, degenerated partitions have local maxima
on their boundary. For such a partition,
one can define the two sets of pixels:
\begin{equation}
\mathbf{J}_1(C)=\{ \mathbf{x}'\in\mathrm{Image}~:~\|\mathbf{x}'-\mathbf{m}(C)\|_\infty=1 \},
\end{equation}
and
\begin{equation}
\mathbf{J}_2(C)=\{ \mathbf{x}' \in \mathbf{J}_1(C)~:~\mathbf{m}(\mathbf{x}') \ne \mathbf{m}(C) \}.
\end{equation}
Let us denote the location of the maximum of a given set $\mathbf{J}$ by
\begin{equation}
\mathbf{M}(\mathbf{J})=\{ \mathbf{x} ~:~ \forall~ \mathbf{x}' \in \mathbf{J} ~ I
(\mathbf{x}')\le I(\mathbf{x}),\}
\end{equation}
where $I(\mathbf{x})$ is the intensity of the pixel $\mathbf{x}$. 
Using the above 
definitions, we can coalesce this degenerated partition 
$C$ with one or more other 
partitions by two ways. Obviously, $\mathbf{n}(\mathbf{m}(C))=\mathbf{m}(C)$,
so we re-define $\mathbf{n}(\mathbf{m}(C))$ by either
\begin{equation}
\mathbf{n}'_1(\mathbf{m}(C)) := \mathbf{M}(\mathbf{J}_1(C)) 
\end{equation}
if and only if $\mathbf{J}_1(C)$ is not the empty set \emph{and}
$\mathbf{m}[\mathbf{M}(\mathbf{J}_1(C))]\ne\mathbf{m}(C)$ 
or
\begin{equation}
\mathbf{n}'_2(\mathbf{m}(C)) := \mathbf{M}(\mathbf{J}_2(C)) 
\end{equation}
if and only if $\mathbf{J}_2(C)$ is not the empty set. Otherwise
we do not affect $\mathbf{n}(\mathbf{m}(C))$. We note that 
the latter expansion may result in a larger amount of coalescing sets, i.e.
in the former case it may happen that the maximum of the neighboring
pixels fall into the same class while in the latter case we definitely
excluded such cases (see the definition of $\mathbf{J}_2(C)$).

\begin{figure*}
\begin{center}
\resizebox{50mm}{!}{\includegraphics{img/star/flwo-blur-img.eps}}\hspace*{5mm}%
\resizebox{50mm}{!}{\includegraphics{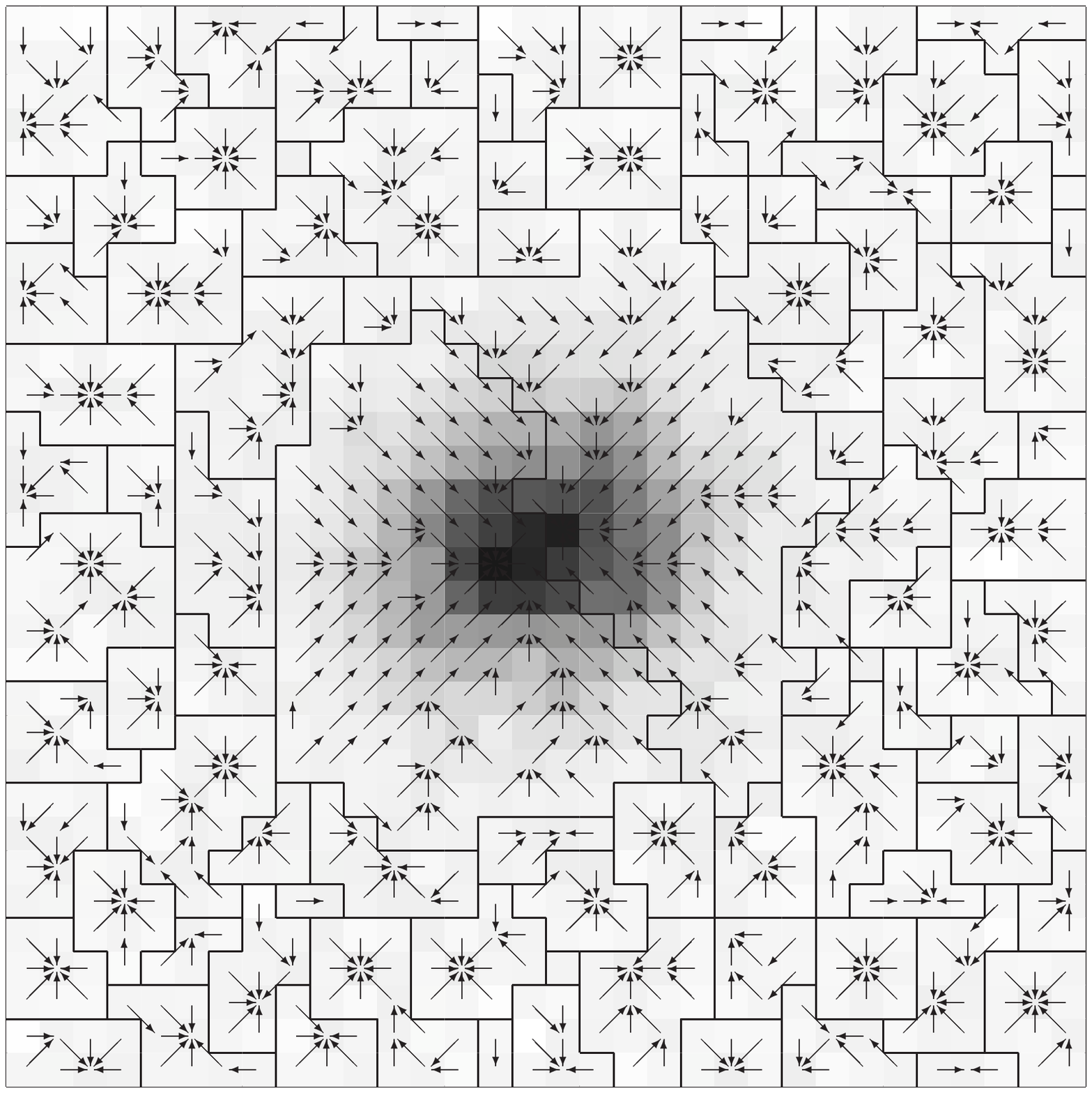}}\hspace*{5mm}%
\resizebox{50mm}{!}{\includegraphics{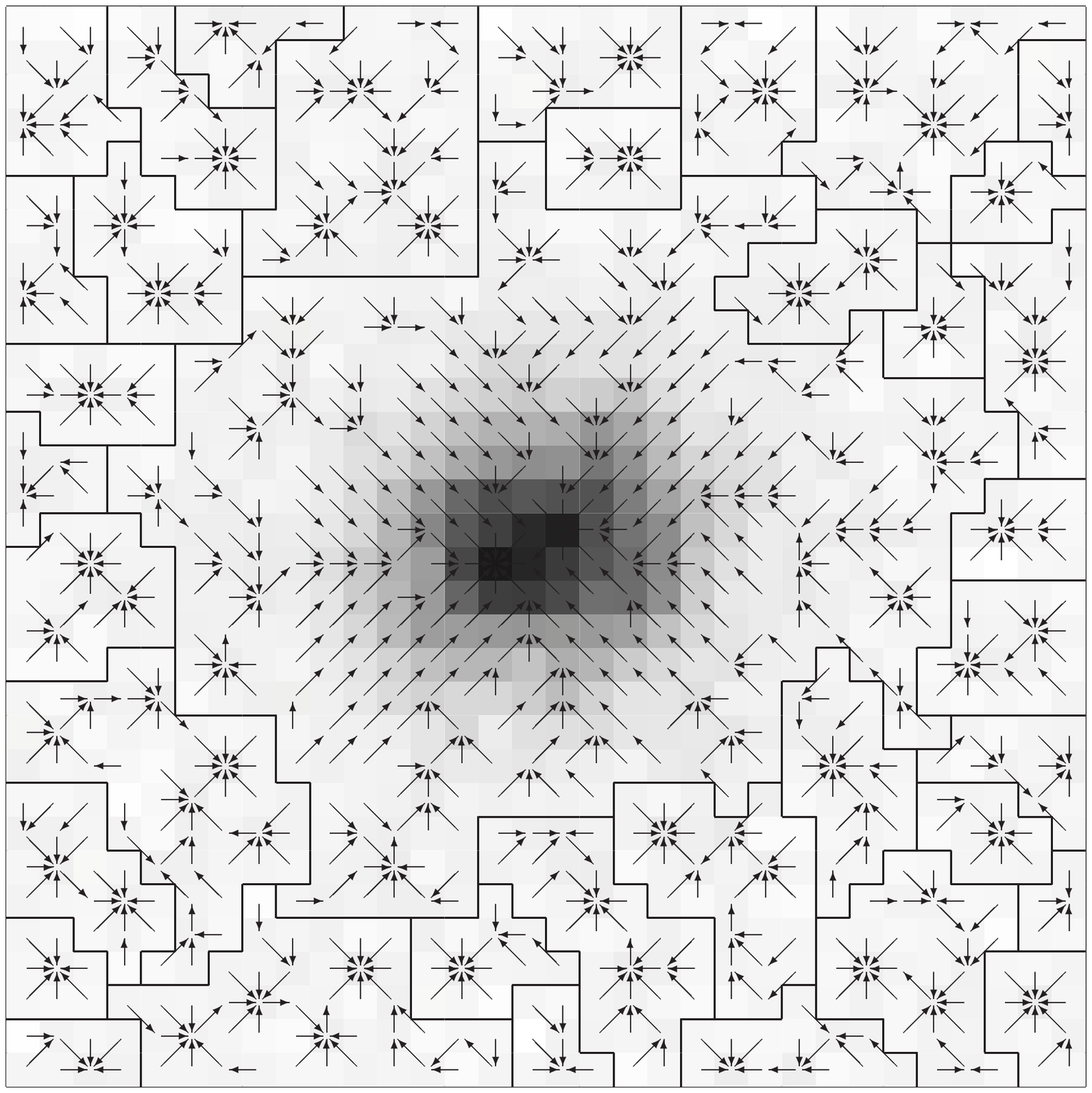}}
\end{center}
\caption{Left panel: a stamp of a star, covering $32\times32$ pixels 
from a typical
blurred KeplerCam image. Middle panel: the links 
and equivalence classes generated from this stamp, using the 
basic algorithm without any coalescing. Right panel: the links 
and equivalence classes generated from the stamp, when the 
partitions with zero prominence are joined to their neighboring partitions.}
\label{fig:starlinkprom0}
\end{figure*}

\phdsubsubsection{Prominence}

In case of highly defocused star images, the PSF
can be donut-shaped and a single star may have separated distinct (and not
degenerated) maxima. To coalesce such equivalence classes, we define
the discrete prominence, with almost the same properties
as it is known from topography. The prominence of a mountain peak in topography
(a.k.a. topographic prominence or autonomous height) is defined as follows.
For every path connecting the peak to higher terrain, find the lowest point 
on that path, that is at a saddle point. The key saddle is defined as 
the highest of these saddles, along all connecting paths. Then the prominence 
is the difference between the elevation of the peak and the elevation of 
the key saddle. This definition cannot be directly applied to our discrete
case, since the number of possible connecting paths between two maxima
is an exponential function of the number of the pixels, i.e. we cannot
get an $\ordo(N)$ algorithm. Thus, we use the following definition for
the key saddle $\mathbf{s}$ of an equivalence class $C$: 
\begin{eqnarray}
\mathbf{s}(C) & = & \big\{\mathbf{x}\in C ~:~ G(\mathbf{x})<K(\mathbf{x}) ~\mathrm{and}~ \\
& & \forall~\mathbf{x}'\in C~G(\mathbf{x}')<K(\mathbf{x}') \Rightarrow I(\mathbf{x}')\le I(\mathbf{x})\big\}. \nonumber
\end{eqnarray}
Thus, the \emph{prominence} of this class is going to be
\begin{equation}
p(C)=I(\mathbf{m}(C))-I(\mathbf{s}(C)).
\end{equation}
Note that $p(C)$ is always non-negative and 
if $C$ is degenerated, $p(C)$ is zero. 
The \emph{related class}es $\mathcal{R}(C)$ of $C$ are defined as 
\begin{equation}
\mathcal{R}(C)=\big\{ C'\in\mathrm{Classes} ~:~ \exists \mathbf{x'}\in C' ~ \|\mathbf{x}'-\mathbf{s}(C)\|= 1\big\}
\end{equation}
We define the set of \emph{parent classes} of $C$ as the set
\begin{eqnarray}
P(C) & = & \big\{ C'\in\mathcal{R}(C) ~:~ \forall~ C''\in\mathcal{R}(C) \\
& & \mathbf{m}(C'')\le\mathbf{m}(C') ~\mathrm{and}~ \mathbf{m}(C)<\mathbf{m}(C'') \big\} \nonumber
\end{eqnarray}
The set of parent classes $P(C)$ can be empty if the class $C$ is the most
prominent one. If at least one parent class exists, 
the \emph{relative prominence} of $C$ is defined as
\begin{equation}
r(C) = \frac{p(C)}{I[\mathbf{m}(P(C))]-\left<\mathbf{B}(P(C))\right>},
\end{equation}
and, by definition, it is always between 0 and 1.
Since the classes with low relative prominences are most likely
parts of a larger object that is dominated by the parent class 
(or, moreover, by the parent of the parent class and so on), we connect
these low-prominence classes to their parents below a critical
relative prominence $r_0$. Namely, we alter 
$\mathbf{n}(\mathbf{s}(C))$ to one point of $P(C)$, say, $\mathbf{x}'\in P(C)$
where $\|\mathbf{s}(C)-\mathbf{x}'\|=1$. Note that this algorithm
for $r_0=0$ yields the same collection of partitions as the
usage of the definition $\mathbf{J}_2(C)$ in the end of the previous 
subsection only for degenerated partitions.

\phdsubsection{Coordinates, shape parameters and analytic models}
\label{sec:stardetection:coordshapemodel}

In the previous sections we have discussed how astronomical images
can be partitioned in order to extract sets of pixels that belong to
the same source. Now we describe how these partitions can be characterized,
i.e. how can one determine the centroid coordinates, total flux of the source and
quantify somehow the shape of the source.

\phdsubsubsection{Weighted mean and standard deviance}

The easiest and fastest way to get some estimation on the centroid coordinates
and the shape parameters of the source is to calculate the statistical
mean and standard deviation of the pixel coordinates, weighted by the 
individual fluxes after background subtraction. 
Let us consider a set of pixels, $C=\{\mathbf{x}_i\}$,
each of them has the flux (ADU value) of $f_i$, while the background level $B$
of this source is calculated by using \eqref{eq:partitionbackground}. 
Then the weighted coordinates are 
\begin{equation}
\left<\mathbf{x}\right>=\frac{\sum\limits_{i} (f_i-B)\mathbf{x}_i}{\sum\limits_{i} (f_i-B)},
\end{equation}
while the statistical standard deviation in the coordinates is the covariance
matrix, defined as
\begin{equation}
\mathbf{S}=\frac{\sum\limits_{i} (f_i-B)(\mathbf{x}_i-\left<\mathbf{x}\right>)\circ(\mathbf{x}_i-\left<\mathbf{x}\right>)}{\sum\limits_{i} (f_i-B)}.
\end{equation}
Let us denote the components of the matrix $\mathbf{S}$ by
\begin{equation}
\mathbf{S}=\dpmatrix{\Sigma+\Delta}{\rm K}{\rm K}{\Sigma-\Delta}. \label{eq:profilecovariance}
\end{equation}
For objects that are not elongated, $\Delta={\rm K}=0$. It can be shown that
for elongated objects, the semimajor axis of the best fit ellipse 
(to the contours) has a position angle of $\varphi=\frac{1}{2}\arg(\Delta,{\rm K})$
and an ellipticity of $\sqrt{\Delta^2+{\rm K}^2}/\Sigma$.
The size of the star profiles are commonly characterized by the ``full width
at half magnitude'' (FWHM), that can be derived from $(\Sigma,\Delta,{\rm K})$ 
as follows. Let us consider an elongated 2 dimensional Gaussian profile
that is resulted by the convolution of a symmetric profile with the matrix
\begin{equation}
\mathbf{s}=\dpmatrix{\sigma+\delta}{\kappa}{\kappa}{\sigma-\delta}.
\end{equation}
It can be shown that such a profile described by $(\sigma,\delta,\kappa)$
has a covariance of 
\begin{equation}
\mathbf{S}=\dpmatrix{\sigma+\delta}{\kappa}{\kappa}{\sigma-\delta}^2,
\end{equation}
i.e. for such profiles, $\mathbf{s}^2=\mathbf{S}$. Since the FWHM of 
a Gaussian profile with $\sigma$ standard deviation is 
$2\sigma\sqrt{2\log 2}\approx2.35\,\sigma$, one can obtain the 
FWHM by calculating the square root of
the matrix defined in \eqref{eq:profilecovariance} and 
multiply the trace of the root (that is $2\sigma$) by the factor 
$1.17$. Therefore, for nearly circular profiles, the FWHM can be 
well approximated by $\sim2.35\,\sqrt{\Sigma}$.

Finally, the total flux of the object is 
\begin{equation}
f=\sum\limits_{i} (f_i-B),
\end{equation}
and the peak intensity is
\begin{equation}
A=\max\limits_{i} (f_i-B).
\end{equation}

\phdsubsubsection{Analytic models}
In order to have a better characterization for the stellar profiles, 
it is common to fit an analytic model function to the pixels.
Such a model has roughly
the same set of parameters: background level, flux (or peak intensity),
centroid coordinates and shape parameters. The most widely used models
are the Gaussian profile (symmetric or elongated) and the Moffat profile. 
In the characterization of stellar profiles, Lorentz profile and/or 
Voight profile are not 
used since these profiles are not integrable in two dimension. 

\begin{figure}
\begin{center}
\resizebox{80mm}{!}{\includegraphics{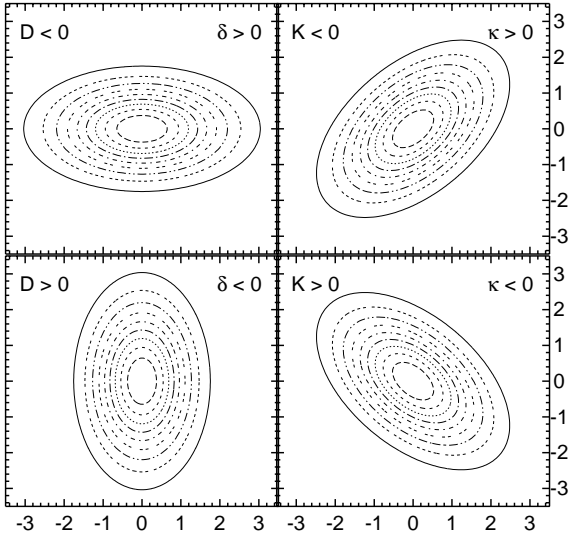}}
\end{center}
\caption{Some analytic elongated Gaussian stellar profiles. 
Each panel shows a contour plot for a profile
where the sharpness parameter $S=1$ and either $|D|=0.5$ or $|K|=0.5$.
Note that if the Gaussian polynomial coefficients $D$ and/or $K$ are positive,
then the respective asymmetric covariance matrix elements $\Delta$ and/or 
${\rm K}$ (and the asymmetric convolution parameters $\delta$ and/or $\kappa$ )
are negative and vice versa.}
\label{fig:stardkbasis}
\end{figure}

In the cases of undersampled images, we found that the profiles can be well
characterized by the Gaussian profiles, therefore in the practical 
implementations (see \texttt{fistar} and \texttt{firandom}, 
Sec.~\ref{sec:prog:fistar}, Sec.~\ref{sec:prog:firandom})
we focused on these models. Namely, these implementations support three
kind of analytic models, both are derivatives of the Gaussian function.
The first model is the symmetric Gaussian profile, characterized by 
five parameters: the background level $B$, the peak intensity $A$,
the centroid coordinates $\mathbf{x}_0=(x_0,y_0)$ and the parameter $S$
that is defined as $S=\sigma^{-2}$, where $\sigma$ is the standard 
deviation of the profile function. Thus, the model for the flux
distribution is
\begin{equation}
f_{\rm sym}(\mathbf{x})=B+A\exp\left[-\frac{1}{2}S(\mathbf{x}-\mathbf{x}_0)^2\right]. \label{eq:shapegausssym}
\end{equation}

The second implemented model is the elongated Gaussian profile that
is characterized by the above five parameters extended with two additional
parameters, resulting a flux distribution of
\begin{eqnarray}
f_{\rm elong}(\mathbf{x})&=&B+A\exp\big\{-\frac{1}{2}\left[S(\Delta x^2+\Delta y^2)+\right. \\
 & & +\left.D(\Delta x^2-\Delta y^2)+K(2\Delta x\Delta y)\right]\big\}, \label{eq:shapegausselong}
\end{eqnarray}
where $\Delta\mathbf{x}=\mathbf{x}-\mathbf{x}_0$ and $D$ and $K$ are the
two additional parameters, that show how the flux deviates
from a symmetric distribution. It is easy to show that the $(S,D,K)$ 
parameters are related to the covariance parameters 
$(\Sigma,\Delta,{\rm K})$ as 
\begin{equation}
\dpmatrix{S+D}{K}{K}{S-D}=\dpmatrix{\Sigma+\Delta}{\rm K}{\rm K}{\Sigma-\Delta}^{-1}. 
\end{equation}

The third model available in the implementations describes a flux distribution
that is called ``deviated'' since the peak intensity is offset from
the mean centroid coordinates. Stellar profiles that can only be 
well characterized by such a flux distribution model are fairly
common among images taken with fast focal ratio instruments due to
the strong comatic aberration. Such a model function can be built
from a Gaussian flux distribution by multiplying the main function
by a polynomial:
\begin{equation}
f_{\rm dev}=B+A\exp\left[-\frac{1}{2}S(\Delta\mathbf{x})^2\right]
\left(1+\sum\limits_{k,\ell} P_{k\ell}\Delta x^{k}\Delta y^{\ell}\right).\label{eq:devprofiledef}
\end{equation}
In the summation of \eqref{eq:devprofiledef}, 
$2\le k+\ell\le M$, where $M$ is the maximal 
polynomial order and $P_{02}+P_{20}$ is constrained to be $0$. Therefore,
for $M=2$, $3$ or $4$ the above function involves $2$, $6$ and $11$ other
parameters in addition to the 5 parameters of the symmetric Gaussian profile. 
If $M=2$, the above polynomial is equivalent to the second order 
expansion of the 
elongated Gaussian model if $P_{20}-P_{02}=-\frac{1}{2}D$ and
$P_{11}=K$. However, for $M=2$ the peak intensity is not offset
from the mean centroid coordinates, therefore in practice $M=2$ is not used.

\begin{figure*}
\begin{center}
\resizebox{50mm}{!}{\includegraphics{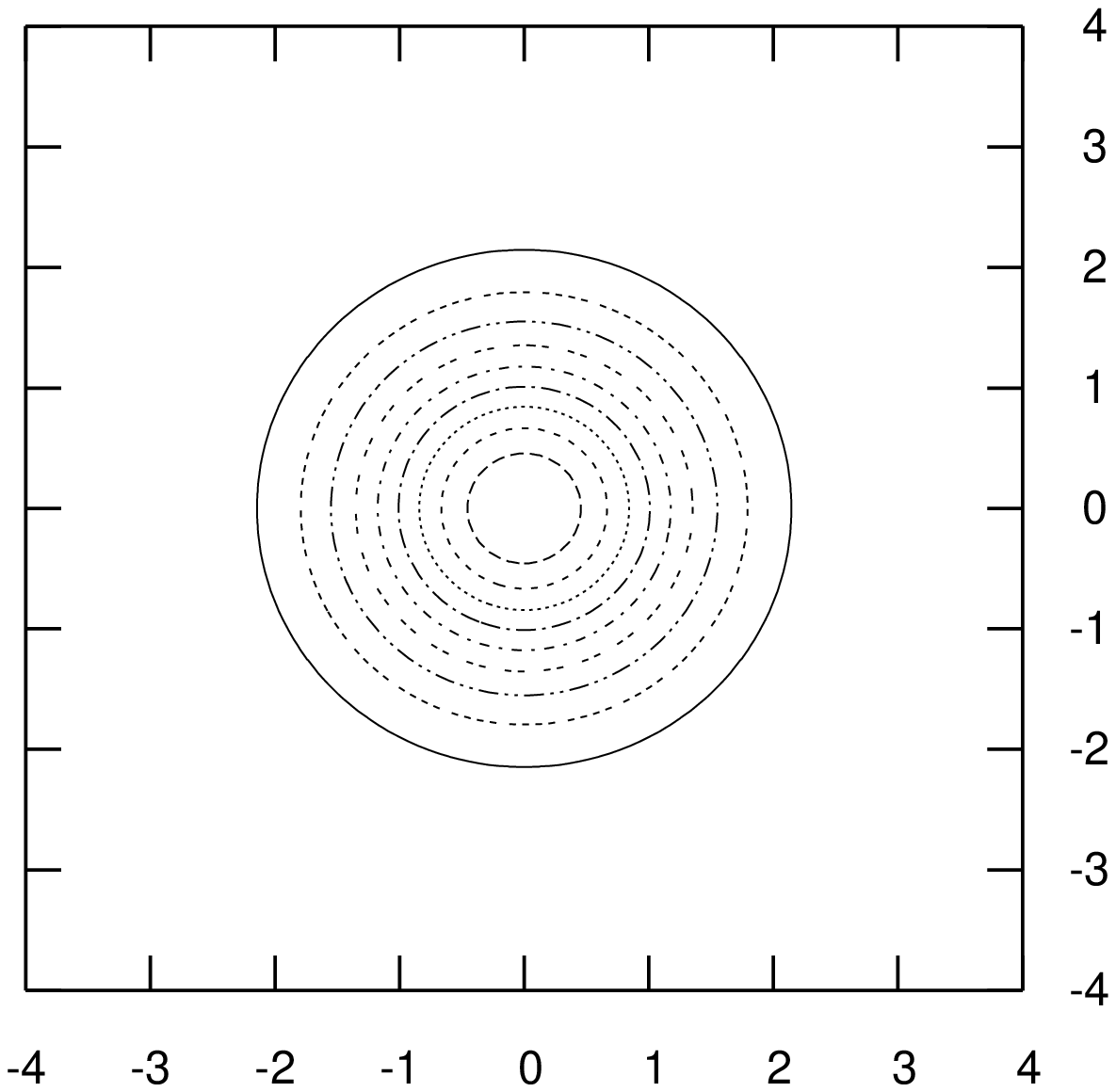}}\hspace*{2mm}%
\resizebox{50mm}{!}{\includegraphics{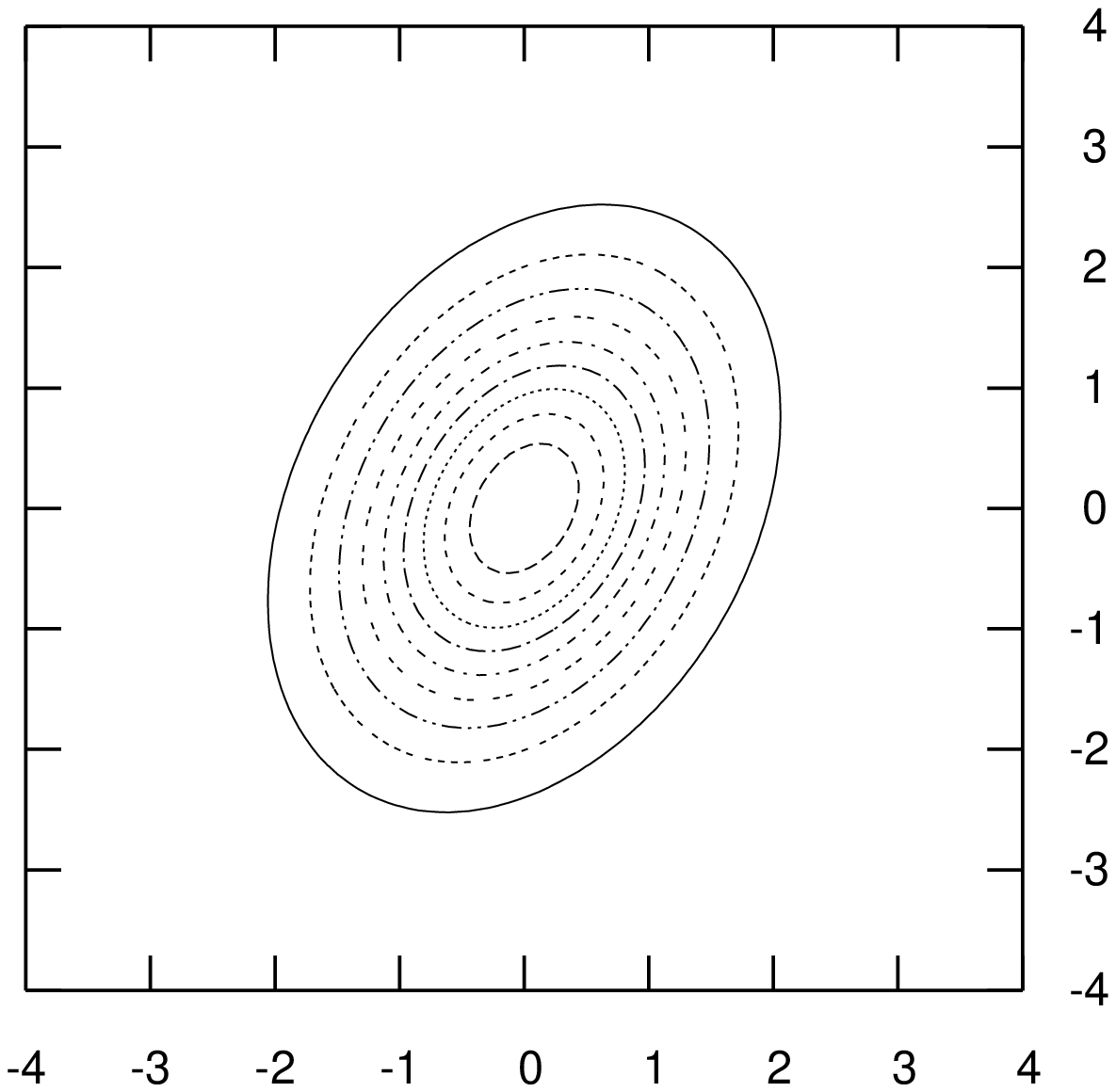}}\hspace*{2mm}%
\resizebox{50mm}{!}{\includegraphics{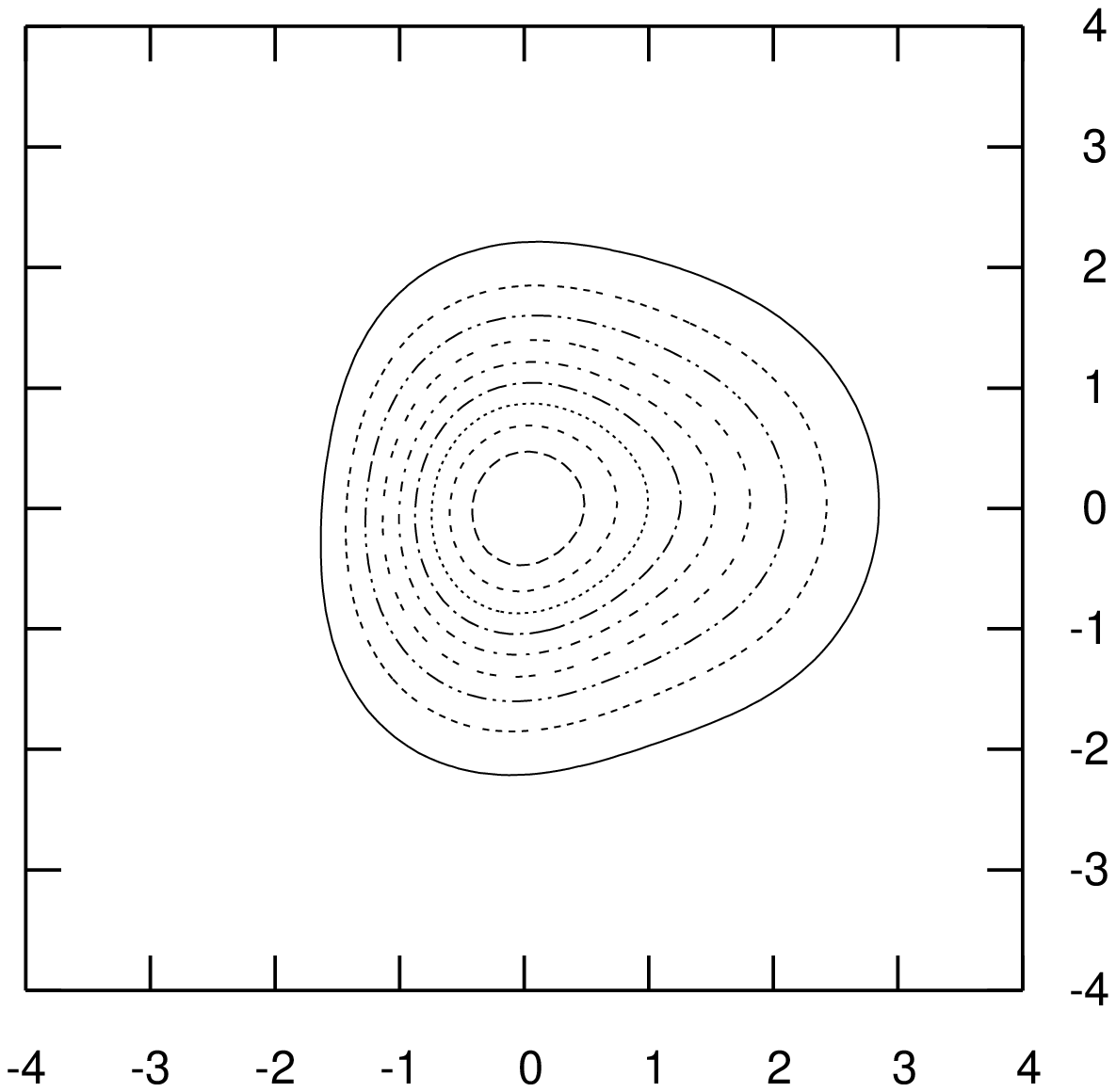}}
\end{center}\vspace*{0mm}
\caption{Analytic models for stellar profiles. From left to right, the three 
panels show the contour plots for a symmetric Gaussian profile, for
an elongated Gaussian profile and a deviated profile model of $M=4$. 
Note that all of the three models have a peak intensity at 
the coordinate $(0,0)$. In the plots the peak intensity is normalized to
unity and the contours show the intensity levels with a step size of $0.1$. 
All of the plotted models have an $S=1$ parameter while the other 
parameters ($D$, $K$ and $P_{k\ell}$) have a value around $\sim0.1 - 0.2$.
Because the choice of $S=1$, all of the models plotted here has a FWHM
of nearly $2.35$.}
\label{fig:starmodels}
\end{figure*}

All of the model functions discussed above are nonlinear
in the centroid coordinates $\mathbf{x}_0$ and the shape parameters $S$,
$D$, $K$ or $P_{k\ell}$. Therefore, in a parameter fit, 
one can use the Levenberg-Marquardt
algorithm \citep{press1992} since the parametric derivatives of the
model functions can easily be calculated and using the parameters of the
statistical mean coordinates and standard deviations as initial values
yields a good convergence. Moreover, if the iterations of 
the Levenberg-Marquardt algorithm fail to converge, it is a good indicator
to discard the source from our list since it is more likely to be a
hot pixel or a structure caused by cosmic ray event\footnote{Both 
cosmic ray events and hot pixels are hard to be modelled with 
these analytic functions.}.

In practice of HATNet and follow-up data reduction, we are
using the above models as follows. In real-time applications, for example
when the guiding correction is based on the astrometric solution,
the derivation of profile centroid coordinates is based on the weighted 
statistical mean of the pixel coordinates (and in this case, we are not 
even interested in the shape parameters, just in the centroid coordinates). 
If more  precise coordinates are needed, for example when one has to derive
the individual astrometric solutions in order to have a list
of coordinates for photometry, the symmetric Gaussian or the elongated
Gaussian models are used. The elongated model is also used when we 
characterize the spatial variations of the stellar profiles. This 
is particularly important when the optics is not adjusted to the detectors:
if the optical axis is not perpendicular to the plane of the CCD chip,
the spatial variations in the $D$ and $K$ parameters show a linear 
trend across the image. If the optical axis is set properly, the linear trend
disappears\footnote{Moreover, \emph{quadratic} trends in the $D$ or
$K$ components may also be there even if the optical axis is aligned properly.
In this case, the magnitude of the quadratic trends is proportional to
the magnitude of comatic aberration or the focal plane curveture.}. 
Finally, if we need to have an analytic description
for the stellar profiles as precise as possible, it is worth to use 
the deviated model.

\phdsubsection{Implementation}

The algorithms for extracting stars and characterizing stellar profiles
have been implemented in a standalone binary program named \texttt{fistar},
part of the \fihat{} package. All of the analytic models described here
are available in the program \texttt{firandom} of which main purpose
is to generate artificial images.
The capabilities of both programs
are discussed in Sec.~\ref{subsec:implementation} in more detail.


\phdsection{Astrometry}
\label{subsec:astrometry}

In the context of reduction of astronomical images, astrometry refers
to basically two things. First, the role of finding the astrometrical
solution is to find the appropriate function that maps the celestial
coordinate system to the image frame and vice versa. Second, the complete
astrometrical solution for any given image should identify the individual
sources (i.e. perform a ``cross-matching''), 
mostly based on a catalog that is assumed to be known in advance.

Theoretically there is no need to have a list from the available sources
found on the image and to have a pre-defined image to find. If one can
use only the pixel intensity information of the current and a
previously analyzed image to determine a relative transformation and supposing 
an astrometrical solution being obtained for another image, the
two mappings can be composed that results the astrometrical transformation
for the current one. This kind of transformations are mostly compositions of 
dilatation, small rotation and shift if the frames have been acquired 
subsequently by the same instrumentation from the same stellar field.
Such attempts of finding the relative transformation based on 
only the pixel intensities have been made by \cite{thiebaut2001}.

In this section, a robust and fast algorithm is presented,
for performing astrometry and
source cross-identification on two dimensional point lists, such as
between a catalogue and an astronomical image, or between two images.
The method is based on minimal assumptions: the lists can be rotated,
magnified and inverted with respect to each other in an arbitrary way. 
The algorithm is tailored to work efficiently on wide fields with large
number of sources and significant non-linear distortions, as long as
the distortions can be approximated with linear transformations
locally, over the scale-length of the average distance between the
points.  The procedure is based on symmetric point matching in a newly
defined continuous triangle space that consists of triangles generated
by an extended Delaunay triangulation. 


\phdsubsection{Introduction}
\label{subsec:astrometry:intro}

Cross-matching two two-dimensional points lists is a crucial step in
astrometry and source identification. The tasks involves finding the
appropriate geometrical transformation that transforms one list into
the reference frame of the other, followed by finding the best matching
point-pairs.
 
One of the lists usually contains the pixel coordinates of sources in
an astronomical image (e.g.~point-like sources, such as stars), while
the other list can be either a reference catalog with celestial
coordinates, or it can also consist of pixel coordinates that originate
from a different source of observation (another image). Throughout this
section we denote the reference (list) as \aref, the image (list) as
\aimg, and the function that transforms the reference to the image as
\atran.

The difficulty of the problem is that in order to find matching pairs,
one needs to know the transformation, and vica versa: to derive the
transformation, one needs point-pairs. Furthermore, the lists may not
fully overlap in space, and may have only a small fraction of sources
in common.

By making simple assumptions on the properties of \atran, however, the
problem can be tackled. A very specific case is when there is
only a simple translation between the lists, and one can use
cross-correlation techniques \citep[see][]{phillips1995} to find the
transformation. 
We note, that a method proposed by \citet{thiebaut2001} uses the
whole image information to derive a transformation (translation and
magnification).

A more general assumption, typical to astronomical applications, is
that $\atran$ is a similarity transformation (rotation, magnification,
inversion, without shear), i.e.~$\atran = \lambda
\mathbf{A}\underline{r} + \underline{b}$, where $\mathbf{A}$ is a
(non-zero) scalar $\lambda$ times the orthogonal matrix,
$\underline{b}$ is an arbitrary translation, and $\underline{r}$ is the
spatial vector of points. Exploiting that geometrical patterns remain
similar after the transformation, more general algorithms have been
developed that are based on pattern matching \citep{groth1986,valdes1995}.
The idea is that
the initial transformation is found by the aid of a specific set of
patterns that are generated from a subset of the points on both \aref\
and \aimg. For example, the subset can be that of the brightest
sources, and the patterns can be triangles.  With the knowledge of this
initial transformation, more points can be cross-matched, and the
transformation between the lists can be iteratively refined.
Some of these methods are implemented as an
\pack{IRAF} task in \prog{immatch} \citep{phillips1995}.

The above pattern matching methods perform well as long as the dominant
term in the transformation is linear, such as for astrometry of narrow
field-of-view (FOV) images, and as long as the number of sources is
small (because of the large number of patterns that can be generated
\--- see later).  In the past decade of astronomy, with the development of
large format CCD cameras or mosaic imagers, many wide-field surveys
appeared, such as those looking for transient events 
(e.g.~ROTSE --- \citealt{akerlof2000}), 
transiting planets (Chapter~\ref{chapter:introduction}), 
or all-sky variability (e.g.~ASAS \--- \citealt{pojmanski1997}). There
are non-negligible, higher order distortion terms in the astrometric
solution that are due to, for instance, the projection of celestial to
pixel coordinates and the properties of the fast focal ratio optical
systems. Furthermore, these images may contain $\sim 10^5$ sources, and
pattern matching is non-trivial.


The presented algorithm is based on, and is a
generalization of the above pattern matching algorithms. It is very
fast, and works robustly for wide-field imaging with minimal
assumptions. Namely, we assume that:
i) the distortions are non-negligible, but small compared to the linear
term,
ii) there exists a smooth transformation between the reference and
image points, 
iii) the point lists have a considerable number of sources in common,
and
iv) the transformation is locally invertible.

This section has the following parts. First we describe symmetrical point
matching in Sec.~\ref{subsec:astrometry:pmatch} before we go on 
to the discussion of finding
the transformation (Sec.~\ref{subsec:astrometry:tran}). 
The software implementation and its
performance on a large and inhomogeneous dataset is demonstrated in
Sec.~\ref{subsec:astrometry:exam}. 

\phdsubsection{Symmetric point matching}
\label{subsec:astrometry:pmatch}

First, let us assume that \atran{} is known. To find point-pairs between
\aref{} and \aimg\, one should first transform the reference points to
the reference frame of the image: $\areft = \atran(\aref)$. Now it is
possible to perform a simple symmetric point matching between \areft\
and \aimg. One point ($R_1\in\areft$) from the first and one point
($I_1\in\aimg$) from the second set are treated as a pair if the
closest point to $R_1$ is $I_1$ \emph{and} the closest point to
$I_1$ is $R_1$. This requirement is symmetric by definition and excludes such
cases when e.g.~the closest point to $R_1$ is $I_1$, but there exists an
$R_2$ that is even closer to $I_1$, etc.

In one dimension, finding the point of a given list nearest to a
specific point ($x$) can be implemented as a binary search. Let us
assume that the point list with $N$ points is ordered in ascending
order.  This has to be done only once, at the beginning, and using the
quicksort algorithm, for example, the required time scales on average
as $\ordo(N \log N)$. Then $x$ is compared to the median of the list:
if it is less than the median, the search can be continued recursively
in the first $N/2$ points, if it is greater than the median, the second
$N/2$ half is used. At the end only one comparison is needed to find
out whether $x$ is closer to its left or right neighbor, so in total
$1+\log_2(N)$ comparisons are needed, which is an $\ordo(\log N)$
function of $N$. Thus, the total time including the initial sorting
also goes as $\ordo(N \log N)$.

As regards a two dimensional list, let us assume again, 
that the points are ordered in
ascending order by their $x$ coordinates (initial sorting $\sim \ordo(N
\log N)$), and they are spread uniformly in a square of unit area.
Finding the nearest point in $x$ coordinate also requires $\ordo(\log
N)$ comparisons, however, the point found presumably is not the
nearest in Euclidean distance. The expectation value of the distance
between two points is $1/\sqrt{N}$, and thus we have to compare points
within a strip with this width and unity height, meaning
$\ordo(\sqrt{N})$ comparisons. Therefore, the total time required by a
symmetric point matching between two catalogs in two dimensions
requires $\ordo(N^{3/2} \log N)$ time.

We note that finding the closest point within a given set of points is
also known as nearest neighbor problem \citep[for a summary see][and
references therein]{gionis2002}.  It is possible to reduce the
computation time in 2 dimensions to $\ordo(N \log N)$ by the aid of
Voronoi diagrams and Voronoi cells, but we have not implemented such an
algorithm in our matching codes.

\phdsubsection{Finding the transformation}
\label{subsec:astrometry:tran}

Let us go back to finding the transformation between \aref{} and \aimg.
The first, and most crucial step of the algorithm is to find an initial
``guess'' \atranone{} for the transformation based on a variant of
triangle matching. Using \atranone, \aref{} is transformed to
\aimg, symmetric point-matching is done, and the paired coordinates are
used to further refine the transformation (leading to \atrani{} in iteration
$i$), and increase the number of matched points iteratively. A major part of
this section is devoted to finding the initial transformation.

\phdsubsubsection{Triangle matching}
\label{subsec:astrometry:trimatch}

It was proposed earlier by \citet{groth1986}, \citet{stetson1989}
and \citep[see][]{valdes1995} to use triangle matching for
the initial ``guess'' of the transformation. 
The total number of triangles that can be formed using $N$ points is
$N(N-1)(N-2)/6$, an $\ordo(N^3)$ function of $N$. As this can
be an overwhelming number, one can resort to using a subset of the points
for the vertices of the triangles to be generated. One can also limit
the parameters of the triangles, such as exclude elongated or large
(small) triangles.

As triangles are uniquely defined by three parameters, for example the
length of the three sides, these parameters (or their appropriate
combinations) naturally span a 3-dimensional triangle space. Because
our assumption is that \atran{} is dominated by the linear term, to
first order approximation there is a single scalar magnification
between \aref{} and \aimg{} (besides the rotation, chirality and
translation). It is possible to reduce the triangle space to a
normalized, two-dimensional triangle space ($(T_x,T_y)\in T$), whereby
the original size information is lost. Similar triangles (with or
without taking into account a possible flip) can be represented by the
same points in this space, alleviating triangle matching between \aref\
and \aimg.

\phdsubsubsection{Triangle spaces}
\label{subsec:astrometry:trispace}

\begin{figure}
\begin{center}
\resizebox{80mm}{!}{\includegraphics{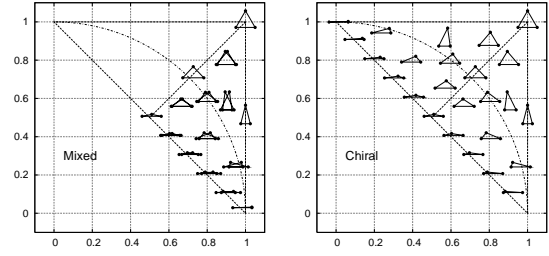}}
\end{center}
\caption{The position of triangles in the mixed and the chiral triangle spaces.
The exact position of a given triangle is represented by its center of
gravity.  Note that in the mixed triangle space some triangles with
identical side ratios but different orientation overlap. The dashed
line shows the boundaries of the triangle space. The dotted-dashed line
represents the right triangles and separates obtuse and acute ones.
\label{fig:trichir}}
\end{figure}

There are multiple ways of deriving normalized triangle spaces. One can
define a ``mixed'' normalized triangle space \tmixed, where the
coordinates are insensitive to inversion between the original
coordinate lists, i.e.~all similar triangles are represented by the
same point irrespective of their chirality \citep{valdes1995}:
\begin{eqnarray}
\tmixed_x & = & p/a, \label{eq:tmix1}\\
\tmixed_y & = & q/a, \label{eq:tmix2}
\end{eqnarray}
where $a$, $p$ and $q$ are the sides of the triangle in descending
order. Triangles in this space are shown on the left panel of
Fig.~\ref{fig:trichir}.
Coordinates in the mixed triangle space are continuous functions of the
sides (and therefore of the spatial coordinates of the vertices of the
original triangle) but the orientation information is lost.  Because we
assumed that \atran{} is smooth and bijective, no local inversions and
flips can occur. In other words, \aref{} and \aimg{} are either flipped
or not with respect to each other, but chirality does not have a
spatial dependence, and there are no ``local spots'' that are mirrored.
Therefore, using mixed triangle space coordinates can yield false
triangle matchings that can lead to an inaccurate initial
transformation, or the match may even fail. Thus, for large sets of
points and triangles it is more reliable to fix the orientation of the
transformation. For example, first assume the coordinates are not
flipped, perform a triangle match, and if this match is unsatisfactory,
then repeat the fit with flipped triangles.

This leads to the definition of an alternative, ``chiral'' triangle
space:
\begin{eqnarray}
\tchir_x & = & b/a, \label{eq:tchir1}\\
\tchir_y & = & c/a, \label{eq:tchir2}
\end{eqnarray}
where $a$, $b$ and $c$ are the sides in counter-clockwise order and $a$
is the longest side. In this space similar triangles with different
orientations have different coordinates. The shortcoming of \tchir{} is
that it is not continuous: a small perturbation of an isosceles
triangle can result in a new coordinate that is at the upper rightmost
edge of the triangle space.

In the following, we show that it is possible to define a
parametrization that is both continuous and preserves chirality. Flip
the chiral triangle space in the right panel of Fig.~\ref{fig:trichir} along
the $T_x + T_y=1$ line. This transformation moves the equilateral
triangle into the origin. Following this, apply radial magnification of
the whole space to move the $T_x + T_y=1$ line to the $T_x^2 + T_y^2=1$
arc (the magnification factor is not constant: $1$ along the direction
of $x$ and $y$-axis and $\sqrt{2}$ along the $T_x=T_y$ line).  Finally,
apply an azimuthal slew by a factor of $4$ to identify the
$T_y=0, T_x>0$ and $T_x=0, T_y>0$ edges of the space. To be more
specific, let us denote the sides as in \tchir: $a$, $b$ and
$c$ in counter-clockwise order where $a$ is the longest, and define
\begin{eqnarray}
\alpha & = & 1-b/a, \\
\beta & = & 1-c/a.
\end{eqnarray}
Using these values, it is easy to prove that by using the definitions
of the following variables:
\begin{eqnarray}
x_1 & = & \frac{\alpha(\alpha+\beta)}{\sqrt{\alpha^2+\beta^2}}, \\
y_1 & = & \frac{\beta(\alpha+\beta)}{\sqrt{\alpha^2+\beta^2}}, \\
x_2 & = & x_1^2-y_1^2, \\
y_2 & = & 2x_1y_1, 
\end{eqnarray}
one can define the triangle space coordinates as:
\begin{eqnarray}
\tcont_x & = & \frac{x_2^2-y_2^2}{(\alpha+\beta)^3} = 
	\frac{(\alpha+\beta)\left(\alpha^4-6\alpha^2\beta^2+\beta^4\right)}
	{(\alpha^2+\beta^2)^2}, \label{eq:tcnt1}\\
\tcont_y & = & \frac{2x_2y_2}{(\alpha+\beta)^3} =
 	\frac{4(\alpha+\beta)\alpha\beta(\alpha^2-\beta^2)}
	{(\alpha^2+\beta^2)^2}\, .\label{eq:tcnt2}
\end{eqnarray}

\begin{figure}
\begin{center}
\resizebox{80mm}{!}{\includegraphics{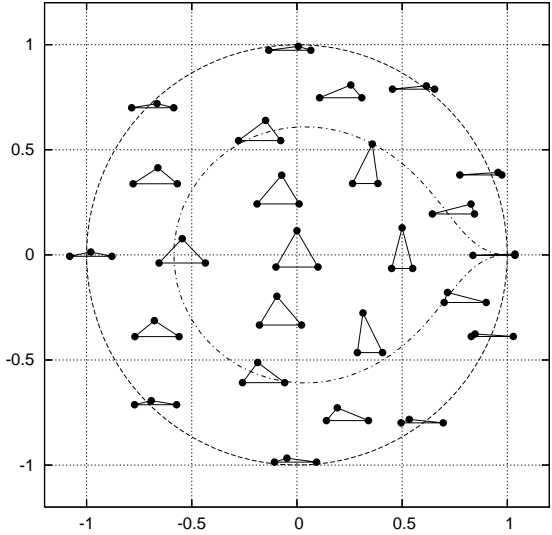}}
\end{center}
\caption{Triangles in the continuous triangle space as defined by
Eqs.~\ref{eq:tcnt1}\---\ref{eq:tcnt2}. 
We show the same triangles as earlier, in Fig.~\ref{fig:trichir}, for the
\tmixed{} and \tchir triangle spaces. Equilateral triangles are centered
in the origin. The dotted-dashed line refers to the right triangles,
and divides the space to acute (inside) and obtuse (outside) triangles.
Isosceles triangles are placed on the $x$-axis (where
$\tcont_y=0$).}
\label{fig:tricont}
\end{figure}

The above defined \tcont{} continuous triangle space has many
advantages. It is a continuous function of the sides for all
non-singular triangles, and also preserves chirality information.
Furthermore, it spans a larger area, and misidentification of triangles
(that may be very densely packed) is decreased. Some triangles in this
space are shown in Fig.~\ref{fig:tricont}.

\phdsubsubsection{Optimal triangle sets}
\label{subsec:astrometry:opttri}

As it was mentioned before, the total number of triangles that can be
formed from $N$ points is $\approx N^3/6$.  Wide-field images typically
contain $\ordo(10^4)$ points or more, and the total number of triangles
that can be generated -- a complete triangle list -- is unpractical for
the following reasons.
First, storing and handling such a large number of triangles with
typical computers is inconvenient. To give an example, a full
triangulation of 10,000 points yields $\sim 1.7\times10^{11}$ triangles.

Second, this complete triangle list includes many triangles that are
not optimal to use. For example large triangles can be significantly
distorted in \aimg{} with respect to \aref, and thus are represented by
substantially different coordinates in the triangle space. The size of
optimal triangles is governed by two factors: the distortion of large
triangles, and the uncertainty of triangle parameters for small
triangles that are comparable in size to the astrometric errors of the
vertices.

To make an estimate of the optimal size for triangles, let us denote
the characteristic size of the image by $D$, the astrometric error by
$\delta$, and the size of a selected triangle as $L$. For the sake of
simplicity, let us ignore the distortion effects of a complex optical
assembly, and estimate the distortion factor $f_d$ in a wide field
imager as the difference between the orthographic and gnomonic
projections
\citep[see][]{calabretta2002}:
\begin{equation}
f_d \approx |(\sin(d)-\tan(d))/d| \approx |1-\cos(d)|\,,
\end{equation}
where $d$ is the radial distance as measured from the center of the
field. For the HATNet frames ($d = D\approx6\arcdeg$ to the corners)
this estimate yields $f_d\approx 0.005$.
The distortion effects yield an error of $f_d L/D$ in the triangle
space \--- the bigger the triangle, the more significant the
distortion. For the same triangle, astrometric errors cause an
uncertainty of $\delta/L$ in the triangle space that decreases with
increasing $L$. Making the two errors equal,
\begin{equation}
\frac{f_d \cdot L}{D} = \frac{\delta}{L},
\end{equation}
an optimal triangle size can be estimated by
\begin{equation}
L_{\rm opt}=\sqrt{\frac{\delta \cdot D}{f_d}}.
\end{equation}
In our case $d=2048$ pixels (or $6\arcdeg$), $f_d=0.005$ and the
centroid uncertainty for an $I=11$ star is $\delta=0.01$, so the
optimal size of the triangles is $L_{\rm opt}\approx 60-70$ pixels.

Third, dealing with many triangles may result in a triangle space that
is over-saturated by the large number of points, and may yield
unexpected matchings of triangles. In all definitions of the previous
subsection, the area of the triangle space is approximately unity.
Having triangles with an error of $\sigma$ in triangle space and
assuming them to have a uniform distribution, allowing a $3\sigma$
spacing between them, and assuming $\sigma=\delta/L_{\rm  opt}$, the
number of triangles is delimited to:
\begin{equation}
T_{\rm max}\approx
\frac{1}{(3\sigma)^2}\approx
\frac{1}{9}\left(\frac{L}{\delta}\right)^2=
\frac{D}{9f_d\delta}\,.
\end{equation}
In our case (see values of $D$, $f_d$ and $\delta$ above) the former
equation yields $T_{\rm opt}\approx 2\times 10^6$ triangles. Note
that this is 5 orders of magnitude smaller than a complete
triangulation ($\ordo(10^{11})$). 

\phdsubsubsection{The extended Delaunay triangulation}
\label{subsec:astrometry:extdel}

\begin{figure}
\begin{center}
\resizebox{80mm}{!}{\includegraphics{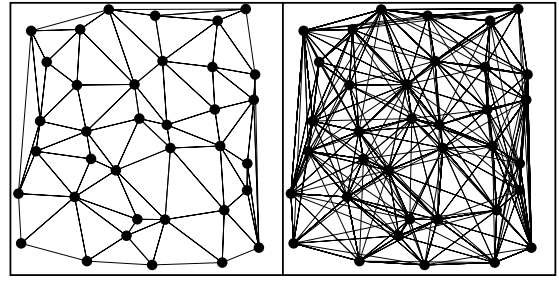}}
\end{center}
\caption{Triangulations of some randomly distributed points: the left panel
shows the Delaunay triangulation (60 triangles in total) 
the right panel exhibits the $\ell=1$ extended triangulation 
(312 triangles) of the same point set.
\label{fig:delaunay}}
\end{figure}

Delaunay triangulation \citep[see][]{shewchuk1996} is a fast and robust
way of generating a triangle mesh on a point-set. The Delaunay
triangles are disjoint triangles where the circumcircle of any triangle
contains no other points from any other triangle. This is also
equivalent to the most efficient exclusion of distorted triangles in a
local triangulation. For a visual example of a Delaunay triangulation
of a random set of points, see the left panel of Fig.~\ref{fig:delaunay}.

Following Euler's theorem (also known as the polyhedron formula), one
can calculate the number of triangles in a Delaunay triangulation of
$N$ points:
\begin{equation}
T_D = 2N-2-C,
\end{equation}
where $C$ is the number of edges on the convex hull of the point set.
For large values of $N$, $T_D$ can be estimated as $2N$, as $2+C$ is
negligible. Therefore, if we select a subset of points (from \aref{} or
\aimg) where neighboring ones have a distance of $L_{\rm opt}$, we
get a Delaunay triangulation with approximately $2D^2/L_{\rm opt}^2$
triangles. The $D$, $\delta$ and $f_d$ values for HAT images
correspond to $\approx 6000$ triangles, i.e.~3000 points. 
In our experience, this yields very fast matching, but it is not robust
enough for general use, because of the following reasons.

Delaunay triangulation is very sensitive for removing a point
from the star list.  According to the polyhedron formula, on the
average, each point has 6 neighboring points and belongs to 6
triangles. Because of observational effects or unexpected events, the
number of points fluctuates in the list. To mention a few examples, it
is customary to build up \aimg{} from the brightest stars in an image,
but stars may get saturated or fall on bad columns, and thus disappear
from the list. Star detection algorithms may find sources depending on
the changing full-width at half maximum (FWHM) of the frames.
Transients, variable stars or minor planets can lead to additional
sources on occasions. In general, if one point is removed, 6 Delaunay
triangles are destroyed and 4 new ones are formed that are totally
disjoint from the 6 original ones (and therefore they are represented
by substantially different points in the triangle space). Removing one
third of the generating points might completely change the
triangulation\footnote{Imagine a honey-bee cell structure where all
central points of the hexagons are added or removed: these two
construction generates disjoint Delaunay triangulations.}.

Second, and more important, there is no guarantee that the spatial {\em
density} of points in \aref{} and \aimg{} is similar. For example, the
reference catalog is retrieved for stars with different magnitude
limits than those found on the image. If the number of points in common
in \aref{} and \aimg{} is only a small fraction of the total number of
points, the triangulation on the reference and image has no common
triangles. 
Third, the number of the triangles  with Delaunay triangulation
($T_{\rm D}$) is definitely smaller than $T_{\rm opt}$; i.e.~the
triangle space could support more triangles without much confusion.

Therefore, it is beneficial to extend the Delaunay triangulation. A
natural way of extension can be made as follows. Define a level $\ell$
and for any given point ($P$) select all points from the point set of
$N$ points that
can be connected to $P$ via maximum $\ell$ edges of the Delaunay
triangulation. Following this, one can generate the \emph{full}
triangulation of this set and append the new triangles to the whole
triangle set. This procedure can be repeated for all points in the point
set at fixed $\ell$. For self-consistence, the $\ell=0$ case is
defined as the Delaunay triangulation itself. 
If all points have 6
neighbors, the number of ``extended'' triangles {\em per data point} is:
\begin{equation}
T_\ell=(3\ell^2+3\ell+1)(3\ell^2+3\ell)(3\ell^2+3\ell-1)/6
\end{equation}
for $\ell>0$, 
i.e.~this extension introduces $\ordo(\ell^6)$ new triangles. 
Because some of the extended triangles are repetitions of other
triangles from the original Delaunay triangulation and from the
extensions of another points, the final dependence only goes as
$\ordo(T_{\rm D}\ell^2)$. We note that our software implementation
is slightly different, and the expansion requires $\ordo(N\ell^2)$ time
and automatically results in a triangle set where each triangle is
unique. To give an example, for $N=10,000$ points the Delaunay
triangulation gives $20,000$ triangles, the $\ell = 1$ extended
triangulation gives $\sim 115,000$ triangles, $\ell = 2$ some $\sim
347,000$ triangles, $\ell = 3$ $875,000$ and $\ell = 4$ $\sim
1,841,000$ triangles, respectively. The extended triangulation is not
only advantageous because of more triangles, and better chance for
matching, but also, there is a bigger variety in size that enhances
matching if the input and reference lists have different spatial
density.

\phdsubsubsection{Matching the triangles in triangle space}
\label{subsec:astrometry:trimatch2}

If the triangle sets for both the reference and the input list are
known, the triangles can be matched in the normalized triangle space
(where they are represented by two dimensional points) using the
symmetric point matching as described in Sec.~\ref{subsec:astrometry:pmatch}.

In the next step we create a ${N_R\times N_I}$ ``vote'' matrix $V$,
where $N_R$ and $N_I$ are the number of points in the reference and
input lists that were used to generate the triangulations,
respectively. The elements of this matrix have an initial value of 0.
Each matched triangle corresponds to 3 points in the reference list
(identified by $r_1$, $r_2$, $r_3$) and 3 points in the input list
($i_1$, $i_2$ and $i_3$). Knowing these indices, the matrix elements
$V_{r_1i_1}$, $V_{r_2i_2}$ and $V_{r_3i_3}$ are incremented. The
magnitude of this increment (the \emph{vote}) can depend on the
distances of the matching triangles in the triangle space: the closer
they are, the higher votes these points get. In our implementation, if
$N_T$ triangles are matched in total, the closest pair gets $N_T$
votes, the second closest pair gets $N_T-1$ votes, and so on.

Having built up the vote matrix, we select the greatest elements of
this matrix, and the appropriate points referring to these row and
column indices are considered as matched sources. We note that not all
of the positive matrix elements are selected, because elements with
smaller votes are likely to be due to misidentifications. We found that
in practice the upper 40\% of the matrix elements yield a robust match.

\phdsubsubsection{The unitarity of the transformations}
\label{subsec:astrometry:unit}

If an initial set of the possible point-pairs are known from
triangle-matching, one can fit a smooth function (e.g.~a polynomial)
that transforms the reference set to the input points.
Our assumption was that the dominant term in our transformation is the
similarity transformation, which implies that the homogeneous linear
part of it should be \emph{almost} unitarity
operator\footnote{Here 
$\mathbf{A}\mathbf{A^+} = I$, where $\mathbf{A^+}$
is the adjoint of $\mathbf{A}$ and $\mathbf{I}$ is the identity,
i.e.~$\mathbf{A}$ is an orthogonal transformation with possible
inversion and magnification.}.
After the transformation is determined, it is useful to measure how much we
diverge from this assumption.
As mentioned earlier (Sec.~\ref{subsec:astrometry:intro}), similarity transformations
can be written as
\begin{equation}
\label{eq:lintr2}
\underline{r^{\prime}} = \lambda \mathbf{A} \underline{r} + \underline{b} 
\, \equiv \,
\lambda \binom{a ~~~ c}{b ~~~ d} \underline{r} +
\underline{b},\,
\end{equation}
where $\lambda\ne0$, and the $a,b,c,d$ matrix components are the sine
and cosine of a given rotational angle, i.e.~$a=d$ and $b=-c$.

If we separate the homogeneous linear part of the transformation, as
described by a matrix similar to that in \eqref{eq:lintr2}, it is a
combination of rotation and dilation with possible inversion if
$|a|\approx |d|$ and $|c|\approx |b|$.
We can define the unitarity of a matrix as:
\begin{equation}
\Lambda^2 := \frac{(a\mp d)^2+(b \pm c)^2}{a^2 + b^2 + c^2 + d^2}\,,\label{eq:defunitarity}
\end{equation}
where the $\pm$ indicates the definition for regular and inverting
transformations, respectively. For a combination of rotation and
dilation, $\Lambda$ is zero, for a distorted transformation
$\Lambda\approx f_d \ll 1$.

The $\Lambda$ unitarity gives a good measure of how well the initial
transformation was determined. It happens occasionally that the
transformation is erroneous, and in our experience, in these cases 
$\Lambda$ is not just larger than the expectational value of $f_d$, but
it is $\approx1$. This enables fine-tuning of the algorithm, such as
changing chirality of the triangle space, or adding further iterations
till satisfactory $\Lambda$ is reached. 

\phdsubsubsection{Point matching in practice}
\label{subsec:astrometry:pract}

In practice, matching points between the \aref{} reference and \aimg\
image goes as the following:
\begin{enumerate}
\item Generate two triangle sets $T_R$ and $T_I$ on \aref{} and \aimg, 
respectively:
	\begin{enumerate}
	\item In the first iteration, generate only Delaunay triangles.
	\item Later, if necessary, extended triangulation can be generated 
		with increasing levels of $\ell$.
	\end{enumerate}
\item Match these two triangle sets in the triangle space using
symmetric point matching.
\item Select some possible point-pairs using a vote-algorithm (yielding 
$N_0$ pairs).
\item Derive the initial smooth transformation \atranone{} 
using a least-squares fit.
	\begin{enumerate}
	\item Check the unitarity of \atranone.
	\item If it is greater than a given threshold ($\ordo(f_d)$), 
		increase $\ell$ and go to step (i)/(b).
		If the unitarity is less than this threshold, proceed to step 5.
	\item If we reached the maximal allowed $\ell$, try the procedure
		with triangles that are flipped with respect to each other
		between the image and reference, i.e.~switch chirality of the
		\tcont{} triangle space. 
	\end{enumerate}

\item Transform \aref{} using this initial transformation 
to the reference frame of the image ($\areft = \atranone(\aref)$).

\item Perform a symmetric point matching between
\areft{} and \aimg{} (yielding $N_1 > N_0$ pairs).

\item Refine the transformation based on the greater number of 
pairs, yielding transformation \atrani, where $i$ is the iteration
number. 

\item If necessary, repeat points 5, 6 and 7 iteratively, increase the
number of matched points, and refine the transformation.
\end{enumerate}

For most astrometric transformations and distortions it holds that
locally they can be approximated with a similarity transformation.  At
a reasonable density of points on \aimg{} and \aimg, the triangles
generated by a (possibly extended) Delaunay triangulation are small
enough not to be affected by the distortions.  The crucial step is the
initial triangle matching, and due to the use of local triangles, it
proves to be robust procedure. It should be emphasized that \atrani\
can be any smooth transformation, for example an affine transformation
with small shear, or polynomial transformation of any reasonable order.
The optimal value of the order depends on the magnitude of the
distortion. The detailed description of fitting such models and
functions can be found in various textbooks
\citep[see e.g.~Chapter 15. in][]{press1992}. It is noteworthy that in
step 7 one can perform a weighted fit with possible iterative rejection
of n-$\sigma$ outlier points.

\phdsubsection{Implementation}
\label{subsec:astrometry:exam}

The coordinate matching and coordinate transforming algorithms are
implemented in two stand-alone binary programs as a part of the
complete data reduction package. The 
program named \prog{grmatch} (Sec.~\ref{sec:prog:grmatch})
matches point sets, including triangle
space generation, triangle matching, symmetric point matching and
polynomial fitting, that is steps 
1 through 4 in Sec.~\ref{subsec:astrometry:pract}. The
other program, \prog{grtrans} (Sec.~\ref{sec:prog:grtrans}), 
transforms coordinate lists using the
transformation coefficients that are output by \prog{grmatch}. The
\prog{grtrans} code is also capable of fitting a general polynomial
transformation between point-pair lists if they are paired or matched
manually or by an external software. We should note that in the 
case of degeneracy, e.g.~when all points are on a perfect lattice,
the match fails.

By combining \prog{grmatch} and \prog{grtrans}, one can easily derive
the World Coordinate System (WCS) information for a FITS data file. 
Output of WCS keywords is now fully implemented in \prog{grtrans},
following the conventions of the package 
\prog{WCSTools}\footnote{http://tdc-www.harvard.edu/wcstools/}
\citep[see][]{mink2002}.
Such information is very useful for manual analysis with well-known FITS
viewers \citep[e.g.~\prog{ds9}, see][]{joye2003}. For a more detailed
description of WCS see \citet{calabretta2002} and on the representation
of distortions see \citet{shupe2005}.

\begin{figure*}
\begin{center}
\resizebox{160mm}{!}{\includegraphics{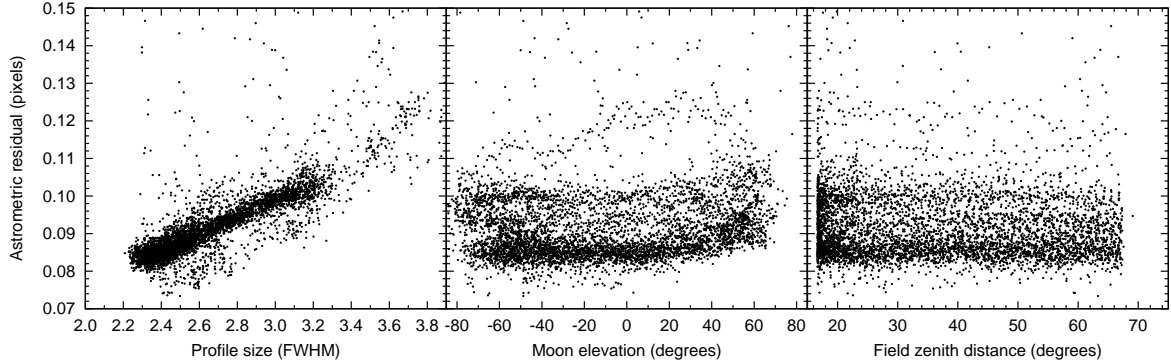}}
\end{center}
\caption{Astrometric residuals as the function of observational conditions.
The left panel shows the strong correlation between the stellar
profile sizes (${\rm FWHM}\approx2.35/\sqrt{S}$): the sharper
the stars are on the image the smaller the astrometric residual is.
The middle panel shows the astrometric residuals as the elevation of the Moon.
Obviously, if the Moon is below the horizon, the residuals are independent
from this ``negative elevation'', however, if the Moon is above
the horizon, the effect of the stronger background illumination 
can be seen well: as the elevation of the Moon increases, the residuals
do also become larger. The right panel shows the residuals
as the function of the field elevation. No correlation between these can
be seen.
\label{fig:astromres}}
\end{figure*}

\phdsection{Registering images}
\label{subsec:imageregistering}

In order to have data ready for image subtraction, the images
themselves have to be transformed to the same reference system (i.e.
the images have to be \emph{registered}). This
transformation is a continuous mapping between the reference coordinate
system and the system of each of the individual images. In practice, all of the
frames are taken by the same instrument so this transformation is 
always nearly identity, affected only by slight rotation, shift and 
small distortions (for instance due to differential refraction as a given
field of the sky is observed at different air masses or small dilations
may occur due to the change of focus). In principle, the 
whole registration process should comply with the following issues.
First, the relative transformations are expected to be as small as possible.
Since dilations are negligible, the combination of rotation and shift
can be described by an affine linear transformation whos determinant is $1$.
However, the distortions resulted by differential refraction
require higher order transformations to be described properly. 
Second, the flux transformation must preserve brightnesses of the sources. 
Namely, any area on
the image referenced by the same absolute (e.g. celestial) coordinates
must contain \emph{exactly} the same amount of flux before and after the
transformation is done. Third, composition of the geometric transformations
must be as ``commutative'' as possible with subsequent image transformations.
Namely, having two $\mathbb{R}^2\to\mathbb{R}^2$ mappings, e.g. $f$ and $g$,
and we denote the transformed version of image $I$ by $T_{f}[I]$, we 
want to keep $\|T_{f\circ g}[I]-T_{f}[T_{g}[I]]\|$ as small as possible. 
Here ``small'' means that the difference between the
images $T_{f\circ g}[I]$ and $T_{f}[T_{g}[I]]$ should be comparable 
with the overall noise level.
In this section the details of this image registration process is discussed. 

\phdsubsection{Choosing a reference image}

In practice, the reference image is chosen to be a ``nice'' image,
with high signal-to-noise ratio and therefore with
small astrometric residual. Since
the signal-to-noise ratio is affected by both the background noise and
the fluxes of the individual stars, images taken near culmination,
after astronomic twilight and when the Moon is below the horizon
are a proper choice in most of the cases. Moreover, in the case of HATNet,
sharper images tend to have smaller astrometric residuals because of
the merging of nearby stars is also smaller and the background noise
affects less pixels. In the panels of Fig.~\ref{fig:astromres},
the astrometric residuals are shown as the function of the
previously discussed observational conditions. As one can expect,
the effect of the image sharpness (characterized by the
stellar profile FWHMs) and the Moon
elevation definitely influence the astrometric residuals. However,
the effect of the field elevation itself is negligible, the 
variation in the airmass between $\sim1.02$ and $\sim2.55$ 
(i.e. $12^\circ\lesssim z\lesssim67^\circ$) causes
no practical fluctuation in the astrometric residuals.

We should note here that the whole process of the 
image subtraction photometry needs not only a specific image
to be an astrometric reference but a couple of images for
photometric reference as well. As we will see later on,
the selection criteria for convolution reference images are
roughly the same as for an astrometric reference. 
Hence, in practice, the astrometric reference
image is always one of the convolution reference frames.

\phdsubsection{Relative transformations}

Once the reference frame for registration has been chosen, the
appropriate geometric transformations between this frame and the 
other frames should be derived (prior to the image transformation itself).
To derive this geometric transformation, one can proceed using one of the 
following methods:
\begin{itemize}
\item Assuming the absolute astrometrical solutions to be known (i.e.
the mappings between the celestial and pixel coordinates), the
solution for the reference frame can be composed with the inverse
of the solution for the current image.
\item Assuming that the sources on both the reference and the current
images are extracted and identified with a previously declared
external catalog, one can match these identifier -- pixel coordinate lists
and fit a geometric transformation involving the matched 
coordinate pairs. 
\item If any kind of astrometric information -- neither absolute
solution nor source identification -- is not known in advance,
one can directly employ the triangulation-based point matching 
algorithm itself, as it was presented earlier (Sec.~\ref{subsec:astrometry}).
\end{itemize}
In practice, the first option is sub-optimal. Since the absolute 
astrometric transformation has higher order distortions than 
in a relative transformation, such composition
does easily lead to numeric round-off errors. Moreover, the direct
composition of two polynomials with an order of 6 (which is needed
for a proper astrometric solution, see Sec.~\ref{sec:difficulties:largefov},
Table~\ref{tab:astromres} or Sec.~\ref{subsec:astrometry}) 
yields
a polynomial with an order of 12, while a relative transformation
between two images needs only a polynomial with a degree of $3-4$ 
(see also Table~\ref{tab:astromres}). The 
naive omission of higher order polynomial coefficients does not result the
``best fit'' and this best fit depends on the domain of the polynomial
therefore this polynomial degradation is always an ambiguous step.

Both the second or third option mentioned above are efficient and
can be used in practice. The last option, involving the point matching
to determine the relative transformation has an advantage: 
on cloudy images where derivation of 
the absolute astrometric solution failed,
the chance to obtain a successful relative transformation is higher.
This is mostly because of both the lack of large-scale distortions and
the smaller polynomial degree required for such transformations.

\begin{figure}
\begin{center}
\noindent
\resizebox{80mm}{!}{\includegraphics{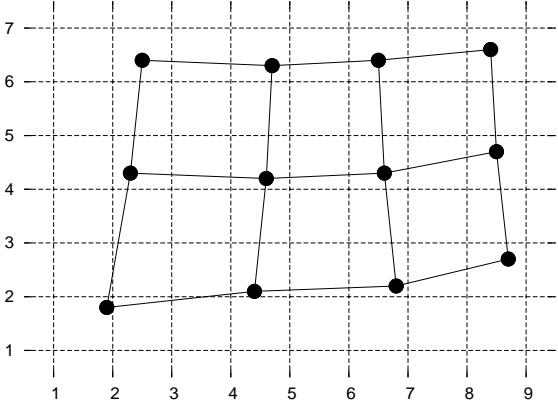}}
\end{center}
\caption{In order to perform a spatial image transformation with 
exact flux conserving, the intensity level of the original image
should be integrated on the quadrilaterals defined by the mapping function.
Each quadrilateral is the projection of one of the pixels in the target image
while the dots represent the projections of the pixel corners. 
The above image shows the pixel grid of the original image and 
the grid of quadrilaterals for a transformation that shrinks the
image by a factor of nearly two.}
\label{fig:transgrid}
\end{figure}

\phdsubsection{Conserving flux}
\label{sec:conservingflux}

Even if the spatial image transformation does not significantly shrink
or enlarge the image, pixels of the target image usually are not mapped exactly
to the pixels of the original image (and vice versa). Therefore,
some sort of interpolation is needed between the adjacent pixel values
in order to obtain an appropriate transformed image. Since the 
spatial transformation is followed by the steps of convolution and 
photometry, exact flux conservation is a crucial issue. If the 
interpolation is performed naively by multiplying the interpolated
pixel values with the Jacobian determinant of the spatial mapping,
the exact flux conservation property is not guaranteed at all. It 
is even more relevant in the cases where the transformation includes
definite dilation or shrinking, i.e. the Jacobian determinant 
significantly differ from unity. 

In order to overcome the problem of the flux conserving transformations,
we have implemented a method based on analytical integration of
surfaces of which are determined by the pixel values. These surfaces
are then integrated on the quadrilaterals whose coordinates
are derived by mapping the pixel coordinates on the target frame
to the system of the original frame. An example 
is shown in Fig.~\ref{fig:transgrid}, where the transformation includes
a shrink factor of nearly two (thus the Jacobian determinant is $\sim 1/4$). 
In practice, two kind of surfaces
are used in the original image. The simplest kind of surface is the 
two dimensional step function, defined explicitly by the discrete pixel values.
Obviously, if the area of the intersections of the quadrilaterals
and the pixel squares is derived, the integration is straightforward:
it is equivalent with a multiplication of this 
intersection area by the actual pixel value. 

A more sophisticated interpolation surface 
can be defined as follows. On each pixel, at the position $(i,j)$, 
we define a biquadratical function of the fractional pixel coordinates 
$(\delta x,\delta y)$, 
namely 
\begin{equation}
f^{ij}(x,y)=\sum\limits_{k=0}^2\sum\limits_{\ell=0}^2 C^{ij}_{k\ell} 
\delta x^k\delta y^\ell.
\end{equation}
For each pixel, we define nine coefficients, $C^{ij}_{k\ell}$. We derive
these coefficients by both constraining the integral of the surface
at the pixel to be equal to the pixel value itself, i.e.
\begin{equation}
\int\limits_{0}^1 \int\limits_{0}^{1} f^{ij}(\delta x,\delta y)
\, \mathrm{d}\delta x\,\mathrm{d}\delta y = P_{ij}, \label{eq:biquadintegral}
\end{equation}
and requiring the joint function $F(x,y)$ describing the surface 
\begin{equation}
F(x,y)=f^{[x][y]}(\{x\},\{y\})
\end{equation}
to be continuous (here $[x]$ denotes the integer part of 
$x$ and $\{x\}$ denotes the
fractional part, i.e. $x=[x]+\{x\}$). This continuity
is equivalent to 
\begin{eqnarray}
f^{[i+1]j}(0,y) & = & f^{ij}(1,y), \label{eq:biquadcont1} \\
f^{[i-1]j}(1,y) & = & f^{ij}(0,y), \label{eq:biquadcont2} \\
f^{i[j+1]}(x,0) & = & f^{ij}(x,1), \label{eq:biquadcont3} \\
f^{i[j-1]}(x,1) & = & f^{ij}(x,0), \label{eq:biquadcont4} 
\end{eqnarray}
for all $0\le x,y \le 1$. Since $f$ is a biquadratical function of 
the fractional pixel coordinates $(x,y)$, it can be shown that the 
above four equations imply $8$ additional constraints for each pixel.
At the boundaries of the image, we can define any feasible boundary condition.
For instance, by fixing the partial derivatives $\partial F/\partial x$
and $\partial F/\partial y$ of the surface $F(x,y)$ to be zero at the 
left/right and the lower/upper edge of the image, respectively. It can be
shown that the integral property of \eqref{eq:biquadintegral},
the continuity constrained by \eqrefs{eq:biquadcont1}{eq:biquadcont4} 
and the boundary conditions define an unique solution for the 
$C^{ij}_{k\ell}$ coefficients. This solution exists for arbitrary values of
the $P_{ij}$ pixel intensities (note that the complete problem of obtaining
the $C^{ij}_{k\ell}$ is a system of linear equations). 
Since the integrals of the $F(x,y)$ surface 
on the quadrilaterals are linear combinations of polynomial integrals,
the pixel intensities on interpolated images can be obtained easily, 
although it is a bit more computationally expensive.

We should note here that if the transformation is a simple shift
(i.e. there are not any dilation, rotation and higher order
distortions at all), the two, previously discussed interpolation 
schemes yield the same results as the classic bilinear and bicubic
\citep{press1992} interpolation. 

In practice, during the above interpolation procedure pixels
that have been marked to be inappropriate\footnote{For instance,
pixels that are saturated or have any other undesired mask.}
are ignored from the determination of the $C^{ij}_{k\ell}$ coefficients,
and any interpolated pixels on the target image 
inherit the underlying masks of the pixels that intersects 
their respective quadrilaterals. Pixels on the target frame that are
mapped off the original image have a special mask which
marks them ``outer'' ones (see also Sec.~\ref{subsec:masking}). It yields
a transparent processing of the images: for instance in the case
of photometry, if the aperture falls completely inside the image
but intersects one or more pixels having this ``outer'' mask yields
the same photometry quality flag as if the aperture is
(partially or completely) off the image. See also 
Sec.~\ref{subsec:photometry} or Sec.\ref{sec:prog:fiphot} 
for additional details.

\phdsubsection{Implementation}
The core algorithms of the interpolations discussed here
are implemented in the program \texttt{fitrans}
(Sec.~\ref{sec:prog:fitrans}). This
program performs the spatial image transformation, involving
both the naive and the integration-based methods and
both the bilinear and bicubic/biquadratical interpolations. 
The transformation itself is the output of the \texttt{grmatch}
or \texttt{grtrans} programs (see also Sec.~\ref{sec:prog:grmatch}
and Sec.~\ref{sec:prog:grtrans}).


\phdsection{Photometry}
\label{subsec:photometry}

The main step in a reduction pipeline intended to measure fluxes 
of objects on the sky is the photometry. All of the steps discussed
before are crucial to prepare the image to be ready for photometry.
Thus at this stage we should have a properly calibrated 
and registered\footnote{Only if we intend to perform image subtraction
based photometry.} image as well as we have to know the positions of the sources of interest.
For each source, the CCD photometry process for a single image yields
only \emph{raw instrumental} fluxes. 
In order to estimate the \emph{intrinsic} flux of a target object, ground-based
observations use nearby comparison objects with known fluxes.
The difference in the raw instrumental fluxes between the target source 
and the source with known flux is then converted involving smooth
transformations to obtain the ratios between the intrinsic flux values.
Such smooth transformation might be the identical transformation 
(this is the simplest of all photometry methods, known as 
single star comparison photometry) or
some higher order transformations for correcting various gradients
(mostly in the transparency: due to the large field of view, the 
airmass and therefore the extinction at the different corners
of the image might significantly differ). Even more
sophisticated transformations can also be performed in order to correct
additional filter- and instrumentation effects yielded by the
intrinsic color (and color differences) between the various sources.
Corrections can also made in order to transform the brightnesses
into standard photometric systems. 
The latter is known as standard transformation and almost in all cases
it requires measurements for standard areas as well \citep{landolt1992}.
Since for all objects, transparency variations cause 
flux increase or decrease proportional
to the intrinsic flux itself, the transformations mentioned above are done on
a logarithmic scale (in practice, magnitude scale). For instance, 
in the case of single-star comparison photometry, the 
difference between the intrinsic magnitudes and the 
raw instrumental magnitudes is constant\footnote{To be precise,
only if the spectra of the two stars are exactly the same \emph{and}
the two objects are close enough to neglect the difference in 
the atmospheric transparency.}.
In this section some aspects of the raw and instrumental photometric methods
are detailed with the exception of topics related to the photometry 
on convolved and/or subtracted images. 
As it was mentioned above, the first step of the photometry is the 
derivation of the raw instrumental magnitudes of the objects or sources
of our interest. 

\phdsubsection{Raw instrumental magnitudes}

In principle, raw magnitudes are derived from two quantities.
First, the total flux of the CCD pixels are determined around the object
centroid. The total flux can be determined in three manners:
\begin{itemize}
\item If a region is assigned to the object of interest, one 
has to count the total flux of the pixels inside this region.
The region is generally defined to be within a fixed distance from
the centroid (so-called \emph{aperture}), but in the case of diffuse or
non-point sources, more sophisticated methods have to be used to
define the boundary of the region. The algorithms implemented in the
program SExtractor \citep{bertin1996} focus on photometry of
such sources. In the following we are interested only in 
stars and/or point-like sources. 
\item If the source profile can be modelled with some kind of analytic
function (see Sec~\ref{sec:stardetection:coordshapemodel}) or 
an empirical model function 
(e.g. the PSF of the image), one can fit such a model surface to the pixels
that are supposed to belong to the object (e.g. to the pixels being inside 
of a previously defined aperture or one of the isophotes). From the 
fitted parameters, the integral of the surface is derived, and 
this integral is then treated as the flux of the object. This method
for photometry is known as PSF photometry. 
\item The previous two methods can be combined as follows. After fitting
the model function, the best fit surface is subtracted from the pixel
values and aperture photometry is performed on this residual. The flux
derived from the residual photometry is then added to the flux derived
from the best fit surface parameters yielding the total flux for the
given object. It is not necessary that the pixels used for surface fitting
are the same as the pixels being inside the aperture.
\end{itemize}
It should be mentioned here that whatever primary method from these
above is used to perform the photometry, estimating the uncertainties
should be done carefully.  

After the total flux of the object has been estimated, one has to 
remove the flux contribution of the background. It is essential in the
case of aperture photometry, however, if a profile function is 
fitted to the pixel values, the contribution of the background
is added to the model function as an additional free parameter. 
If the photometric aperture is a circular region, the background is
usually defined as a concentric annulus, whose inner radius is 
larger than the radius of the aperture. If the field is not 
crowded, the background level is simply the mean or median 
of the pixel values found in the annulus. On the other hand, 
if the field is extremely crowded, the determination of the 
background level might even be impossible. A solution for this 
issue can be either profile (PSF) fitting or photometry based
on differential images (see Sec.~\ref{subsec:subtractedphotometry}).
Note that on highly crowded fields, apertures significantly overlap.
One advantage of the profile/PSF model fitting method is the ability
to fit adjacent profiles simultaneously.

In practice, additional data are obtained and reported for a single 
raw instrumental photometry measurement, such as:
\begin{itemize}
\item Noise estimations, based on the Poisson statistics of the 
flux values, the uncertainty of the background level determination,
and optionally scintillation noise can also be estimated
\citep{young1967};
\item Characteristics of the background: total number of pixels
used to derive the background level, number of outlier pixels -- 
such as pixels of nearby stars or cosmics events -- rejected from the
background determination procedure and so.
\item Quality flags, such as various pixel masks happen to fall in the 
aperture.
\end{itemize}

\begin{figure}
\begin{center}
\noindent
\resizebox{80mm}{!}{\includegraphics{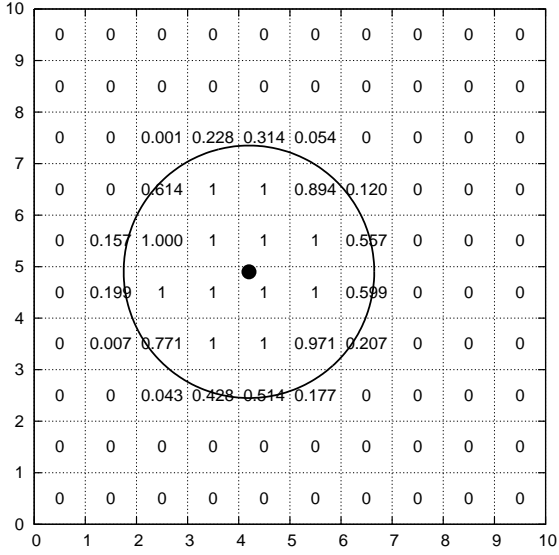}}
\end{center}
\caption{Weight matrix for a circular aperture centered at 
$(x_0,y_0)=(4.2,4.9)$ and having a radius of $r_0=2.45$ pixels. The numbers
written in the squares show the area of the intersection of the given
square and the circle.}
\label{fig:aperture}
\end{figure}

\phdsubsection{Formalism for the aperture photometry}
\label{sec:apertureformalism}

In practice, aperture photometry derives the raw instrumental magnitudes 
as follows.
Let us consider an image $I$ with the pixel intensities $I(x,y)\equiv I_{xy}$
where $(x,y)$ are the respective pixel coordinates. Let us define the
weight matrix for the circular aperture centered at $(x_0,y_0)$ and having
a radius of $r_0$ as 
\begin{eqnarray}
A_{xy} & \equiv & A^{x_0,y_0,r_0}_{xy} = \\
  & = & \int\limits_{x_0}^{x_0+1}\mathrm{d}x
	\int\limits_{y_0}^{y_0+1}\mathrm{d}y \,\Theta \left[ r_0^2 - (x-x_0)^2-(y-y_0)^2 \right], \nonumber
\end{eqnarray}
where $\Theta(\cdot)$ is the Heaviside step function 
(see also Fig.~\ref{fig:aperture}). Due to the definition
of $A_{xy}$, it is unity inside the aperture, has some value 
between $0$ and $1$ at the boundary (depending on
the area of overlap), and it is zero further outside
from the aperture centroid. The total raw instrumental 
flux $f_{\rm total}$ is then simply derived as
\begin{equation}
f_{\rm total}=\sum\limits_{x,y} A_{xy}I_{xy}. \label{eq:apphot}
\end{equation}
The background level in the annulus having inner and outer radii of $r_1$
and $r_2$ respectively, around the centroid $(x_0,y_0)$ can be derived as
\begin{equation}
B = \frac{\sum\limits_{x,y} I_{xy}\left(A^{x_0,y_0,r_2}_{xy}-A^{x_0,y_0,r_1}_{xy}\right)}{r_2^2-r_1^2}.
\end{equation}
The raw flux of the object in the aperture after the background level
removal is 
\begin{equation}
f=\sum\limits_{x,y} A_{xy}(I_{xy}-B) = f_{\rm total} - Br_0^2.
\end{equation}
Albeit this discussion seems to be rather trivial, the same formalism
will be used later on in Sec.~\ref{subsec:subtractedphotometry} 
while considering the details of photometry performed on subtracted images.

\phdsubsection{Magnitude transformations}

As it was mentioned earlier,
raw magnitude lists on subsequent frames yielded by the photometry 
have to be transformed to the same reference system in order to have
instrumental and/or standard magnitudes for our objects. For a given frame, 
let us assume to have a list of stars with $m^{(i)}$ raw
magnitudes, located at the $(x_i,y_i)$ position on the image. Let us
denote the raw magnitudes of these objects on a certain reference 
frame by $m_0^{(i)}$.
For images obtained by small field-of-view instrumentation, the 
$m^{(i)}-m_0^{(i)}$ difference depends only on the color of the 
star, due to the wavelength dependence of the atmospheric extinction. For
images obtained by larger field-of-view optics, the difference between
the instrumental magnitudes depend also on the $(x,y)$ centroid positions
due to the gradient in the extinction level throughout the image. 
In practice, both the spatial and color dependence of the differential
magnitudes can be well characterized by polynomials. Such a transformation
is quantified as 
\begin{equation}
m^{(i)}-m_0^{(i)} = \sum\limits_{c=0}^{N} 
\left( (C^{(i)})^c \sum\limits_{0\le k+\ell \le N_c} K_{ck\ell} x_i^{k}y_i^{\ell}\right),
\end{equation}
where $C^{(i)}$ is some color index (e.g. $V-I$ or 
$J-K$) of the star, and $N$ and $N_c$ are the maximal 
polynomial orders in the color 
and in the spatial coordinates, respectively. The $K_{ck\ell}$ coefficients
can be obtained by involving the linear least squares method, if each 
of the stars are weighted appropriately. The weights assigned to the stars
can be derived from both the photon noise and the light curve residuals.
In practice, the above mentioned magnitude transformation is done
iteratively. First, instrumental magnitude lists for each frame are transformed
to the instrumental system of one of the frames. This reference frame
is usually selected from the ``best'' frames, i.e. that has been obtained 
at low airmass and good generic atmospheric conditions, 
has small astrometric residuals and the illumination
of the Moon and/or sky background (due to twilight) is the smallest. 
After each magnitude list have been transformed, light curves are 
gathered and the individual scatters are derived for each star. 
The transformation is then repeated while the contribution of each
star is weighted by the light curve scatters. This kind of weighting
gives lower weight for stars whose scatter have been underestimated
(due to unresolved remaining systematics, for instance) or have
intrinsic but not known variability. Of course, stars with known variability
should be excluded from the fit, including our target stars as well.

\phdsubsection{Implementation}

In the \fihat{} package, the above discussed photometric algorithms
are implemented as follows. The aperture photometry and the 
related features -- such as background level determination, noise estimations,
assignment of quality flags, conversion of fluxes to instrumental
magnitudes -- are implemented in the program \texttt{fiphot} (see
Sec.~\ref{sec:prog:fiphot}). The point-spread functions are derived
by the program \texttt{fistar} (Sec.~\ref{sec:prog:fistar}). Moreover,
this program is also capable to fit the derived PSFs to the individual
detected profiles. Currently, none of these programs deals explicitly with profile 
fit residuals, however, the output of \texttt{fistar} can be used
as an input for \texttt{firandom} (both for analytical profile models
and PSFs) to create model images. Such model images are suitable
to subtract from the original images yielding complete residual images.
The program \texttt{lfit} is another alternative for fitting analytic
stellar profile models that are not supported by 
\texttt{fistar}/\texttt{firandom}. Magnitude transformations 
between two frames can also be performed with the program \texttt{lfit}.
See also Sec.~\ref{subsec:photometricanalysis} and Fig.~\ref{fig:apphotexample}
about the practical details about how these programs can be applied
for real observations.


\phdsection{Image convolution and subtraction}
\label{subsec:convolution}

In a generic variability survey, such as the HATNet project, 
we are primarily focusing on the detection and the quantifications
of source brightness variations. The idea behind the photometry methods 
involving image subtraction is to derive the part of the flux that
varies from image to image.
It is rather easy to see that simple per-pixel
arithmetic subtraction is not 
sufficient to derive the difference between two images. First, the 
centroid positions of the stars are different for each image. The magnitude
of this difference depends on the precision and the systematic
variations in the mount tracking, as well as other side effects such
as field rotation and the intrinsic differential refraction. 
However, it is rather easy to overcome this problem by registering the 
images to the same reference system 
(Sec.~\ref{subsec:imageregistering}). Second, background level may
vary from image to image. Changes in the background can be
modelled by adding a constant or some slowly varying function 
to the (convolved) image.
Third, the stellar profiles
are also vary from frame to frame, due to the variations in the 
seeing or in the focus. In order to have the smallest residual between
two images, one should not only register these to the same reference
system but on at least one of the images, the profiles should be 
transformed to match the profiles of the other image. This profile
transformation is performed as a convolution, namely
the image $R$ is transformed to $R'$ as 
\begin{equation}
R'=B+R\star K,
\end{equation}
where $K$ is the convolution kernel and the operator $(\cdot)\star(\cdot)$ 
denotes the convolution. For (astronomical) images that are
sampled on discrete pixels, the operation of convolution is defined as
\begin{equation}
R'_{xy}=\sum\limits_{-B_{\rm K} \le i,j \le +B_{\rm K}} R_{(x-i)(y-j)} K_{ij}.
\end{equation}
Here, the convolution kernel $K_{ik}$ is sampled on a grid of
$(2B_{\rm K}+1)\times(2B_{\rm K}+1)$ pixels and $I_{xy}$ refers to
the intensity of the pixel at $(x,y)$. If the difference of 
FWHMs of the image $R$ and $R'$ are small, the kernel can be sampled on
a smaller grid. In general, a kernel function with an FWHM of $F_{\rm K}$
yields a profile FWHM $F'$ on the convolved image of
\begin{equation}
F'\approx\sqrt{F^2+F_{\rm K}^2},
\end{equation}
where $F$ is the FWHM of the profiles on the image $R$.

Supposing two images, $I$ and $R$, the main problem of the
image convolution and subtraction method is to find the appropriate kernel $K$
with which the image $R$ convolved, the resulting image
is nearly identical to $I$. The first attempt to find this
optimal kernel \citep{tomaney1996} was based on an inverse Fourier
transformation between the two PSFs of the images. Theoretically,
inverse Fourier transformation yields the appropriate kernel,
however, the practical usage of this method is limited due to
the high signal-to-noise ratio that is needed by a Fourier inversion.
\cite{kochanski1996} attempted to find the kernel $K$ by minimizing
the merit function
\begin{equation}
\chi^2_{\infty}=\sum\limits_{xy}\left|I_{xy}-(R\star K)_{xy}\right|.
\end{equation}
This minimization yields a non-linear equation for the kernel $K$ and
therefore it is not computationally efficient. The most cited algorithm
related to image subtraction was given by \cite{alard1998}. In this
work, an additional term was added to the convolution transformation,
which allows to fit not only the convolution transformation but
the background variations:
\begin{equation}
I=B+R\star K.		\label{eq:convolutionbasic}
\end{equation}
The basic idea of \cite{alard1998} was to minimize the function 
\begin{equation}
\chi^2=\sum\limits_{xy}\left(I_{xy}-[B_{xy}+(R\star K)_{xy}]\right)^2 \label{eq:lqkernelfit}
\end{equation}
and search the kernel solution $K$ in the form of
\begin{equation}
K=\sum\limits_i C_iK^{(i)}.	\label{eq:kernelcoeffs}
\end{equation}
In their work, the kernels $K^{i}$ were two dimensional Gaussian functions
with variable FWHMs multiplied by polynomials. Assuming the 
background variations to be constant, i.e. $B_{xy}\equiv B$, 
minimizing \eqref{eq:lqkernelfit} yields a linear set of equations
for the parameters $B$ and $C_i$, thus its solution is straightforward
(and efficient). Shortly after, \cite{alard2000} gave a more
sophisticated method that allows the kernel parameters as well as
the background level to vary across the image:
\begin{equation}
I_{xy}=B(x,y)+[R\star K(x,y)]_{xy}.
\end{equation}
Both the background variations and the kernel coefficients were
searched as a polynomial function of the pixel coordinates, namely
\begin{equation}
B(x,y)=\sum\limits_{0\le k+\ell \le N_{\rm bg}}B_{k\ell}x^ky^\ell	\label{eq:backgroundcoeffs}
\end{equation}
and
\begin{equation}
K(x,y)=\sum\limits_i \sum\limits_{0\le k+\ell \le N^{(i)}_{\rm K}}C_{ik\ell}K^{(i)}x^ky^\ell.
\end{equation}
It is easy to show that finding the optimal $B_{k\ell}$ and $C_{ik\ell}$ 
coefficients still requires only linear least squares minimization.
\cite{alard1998} also discuss how the individual pixels used in the
fit must be weighted by the Poisson noise level in order to have
a consistent result. Recently, \cite{bramich2008} searched the 
optimal kernel $K$ by assuming an alternate set of kernel base 
functions $K^{(i)}$, involving discrete kernels
instead of Gaussian functions. These discrete kernels are defined as
\begin{equation}
K^{(u,v)}=\delta^{(uv)},
\end{equation}
where
\begin{equation}
(\delta^{(uv)})_{xy}=\left\{
        \begin{tabular}{ll}
                 $1$ & \hspace*{2mm} if $u=x$ and $v=y$, \\
                 $0$ & \hspace*{2mm} otherwise.
        \end{tabular}\right.
\end{equation}
The total number of base kernels is then $N_{\rm kernels}=(2B_{\rm K}+1)^2$.
\cite{yuan2008} attempted to find the 
solution $K_i$, $B$ and $K_r$ of the equation 
\begin{equation}
I\star K_i = B+R\star K_r.	\label{eq:crossconvolutionbasic}
\end{equation}
This method is known as \emph{cross-convolution} and works properly in the 
cases when there is no suitable solution for \eqref{eq:convolutionbasic}.
For instance, on the image $R$ the profiles have such shape parameters
where $K>0$ and $D=0$ while on the image $I$ these parameters 
are $K<0$ and $D=0$. The method of cross-convolution has a disadvantage,
namely if one finds a solution $K_i$ and $K_r$ for 
\eqref{eq:convolutionbasic}, $K_i\star G$ and $K_r\star G$ is also a 
solution (where $G$ is an arbitrary convolution kernel). Therefore
\eqref{eq:convolutionbasic} is degenerated unless additional constraints
are introduced (e.g. by minimizing the $\|K_i-K_r\|$ difference 
simultaneously).

\phdsubsection{Reference frame}

The noise characteristics of the subtracted image is determined by 
both the reference image $R$ and the target image $I$. If both images
are individual frames, the generic noise level is approximately
$\sqrt{2}$ times larger than that of on the individual frames. In order
to reduce the noise level on the subtracted frames, the reference 
image $R$ is created from several individual frames. If the number of
such frames is $N$, the noise level of the subtracted images is  
$\sqrt{1+1/N}\approx{1+1/(2N)}$ (supposing that both the reference
frames and the target image have the same noise level). Thus, 
a number of $N\approx 20-25$ frames are sufficient to increase the
noise level on the subtracted image only by a few 
percent\footnote{Strictly speaking, a noisy reference frame implies
a \emph{correlated} noise on the subtracted frames since the 
same image (or its versions derived by convolution) is subtracted
from the original frames. Therefore, it is an upper limit
for the noise increment in the final light curves. However, the 
scatter in the convolution parameters also increase the light
curve noise, but this cannot be quantified in a simple way.}.

\phdsubsection{Registration}

As it was seen related to the difficulties of the photometry on 
undersampled images (Sec.~\ref{sec:undersampledproblems}), the 
interpolation of such images with sharp profiles is likely to yield artifacts,
``spline undershoots'' and therefore systematic residuals
(Fig.~\ref{fig:samplespline}). Since the
FWHM of the HATNet frames is too small to clearly remove such residuals,
we have used the following sophisticated registration process. First,
using the stellar profile parameters and flux estimations yielded 
by the modelling described in Sec.~\ref{sec:stardetection:coordshapemodel},
a model for the images is created, involving the program \texttt{firandom}
(Sec.~\ref{sec:prog:firandom}). This image model is then subtracted
from the original image, yielding a residual with no sharp structures.
The residual image is then transformed to the reference system, simultaneously
with the transformation of the centroid coordinates found
in the stellar profile parameter list. Using the transformed 
stellar profile parameters, another model image is created that is
added to the transformed residual image. Since the stellar profiles
can be well modelled by an analytic function, this way of image registration
yields no artifacts on the transformed images, even for highly 
undersampled profiles. Additionally, we do not have to involve \emph{all}
of the stars on the image, only the brighter ones, since for
fainter stars the amplitude of spline undershoots are comparable to 
or less than the noise level.

This kind of transformation is even more relevant during the creation
of the reference image $R$ since this image is created by
averaging some of the most sharpest images.

\phdsubsection{Implementation}

Those methods discussed above that are based on the technique of 
linear least squares are implemented in the program \texttt{ficonv} 
(see Sec.~\ref{sec:prog:ficonv}). The practical details of the
photometry based on the method of image subtraction are
explained in Chapter~\ref{chapter:hatnetdiscovery},
related to the HAT-P-7(b) planetary system.


\phdsection{Photometry on subtracted images}
\label{subsec:subtractedphotometry}

As it was discussed in the previous section (Sec.~\ref{subsec:convolution}),
the method of image convolution and subtraction aids the photometry
process by both decreasing the fluctuations in the background level
and reducing the influence of the nearby stars on the background area level. 
A great advantage of the image subtraction method is that it does not
need to know about stars (initially), can use all of the pixels 
and works in extremely crowded images. 
In the simplest case when both
the reference image $R$ and the target image $I$ have exactly the
same intensity level and the stellar profiles are nearly the same, 
the flux of a given
star on image $I$ can be obtained by simply adding the reference flux
and the flux measured on the residual image. 

However, even in the
cases where the stellar profiles are nearly the same but the images $R$
and $I$ have different intensity levels (for example, image $I$ was 
acquired at higher airmass or lower transparency while the reference
was chosen to be one of the high signal-to-noise images, acquired
at high horizontal altitudes), the photometry on the subtracted images
is not as simple as before. Let us consider the following situation.
The flux of a given isolated star on the reference image is found to
be 1000\,ADUs. In the target image, this star has an intrinsic flux decrease
of 1\%, thus if this image had been acquired under the 
same conditions as the reference image, the flux of the star would be 
990\,ADUs. Let us suppose now that due to the low sky transparency,
all of the stars have a flux decrease of 50\%, thus our star is
measured to have a flux of 495\,ADUs. The best fit kernel solution that
transforms the reference image to the target image is then 
$K=\frac12\delta$. Therefore, the residual flux of the target star 
would be -5\,ADUs. If this residual flux is simply added to the 
reference flux, the obtained flux is only 995\,ADUs, thus
the measured flux decrease (the signal itself) is significantly
underestimated. Moreover, if the kernel solution of \eqref{eq:convolutionbasic}
implies significant difference between the FWHMs of the stellar
profiles in the reference and target image, both the methods of
PSF and aperture photometry should be tweaked.

Using the formalism shown in Sec.~\ref{sec:apertureformalism},
aperture photometry on subtracted images can be performed as follows.
It is easy to show that for any weight matrix of $A_{xy}$, the 
relation
\begin{equation}
\sum\limits_{x,y} (R\star K)_{xy} (A\star K)_{xy} =
\sum\limits_{x,y} (R_{xy} A_{xy}) \|K\|_1^2 \label{eq:ismphotbase}
\end{equation}
is true if the aperture $A$ supports the convolved profile of $R\star K$
and it is a rather good approximation if the aperture has a size 
that is comparable to the profile FWHM. The norm $\|K\|_p$ is defined as 
\begin{equation}
\|K\|_p:=\sqrt[p]{\sum\limits_{x,y}|K_{xy}|^p}.
\end{equation}
Moreover, the ratio of the two sides in \eqref{eq:ismphotbase} is
independent from $\|K\|_1$ (even if the aperture $A$ does not support
completely the convolved profile on $R\star K$) or in other words, 
this ratio does not
change if $K$ is multiplied by an arbitrary positive constant.
Therefore, involving an aperture of $A_{xy}$, the flux 
of a source found on the convolved image $C=R\star K$ can be obtained as
\begin{equation}
f_{\rm C}=\frac{\sum\limits_{x,y}C_{xy} (A\star K)_{xy}}{\|K\|_1^2} \label{eq:convflux}
\end{equation}
and this raw flux is independent from the large
scale flux level variations that are quantified by $\|K\|$. 
The total flux $f$ of the source can be derived from the flux on the 
reference image and the flux of the target image. Since 
the method of image subtraction tries to find the optimal kernel $K$, that
minimizes $\|I-B-R\star K\|_2$, combining \eqref{eq:convflux}
and \eqref{eq:apphot} from Sec.~\ref{sec:apertureformalism}, $f$
is obtained as
\begin{equation}
f=\frac{\sum\limits_{x,y}S_{xy} (A\star K)_{xy}}{\|K\|_1^2} + \sum\limits_{x,y}R_{xy} A_{xy}. \label{eq:subflux}
\end{equation}
Here $S$ is $I-B-R\star K$, the subtracted image.
Of course, one can derive a background level around the target object
on the subtracted images, but in most of the cases this 
background level is zero within reasonable uncertainties. However,
it is worth to include such a background correction even on the 
subtracted images since unpredictable small-scale background 
variations\footnote{For instance, variations yielded by thin clouds or 
scattered light, that cannot be characterized by a function 
like in \eqref{eq:backgroundcoeffs}} can occur at any time.


\begin{figure*}
\begin{center}
\resizebox{160mm}{!}{\includegraphics{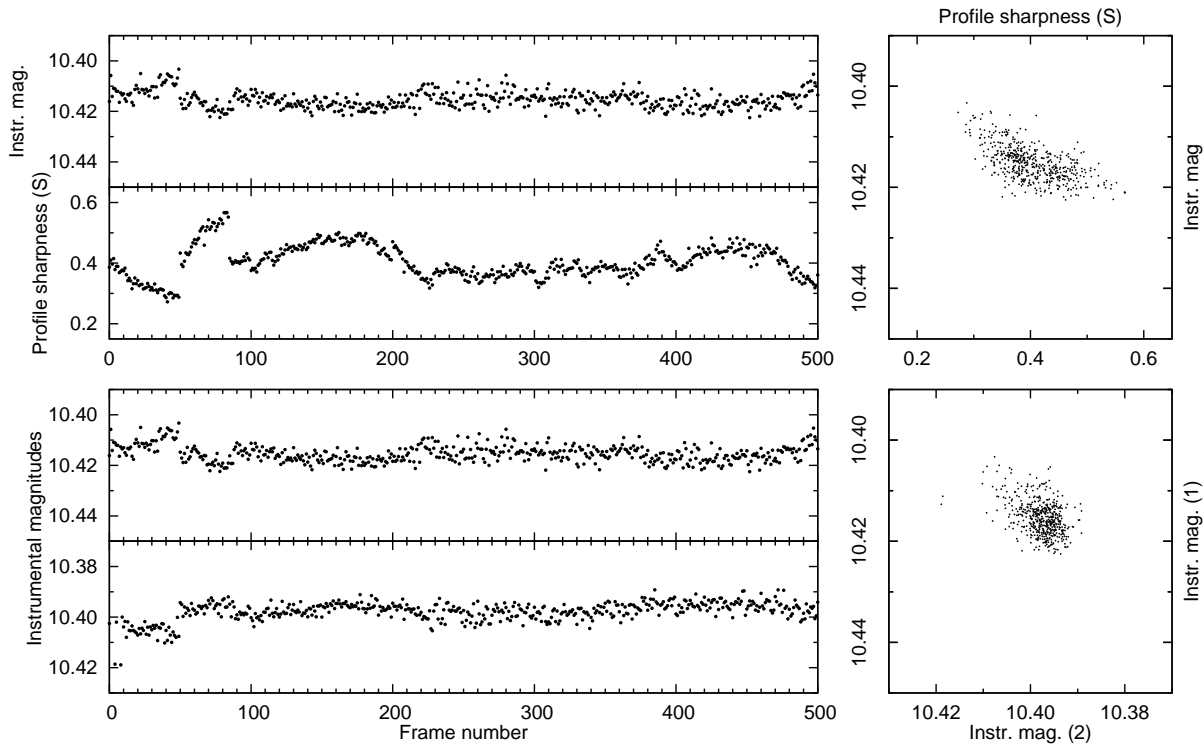}}
\end{center}
\caption{Typical examples of trends. The upper panels display the 
primary concepts of the External Parameter Decorrelation: for a
particular star, the lower inset shows the variance in the 
profile sharpness parameter ($S$) throughout the night while
the upper inset shows the instrumental magnitude. The panel in
the upper-right corner shows the distribution of the individual
measurements in the $S - {\rm magnitude}$ parameter space. The 
correlation between these two parameters can be seen clearly.
The lower panels display light curves for two given stars 
in the same instrumental photometric system. The insets on the
left show the two light curves while the plot in the lower-right
corner shows the ${\rm magnitude} - {\rm magnitude}$ distribution.
The correlation between the two magnitudes is quite clear also in this case.
\label{fig:trends}}
\end{figure*}

\phdsection{Trend filtering}
\label{subsec:trendfiltering}

Photometric time series might show systematic variations due to
various effects. Of course, if a certain star is indeed a variable,
the main source of photometric variations should be the intrinsic changes
in the stellar brightness. However, there are various other effects
that yield unexpected trends in the light curves, which still present
after the magnitude transformation and even if sophisticated
algorithms are involved in the data reduction (such as image subtraction
based photometry). The primary 
reasons for such trends are the following. Observational conditions
might vary (even significantly) throughout the night, for instance
clouds are blocking the light at some regions of the field, or the
background level is increasing due to the twilight or the proximity 
of the Moon. Additionally, instrumental effects, such as variations
in the focal length or drops or increases in the detector temperature 
can result in various trends. And finally, lack of the proper data
reduction is also responsible for such effects. For instance,
faults in the calibration process, insufficiently large polynomial 
orders in the astrometric or magnitude transformations, 
underestimated or overestimated aperture sizes, badly determined PSFs, 
inappropriate reference frames; all of these are plausible reasons
for unexpected systematic variations. In this section the 
efforts are summarized intended to reduce the remaining trends
in light curves.

The basic concepts of trend removal are the following. First, one
can assume that instrumental magnitudes have some remaining dependence on
additional quantities that are also derived during the data reduction.
Such external parameters can be the profile shape parameters, 
centroid coordinates, celestial positions (such as elevation 
or hour angle of the target field or
object), or environmental parameters (external temperature). The dependence
on these parameters therefore results in a definite correlation. 
Assuming some qualitative dependence, these correlations can then
be removed, yielding light curves with smaller scatter. The type
of the qualitative dependence is related to certain parameters
against which the de-correlation is performed (see later on some
examples). In general, this method of the External Parameter Decorrelation
\citep[EPD; see e.g.][]{bakos2007hatp2} yields a linear least squares fit. 
Second, either if we have no information about all of the external
parameters or there are other sources for the trends that cannot be
quantified by any specific external parameters (for instance,
there are thin clouds moving across the subsequent images), one
can involve the method of Trend Filtering Algorithm 
\citep[TFA;][]{kovacs2005}.
This algorithm is based on the experience that there are stars
with no intrinsic variability showing the same features in their light
curves. TFA removes these trends by using a set of \emph{template}
stars (preferably none of them are variables) and searching for
coefficients that can be used to perform a linear combination between
the template light curves and then this best fit linear combination is 
subtracted from the original signal. 
Fig.~\ref{fig:trends} displays
these two primary sources of the trends, in the case of some 
non-variable stars\footnote{These stars are suspected not to be variables above
the noise limits of the measurements. The data displayed here 
originate from the first follow-up transit measurements of the HAT-P-7(b) 
planetary system on 2007 November 2. See Chapter~\ref{chapter:hatnetdiscovery}
for further details about the related data reductions.}. In the cases
when analysis is performed on a photometric data set which does have
only time series information about the magnitudes, the method of EPD
cannot be applied while TFA still can be very effective
\citep[for a recent application, see e.g.][]{szulagyi2009}.

Of course, there are several other methods found
in the literature that are intended to remove or at least, decrease
the amplitude of unexpected systematic variations in the light curves.
The concept of the SysRem method \citep{tamuz2005} can be summarized
shortly as an algorithm that searches decorrelation coefficients 
similar to the ones used in the EPD simultaneously to all of the 
light curves then repeats this procedure by assuming the external parameters
themselves to be unknowns. This method of SysRem has been improved
by \cite{cameron2006} in order to have a more robust and reliable generic
transit search algorithm. The ad-hoc template selection of the TFA
has been replaced by a hierarchical clustering algorithm by 
\cite{kim2008}, assuming that stars showing similar trends are
somehow localized. In the following, we are focusing on the EPD 
and TFA algorithms, since in the HATNet data reductions these
algorithms play a key role. 

\phdsubsection{Basic equations for the EPD and TFA}

Let us assume having a photometric time series for a particular
star and denote the instrumental magnitudes by $m_i$ ($i=1,\dots,N$ where
$N$ is the total number of data points). The external parameters
involved in the decorrelation are denoted by $p^{(k)}_i$ ($k=1,\dots,P$, where
$P$ is the number of the independent external parameters)
while the magnitudes template stars are $m^{(t)}_i+\bar{m}^{(t)}$ 
($t=1,\dots,T$, where
$T$ is the total number of template stars and $\bar{m}^{(t)}$ is
the mean magnitude for the template star $t$). The method of EPD then 
minimizes the merit function
\begin{equation}
\chi^2_{\rm EPD} = \sum\limits_i w_i\left( m_i - m_0 - \sum\limits_k E_k p^{(k)}_i \right)^2, \label{eq:epdbase}
\end{equation}
where $E_k$'s are the appropriate EPD coefficients, $m_0$ is the 
mean brightness of the star and the weight of the given photometric point
$i$ is $w_i$, usually $w_i=\sigma_i^{-2}$ ($\sigma_i$ is the 
individual photometric uncertainty for the measurement $i$). 
One of the most frequently used $\mathbf{p}_i$ parameter vector used in the
EPD of HATNet light curves is $\mathbf{p}_i=\{ x_i-\bar{x},y_i-\bar{y},
S_i,D_i,K_i,1/\cos(z_i),\tau_i \}$, where $x_i$ and $y_i$ are the centroid coordinates
on the original frames, $S_i$, $D_i$ and $K_i$ are the stellar profile
shape parameters defined in \eqref{eq:shapegausselong}, $z_i$ is the
zenith distance (thus, $1/\cos(z_i)$ is the airmass) and $\tau_i$ is the hour angle.
The $\bar{q}$ refers to the average of the quantity $q$. Although the 
EPD method yields a linear equation for the coefficients $E_{k}$, omitting
the subtraction of the average centroid coordinates might significantly
offset the value of $m_0$ from the real mean magnitude. Due to the
linearity of the problem, this is not relevant unless one wants to 
rely on the value of $m_0$ in some sense\footnote{For instance,
light curves from the same source might have different average magnitudes
in the case of multi-station observations. The average magnitudes are
then shifted to the same level prior to the joint analysis of this
photometric data. Either $m_0$ or the median value of the light curve
magnitudes can be used as an average value.}
The function that is minimized by TFA is 
\begin{equation}
\chi^2_{\rm TFA} = \sum\limits_i w_i\left( m_i - m_0 - \sum\limits_t F_t m^{(t)}_i \right)^2, \label{eq:tfabase}
\end{equation}
where the appropriate coefficient for the template star $t$ is $F_t$. 
The similarities between \eqref{eq:epdbase} and \eqref{eq:tfabase} are
obvious. Indeed, one can perform the two algorithms simultaneously,
by minimizing the joint function of 
\begin{equation}
\chi^2_{\rm E+T} = \sum\limits_i w_i\left( m_i - m_0 - \sum\limits_k E_k p^{(k)}_i - \sum\limits_t F_t m^{(t)}_i\right)^2. \label{eq:jointepdtfa}
\end{equation}
The de-trended light curve is then 
\begin{eqnarray}
m^{\rm (EPD)}_i & = & m_i - \sum\limits_k E_k p^{(k)}_i, \\
m^{\rm (TFA)}_i & = & m_i - \sum\limits_t F_t m^{(t)}_i \hspace*{3ex}{\rm or}\\
m^{\rm (EPD+TFA)}_i & = & m_i - \sum\limits_k E_k p^{(k)}_i- \sum\limits_t F_t m^{(t)}_i,
\end{eqnarray}
for EPD, TFA and the joint trend filtering, respectively.

\phdsubsection{Reconstructive and simultaneous trend removals}

Of course, we are not really interested in the de-trending of 
non-variable stars. Unless one wants to quantify the generic quality
of a certain photometric pipeline, the importance of any trend 
removal algorithm are relevant only in the cases where the
stars \emph{have} intrinsic brightness variations. In the following,
we suppose that the physical variations can be quantified by a small
set of parameters $\{A_r\}$, namely the fiducial signal of a particular 
star can be written as
\begin{equation}
m_i^{0} = m_0 + F(t_i,A_1,\dots,A_{\rm R})
\end{equation}
where $F$ is some sort of model function. 

In principle, one can manage variable stars by four considerations.
First, even stars with physical brightness variations are treated
as non-variable stars. This naive method is likely to distort
the signal shape by treating the intrinsic changes in the brightness
to be unexpected. In the cases where the periodicity of these intrinsic
variations are close to the periodicity of the generic
trends\footnote{For instance, trends with a period of a day are 
generally very strong.} or when the period is comparable or longer
with the observation window, either EPD or TFA tend to kill the real
signal itself. Second, one can involve the method of \emph{signal reconstruction},
as it was implemented by \cite{kovacs2005}. In this method, the 
signal model parameters $\{A_r\}$ are derived using the noisy signal,
and then the fit residuals undergo either the EPD or TFA. The model
signal $F(t_i,\dots)$ is added to the de-trended residuals, yielding
a complete signal reconstruction. The steps can be repeated until convergence 
is reached. Third, one can involve the \emph{simultaneous}
derivation of the $A_r$ model parameters and the $E_k$/$F_t$ coefficients
by minimizing the merit function 
\begin{equation}
\chi^2 = \sum\limits_i w_i\left[ m_i - m_0 - F\left(t_i,\{A_r\}\right) - \sum\limits_k E_k p^{(k)}_i \right]^2. \label{eq:simultaneoustrend}
\end{equation}
(This merit function shows the simultaneous trend removal for EPD. The 
TFA and the joint EPD+TFA can be applied similarly.)
The fourth method derives the $E_k$ and/or $T_f$ coefficients on
sections of the light curve where the star itself shows no real
variations. This is a definitely useful method in the analysis of 
planetary transit light curves, since the star itself can be assumed
to have constant brightness within 
noise limitations\footnote{At least, in the most of the cases. A famous
counter-example is the star CoRoT-Exo-2 of \cite{alonso2008}.} 
and therefore the
light curve should show no variations before and after the transit. 
If these out-of-transit sections of the light curves are sufficiently
long, the trend removal coefficients $E_k$ and/or $T_f$
can safely be obtained.

There are some considerations regarding to the 
$F(t_i,A_1,\dots,A_{\rm R})$ function and its parameters $\{A_r\}$
that should be mentioned here. In principle, one can use a
model function that is related to the \emph{physics} of
the variations. For instance, a light curve of a transiting 
extrasolar planet host star can be well modelled by $5$ 
parameters\footnote{Other parameters might be present if we do not have
\emph{a priori} assumptions for the limb darkening and/or the planetary orbit
is non-circular and the signal-to-noise of the light curve is sufficiently
large to see the asymmetry.}:
period ($P$), epoch ($E$), depth of the transit ($d$), duration of the transit
($\tau_{14}$) and the duration of the ingresses/egresses ($\tau_{12}$)
\citep[see e.g.][about how these parameters are related to the physical
parameters of the system, such as normalized semimajor axis, planetary
radius and orbital inclination]{carter2008}. Although 
the respective model function,
$F_{\rm transit}(t_i,P,E,d,\tau_{14},\tau_{12})$ is highly non-linear
in its parameters, the simultaneous signal fit and trend removal
of \eqref{eq:simultaneoustrend} can be performed, and the fit yields
reliable results in general\footnote{Only if the transit instances 
inter/extrapolated from the initial guess for the epoch $E$ and 
period $P$ sufficiently cover the observed transits. Otherwise, 
all of the parametric derivatives of $F$ will be zero and only methods
based on systematic grid search (e.g. BLS) yield reliable results.}. 
In the cases where we do not have any \emph{a priori} knowledge of
the source of the variations, but the signal can be assumed to 
be periodic, one can use a periodic model for $F$, that is, 
for instance, a linear combination of step functions. Although the number of 
free parameters (which must be involved in such a fit) are significantly
larger, in the cases of HATNet light curves, the fit can be achieved properly.
The signal reconstruction algorithm of \cite{kovacs2005} use a
step function (also known as ``folded and binned light curve models'')
for this purposes. Like so, $F$ can also be written as a Fourier 
series with finite terms. If the period and epoch are kept fixed,
both assumptions for the function $F$
(i.e. step function or Fourier expansion) yield a linear fit 
for both the model parameters and the EPD/TFA coefficients.

It should be mentioned here that the signal reconstruction mode
and the simultaneous trend removal yields roughly the same results.
However, a prominent counter-example is the case of HAT-P-11(b)
\citep{bakos2009hatp11}, where the reconstruction mode yielded an unexpectedly
high impact parameter for the system. In this case, only the 
method of simultaneous EPD and TFA was able to reveal a refined
set of light curve parameters that are expected to be more accurate on an
absolute scale. Further discussion of this problem 
can be found in \cite{bakos2009hatp11}.

\phdsubsection{Efficiency of these methods}

It is important to emphasize that both the EPD and TFA algorithms 
(independently from their native, reconstructive or simultaneous 
applications) reduce the effective degrees of freedom and therefore
the light curve scatter \emph{always} decreases. In order to 
determine whether the application of any of these algorithms 
is effective, one should compute the unbiased residuals of the 
fit after the derivation of the de-correlation coefficients. Alternatively,
one can increase the scatter of a particular light curve by the factor
$\sqrt{N/(N-P)}$ where $N$ is the number of total data points in the light
curve and $P$ is the number of parameters involved in the EPD 
or TFA. We should keep in mind that both during the selection 
of the appropriate external parameters and during the template selection,
the unbiased residuals must be checked carefully, otherwise 
the efficiency of these algorithms can easily be overrated.


\begin{table*}
\caption{Comparison of various data storage schemes. In this list,
``blobs'' are used as an acronym for  ``binary large objects''
(a collection of purely binary data in a single file).
.\label{tab:datastorage}}
\begin{center}
\footnotesize
\begin{tabularx}{160mm}{XX}
\hline
Pros	& Cons \\
\hline
{\bf\small FITS}$^{1}$

\footnotesize
\begin{itemize}
\xitem Flexible Image Transport System. The most common format 
and standard for astronomical data storage, especially for images 
(either raw or calibrated) and spectra.

\xitem Extensible format, supports not only multidimensional numeric 
arrays but in addition, structured flat tables and ASCII tables 
can also be stored inside a single FITS file.

\xitem Metadata storage is also available in a form of
keywords and their associated values. For instance, the location
of observation, information about the observer, date and time of the 
observation, instrumental details (such as filters, exposure time, 
optical data for the telescope) are stored almost always 
in FITS files involving consensual keywords.
\end{itemize}
&
\footnotesize
\begin{itemize}
\xitem Although some parts of the FITS files are stored in ASCII 
form (such as keywords and their values and textual tables),
extracting data from FITS files requires special tools. Moreover,
to access to just a smaller segment of the FITS data, one likely 
has to parse the whole file. For instance, if in a single file there
are 100 stored tables and one needs data only from the last table,
all of the other tables (at least their headers) have to be read and parsed,
since there is no pre-defined location of the last table.

\xitem Inserting or removing some keywords to/from a FITS header likely
results in an update of the whole file.
\end{itemize}
\\

\hline
{\bf\small Binary (large) objects}

\footnotesize
\begin{itemize}
\xitem Binary large object files (also known as ``blobs'') provide
the fastest way for both accessing (reading) and writing data.

\xitem Various indexing algorithms are available to make the data
access more efficient. Such algorithms can be optimized for 
any kind of data structure and access mode, 
including sequential access or two- or 
multidimensional hierarchy of data records.
\end{itemize}
&
\footnotesize
\begin{itemize}
\xitem Such blobs are \emph{not} human-readable, special programs are required
for accessing, reading or modifying the data. Basic tools found in UNIX-like
systems are not capable for generic manipulation of binary data.

\xitem Binary representation of integers and floating point numbers depends
on the actually used computer/processor architecture. Unless special
attention is given, such blobs cannot be copied from one computer
to another if they are using different architectures. Involving an architecture
independent storage format reduces some advantages of 
blobs (such as fast access).
\end{itemize}
\\
\hline
{\bf\small Linear ASCII/text files}

\footnotesize
\begin{itemize}
\xitem Human-readable format, easy to interpret.

\xitem Basic tools found in UNIX-like systems are capable to view or manipulate
plain textual data.

\xitem All of the programming languages, including data processing environments
and plotting tools support to read and parse numeric data from textual formats.

\xitem Modifications are easy to implement. Any kind of text editors 
or word processors are appropriate for manual manipulating of the data.
\end{itemize}
&
\footnotesize
\begin{itemize}
\xitem Access to massive numeric data in stored in textual format can
significantly be slower than access to the same amount of data that are 
stored in blobs. 

\xitem Random or even non-sequential access of small chunks of data stored in
a single text file requires the reading and parsing of the whole file.

\xitem The same type of data require $5-8$ times larger storage space 
than if these data were stored in blobs (depending on the actual data types
and/or our needs for a well structured file).
\end{itemize}
\\
\hline
{\bf\small Third-party applications: database servers, external storage systems}

\footnotesize
\begin{itemize}
\xitem Easy to maintain. Database solutions support various methods
for management, access control and such servers come with 
programming interfaces for many languages and environments. 

\xitem The underlying database engines involve large number of algorithms
for optimal data storage and allow efficient queries (using various
indexing methods). The engines can be fine-tuned in order to optimize for
our particular problem.
\end{itemize}
&
\footnotesize
\begin{itemize}
\xitem The indexing and therefore the access to the data is optimized
for one dimensional arrays of records (i.e. ``flat'' tables). 
Therefore storing images or other two- or multidimensional data structures
(such as astronomical catalogs, long photometric time series of enormous
amount of objects) cannot be implemented efficiently using classical
database engines.
\end{itemize}
\\
\hline
\end{tabularx}
\end{center}
\scriptsize
\begin{flushleft}
\noindent $^{1}$ The detailed documentation about the FITS file format 
is available from \texttt{http://fits.gsfc.nasa.gov/}.
\end{flushleft}
\end{table*}

\phdsection{Major concepts of the software package}
\label{subsec:majorconcepts}

Continuous monitoring of the sky yields
enormous amount of data. In the HATNet project, 6 telescopes expose 
images with a cadence of 5.5 minutes. Each image is a 
$2\,{\rm k}\times 2\,{\rm k}$ (up to August 2007) or
$4\,{\rm k}\times 4\,{\rm k}$ array of pixels, thus the amount of data
gathered on each clean night is $\sim 80-120$ scientific frames for 
a single telescope, equivalent to $7-11$ or $30-45$ gigabytes of
uncompressed calibrated images (assuming frames with the size of
$2\,{\rm k}\times 2\,{\rm k}$ or $4\,{\rm k}\times 4\,{\rm k}$ pixels,
respectively). In other words, if a single field is monitored for 2 months
by two of the telescopes \citep[see e.g.][for a description of the
actual observational principles]{bakos2007hatp2}, yielding
$\sim 5000$ individual scientific frames. The amount of 
data associated to this certain field is $\sim 300-350$\,gigabytes in a form
of calibrated images (assuming $4\,{\rm k}\times 4\,{\rm k}$ images).
If photometry is performed on these frames, the 
amount of associated information for 10~000 stars and for a single frame 
is $\sim3$\,megabytes of data, therefore one needs
hundreds of gigabytes storage space just for the photometric results.
All in all, the total amount of data that can be associated to
the reduction of a single monitored field can be even be close to one terabyte,
including all of the results of previously mentioned data types as well as other
ones, for instance astrometrical information, subtracted images,
or light curves with some sort of de-trending.

The components of the software package must be appropriate to 
manage such a huge amount of data. Thus, before going into the
details of the practical implementation, two issues should be clarified.
First, what kinds of data structures do appear during the reduction
of the images? This is a rather important question since the programs
not only have to access and manipulate these data but the resource limitations
of the computers do also constrain the available solutions. Second,
what are the existing software solutions which can efficiently
be exploited? We are especially focusing on such operating systems
and the related tools that are supported by larger communities 
and have a free and portable implementation.

\phdsubsection{Data structures}

At a first glance, data associated with image reduction can be 
classified into two major groups. The first group, that requires
the most of the storage space is in the form of massive linear data,
such as sequences of records, arrays of basic types or other multidimensional
arrays. Astronomical images, processed images (such as registered
or subtracted ones), instrumental photometric information,
light curves, de-trended light curves, Fourier or other kind of
spectra of the light curves belong to this group. All of these
data are a set of records with the same structure. For instance, 
\begin{itemize}
\item an image is a two dimensional array of integer or real numbers;
\item the list of extracted sources, where each record contains information
on the source's coordinates, brightness, shape parameters and possible
catalogue identifiers;
\item a light curve
is a series of individual photometric measurements, where each
measurement has a time, some sort of quality flag, magnitudes
for various apertures and/or various photometric methods, uncertainty
estimations; or
\item instrumental photometry, where the records contain
the same kind of information as the records of light curves, but
one set of records is associated not to a particular object covering
a long timebase, but to a single frame and numerous individual objects;
\item additional catalogue information for each star, that can be
useful in the interpretation of the photometric time series: 
such as brightness, color, spectral type, evolutionary state, parallax 
(if known), variability (if known).
\end{itemize}
These data types in the following are referred to as simply ``data'' 
in a general context.

The second major group of data types is the ``metadata'', that do not
have linear structure like the data types discussed above, and represent
definitely smaller amount of information. For instance,
\begin{itemize}
\item observational conditions for each image, such as date and time
of the observation, location, instrument description, primary target object
or field;
\item astrometric solution, where the information itself is the
transformation that maps a reference catalogue to the frame 
of the image;
\item point-spread function for a single image;
\item kernel solution, that describes the convolution function
used in the process of image subtraction.
\end{itemize}
Table~\ref{tab:datafiles} summarizes the above mentioned
various data types and their expected storage space requirements
appearing in the photometric analysis. 

\begin{table*}
\caption{An overview of data files used to store information
required by the image reduction process or created during 
the reduction. Each file type is referred by its extension.
\label{tab:datafiles}}
\begin{center}
\sloppy
\footnotesize
\begin{tabularx}{164mm}{|l|X|X|X|}
\hline
File size ($\rightarrow$)
	& $\mathcal{O}(1)$ 
	& $\propto N_{\rm frame}$
	& $\propto N_{\rm object}$ \\
\# of files($\downarrow$) & & & \\
\hline
$\mathcal{O}(1)$ 
&
\texttt{*.config}: generic information about the whole reduction
and observational conditions 
(name and coordinates of the target field, involved
reduction algorithms and their fine-tune parameters)

&
\texttt{*.list}: list of frames to be processed during the reduction

\texttt{*.stat}: basic statistics for each frame (both image
statistics such as number of detected objects and information on
the observational circumstances, e.g. zenith distance, airmass,
elevation of the Moon, stellar profile FWHM)
&
\texttt{*.cat}: list of objects and some catalogue information
that is used during the reduction

\texttt{*.lcstat}: light curve statistics (also known as ``magnitude-rms''
statistics)
\\
\hline
$\propto N_{\rm frame}$ 
&
\texttt{*.trans}: astrometric solution (the transformation that
maps the reference catalogue to the coordinate system of the image)

\texttt{*.kernel}: kernel solution (the convolution function
used in the image subtraction process)

\texttt{*-psf.fits}: best fit point-spread function for a given image
&
~
\vspace*{8mm}
	\begin{center} 
	\begin{picture}(20,20)
	\put(0,0){\line(1,1){20}}
	\put(0,20){\line(1,-1){20}}
	\end{picture}
	\end{center}
\vspace*{\fill}
~
&
\texttt{*.fits}: calibrated images$^{1}$

\texttt{*-sub.fits}: convolved and subtracted images$^{1}$

\texttt{*.stars}: list of detected sources and their properties
(coordinates, shape parameters, brightness estimation)

\texttt{*.phot}: instrumental photometric measurements
\\
\hline
$\propto N_{\rm object}$ 
&
%
%
\texttt{*.xmmc}: best fit and Monte-Carlo distribution of
the parameters of the light curve model function (if the object is turned
out to be interesting)

\texttt{*.info}: summary information of the planetary, orbital
and stellar data for the actual object (if the object is indeed a 
planet-harboring star)
&
\texttt{*.lc}: light curves

\texttt{*.epdlc}: de-trended light curves involving only the 
External Parameter Decorrelation algorithm

\texttt{*.tfalc}: de-trended light curves involving the 
Trend Filter Algorithm
&
~
\vspace*{9mm}
	\begin{center} 
	\begin{picture}(20,20)
	\put(0,0){\line(1,1){20}}
	\put(0,20){\line(1,-1){20}}
	\end{picture}
	\end{center}
\vspace*{\fill}
~
\\
\hline
\end{tabularx}
\end{center}
\begin{flushleft}
\scriptsize
\noindent $^{1}$ Strictly speaking, the size of these files does
not depend on the number of objects that are extracted from
the image and/or targets for further photometry. However, larger
images tend to have greater number of sources of interest.
\end{flushleft}
\end{table*}

Of course, both linear data and metadata that are created
during the image reduction process should be stored in some format.
There are various concepts for data formats available in modern computers and
operating systems, so one can choose the most suitable format for
each purpose. In astronomy, people commonly store and share data in FITS format.
Many programs use human-readable (ASCII or text) files both for 
input and output. Some other programs store their information
in binary format, where the contents of the files cannot even be viewed
without a special program. And there are robust database systems,
that hide the details of the actual storage and give a relatively
lightweight interface to access or manipulate the data.
Each type of the above mentioned data representations has its own advantages
and disadvantages. In Table~\ref{tab:datastorage} these properties
are summarized for these four major representation schemes. 
During the reduction of HATNet data, we have chosen a mixed form of data 
representation as follows. The images, including the raw, calibrated
and processed ones are stored in FITS format. Moreover, we use three
dimensional FITS images to store the spatial variations 
of the point-spread function. Other metadata, such as astrometrical
solutions, kernel solutions, catalogue information are stored in text
files. Instrumental photometric measurements and light curves are 
also stored in the form of text files. Temporary data 
(needed for intermediate steps of the reduction) are stored in binary form,
since such data are not needed to be portable and an advantage
of the binary format is the significantly smaller storage space
requirement.

\phdsubsection{Operating environment}

In order to both have a portable and robust set of tools, one has 
to build a software package on the top of widely standardized 
and documented environment. The most widespread and approved
standard is the ``Portable Operating System Interface'' or 
POSIX\footnote{http://en.wikipedia.org/wiki/POSIX},
that intended to standardize almost all layers of the operating system,
from the system-level application program interfaces (APIs, such 
as file manipulation or network access) up to the highest level of 
programs such as shell environments, related scripting languages and 
other basic utilities. 

The actual development of the package \fihat{} was done under 
GNU\footnote{http://www.gnu.org/}/Linux\footnote{http://www.kernel.org/}
systems, that is one of the most frequently used POSIX compliant, 
UNIX-like\footnote{http://en.wikipedia.org/wiki/Unix-like} 
free operating system. The main code was written in ANSI C
(featured with some GNU extensions) and intended to be compiled without
any difficulties on various other UNIX systems such as SunOS/Sparc
and Mac OSX. The compilation of the package does not require additional
packages or libraries, only the GNU C Compiler 
(\texttt{gcc}\footnote{http://gcc.gnu.org/}), its 
standard library (\texttt{glibc}\footnote{http://www.gnu.org/software/libc/}), 
the associated standard header files
and some related development utilities.
(Such as \texttt{make}\footnote{http://www.gnu.org/software/make/} or 
the \texttt{ar}\footnote{http://www.gnu.org/software/binutils/} 
object archived. In almost
all of the systems these come with \texttt{gcc} as its dependencies)
Therefore, all of the requirements of the package include only
free and open source software (F/OSS).

In practice, to have a complete data reduction environment the users 
of the package might have to use additional text processing utilities
such as an implementation of the AWK programming language
(for instance, \texttt{gawk}\footnote{http://www.gnu.org/software/gawk/},
that is included in all of the free GNU/Linux systems)
and basic text processing utilities (such as \texttt{paste}, \texttt{cat},
\texttt{sort}, \texttt{split}, included in the 
\texttt{textutils}/\texttt{coreutils}\footnote{http://www.gnu.org/software/coreutils/}
GNU package). And finally, for visualization purposes, the 
SAOImage/DS9 utility\footnote{http://hea-www.harvard.edu/RD/ds9/} 
\citep{joye2003} is highly recommended.


\begin{table*}
\caption{An overview of the standalone binary programs included 
in the package, displaying their main purposes and the types of input and
output data.
\label{tab:progsummary}}
\begin{center}
\scriptsize
\sloppy
\begin{tabularx}{164mm}{|l|p{50mm}|X|X|}
\hline 
Program & Main purpose & Type of input & Type of output \\
\hline
\texttt{fiarith}
&
Evaluates arithmetic expressions on images as operands.
&
A set of FITS images.
& 
A single FITS image.
\\
\texttt{ficalib}
&
Performs various calibration steps on the input images.
&
A set of raw FITS images.
& 
A set of calibrated FITS image.
\\
\texttt{ficombine}
&
Combines (most frequently averages) a set of images.
&
A set of FITS images.
& 
A single FITS image.
\\
\texttt{ficonv}
&
Obtains an optimal convolution transformation between two images
or use an existing convolution transformation to convolve an image.
&
Two FITS images or a single image and a transformation.
&
A convolution transformation or a single image.
\\
\texttt{fiheader}
&
Manipulates, i.e. reads, sets, alters or removes some FITS header keywords
and/or their values.
&
A single FITS image (alternation) or more FITS images (if header contents
are just read).
&
A FITS image with altered header or a series of keywords/values from the
headers.
\\
\texttt{fiign}
&
Performs low-level manipulations on masks associated to FITS images.
&
A single FITS image (with some optional mask).
&
A single FITS image (with an altered mask).
\\
\texttt{fiinfo}
&
Gives some information about the FITS image in a human-readable form
or creates image stamps in a conventional format.
&
A single FITS image.
&
Basic information or PNM images.
\\
\texttt{fiphot}
&
Performs photometry on normal, convolved or subtracted images.
&
A single FITS image (with additional reference photometric information
if the image is a subtracted one).
&
Instrumental photometric data.
\\
\texttt{firandom}
&
Generates artificial object lists and/or artificial (astronomical) images.
&
List of sources to be drawn to the image or an arithmetic expression that 
describes how the list of sources is to be created.
&
List of sources and/or a single FITS image.
\\
\texttt{fistar}
&
Detects and characterizes point-like sources from astronomical images.
&
A single FITS image.
&
List of detected sources and an optional PSF image (in FITS format).
\\
\texttt{fitrans}
&
Performs generic geometric (spatial) transformations on the input image.
&
A single FITS image.
&
A single, transformed FITS image.
\\
\texttt{fi[un]zip}
&
Compresses and decompresses primary FITS images.
&
A single uncompressed or compressed FITS image file.
&
A single compressed or uncompressed FITS image file.
\\
\texttt{grcollect}
&
Performs data transposition on the input tabulated data or do some sort
of statistics on the input data.
&
A set of files containing tabulated data.
&
A set of files containing the transposed tabulated data or a single
file for the statistics, also in a tabulated form.
\\
\texttt{grmatch}
&
Matches lines read from two input files of tabulated data, using
various criteria (point matching, coordinate matching or identifier matching).
&
Two files containing tabulated data (that must be two point sets in the
case of point or coordinate matching).
&
One file containing the matched lines and in the case of point matching,
an additional file that describes the best fit geometric transformation between
the two point sets.
\\
\texttt{grselect}
&
Selects lines from tabulated data using various criteria.
&
A single file containing tabulated data.
&
The filtered rows from the input data.
\\
\texttt{grtrans}
&
Transforms a single coordinate list or derives a best-fit transformation 
between two coordinate lists.
&
A single file containing a coordinate list and
a file that describes the transformation or two files, each one is 
containing a coordinate list.
&
A file with the transformed coordinate list in tabulated from or
a file that contains the best-fit transformation.
\\
\texttt{lfit}
&
General purpose arithmetic evaluation, regression and data analysis tool.
&
Files containing data to be analyzed in a tabulated form.
&
Regression parameters or results of the arithmetic evaluation.
\\
\hline
\end{tabularx}
\end{center}
\begin{flushleft}
\end{flushleft}
\end{table*}

\phdsection{Implementation}
\label{subsec:implementation}

In this subsection I summarize the standalone programs that are implemented
as distinct binary executables. The programs can be divided into two
well separated groups with respect to the main purposes. In the first
group there are the programs that manipulate the (astronomical) images
themselves, i.e. read an image, generate one or do a specific transformation
on an image. In the second group, there are the programs that manipulate
textual data, mostly numerical data presented in a tabulated form. 

Generally, all of these programs are capable to the following.
\begin{itemize}
\item The codes give release and version information as well as the invocation
can be logged on demand. The version information can be reported by a single
call of the binary, moreover it is logged along with the invocation arguments
in the form of special FITS keywords (if the main output of the actual
code is a processed FITS image) and in the form of textual comments
(if the main output of the code is text data). Preserving the version 
information along with the invocation arguments makes any kind of output
easily reproducible. 
\item All of the codes are capable to read their data to be processed from the 
standard input and write the output data to the standard output.
Since many of these programs manipulate relatively large amount of data,
the number of unnecessary hard disk operations should be 
reduced as small as possible.
Moreover, in many cases the output of one of the programs is the input 
of the another one.
Pipes, available in all of the modern UNIX-like operating systems,
are basically designed to perform such bindings between the output and
input of two programs. Therefore, such a capability of redirecting the
input/output data streams significantly reduce the overhead
of background storage operations.
\item The programs that deal with symbolic operations and
functions, a general back-end library\footnote{available from {http://libpsn.sf.net}}
is provided to make a user-friendly
interface to specify arithmetic expressions. This kind of approach 
in software systems is barely used, since such a symbolic
specification of arithmetic expressions does not provide a standalone language.
However, it allows an easy and transparent way for arbitrary operations,
and turned out to be very efficient in higher level data reduction scripts.
\item The programs that manipulate FITS images are capable to 
handle files with multiple extensions. The FITS standard allows the user
to store multiple individual images, as well as (ASCII or binary) tabulated
data in a single file. The control software of some detectors 
produces images that are stored in this extended format, for 
example, such detectors where 
the charges from the CCD chip are read out in multiple directions
(therefore the camera electronics utilizes more than one amplifier 
and A/D converter, thus yield different bias and noise levels). 
Other kind of detectors (which acquire individual images with a very
short exposure time) might store the data in 
the three dimensional format called ``data cube''. The developed
codes are also capable to handle such data, therefore it is possible
to do reductions on images obtained by the Spitzer Space Telescope,
that optionally uses such data structures for image storage.
\end{itemize}
The list of standalone binaries and their main purposes 
that come with the package are shown in Table~\ref{tab:progsummary}. 

\phdsubsection{Basic operations on FITS headers and keywords -- \texttt{fiheader}}

The main purpose of the \texttt{fiheader} utility is to read specific
values from the headers of FITS files and/or alter them on demand. 

Although most of the information about the observational conditions
is stored in the form of FITS keywords, image manipulation programs
use only the necessary ones and most of the image processing
parameters are passed as command line arguments (such keywords and data
are, for example, the gain, the image centroid coordinates, 
astrometrical solutions). The main reasons why this kind of approach
was chosen are the following. 
\begin{itemize}
\item First, interpreting many of the standard keywords
leads to false information about the image in the cases of wide-field
or heavily distorted images. Such a parameter is the gain that 
can be highly inhomogeneous for images acquired by an optical system
with non-negligible vignetting and the gain itself cannot be described
by a single real number\footnote{For which the \emph{de facto} standard is  
the \texttt{GAIN} keyword.}, 
rather a polynomial or some equivalent function. Similarly, the standard
World Coordinate System information, describing the astrometrical
solution of the image, has been designed for small field-of-view images,
i.e. the number of coefficients are insufficiently few to properly
constrain the astrometry of a distorted image.
\item Second, altering the meanings of standard keywords leads to 
incompatibilities with existing software. For example, if
the format of the keyword \texttt{GAIN} was changed to be a string of 
finite real numbers (describing a spatially varied gain), other
programs would not be able to parse this redefined keyword.
\end{itemize}
Therefore, our conclusion was not altering the syntax of the existing keywords,
but 
to define some new (wherever it was necessary). The \texttt{fiheader}
utility enables the user to read any of the keywords, and allows
higher level scripts to interpret the values read from the headers
and pass their values to other programs in the form of command line
arguments.

\phdsubsection{Basic arithmetic operations on images -- \texttt{fiarith}}
\label{sec:prog:fiarith}

The program \texttt{fiarith} allows the user to perform simple
operations on one or more astronomical images. Supposing all of the
input images have the same size, the program allows the user to do
\emph{per pixel} arithmetic operations as well as manipulations 
depend on the pixel coordinates themselves. 

The invocation syntax simply reflects the desired operations. For 
example the common way of calibrating image $I$, using bias ($B$), dark ($D$) 
and flat ($F$) images, which can be written as
\begin{equation}
C=\frac{I-B-D}{F/\|F\|},
\end{equation}
where $C$ denotes the calibrated image (see also 
equation~\ref{eq:calibgeneral}). Thus, the computation of the calibrated 
image $C$ can be written as 
\begin{scmd}
fiarith "('I'-'B'-'D')/('F'/norm('F'))" -o C%
\end{scmd}

\phdsubsection{Basic information about images -- \texttt{fiinfo}}
\label{sec:prog:fiinfo}

The aim of the program \texttt{fiinfo} is twofold. First, 
this program is capable to gather some statistics and masking 
information of the image. These include
\begin{itemize}
\item general statistics, such as mean, median, minimum, maximum,
standard deviation of the pixel values;
\item statistics derived after rejecting the outlier pixels;
\item estimations for the background level and its spatial variations;
\item estimations for the background noise; and
\item the number of masked pixels, detailing for all occurring mask types.
\end{itemize}
The most common usage of \texttt{fiinfo} in this statistical mode is 
to deselect those calibration frames that seem to be faulty 
(e.g. saturated sky flats, aborted images or so).

Second, the program
is capable to convert astronomical images into widely used 
graphics file formats. Almost all of the
scaling options available in the well known \texttt{DS9} program
\citep[see][]{joye2003} have been implemented in \texttt{fiinfo},
moreover, the user can define arbitrary color palettes as well.
In practice, \texttt{fiinfo} creates only images in PNM (portable 
anymap) format. Images stored in this format can then be converted to any of 
the widely used graphics file formats (such as JPEG, PNG), using existing 
software (e.g. \texttt{netpbm}, \texttt{convert}/ImageMagick).
Figures in this thesis displaying stamps from real (or mock)
astronomical images have also been created using this mode of the program.

\phdsubsection{Combination of images -- \texttt{ficombine}}
\label{sec:prog:ficombine}

The main purpose of image combination is to create a single image
with good signal-to-noise ratio from individual images with 
lower signal-to-noise ratio. The program \texttt{ficombine} is
intended to perform averaging of individual images.
In practice, the usage of this program is twofold.
First, it is used to create the master calibration frames, as
it is defined by \eqref{eq:avgbias}, \eqref{eq:avgdark} and
\eqref{eq:avgflat}. Second, the reference frame required by the
method of image subtraction is also created by averaging individual
registered object frames (see also Sec.~\ref{sec:prog:fitrans} about
the details of image registration). 

In the actual implementation, such combination is employed as a 
\emph{per pixel} averaging, where the method of averaging and its
fine tune parameters can be specified via command line arguments. The
most frequently used ``average values'' are the mean and median values.
In many applications, rejection of outlier values are required,
for instance, omitting pixels affected by cosmic ray events. The
respective parameters for tuning the outlier rejection are also
given as command line options. See Sec.~\ref{sec:prog:ficalib} 
for an example about the usage of \texttt{ficombine}, demonstrating
its usage in the calibration pipeline.

\begin{figure*}
\begin{lcmd}
\small
\#!/bin/sh \\
~ \\
\# \textit{Names of the individual files storing the raw bias, flat and object frames are stored here:} \\
BIASLIST=(\$SOURCE/[0-9]*.BIAS.fits) \\
FLATLIST=(\$SOURCE/[0-9]*.FLAT.fits) \\
OBJLIST=(\$SOURCE/[0-9]*.TARGET.fits) \\
~ \\
\# \textit{Calibrated images: all the images are got an 'R' prefix and put in the appropriate directory:} \\
R\_BIASLIST=(\$(for f in \$\{BIASLIST[*]\} ; do echo \$MSTTMP/bias/R`basename \${f}` ; done)) \\
R\_FLATLIST=(\$(for f in \$\{FLATLIST[*]\} ; do echo \$MSTTMP/flat/R`basename \${f}` ; done)) \\
R\_OBJLIST=( \$(for f in \$\{OBJLIST[*]\}  ; do echo \$TARGET/R`basename \${f}` ; done)) \\
~ \\
\# \textit{These below are KeplerCam specific data, defining the topology and geometry of the CCD itself.} \\
\# \textit{The camera has four readout registers and therefore four amplifiers and A/D converters as well.} \\
MS\_NAME=(IM1 IM2 IM3 IM4) \\
MS\_OPAR=spline,order=3,iterations=2,sigma=3 \\
MS\_OVER=(area=\{2:0:7:1023,1034:0:1039:1023,2:0:7:1023,1034:0:1039:1023\},\$\{MS\_OPAR\}) \\
MS\_OFFS=(1024,1024 0,1024 1024,0 0,0) \\
MS\_TRIM=image=[8:0:1031:1023] \\
~ \\
M\_ARGS="--mosaic size=[2048,2048]" \\
M\_ARGS="\$M\_ARGS --mosaic [name=\$\{MS\_NAME[0]\},\$MS\_TRIM,overscan=[\$\{MS\_OVER[0]\}],offset=[\$\{MS\_OFFS[0]\}]]" \\
M\_ARGS="\$M\_ARGS --mosaic [name=\$\{MS\_NAME[1]\},\$MS\_TRIM,overscan=[\$\{MS\_OVER[1]\}],offset=[\$\{MS\_OFFS[1]\}]]" \\
M\_ARGS="\$M\_ARGS --mosaic [name=\$\{MS\_NAME[2]\},\$MS\_TRIM,overscan=[\$\{MS\_OVER[2]\}],offset=[\$\{MS\_OFFS[2]\}]]" \\
M\_ARGS="\$M\_ARGS --mosaic [name=\$\{MS\_NAME[3]\},\$MS\_TRIM,overscan=[\$\{MS\_OVER[3]\}],offset=[\$\{MS\_OFFS[3]\}]]" \\
~ \\
\# \textit{The calibration of the individual bias frames, followed by their combination into a single master image:} \\
ficalib	~~-i \$\{BIASLIST[*]\} --saturation 50000 \$M\_ARGS -o \$\{R\_BIASLIST[*]\}  \\
ficombine \$\{R\_BIASLIST[*]\} --mode median -o \$MASTER/BIAS.fits \\

\# \textit{The calibration of the individual flat frames, followed by their combination into a single master image:} \\
ficalib	-i \$\{FLATLIST[*]\} --saturation 50000 \$M\_ARGS -o \$\{R\_FLATLIST[*]\} $\backslash$ \\
\ptab	--input-master-bias \$MASTER/BIAS.fits --post-scale 20000 \\
ficombine \$\{R\_FLATLIST[*]\} --mode median -o \$MASTER/FLAT.fits \\
~ \\
\# \textit{The calibration of the object images:} \\
ficalib	-i \$\{OBJLIST[*]\} --saturation 50000 \$M\_ARGS -o \$\{R\_OBJLIST[*]\} $\backslash$ \\
\ptab	--input-master-bias \$MASTER/BIAS.fits --input-master-flat \$MASTER/FLAT.fits
\end{lcmd}
\caption{A shell script demonstrating the proper usage of the 
\texttt{ficalib} and \texttt{ficombine} programs on the example of the
calibration of the KeplerCam mosaic images. The names for the 
files containing the input raw frames (both calibration frames and
object frames) are stored in the arrays \texttt{\$BIASLIST[*]}, \texttt{\$FLATLIST[*]}
and \texttt{\$OBJLIST[*]}. The variable \texttt{\$M\_ARGS} contains 
all necessary information related to the specification of the mosaic 
topology and geometry as well as the overscan areas associated to each
readout direction. The individual calibrated bias and flat frames are 
stored in the subdirectories of the \texttt{\$MSTTMP} directory.
These files are then combined to a single master bias and flat frame,
that are used in the final step of the calibration, when the object
frames themselves are calibrated. The final calibrated scientific images
are stored in the directory \texttt{\$TARGET}. Note that each flat
frame is scaled after calibration to have a mean value of 20,000\,ADU. 
In the case of dome flats, this scaling is not necessary, but in the
case of sky flats, this steps corrects for the variations in the
sky background level (during dusk or dawn). 
}
\label{fig:ficalibexample}
\end{figure*}

\phdsubsection{Calibration of images -- \texttt{ficalib}}
\label{sec:prog:ficalib}

In principle, the program \texttt{ficalib} implements 
the evaluation of \eqref{eq:calibgeneral}
in an efficient way. It is optimized for the assumption that
all of the master calibration frames are the same for all of the input images.
Because of this assumption, the calibration process is much more
faster than if it was done independently on each image, using 
the program \texttt{fiarith}.

Moreover, the program \texttt{ficalib} automatically performs
the overscan correction (if the user specifies overscan regions),
and also trims the image to its designated size (by clipping these 
overscan areas). The output images inherit the masks from
the master calibration images, as well as additional pixels might
be masked from the input images if these were found to be saturated
and/or bloomed. 
When a single chip camera uses multiple readout gates, amplifiers 
and A/D converters the images are stored in a so-called mosaic format
(such as KeplerCam). The program \texttt{ficalib} is capable to
combine these mosaic image regions into one single image. 

In Fig.~\ref{fig:ficalibexample} a shell script is shown that demonstrates
the usage of the programs \texttt{ficalib} and \texttt{ficombine} on
a real-life application, namely how the images acquired by the FLWO 
KeplerCam\footnote{See: http://www.sao.arizona.edu/FLWO/48/kep.primer.html}
are completely calibrated. 

\phdsubsection{Rejection and masking of nasty pixels -- \texttt{fiign}}
\label{sec:prog:fiign}

The aim of the program \texttt{fiign} is twofold. 
First, it is intended to perform 
low-level operations on masks associated to FITS images, such as
removing some of the masks, converting between layers of the masks 
and merging or combining masks from separate files. Second, various
methods exist with which the user can add additional masks based 
on the image itself. These additional masks can be used to mark 
saturated or blooming pixels, pixels with unexpectedly low and/or high 
values or extremely sharp structures, especially pixels that are 
resulted by cosmic ray events. 

This program is a crucial piece in the calibration pipeline if 
it is implemented using purely the \texttt{fiarith} program. However,
most of the functionality of \texttt{fiign} is also integrated in
\texttt{ficalib} (see Sec.~\ref{sec:prog:ficalib}). Since
\texttt{ficalib} much more efficiently implements the operations
of the calibration than if these were implemented by individual calls 
of \texttt{fiarith}, \texttt{fiign} is used only occasionally in practice.

\begin{figure*}
\begin{center}
\resizebox{50mm}{!}{\includegraphics{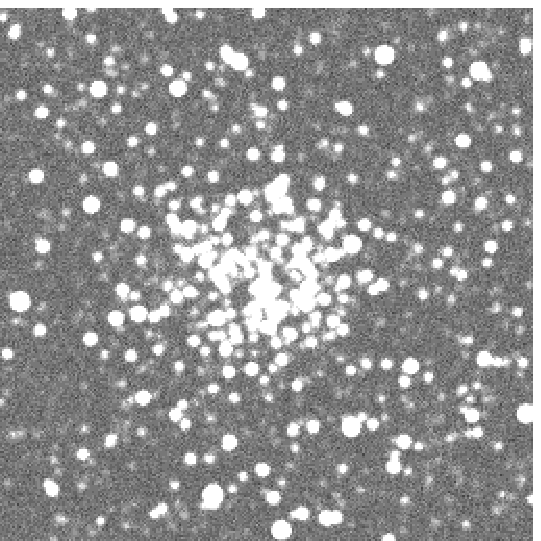}}\hspace*{5mm}%
\resizebox{50mm}{!}{\includegraphics{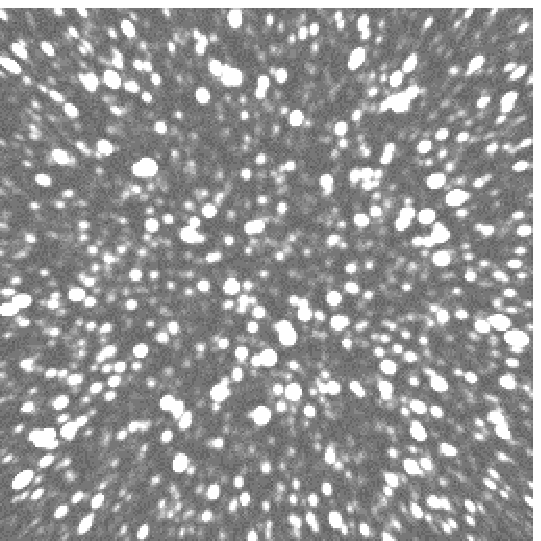}}\hspace*{5mm}%
\resizebox{50mm}{!}{\includegraphics{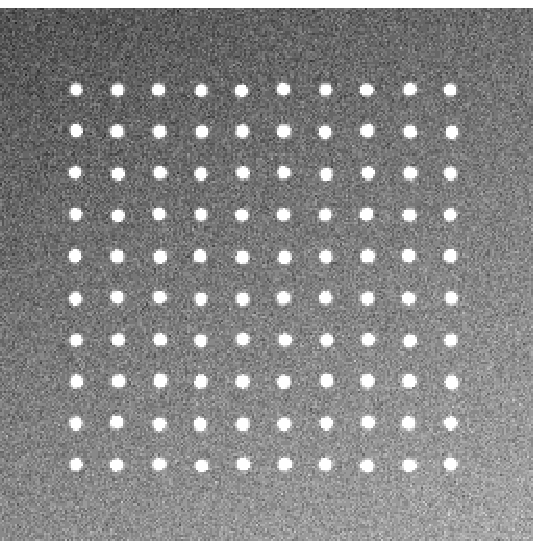}}
\begin{lcmd}
\small
\#!/bin/sh \\
~ \\
firandom --size 256,256 $\backslash$ \\
\ptab	~--list "f=3.2,500*[x=g(0,0.2),y=g(0,0.2),m=15-5*r(0,1)\^{ }2]" $\backslash$ \\
\ptab	~--list "f=3.2,1400*[x=r(-1,1),y=r(-1,1),m=15+1.38*log(r(0,1))]" $\backslash$ \\
\ptab	~--sky 100 --sky-noise 10 --integral --photon-noise --bitpix -32 --output globular.fits \\
~ \\
firandom --size 256,256 $\backslash$ \\
\ptab	~--list "5000*[x=r(-1,1),y=r(-1,1),s=1.3,d=0.3*(x*x-y*y),k=0.6*x*y,m=15+1.38*log(r(0,1))]" $\backslash$ \\
\ptab	~--sky 100 --sky-noise 10 --integral --photon-noise --bitpix -32 --output coma.fits \\
~ \\
firandom --size 256,256 $\backslash$ \\
\ptab	~--list "f=3.0,100*[X=36+20*div(n,10)+r(0,1),Y=36+20*mod(n,10)+r(0,1),m=10]" $\backslash$ \\
\ptab	~--sky "100+x*10-y*20" --sky-noise 10 --integral --photon-noise --bitpix -32 --output grid.fits \\
~ \\
for base in globular coma grid ; do \\
\ptab	fiinfo	 \$\{base\}.fits --pgm linear,zscale --output-pgm - | pnmtoeps -g -4 -d -o \$\{base\}.eps  \\
done
\end{lcmd}
\end{center}
\caption{Three mock images generated using the program
\texttt{firandom}. The first image (\texttt{globular.fits}) on the left 
shows a ``globular cluster'' with some field stars as well. For simplicity,
the distribution of the cluster stars are Gaussian and the magnitude
distribution is quadratic while the field stars distribute uniformly and
their magnitudes is derived from assuming uniformly distributed stars of 
constant brightness. The second image (\texttt{coma.fits}) simulates
nearly similar effect on the stellar profiles what comatic aberration 
would cause. The shape parameters $\delta$ and $\kappa$ (referred 
as \texttt{d} and \texttt{k} in the command line argument of the program,
see also Sec.~\ref{sec:stardetection:coordshapemodel})
are specific functions of the spatial coordinates. The magnitude 
distribution of the stars is the same as for the field stars in
the previous image. The third image (\texttt{grid.fits})
shows a set of stars positioned
on a grid. The background of this image is not constant.
The shell script below the image stamps is used to create these FITS files.
The body of the last iterator loop in the script converts the FITS files 
into PGM format, using the \texttt{fiinfo} utility (see 
Sec.~\ref{sec:prog:fiinfo}) and the well-known 
\texttt{zscale} intensity scaling algorithm \citep[see DS9,][]{joye2003}.
The images yielded by \texttt{fiinfo} are instantly converted to EPS 
(encapsulated Postscript) files, that is the preferred format for many
typesetting systems, such as \LaTeX.}
\label{fig:firandomexamples}
\end{figure*}

\phdsubsection{Generation of artificial images -- \texttt{firandom}}
\label{sec:prog:firandom}

The main purpose of the program \texttt{firandom} is to create
artificial images. These artificial images can be used either to 
create \emph{model images} for real observations 
(for instance, to remove fitted stellar PSFS) or \emph{mock images}
that are intended to simulate some of the influence related to
one or more observational artifacts and realistic effects. 
In principle, \texttt{firandom}
creates an image with a given background level on which sources are drawn.
Additionally, \texttt{firandom} is capable to add noise to the images,
simulating both the effect of readout and background noise as well
as photon noise. In the case of mock images, \texttt{firandom} is 
also capable to generate the object list itself. The stellar 
profile models that are supported by \texttt{firandom} and therefore
available for artificial images are the same set of functions described
in Sec.\ref{sec:stardetection:coordshapemodel}. Moreover, \texttt{firandom}
is capable to draw stellar profiles derived from PSFs (by
the program \texttt{fistar}, see also Sec.~\ref{sec:prog:fistar}).

The program features symbolic input processing, i.e. the variations
in the background level, the spatial distribution of the object 
centroids (in the case of mock images), the profile shape parameters,
fluxes for individual objects and the noise level can be specified
not only as a tabulated dataset but in the form of arithmetic
expressions. In these expressions one can involve various built-in 
arithmetic operators and functions, including random number generators.
Of course, the generated mock coordinate lists can also be saved
in tabulated form. 
The mock images used during the generation of Fig.~\ref{fig:subpixphot},
Fig.~\ref{fig:optaper} or Fig.~\ref{fig:modelfwhm} have been created
by \texttt{firandom}.

In Fig.~\ref{fig:firandomexamples}, some examples are shown that demonstrate
the usage of the program \texttt{firandom}. 

\phdsubsection{Detection of stars or point-like sources -- \texttt{fistar}}
\label{sec:prog:fistar}

The star detection and stellar profile modelling algorithms 
described in Sec.~\ref{subsec:stardetection}
are implemented in the program \texttt{fistar}. The main purpose of this
program is therefore to search for and characterize point-like sources.
Additionally, the program is capable to derive the point-spread function
of the image, and spatial variations of the PSF can also be fitted
up to arbitrary polynomial order. 

The list of detected sources, their centroid coordinates, shape
parameters (including FWHM) and flux estimations are written to a
previously defined output file. This file can have arbitrary format,
depending on our needs. The best fit PSF is saved in FITS format. If
the PSF is supposed to be constant throughout the image, 
the FITS image is a normal two-dimensional image. Otherwise,
the PSF data and the associated polynomial coefficients are stored
in ``data cube'' format, and the size of the $z$ (\texttt{NAXIS3}) axis
is $(N_{\rm PSF}+1)(N_{\rm PSF}+2)/2$, where $N_{\rm PSF}$
is the polynomial order used for fitting the spatial variations.

\phdsubsection{Basic coordinate list manipulations -- \texttt{grtrans}}
\label{sec:prog:grtrans}

The main purpose of the program \texttt{grtrans} is to perform
coordinate list transformations, mostly related to stellar profile
centroid coordinates and astrometrical transformations. 
Since this program is used exhaustively with the program
\texttt{grmatch}, examples and further discussion of this
program can be found in the next section, Sec.~\ref{sec:prog:grmatch}.

\phdsubsection{Matching lists or catalogues -- \texttt{grmatch}}
\label{sec:prog:grmatch}

\begin{figure}
\begin{center}
\resizebox{80mm}{!}{\includegraphics{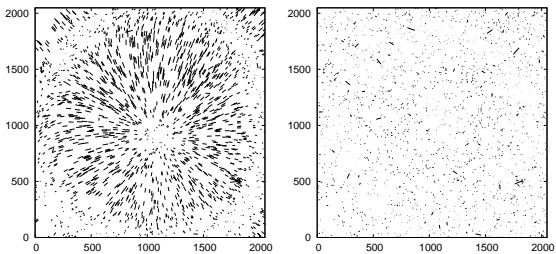}}
\end{center}
\caption{Vector plots of the difference between the transformed reference and
the input star coordinates for a typical HAT field. The left panel shows
the difference for second-order, the right panel for fourth-order
polynomial fits.
\label{fig:distvect}}
\end{figure}

\begin{figure}
\begin{center}
\resizebox{80mm}{!}{\includegraphics{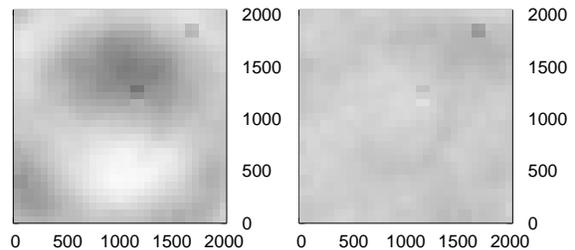}}
\end{center}
\caption{The difference between the $Y$ coordinates of the transformed 
reference and the input star coordinates for a typical HAT field. The left 
panel shows the difference for fourth-order, the right panel for sixth-order
polynomial fits.
\label{fig:distshift}}
\end{figure}

The main purpose of the \texttt{grmatch} code is to implement 
the point matching algorithm that is the key point in the derivation
of the astrometric solution and source identification. See 
Section~\ref{subsec:astrometry} about more details on the algorithm 
itself. We note here that although the program \texttt{grmatch} is 
sufficient for point matching and source identification purposes,
but one needs other codes to interpret or use the outcome of this program.
For instance, tabulated list of coordinates can be transformed 
from one reference frame to another, using the program \texttt{grtrans} 
while the program \texttt{fitrans} is capable to apply these
transformations (yielded by \texttt{grmatch}) on FITS images, in order to,
for instance, register images to the same reference frame.


\begin{figure*}
\begin{lcmd}
\small
\ptab \\
for base in \$\{LIST\_OF\_FRAMES[*]\} ; do \\
\ptab \\
\ptab	grmatch --reference \$CATALOG --col-ref \$COL\_X,\$COL\_Y --col-ref-ordering -\$COL\_MAG $\backslash$ \\
\ptab	\ptab	--input \$AST/\$base.stars --col-inp 2,3 --col-inp-ordering +8 $\backslash$ \\
\ptab	\ptab	--weight reference,column=\$COL\_MAG,magnitude,power=2 $\backslash$ \\
\ptab	\ptab	--order \$AST\_ORDER --max-distance \$MAX\_DISTANCE $\backslash$ \\
\ptab	\ptab	--output-transformation \$AST/\$base.trans --output \$AST/\$base.match || break \\
\ptab	\ptab \\
\ptab   grtrans \$CATALOG $\backslash$ \\
\ptab	\ptab	--col-xy  \$COL\_X,\$COL\_Y --input-transformation \$AST/\$base.trans $\backslash$ \\
\ptab	\ptab	--col-out \$COL\_X,\$COL\_Y --output - | $\backslash$ \\
\ptab	grmatch --reference - --col-ref \$COL\_X,\$COL\_Y --input \$AST/\$base.stars --col-inp 2,3 $\backslash$ \\
\ptab	\ptab	--match-coords --max-distance \$MAX\_MATCHDST --output - | $\backslash$ \\
\ptab	grtrans --col-xy  \$COL\_X,\$COL\_Y --input-transformation \$AST/\$base.trans --reverse $\backslash$ \\
\ptab	\ptab	--col-out \$COL\_X,\$COL\_Y --output \$AST/\$base.match \\
\ptab \\
done \\
\ptab 
\end{lcmd}
\caption{A typical application for the 
\texttt{grmatch} -- \texttt{grtrans} programs, for the cases where a few of
the stars have high proper motion thus have significant offsets from
the catalogue positions. For each frame (named \texttt{\$base}), the input 
catalogue (\texttt{\$CATALOG}) is matched with the respective list
of extracted stars (found in the \texttt{\$AST/\$base.stars} file), keeping a
relatively large maximum distance between the nominal and detected stellar
positions (\texttt{\$MAX\_DISTANCE}, e.g. $4-6$ pixels, derived from the
expected magnitude of the proper motions from the catalogue epoch and the
approximate plate scale). This first initial match identifies all of the
sources (including the ones with large proper motion),
stored in \texttt{\$AST/\$base.match} file in the form of matched
detected source and catalogue entries. However, the astrometric 
transformation (stored in \texttt{\$AST/\$base.trans}) is systematically 
affected by these high proper motion stars. In order to get rid of this
effect, the match is performed again by excluding the stars 
with higher residual distance (by setting \texttt{\$MAX\_MACHDIST} to
e.g. $1-2$ pixels). 
The procedure is then repeated for all frames (elements of the 
\texttt{\$LIST\_OF\_FRAMES[]} array) in the similar manner.
}
\label{fig:grmatchgrtransexample}
\end{figure*}

\phdsubsubsection{Typical applications}
\label{sec:grmatch:applications}

As it was discussed before, the programs \texttt{grmatch} and \texttt{grtrans}
are involved in the photometry pipeline, following the
star detection. If the accuracy of the coordinates in the 
reference catalogue is sufficient to yield a consistent plate solution,
one can obtain the photometric centroids by simply invoking 
these programs. A more sophisticated example for these program is 
shown in Fig.~\ref{fig:grmatchgrtransexample}. In this example these programs 
are invoked twice in order to both derive a proper astrometric 
solution\footnote{By taking into account only the stars with negligible
proper motion.} and properly identify the stars with larger proper 
motions\footnote{That would otherwise significantly distort the 
astrometric solution.}. Such iterative invocation scheme is used
frequently in case of the reduction of follow-up photometry data
(see Chapter~\ref{chapter:followup} and Sec.~\ref{sec:hatp2:photometric} 
for some other practical details). The simple direct application
of \texttt{grmatch} and \texttt{grtrans} as a part of 
a complete photometric pipeline is displayed in 
Fig.~\ref{fig:apphotexample}.

\phdsubsection{Transforming and registering images -- \texttt{fitrans}}
\label{sec:prog:fitrans}

As it was discussed earlier (Sec.~\ref{subsec:convolution}), the 
image convolution and subtraction process requires the images
to be in the same spatial reference system. The details of this
registration process have been explained already in 
Sec.~\ref{subsec:imageregistering}. The purpose of the 
program \texttt{fitrans} is to implement these various image interpolation
methods.

In principle, \texttt{fitrans} reads an image and a transformation
file, performs the spatial transformation and writes the output image
to a separate file. Image data are read from FITS files while
the transformation files are presumably derived from the appropriate 
astrometric solutions. The output of the \texttt{grmatch} and 
\texttt{grtrans} programs can be directly passed to \texttt{fitrans}.
Of course, \texttt{fitrans} takes into account the masks associated
to the given image as well as derive the appropriate mask for the
output file. Pixels which cannot be mapped from the original image
have always a value of zero and these are marked
as \emph{outer pixels} (see also Sec.~\ref{subsec:masking}).

In the HATNet data reduction, this spatial transformation requires
significant amount of CPU time since the exact integration on 
biquadratic interpolation surfaces is a computationally expensive 
process (Sec.~\ref{sec:conservingflux}). However, distinct image 
transformations can be performed independenlty (i.e. a given transformation 
does not have any influence on another transformations), thus the 
complete registration process can easily be performed in parallel. 

\phdsubsection{Convolution and image subtraction -- \texttt{ficonv}}
\label{sec:prog:ficonv}

This member of the \fihat{} package is intended to implement the
tasks related to the kernel fit, image convolution and subtraction. 
In principle, \texttt{ficonv} has two basic modes. First, assuming an existing 
kernel solution, it evaluates \eqref{eq:convolutionbasic} on 
an image and writes the convolved result to a separate image file.
Second, assuming a base set of kernel functions 
(equation~\ref{eq:kernelcoeffs}) and some model for the 
background variations (equation~\ref{eq:backgroundcoeffs})
it derives the best fit kernel solution for \eqref{eq:convolutionbasic},
described by the coefficients $C_{ik\ell}$ and $B_{k\ell}$, respectively.
Since this fit yields a linear equation for these coefficients, the
method of classic linear least squares minimization can be efficiently
applied. However, the least squares matrix can have a relatively
large dimension in the cases where the kernel basis is also large and/or
higher order spatial variations are allowed. In the fit mode,
the program yields the kernel solution, and optionally
the convolved ($C=B+R\star K$) and the subtracted residual image ($S=I-C$)
can also be saved into separate files
without additional invocations of \texttt{ficonv}
and/or \texttt{fiarith}.

The program \texttt{ficonv} also implements the fit for cross-convolution
kernels (equation~\ref{eq:crossconvolutionbasic}). In 
this case, the two kernel solutions are saved to two distinct files.
Subsequent invocations of \texttt{ficonv} and/or \texttt{fiarith} 
can then be used to analyze various kinds of outputs.

In Sec.~\ref{subsec:subtractedphotometry} we were discussing the
relevance of the kernel solution in the case when the photometry
is performed on the residual (subtracted) images. The best fit 
kernel solution obtained by \texttt{ficonv} has to be directly passed
to the program \texttt{fiphot} (Sec.~\ref{sec:prog:fiphot}) in order
to properly take into account the convolution information during the 
photometry (equation~\ref{eq:subflux}). 

\begin{figure*}
\begin{lcmd}
\small
SELF=\$0; base="\$1" \\
if [ -n "\$base" ] ; then \\
\ptab	fitrans	\$\{FITS\}/\$base.fits $\backslash$ \\
\ptab	\ptab	--input-transformation \$\{AST\}/\$base.trans --reverse -k -o \$\{REG\}/\$base-trans.fits  \\
else \\
\ptab	pexec -f BASE.list -e base -o - -u - -c -- "\$SELF $\backslash$\$base" \\
fi
\end{lcmd}
\begin{lcmd}
\small
SELF=\$0; base="\$1" \\
if [ -n "\$base" ] ; then \\
\ptab	KERNEL="i/4;b/4;d=3/4" \\
\ptab	ficonv ~--reference ./photref.fits $\backslash$ \\
\ptab	\ptab	--input \$\{REG\}/\$base-trans.fits --input-stamps ./photref.reg --kernel "\$KERNEL" $\backslash$ \\
\ptab	\ptab	--output-kernel-list \$\{AST\}/\$base.kernel --output-subtracted \$\{REG\}/\$base-sub.fits
else \\
\ptab	pexec -f BASE.list -e base -o - -u - -c -- "\$SELF $\backslash$\$base" \\
fi
\end{lcmd}
\caption{Two shell scripts demonstrating the invocation syntax of
the \texttt{fitrans} and \texttt{ficonv}. Since the computation of
the transformed and convolved images require significant amount of CPU
time, the utility \texttt{pexec} (\texttt{http://shellpexec.sf.net})
is used to run the jobs in parallel on multiple CPUs.}
\label{fig:imgsubexample}
\end{figure*}

\phdsubsection{Photometry -- \texttt{fiphot}}
\label{sec:prog:fiphot}

The program \texttt{fiphot} is the main code in the \fihat{} package
that performs the raw and instrumental photometry. In the
current implementation, we were focusing on the aperture photometry,
performed on normal and subtracted images. Basically, \texttt{fiphot}
reads an astronomical image (FITS file) and a centroid list file, 
where the latter
should contain not only the centroid coordinates but the individual 
object identifiers as well\footnote{If the proper object identification
is omitted, \texttt{fiphot} assigns some arbitrary (but indeed unique)
identifiers to the centroids, however, in practice it is almost useless.}.

In case of image subtraction-based photometry, \texttt{fiphot}
requires also the kernel solution (derived by \texttt{ficonv}). Otherwise,
if this information is omitted, the results of the photometry are not
reliable and consistent. See also Sec.~\ref{subsec:subtractedphotometry}
for further details about this issue.

In Fig.~\ref{fig:apphotexample}, a complete shell script
is displayed, as an example of various \fihat{} programs related
to the photometry process.

Currently, PSF photometry is not implemented directly in the 
program \texttt{fiphot}. However, the program \texttt{fistar} 
(Sec.~\ref{sec:prog:fistar}) is capable to do PSF fitting on 
the detected centroids, although its output is not compatible
with that of \texttt{fiphot}. Alternatively, \texttt{lfit}
(see Sec.~\ref{sec:prog:lfit}) can be used to perform profile fitting,
if the pixel intensities are converted to ASCII tables in 
advance\footnote{The program
\texttt{fiinfo} is capable to produce such tables with three columns:
a list of $x$ and $y$ coordinates followed by the respective pixel 
intensitie.}, however, it is not computationally efficient. 

\begin{figure*}
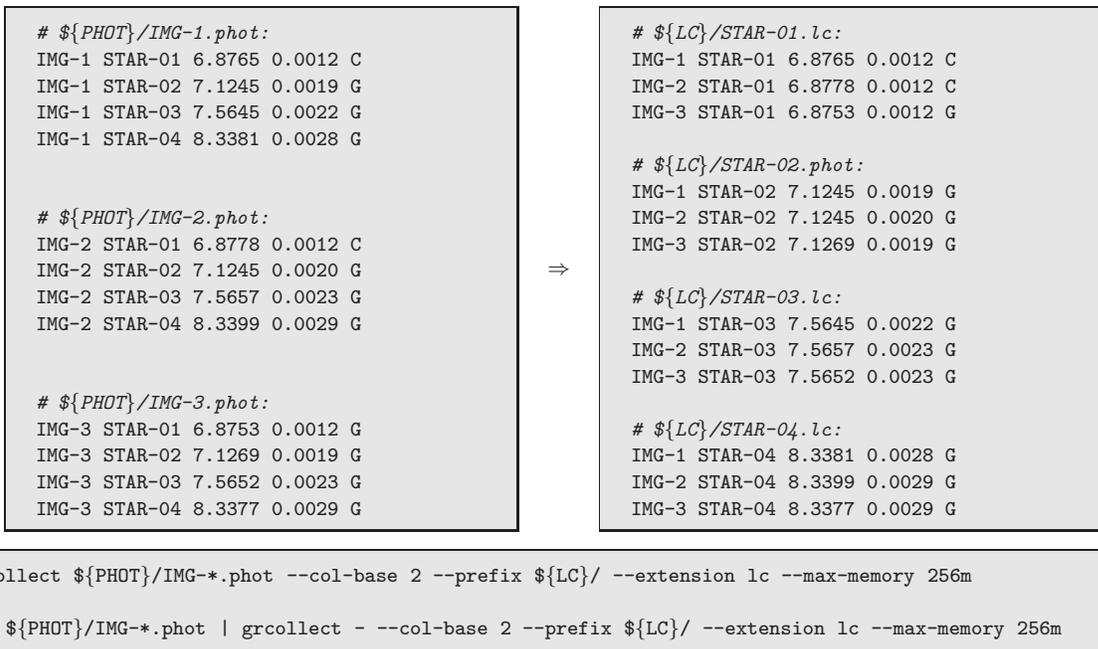

\begin{center}
\begin{ccmd}
\small
\textit{\# \$\{PHOT\}/IMG-1.phot:} \\
IMG-1 STAR-01 6.8765 0.0012 C \\
IMG-1 STAR-02 7.1245 0.0019 G \\
IMG-1 STAR-03 7.5645 0.0022 G \\
IMG-1 STAR-04 8.3381 0.0028 G \\
~ \\
~ \\
\textit{\# \$\{PHOT\}/IMG-2.phot:} \\
IMG-2 STAR-01 6.8778 0.0012 C \\
IMG-2 STAR-02 7.1245 0.0020 G \\
IMG-2 STAR-03 7.5657 0.0023 G \\
IMG-2 STAR-04 8.3399 0.0029 G \\
~ \\
~ \\
\textit{\# \$\{PHOT\}/IMG-3.phot:} \\
IMG-3 STAR-01 6.8753 0.0012 G \\
IMG-3 STAR-02 7.1269 0.0019 G \\
IMG-3 STAR-03 7.5652 0.0023 G \\
IMG-3 STAR-04 8.3377 0.0029 G 
\end{ccmd}\hspace*{3.20ex}$\Rightarrow$\hspace*{3.20ex}%
\begin{ccmd}
\small
\textit{\# \$\{LC\}/STAR-01.lc:} \\
IMG-1 STAR-01 6.8765 0.0012 C \\
IMG-2 STAR-01 6.8778 0.0012 C \\
IMG-3 STAR-01 6.8753 0.0012 G \\
~ \\
\textit{\# \$\{LC\}/STAR-02.phot:} \\
IMG-1 STAR-02 7.1245 0.0019 G \\
IMG-2 STAR-02 7.1245 0.0020 G \\
IMG-3 STAR-02 7.1269 0.0019 G \\
~ \\
\textit{\# \$\{LC\}/STAR-03.lc:} \\
IMG-1 STAR-03 7.5645 0.0022 G \\
IMG-2 STAR-03 7.5657 0.0023 G \\
IMG-3 STAR-03 7.5652 0.0023 G \\
~ \\
\textit{\# \$\{LC\}/STAR-04.lc:} \\
IMG-1 STAR-04 8.3381 0.0028 G \\
IMG-2 STAR-04 8.3399 0.0029 G \\
IMG-3 STAR-04 8.3377 0.0029 G   
\end{ccmd}
\begin{lcmd}
\small
grcollect \$\{PHOT\}/IMG-*.phot --col-base 2 --prefix \$\{LC\}/ --extension lc --max-memory 256m \\
~ \\
cat \$\{PHOT\}/IMG-*.phot | grcollect - --col-base 2 --prefix \$\{LC\}/ --extension lc --max-memory~256m
\end{lcmd}
\end{center}
\caption{The schematics of the data transposition. Records for
individual measurements are written initially to photometry files
(having an extension of \texttt{*.phot}, for instance). These records
contain the source identifiers. During the transposition, photometry 
files are converted to light curves. In principle, these light curves
contain the same records but sorted into distinct files by 
the object names, not the frame identifiers. The command lines on
the lower panel show some examples how this data transposition
can be employed involving the program \texttt{grcollect}.}
\label{fig:transpositionbasics}
\end{figure*}

\phdsubsection{Transposition of tabulated data -- \texttt{grcollect}}
\label{sec:prog:grcollect}

Raw and instrumental photometric data obtained for each frame are 
stored in separate files by default as it was discussed earlier 
(see Sec.~\ref{subsec:photometry}, Sec.~\ref{subsec:subtractedphotometry} 
and Sec.~\ref{sec:prog:fiphot}).
We refer to these files as \emph{photometric files}. In order to analyze 
the per-object outcome of our data reductions, one has to have
the data in the form of \emph{light curve files}. Therefore,
the step of photometry (including the magnitude transformation)
is followed immediately by the step of \emph{transposition}. See
Fig.~\ref{fig:transpositionbasics} about how this step looks like in
a simple case of 3 photometric files and 4 objects. 

The main purpose of the program \texttt{grcollect} is to perform
this transposition on the photometric data in order to have 
the measurements being stored in the form of light curves and therefore
to be adequate for further per-object analysis (such as 
light curve modelling). The invocation syntax
of \texttt{grcollect} is also shown in Fig.~\ref{fig:transpositionbasics}.
Basically, small amount of information is needed 
for the transposition process: the name of the input files, the 
index of the column in which the object identifiers are stored and the optional
prefixes and/or suffixes for the individual light curve file names.
The maximum memory that the program is allowed to use is also specified
in the command line argument. In fact, \texttt{grcollect} does not 
need the original data to be stored in separate files. The second example
on Fig.~\ref{fig:transpositionbasics} shows an alternate
way of performing the transposition, namely 
when the whole data is read from the standard input (and the 
preceding command of \texttt{cat} dumps all the data to the standard output,
these two commands are connected by a single uni-directional pipe).

\begin{figure}
\begin{center}
\noindent
\resizebox{80mm}{!}{\includegraphics{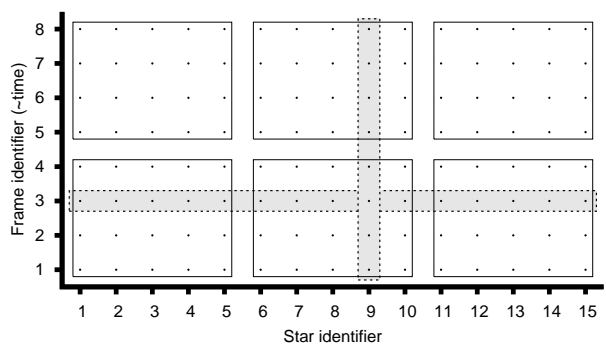}}
\end{center}
\caption{Storage schemes for photometric data. Supposing a series of 
frames, on which nearly the same set of stars have individual photometric
measurements, the figure shows how these data can be arranged 
for practical usages. The target stars (their identifiers) are 
arranged along the abscissa while the ordinate shows the frame identifiers
to which individual measurements (symbolized by dots) belong.
Raw and instrumental photometric data are therefore represented here
as rows (see the marked horizontal stripe for frame \#3, for instance) while 
the columns refer to light curves. In practice, native ways of transposition 
are extremely ineffective if the total amount of data does not fit into
the memory. The transposition can be speeded up by using an intermediate
stage of data storage, so-called macroblocks. In the figure, each macroblock
is marked by an enclosing rectangle. See text for further details. }
\label{fig:macroblock}
\end{figure}

The actual implementation of the transposition inside \texttt{grcollect}
is very simple: it reads the data from the individual files (or from
the standard input) until the data fit in the available memory. 
If this temporary memory is full of records, this array is sorted by 
the object identifier and 
the sorted records are written/concatenated to distinct files.
The output files are named based on the appropriate object identifiers. 
This procedure is repeated until there are available data. 
Although this method creates the light curve files, it means
that neither the whole process nor the access 
to these light curve files is effective. In case of HATNet, 
when we have thousands of frames in a single reduction and there
are several tens or hundreds of thousands individual stars that are
intended to have photometric measurements and each 
record is quite long\footnote{A record for a single photometric
measurement is several hundreds of bytes long since it contains
information for multiple apertures (including flux error estimations
and quality flags) as well as there are additional
fields for the stellar profile parameters and other observational
quantities used in further trend filtering.}, the total amount 
of data is in the order of hundreds of gigabytes. For even 
modern present-day computers, such a large
amount of data does not fit in the memory. Therefore, referring to
the simple process discussed above, light curve files are not written
to the disk at once but in smaller chunks. These chunks are located on
different cylinders of the disk: files are therefore extremely fragmented.
Both the creation and the access of these fragmented files are
extremely inefficient, since fragmented files require additional
highly time-consuming disk operations such as random seeks between cylinders. 
In practice, even on modern computers (being used by
the project), the whole
process requires a day or so to be completed, although the sequential
access to some hundreds of gigabytes of data would require only an
hour or a few hours (with a plausible I/O bandwidth of $\sim 50$\,MB/sec).
In order to overcome this problem, one 
can either use an external database engine that features optimizations
for such two-dimensional queries or tweak the above transposition 
algorithm to avoid unexpected and/or expensive disk operations. 
Now we briefly summarize an approach how the transposition can be made
more effective if we consider some assumptions for the data structure.
The program \texttt{grcollect} is capable to do transpositions
even if some of the keys (stellar identifiers) are missing or if
there are more than one occurrences for a single key in a given file. 
Let us assume that 1) in each input file every stellar identifier is unique
and 2) the number of missing keys is negligible compared to the total
number of photometric data records\footnote{Each record
represents a single photometric measurement for a single instant, including
all additional relevant data (such as the parameters involved in
the EPD analysis, see earlier)}. 
Assuming a total of $N_{\rm F}$ frames and $N_{\star}$
unique stellar identifiers (in the whole photometric data), the total number
of records is $N_{\rm R}\lesssim N_{\rm F}N_\star$. The total memory capacity of 
the computer is able to store $M$ records simultaneously. Let us
denote the average disk seek time by $\tau$ and the sequential
access speed by $\omega$ (in the units of records per second).
The transposition can then be performed effectively in two stages.
In the first stage the photometry files are converted to individual 
files, so-called macroblocks, where each of them is capable to
store $(M/N_{\rm F})\times (M/N_\star)$ records, each macroblock
represent a continuous rectangle in the stellar identifier -- frame
space (see Fig.~\ref{fig:macroblock}).
In the second stage, macroblock files are converted into light curves. 
Due to the size of the macroblock, $MN_{\rm F}/N_\star$ photometric
files can be read up sequentially and stored in the memory at the same time. 
If the relation 
\begin{equation}
1 \ll \frac{M^2}{\tau N_{\rm f}N_\star\omega} \label{eq:macroblockefficiency}
\end{equation}
is true for the actual values of $M$, $N_{\rm f}$, $N_\star$, $\omega$ 
and $\tau$, the macroblocks can be accessed randomly after the first stage
(independently from the order in which they have been written to the disk),
without too much dead time due to the random seeks.
Therefore, at the second stage when macroblocks are read in the appropriate
order of the stellar identifiers, $MN_\star/N_{\rm F}$ light curves
can be flushed simultaneously without any additional disk 
operations beyond sequential writing. 

\begin{table*}
\caption{Algorithms supported 
by \texttt{lfit} and their respective 
requirements for the model function. The first column refers
to the internal and command line identifier of the algorithms. The second
column shows whether the method requires the parametric derivatives of the 
model functions in an analytic form or not. The third column indicates
whether in the cases when the method requires parametric derivatives, 
should the model function be linear in all of the parameters.
}\label{table:lfitmethods}
\begin{center}
\footnotesize
\begin{tabularx}{160mm}{lllX}
\hline
Code		& derivatives		& linearity		& Method or algorithm	\\
\hline
\texttt{L/CLLS}	& yes			& yes			& Classic linear least squares method	\\
\texttt{N/NLLM}	& yes			& no			& (Nonlinear) Levenberg-Marquardt algorithm \\
\texttt{U/LMND}	& no			& no			& Levenberg-Marquardt algorithm employing numeric parametric derivatives \\
\texttt{M/MCMC}	& no			& no			& Classic Markov Chain Monte-Carlo algorithm$^{1}$	\\
\texttt{X/XMMC}	& yes			& no			& Extended Markov Chain Monte-Carlo$^{2}$ \\
\texttt{K/MCHI}	& no			& no			& Mapping the values $\chi^2$ on a grid	(a.k.a. ``brute force'' minimization) \\
\texttt{D/DHSX}	& optional$^{3}$	& no			& Downhill simplex \\
\texttt{E/EMCE}	& optional$^{4}$	& optional$^{4}$	& Uncertainties estimated by refitting to synthetic data sets	\\
\texttt{A/FIMA}	& yes			& no			& Fisher Information Matrix Analysis \\
\hline
\end{tabularx}
\end{center}
\begin{minipage}{\textwidth}
\scriptsize
\noindent \hspace*{1ex} $^{1}$
	The implemented transition function is based on the
	Metropolitan-Hastings algorithm and the optional Gibbs sampler.
	The transition amplitudes must be specified initially. Iterative
	MCMC can be implemented by subsequent calls of \texttt{lfit}, involving
	the previous inverse statistical variances for each parameters as 
	the transition amplitudes for the next chain.

\noindent \hspace*{1ex} $^{2}$
	The also program reports the summary related to the sanity 
	checks (such as correlation lengths, Fisher covariance, statistical
	covariance, transition probabilities and the best fit value obtained
	by an alternate /usually the downhill simplex/ minimization).

\noindent \hspace*{1ex} $^{3}$
	The downhill simplex algorithm may use the parametric derivatives to
	estimate the Fisher/covariance matrix for the initial conditions
	in order to define the control points of the initial simplex.
	Otherwise, if the parametric derivatives do not exist, the 
	user should specify the ``size'' of the initial simplex somehow
	in during the invocation of \texttt{lfit}.

\noindent \hspace*{1ex} $^{4}$
	Some of the other methods (esp. CLLS, NLLM, DHSX, in practice)
	can be used during the minimization process of the orignal data
	and the individual synthetic data sets.
\end{minipage}
\end{table*}

In the case of the computers used in HATNet data reduction, 
$M\approx 10^7$, $N_{\rm f}\approx 10^4$,
$N_\star\approx 10^5$, $\omega \approx 10^5$\,records/sec and $\tau \approx
10^{-2}$\,sec, the right-hand side of \eqref{eq:macroblockefficiency}
is going to be $\approx 10^2$, so the discussed way of two-stage transposition
is very efficient. Indeed, the whole operation
can be completed within $3-5$\,hours, instead of a day or few days 
that is needed by the normal one-stage transposition. Moreover, due
to the lack of random seeks, the computer itself remains responsible
for the user interactions. In the case of one-stage transposition,
the extraordinary amount of random seeks inhibit almost any 
interactive usage.

\phdsubsection{Archiving -- \texttt{fizip} and \texttt{fiunzip}}
\label{sec:prog:fizip}

Due to the large disk space required to store the raw, calibrated
and the derived (registered and/or subtracted) frames,
it is essential to compress and archive the image files that are barely used.
The purpose of the \prog{fizip} and \prog{fiunzip} programs is to compress and 
decompress primary FITS data, by keeping the changes in the primary FITS
header to be minimal. The compressed data is stored in a one-dimensional
8 bit (\texttt{BITPIX=8}, \texttt{NAXIS=1}) array, therefore these 
keywords does not reflect the original image dimension or data type. 

All of the other keywords
are untouched. Some auxiliary information on the compression is stored in
the keywords starting with ``\texttt{FIZIP}'', the contents of these keywords
depend on the involved compression method. \prog{fizip} rejects compressing
FITS file where such keywords exist in the primary header.

In practice, \prog{fizip} and \prog{fiunzip} refer to the same
program (namely, \prog{fiunzip} is a symbolic link to \prog{fizip})
since the algorithms involved in the compression and decompression 
refer to the same codebase or external library. \prog{fizip} and \prog{fiunzip}
support well known compression algorithms, such as the GNU zip (``gzip'')
and the block-sorting file compressor (also known as ``bzip2'') algorithm.

These compression algorithms are lossless. However, \prog{fizip}
supports rounding the input pixel values to the nearest integer 
or to the nearest fraction of some power of 2. Since the common representation
of floating-point real numbers yields many zero bits if the number itself
is an integer or a multiple of power of 2 (including fractional 
multiples), the compression is more effective if this kind of 
rounding is done before the compression. This ``fractional rounding''
yields data loss. However, if the difference between the original and 
the rounded values are comparable or less than the readout noise
of the detector, such compression does not affect the quality of the further
processing (e.g. photometry).

\phdsubsection{Generic arithmetic evaluation, regression and data analysis -- \texttt{lfit}}
\label{sec:prog:lfit}

Modeling of data is a prominent step in the analysis 
and interpretation of astronomical observations. In 
this section, a standalone command line driven tool, 
named \texttt{lfit} is introduced, designed for both interactive 
and batch processed regression analysis as well as generic 
arithmetic evaluation. 

This tool is built on the top of the \texttt{libpsn}
library\footnote{http://libpsn.sf.net, developed by the author}, 
a collection of functions managing symbolic arithmetic expressions. 
This library provides 
both the back-end for function evaluation as well as analytical calculations
of partial derivatives. Partial derivatives are required by most of 
the regression methods 
(e.g. linear and non-linear least squares fitting) and uncertainty 
estimations (e.g. Fisher analysis). The program features many built-in
functions related to special astrophysical problems. Moreover, it allows
the end-user to extend the capabilities during run-time using dynamically
loaded libraries.

In general, \texttt{lfit} is used extensively in the data reduction
steps of the HATNet project. The program acts both in the main 
``discovery'' pipeline and it is involved in the  
characterization of follow-up data, including photometric and radial 
velocity measurements. Currently, \texttt{lfit} implements
executively the EPD algorithm (including the normal, the 
reconstructive and the simultaneous modes) as well as the
simultaneous TFA algorithm \citep[see e.g.][]{bakos2009hatp11}.

\phdsubsubsection{User interface and built-in regression methods}
Due to the high modularization and freedom in its user interface,
the program \texttt{lfit} allows the user to compare the results of different 
regression analysis techniques. The program features 9 built-in algorithms 
at the moment, including the classic linear least squares minimization
\citep{press1992}, the non-linear methods 
\citep[Levenberg-Marquard, downhill simplex, see also][]{press1992},
various methods providing an \emph{a posteriori} distribution
for the adjusted parameters, such as Markov Chain Monte-Carlo
\citep{ford2004}, or the method of refitting to synthetic data sets
\citep{press1992}. The program is also capable to derive the 
covariance or correlation matrix of the parameters involving
the Fisher information analysis \citep{finn1992}. The comprehensive
list of the supported algorithms can be found in Table~\ref{table:lfitmethods}.

The basic concepts of \texttt{lfit} is shown in 
Fig.~\ref{fig:lineexample} in a form of a complete example for
linear regression.

\begin{table*}
\caption{Basic functions found in the built-in astronomical 
extension library. These
functions cover the fields of simple radial velocity analysis,
some aspects of light curve modelling and data reduction. These functions
are a kind of ``common denominators'', i.e. they do not provide a direct
possibility for applications but complex functions can be built on the
top of them for any particular usage. All of the functions below
with the exception of \texttt{hjd()} and \texttt{bjd()} 
have partial derivatives that can be evaluated 
analytically by \texttt{lfit}.}
\label{table:lfitastromext}
\begin{center}
\footnotesize
\begin{tabularx}{\textwidth}{lX}
\hline
Function	& Description	\\
\hline
${\tt hjd}({\rm JD},\alpha,\delta)$	&
	Function that calculates the 
	heliocentric Julian date from the Julian day $J$ and
	the celestial coordinates $\alpha$ (right ascension)
	and $\delta$ (declination). \\

${\tt bjd}({\rm JD},\alpha,\delta)$	&
	Function that calculates the 
	barycentric Julian date from the Julian day $J$ and
	the celestial coordinates $\alpha$ (right ascension)
	and $\delta$ (declination). \\
${\tt ellipticK}(k)$		&
	Complete elliptic integral of the first kind.	 \\
${\tt ellipticE}(k)$		&
	Complete elliptic integral of the second kind.	 \\
${\tt ellipticPi}(k,n)$		&
	Complete elliptic integral of the third kind.	 \\
${\tt eoq}(\lambda,k,h)$	&
	Eccentric offset function, `q` component. The arguments
	are the mean longitude $\lambda$, in radians and the 
	Lagrangian orbital elements $k=e\cos\varpi$, $h=e\sin\varpi$. \\
${\tt eop}(\lambda,k,h)$	&
	Eccentric offset function, `p` component. \\
${\tt ntiu}(p,z)$		&
	Normalized occultation flux decrease. This function calculates the
	flux decrease during the eclipse of two spheres when one of the
	spheres has uniform flux distribution and the other one by which
	the former is eclipsed is totally dark. The bright source is assumed
	to have a unity radius while the occulting disk has a radius of $p$.
	The distance between the centers of the two disks is $z$.	\\
${\tt ntiq}(p,z,\gamma_1,\gamma_2)$		&
	Normalized occultation flux decrease when eclipsed sphere has a 
	non-uniform flux distribution modelled by quadratic limb darkening law. 
	The limb darkening is characterized 
	by $\gamma_1$ and $\gamma_2$. \\
\hline
\end{tabularx}
\end{center}
\end{table*}

\phdsubsubsection{Built-in functions related to astronomical data analysis}
The program \texttt{lfit} provides various built-in functions related
to astronomical data analysis, especially ones that are required by
exoplanetary research. All of these functions are some sort of 
``base functions'', with a few parameters from which one can
easily form more useful ones using these capabilities of \texttt{lfit}.
Good examples are the eccentric offset functions $\eop(\lambda,k,h)$
and $\eoq(\lambda,k,h)$ (Sec.~\ref{sec:hatp2:eof}), that 
have only three parameters but the functions
related to the radial velocity analysis can easily be defined
using these two functions. The full list of these special 
functions can be found in Table~\ref{table:lfitastromext}. The 
actual implementation of the above mentioned radial velocity
model functions can be found in Chapter~\ref{chapter:followup},
in Fig.~\ref{fig:lfitrvmacros}.

\phdsubsubsection{Extended Markov Chain Monte-Carlo}
In this section we discuss in more details 
one of the built-in methods, 
that combines a Markov Chain Monte-Carlo algorithm with the parametric 
derivatives of the model functions in order to yield faster convergence 
and more reliable results, especially in the cases of highly correlated 
parameters. 

The main concept of the MCMC algorithm \citep[see e.g.][]{ford2004},
is to generate an \emph{a posteriori} probability distribution of the
adjusted parameters. It is based on random walks in the parameter
space as follows. In each step, one draws an alternate parameter vector from
an \emph{a priori} distribution and then evaluates the merit function
$\chi^2$. If the value of the $\chi^2$ decreases, we accept the
transition (since the newly drawn parameter vector represents 
a better fit), otherwise the transition is accepted by a certain
probability (derived from the increment in $\chi^2$).
The final distribution of the parameters depends on both the 
\emph{a priori} distribution and the probability function used 
when the value of the $\Delta \chi^2$ is positive. The main problem
of the MCMC method is that the \emph{a posteriori} probability
distribution can only be estimated if the \emph{a priori} 
distribution is chosen well, but initially we do not have any hint
for both distributions. The idea behind MCMC is to derive multiple
chains, by taking the \emph{a posteriori} distribution of the
previous chain as the input (\emph{a priori}) distribution for the upcoming
chain. In regular cases, the chains converge to a final distribution
after some iterations and therefore the last one can be accepted as
a final result. In the literature, several
attempts are known to define an \emph{a priori} transition
function \citep[see also][]{ford2004}. 
Here we give a simple method that not only provides a 
good hint for the \emph{a priori} distribution but yields several
independent sanity checks that are then used to verify the convergence
of the chain. The transition function used by this extended Markov Chain
Monte-Carlo algorithm (XMMC) is a Gaussian distribution of which 
covariances are derived from the Fisher covariance matrix \citep{finn1992}. 
The sanity checks are then the following:
\begin{itemize}
\item The resulted parameter
distribution should have nearly the same statistical covariance as the 
analytical covariance\footnote{In practice,
the program \texttt{lfit} reports the individual uncertainties of the 
parameters and the correlation matrix. Of course, this information
can easily be converted to a covariance matrix and vice versa.}.
\item The autocorrelation lengths of the chain parameters have to be 
small, i.e. nearly $\sim 1 - 2$ steps. Chains failed to converge have
significantly larger autocorrelation lengths. 
\item The transition probability has to be consistent with the
theoretical probabilities. This theoretical probability depends only
on the number of adjusted parameters. 
\item The statistical centroid (mode) of the distribution must agree with
both the best fit parameter derived from alternate methods (such as
downhill simplex) as well as the chain element with the smallest $\chi^2$.
\end{itemize}
The method of XMMC has some disadvantages. First, the 
transition probabilities exponentially
decrease as the number of adjusted parameters increases, therefore,
the required computational time can be exceptionally high in some cases.
The Gibbs sampler (used in the classic MCMC) 
provides roughly constant transition probability. 
Second, the derivation of the Fisher covariance matrix 
requires the knowledge of the parametric derivatives of the merit 
function. In the actual implementation of \texttt{lfit}, XMMC
one can use the method of XMMC if the parametric derivatives
are known in advance in an analytical form. Otherwise, the
XMMC algorithm cannot be applied at all. 

\begin{figure*}
\begin{lcmd}
\small
	\# This command just prints the content of the file ``line.dat'' to the standard output: \\
	\$ \textit{cat line.dat} \\
	2 ~8.10 \\
	3 10.90 \\
	4 14.05 \\
	5 16.95 \\
	6 19.90 \\
	7 23.10 \\
	\# Regression: this command fits a ``straight line'' to the above data: \\
	\$ \textit{lfit -c x,y -v a,b -f "a*x+b" -y y line.dat} \\
	\hspace*{0ex}~~~~~2.99714~~~~~~2.01286 \\
	\# Evaluation: this command evaluates the model function assuming the parameters to be known: \\
	\$ \textit{lfit -c x,y -v a=2.99714,b=2.01286 -f "x,y,a*x+b,y-(a*x+b)" -F \%6.4g,\%8.2f,\%8.4f,\%8.4f line.dat} \\
	\hspace*{0ex}~~~~~  2~~~~~~~~~8.10~~~~~~8.0071~~~~~~0.0929 \\
	\hspace*{0ex}~~~~~  3~~~~~~~~10.90~~~~~11.0043~~~~~-0.1042 \\
	\hspace*{0ex}~~~~~  4~~~~~~~~14.05~~~~~14.0014~~~~~~0.0486 \\
	\hspace*{0ex}~~~~~  5~~~~~~~~16.95~~~~~16.9986~~~~~-0.0486 \\
	\hspace*{0ex}~~~~~  6~~~~~~~~19.90~~~~~19.9957~~~~~-0.0957 \\
	\hspace*{0ex}~~~~~  7~~~~~~~~23.10~~~~~22.9928~~~~~~0.1072 \\
	\$ \textit{lfit -c x,y -v a,b -f "a*x+b" -y y line.dat --err} \\
	\hspace*{0ex}~~~~~2.99714~~~~~~2.01286 \\
	\hspace*{0ex}~~~0.0253144~~~~~0.121842 
\end{lcmd}
\caption{These pieces of commands show the two basic operations of 
\texttt{lfit}: the first invocation of \texttt{lfit} fits a straight
line, i.e. a model function with the form of $ax+b=y$ to the data
found in the file \texttt{line.dat}. This file is supposed to contain
two columns, one for the $x$ and one for the $y$ values.
The second invocation of \texttt{lfit} evaluates the model function.
Values for the model parameters ($a$, $b$) are taken from the command line
while the individual data points ($x$, $y$) are still read from the 
data file \texttt{line.dat}. The evaluation mode allows the user to 
compute (and print) arbitrary functions of the model parameters \emph{and}
the data values. In the above example, the model function itself
and the fit residuals are computed and printed, following the read
values of $x$ and $y$. Note that the printed values are formatted for
a minimal number significant figures (\%6.4g) or for a fixed number of  
decimals (\%8.2f or \%8.4f).
The last command is roughly the same as the first command for regression,
but the individual uncertainties are also estimated by normalizing
the value of the $\chi^2$ to unity.}
\label{fig:lineexample}
\end{figure*}

However, in the case of HATNet data analysis, we found the method of 
XMMC to be highly efficient and we used it in several analyses
related to the discoveries. Moreover, the most important functions
concerning to this analysis, such as light curve and radial velocity
model functions have known analytic partial derivatives. These
derivatives for transit light curve model functions can be 
found in \cite{pal2008lcdiff}. An analytic formalism for radial velocity
modelling is discussed in Sec.~\ref{sec:hatp2:eof} and some
additional related details and applications 
are presented in \cite{pal2009kepler}.
In this thesis (in Chapter~\ref{chapter:hatnetdiscovery})
a detailed example is given on the application of the XMMC algorithm
in the analysis of the HAT-P-7(b) planetary system.


\begin{figure}
\begin{center}
\scriptsize
\setlength{\unitlength}{1mm}
\setlength{\fboxsep}{0pt}
\begin{picture}(80,150)(0,-150)

\thicklines
\put(20,-10){\colorbox[gray]{0.9}{\framebox(40,5){\textbf{raw images}}}}
\put(40,-10){\vector(0,-1){5}}
\put(25,-20){\framebox(30,5){Calibration}}
\put(40,-20){\vector(0,-1){5}}
\put(20,-30){\colorbox[gray]{0.9}{\framebox(40,5){\textbf{calibrated images}}}}
\put(40,-30){\vector(0,-1){5}}
\put(25,-40){\framebox(30,5){Star detection}}
\put(40,-40){\vector(0,-1){5}}
\put(25,-50){\framebox(30,5){Astrometry}}

\put(30,-50){\vector(-1,-1){10}}
\put(5,-65){\framebox(30,5){Photometry}}
\put(20,-65){\vector(0,-1){5}}
\put(5,-75){\framebox(30,5){Magnitude transf.}}
\put(20,-75){\vector(1,-1){10}}

\put(50,-50){\vector(1,-1){5}}
\put(45,-60){\framebox(30,5){Image transf.}}
\put(60,-60){\vector(0,-1){5}}
\put(45,-70){\framebox(30,5){Convolution}}
\put(60,-70){\vector(0,-1){5}}
\put(45,-80){\framebox(30,5){Photometry}}
\put(55,-80){\vector(-1,-1){5}}

\put(20,-90){\colorbox[gray]{0.9}{\framebox(40,5){\textbf{instrumental photometry}}}}
\put(40,-90){\vector(0,-1){5}}
\put(25,-100){\framebox(30,5){Transposition}}
\put(40,-100){\vector(0,-1){5}}
\put(20,-110){\colorbox[gray]{0.9}{\framebox(40,5){\textbf{instrumental light curves}}}}
\put(40,-110){\vector(0,-1){5}}
\put(25,-120){\framebox(30,5){Trend filtering}}
\put(40,-120){\vector(0,-1){5}}
\put(20,-130){\colorbox[gray]{0.9}{\framebox(40,5){\textbf{light curves}}}}
\put(40,-130){\vector(0,-1){5}}
\put(25,-140){\framebox(30,5){Analysis${}^\ast$}}
\put(40,-140){\vector(0,-1){5}}
\put(20,-150){\colorbox[gray]{0.9}{\framebox(40,5){\textbf{results}}}}
\end{picture}
\end{center}
\caption{
Flowchart of the typical photometric reduction pipeline. Each empty box represents
a certain step of the data processing that requires non-negligible
amount of computing resources. Filled boxes represent the type of data
that is only used for further processing, thus the four major steps
of the reduction are clearly distinguishable. See text for further details.}
\label{fig:photometrypipeline}
\end{figure}

\phdsection{Analysis of photometric data}
\label{subsec:photometricanalysis}

In this section we describe briefly how the previously discussed algorithms
and the respective implementations are used in the practice of photometric
data reduction. The concepts for the major steps in the photometry are roughly 
the same for the HATNet and follow-up data, however, the latter has 
two characteristics that make the processing more convenient. First, 
the total amount of frames are definitely smaller, a couple of hundred frames
for a singe night or event, while there are thousands or tens of thousands
of frames for a typical observation of a certain HATNet field. 
Second, the number of stars on each individual frame is also smaller
(a few hundred instead of tens or hundreds of thousands).
Third, during the reduction of follow-up photometric data, we 
have an expectation for the signal shape. The signal can be easily 
obtained even by lower quality of data and/or when some of the 
reduction steps are skipped (e.g. trend filtering or a higher 
order magnitude transformation). 

The schematics of a typical photometric pipeline (as used
for HATNet data reductions) is shown in 
Fig.~\ref{fig:photometrypipeline}. It is clear from the figure that
the steps of the reduction are the same up to astrometry both in cases
when the fluxes are derived either by normal (aperture) photometry 
or image subtraction method. In the first case, the astrometric
solution is directly used to compute the aperture centroids for
all objects of interest, while in case of image subtraction, the
image registration parameters are based on astrometry. After the 
instrumental magnitudes are obtained, the process of the photometric
files (including transposition, trend filtering and 
per-object light curve analysis) are the same again. In practice,
both primary photometric methods yield fluxes for several apertures.
Therefore, joint processing of various photometric data is also feasible
since the subsequent steps do not involve additional information beyond
the instrumental magnitudes. The only exception is that additional
data can be involved in the 
EPD algorithm in case of image subtraction photometry. Namely,
the kernel coefficients $C_{ik\ell}$ can be added to the set of EPD
parameters $p^{(i)}$ (see equation~\ref{eq:epdbase}), by evaluating 
for the spatial variations of each object:
\begin{equation}
p^{(i)} = \sum\limits_{0\le k+\ell \le N^{(i)}_{\rm K}}C_{ik\ell}x^ky^\ell,
\end{equation}
where $(x,y)$ is the centroid coordinate of the actual object of interest.
In the following two chapters, I discuss how the above outlined
techniques are applied in the case of HATNet and follow-up data
reductions.

\begin{figure*}
\begin{lcmd}
\small
\#!/bin/sh \\
~ \\
CATALOG=input.cat~ \# name of the reference catalog \\
COLID=1~ ~ ~ ~ ~ ~ \# column index of object identifie (in the \$CATALOG file) \\
COLX=2 ~ ~ ~ ~ ~ ~ \# column index of the projected X coordinate (in the \$CATALOG file) \\
COLY=3 ~ ~ ~ ~ ~ ~ \# column index of the projected Y coordinate (in the \$CATALOG file) \\
COLMAG=4 ~ ~ ~ ~ ~ \# column index of object magnitude (in the \$CATALOG file) \\
COLCOLOR=5 ~ ~ ~ ~ \# column index of object color (in the \$CATALOG file) \\
THRESHOLD=4000 ~ ~ \# threshold for star detection \\
GAIN=4.2 ~ ~ ~ ~ ~ \# combined gain of the readout electronics and the A/D converter in electrons/ADU \\
MAGFLUX=10,10000 ~ \# magnitude/flux conversion \\
APERTURE=5:8:8 ~ ~ \# aperture radius, background area inner radius and thickness (all in pixels) \\
~ \\
mag\_param=c0\_00,c0\_10,c0\_01,c0\_20,c0\_11,c0\_02,c1\_00,c1\_01,c1\_10 \\
mag\_funct="c0\_00+c0\_10*x+c0\_01*y+0.5*(c0\_20*x\^{ }2+2*c0\_11*x*y+c0\_02*y\^{ }2)+color*(c1\_00+c1\_10*x+c1\_01*y)" \\
~ \\
for base in \$\{LIST[*]\} ; do \\
\ptab	fistar ~\$\{FITS\}/\$base.fits --algorithm uplink --prominence 0.0 --model elliptic $\backslash$ \\
\ptab	\ptab	--flux-threshold \$THRESHOLD --format id,x,y,s,d,k,amp,flux -o \$\{AST\}/\$base.stars \\
\ptab	grmatch --reference \$CATALOG --col-ref \$COLX,\$COLY --col-ref-ordering -\$COLMAG $\backslash$ \\
\ptab	\ptab	--input \$\{AST\}/\$base.stars --col-inp 2,3 --col-inp-ordering +8 $\backslash$ \\
\ptab	\ptab	--weight reference,column=\$COLMAG,magnitude,power=2 $\backslash$ \\
\ptab	\ptab	--triangulation maxinp=100,maxref=100,conformable,auto,unitarity=0.002 $\backslash$ \\
\ptab	\ptab	--order 2 --max-distance 1 $\backslash$ \\
\ptab	\ptab	--comment --output-transformation \$\{AST\}/\$base.trans || continue \\
\ptab	grtrans \$CATALOG --col-xy \$COLX,\$COLY --col-out \$COLX,\$COLY $\backslash$ \\
\ptab	\ptab	--input-transformation \$\{AST\}/\$base.trans --output - | $\backslash$ \\
\ptab	fiphot ~\$\{FITS\}/\$base.fits --input-list - --col-xy \$COLX,\$COLY --col-id \$COLID $\backslash$ \\
\ptab	\ptab	--gain \$GAIN --mag-flux \$MAGFLUX --aperture \$APERTURE --disjoint-annuli $\backslash$ \\
\ptab	\ptab	--sky-fit mode,iterations=4,sigma=3 --format IXY,MmBbS $\backslash$ \\
\ptab	\ptab	--comment --output \$\{PHOT\}/\$base.phot \\
\ptab	paste ~	\$\{PHOT\}/\$base.phot \$\{PHOT\}/\$REF.phot \$CATALOG | $\backslash$ \\
\ptab	lfit ~ ~--columns mag:4,err:5,mag0:12,x:10,y:11,color:\$((2*8+COLCOLOR)) $\backslash$ \\
\ptab	\ptab	--variables \$mag\_param --function "\$mag\_funct" --dependent mag0-mag --error err $\backslash$ \\
\ptab	\ptab	--output-variables \$\{PHOT\}/\$base.coeff \\
\ptab	paste ~ \$\{PHOT\}/\$base.phot \$\{PHOT\}/\$REF.phot | $\backslash$ \\
\ptab	lfit ~ ~--columns mag:4,err:5,mag0:12,x:10,y:11,color:\$((2*8+COLCOLOR)) $\backslash$ \\
\ptab	\ptab	--variables \$(cat \$\{PHOT\}/\$base.coeff) $\backslash$ \\
\ptab	\ptab	--function "mag+(\$mag\_funct)" --format \%9.5f --column-output 4 | $\backslash$ \\
\ptab	awk ~ ~ '\{ print \$1,\$2,\$3,\$4,\$5,\$6,\$7,\$8; \}' > \$\{PHOT\}/\$base.tphot \\
done \\
for base in \$\{LIST[*]\} ; do test -f \$\{PHOT\}/\$base.tphot \&\& cat \$\{PHOT\}/\$base.tphot ; done | $\backslash$ \\
grcollect - --col-base 1 --prefix \$LC/ --extension .lc
\end{lcmd}
\caption{A shell script demonstrating a complete working pipeline for
aperture photometry. The input FITS files are read from the directory
\texttt{\$\{FITS\}} and their base names (without the \texttt{*.fits} extension)
are supposed to be listed in the array \texttt{\$\{LIST[*]\}}. These
base names are then used to name the files storing data obtained during
the reduction process. Files created by the subsequent calls of 
the \texttt{fistar} and \texttt{grmatch} programs are related to the
derivation of the astrometric solution and the respective files
are stored in the directory \texttt{\$\{AST\}}. The photometry centroids
are derived from the original input catalog (found in the file \texttt{\$CATALOG})
and the astrometric transformation (plate solution, stored in the \texttt{*.trans})
files. The results of the photometry are put into the directory \texttt{\$\{PHOT\}}.
Raw photometry is followed by the magnitude transformation. This branch
involves additional common UNIX utilities such as \texttt{paste} and \texttt{awk}
in order to match the current and the reference photometry as well as 
to filter and resort the output after the magnitude transformation. 
The derivation of the transformation coefficients is done by the
\texttt{lfit} utility, that involves \texttt{\$mag\_funct} with the
parameters listed in \texttt{\$mag\_param}. This example features a quadratic
magnitude transformation and a linear color dependent correction 
(to cancel the effects of the differential refraction). The final light curves
are created by the \texttt{grcollect} utility what writes the individual
files into the directory \texttt{\$\{LC\}}.}
\label{fig:apphotexample}
\end{figure*}


\phdchapter{HATNet discoveries}
\label{chapter:hatnetdiscovery}

\newcommand{\hatcurA}{HAT-P-7}
\newcommand{\hatcurAb}{HAT-P-7b}

\newcommand{\hatcurACCra}{\ensuremath{19^{\mathrm{h}}28^{\mathrm{m}}59^{\mathrm{s}}.35}}	%
\newcommand{\hatcurACCdec}{\ensuremath{+47^{\circ}58'10''.2}}		%
\newcommand{\hatcurACCmag}{\ensuremath{9.85}}
\newcommand{\hatcurACCtwomass}{2MASS~19285935+4758102}
\newcommand{\hatcurACCgsc}{GSC~03547-01402}
\newcommand{\hatcurACCtassmv}{10.51}
\newcommand{\hatcurACCtassmvshort}{10.5}
\newcommand{\hatcurACCtassvi}{\ensuremath{0.60\pm0.07}}

\newcommand{\hatcurALCdip}{\ensuremath{7.0}}				
\newcommand{\hatcurALCrprstar}{\ensuremath{0.0761\pm0.0009}}		%
\newcommand{\hatcurALCimp}{\ensuremath{0.44^{+0.10}_{-0.15}}}		%
\newcommand{\hatcurALCdur}{\ensuremath{0.1625\pm0.0029}}			%
\newcommand{\hatcurALCingdur}{\ensuremath{0.0141\pm 0.0020}}		%
\newcommand{\hatcurALCP}{\ensuremath{2.2047298\pm0.0000024}}		%
\newcommand{\hatcurALCPprec}{\ensuremath{2.2047298}}			%
\newcommand{\hatcurALCPshort}{2.2047}					%
\newcommand{\hatcurALCT}{\ensuremath{2,453,785.8503\pm0.0008}}		
\newcommand{\hatcurALCMT}{\ensuremath{53,785.8503\pm0.0008}}		%

\newcommand{\hatcurASMEteff}{\ensuremath{6350\pm80}}			%
\newcommand{\hatcurASMEzfeh}{\ensuremath{+0.26\pm0.08}}			%
\newcommand{\hatcurASMElogg}{\ensuremath{4.06\pm0.10}}			%
\newcommand{\hatcurASMEvsin}{\ensuremath{3.8\pm0.5}}			%

\newcommand{\hatcurAYYm}{\ensuremath{1.49^{+0.06}_{-0.05}}}		%
\newcommand{\hatcurAYYmshort}{\ensuremath{1.49}}				%
\newcommand{\hatcurAYYmlong}{\ensuremath{1.497^{+0.064}_{-0.051}}}	%
\newcommand{\hatcurAYYr}{\ensuremath{1.92^{+0.17}_{-0.11}}}		%
\newcommand{\hatcurAYYrshort}{\ensuremath{1.92}}				%
\newcommand{\hatcurAYYrlong}{\ensuremath{1.921^{+0.171}_{-0.110}}}	%
\newcommand{\hatcurAYYrho}{\ensuremath{0.298\pm0.054}}			%
\newcommand{\hatcurAYYlogg}{\ensuremath{4.05^{+0.04}_{-0.06}}}		%
\newcommand{\hatcurAYYlum}{\ensuremath{5.3^{+1.1}_{-0.6}}}		%
\newcommand{\hatcurAYYmv}{\ensuremath{2.91\pm0.16}}			%
\newcommand{\hatcurAYYage}{\ensuremath{2.1\pm1.0}}			%
\newcommand{\hatcurAYYspec}{F6}

\newcommand{\hatcurARVK}{\ensuremath{213.2\pm1.9}}				%
\newcommand{\hatcurARVgamma}{\ensuremath{-37.0\pm1.5}}

\newcommand{\hatcurAPPi}{\ensuremath{84\fdg1^{+2.2}_{-2.0}}}		%
\newcommand{\hatcurAPPg}{\ensuremath{22.1\pm3.0}}			%
\newcommand{\hatcurAPPlogg}{\ensuremath{3.34\pm0.07}}			%
\newcommand{\hatcurAPPar}{\ensuremath{4.25^{+0.24}_{-0.28}}}			%
\newcommand{\hatcurAPParel}{\ensuremath{0.0379\pm0.0004}}		%
\newcommand{\hatcurAPPrho}{\ensuremath{0.78\pm0.16}}			%

\newcommand{\hatcurAPPm}{\ensuremath{1.80\pm0.06}}		%
\newcommand{\hatcurAPPmshort}{\ensuremath{1.80}}				%
\newcommand{\hatcurAPPmlong}{\ensuremath{1.800^{+0.063}_{-0.059}}}	%
\newcommand{\hatcurAPPr}{\ensuremath{1.42^{+0.14}_{-0.10}}}		%
\newcommand{\hatcurAPPrshort}{\ensuremath{1.42}}				%
\newcommand{\hatcurAPPrlong}{\ensuremath{1.421^{+0.144}_{-0.097}}}	%
\newcommand{\hatcurAPPmrcorr}{\ensuremath{0.81}}				%
\newcommand{\hatcurAPPteff}{\ensuremath{2175_{-60}^{+85}}}		%
\newcommand{\hatcurAPPtheta}{\ensuremath{0.064\pm0.005}}

\newcommand{\hatcurAXdist}{\ensuremath{320^{+30}_{-20}}}			%

In the past few years, the HATNet project announced $11$ discoveries
and became one of the most successful initiatives 
searching for transiting extrasolar planets. In this chapter the procedures 
of the photometric measurements and analysis of spectroscopic data 
(including radial velocity are explained, emphasizing how the algorithms
and programs were used in the data reduction and analysis. 
The particular example of the planetary system HAT-P-7(b)
clearly demonstrates all of the necessary steps that are generally 
required by the detection and confirmation of transiting extrasolar
planets. In Sec.~\ref{sec:hatp7:detection}, the issues 
related to the primary photometric detection are explained. 
Sec~\ref{sec:hatp7:followup} summarizes the follow-up observations,
which are needed by the proper confirmation of the planetary nature.
Mainly, the roles of these photometric follow-up observations are treefold. 
First, it provides additional data in order to have a better
estimation of the planetary parameters whose are derived from
the light curve of the system. Like so, spectroscopic analysis yields 
additional information from which the planetary mass or the properties
and physical parameters of the host star can be deduced. Third,
analysis of follow-up data helps to exclude other
scenarios that are likely to show similar photometric or
spectroscopic variations what a transiting extrasolar planet shows.
In Sec~\ref{sec:hatp7:analysis}, the methods are explained 
that we were using to obtain the final planetary parameters.

\begin{figure*}
\begin{center}
\noindent
\resizebox{160mm}{!}{\includegraphics{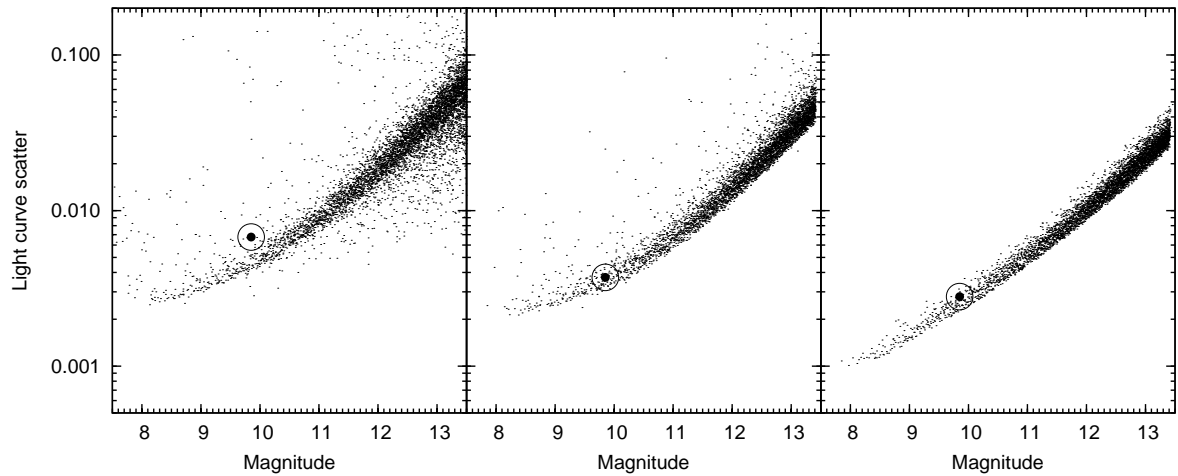}}
\end{center}
\caption{Light curve statistics for the field ``G154'', obtained by 
aperture photometry (left panel) and photometry based on 
the method of image subtraction (middle panel). The right panel
shows the lower noise limit estimation derived from the Poisson- and 
background noise. Due to the strong vignetting of the optics,
the effective gain varies across the image. Therefore, the 
distribution of the points on the right panel is not a clear thin
line. Instead, the thickness of the line is approximately equivalent to
a factor of $\sim 2$ between the noise level, indicating a highly varying
vignetting of a factor of $\sim 4$. The star HAT-P-7 (GSC~03547-01402)
is represented by the thick dot. The light curve scatter for this star 
has been obtained involving only out-of-transit data. This star is
a prominent example where the method of image subtraction photometry 
significantly improves the light curve quality.}
\label{fig:g154lcstat}
\end{figure*}

\phdsection{Photometric detection}
\label{sec:hatp7:detection}

The HATNet telescopes \mbox{HAT-7} and \mbox{HAT-8}
\citep[HATNet;][]{bakos2002,bakos2004} observed HATNet field G154,
centered at $\alpha = 19^{\rm h} 12^{\rm m}$, $\delta = +45^\circ00'$, 
on a near-nightly basis from 2004 May 27 to 2004 August 6.
Exposures of 5 minutes were obtained at a 5.5-minute cadence whenever
conditions permitted; all in all 5140 exposures were secured, each
yielding photometric measurements for approximately $33,000$ stars in
the field down to $I\sim13.0$. The field was observed in network mode,
exploiting the longitude separation between \mbox{HAT-7}, stationed at
the Smithsonian Astrophysical Observatory's (SAO) Fred Lawrence Whipple
Observatory (FLWO) in Arizona ($\lambda=111^\circ$\,W), and
\mbox{HAT-8}, installed on the rooftop of SAO's Submillimeter Array
(SMA) building atop Mauna Kea, Hawaii ($\lambda=155^\circ$\,W). We note that
each \lc{} obtained by a given instrument was shifted to
have a median value to be the same as catalogue magnitude of the 
appropriate star, allowing to merge \lc{}s acquired by different 
stations and/or detectors.

Following standard frame calibration procedures, astrometry was
performed as described in Sec.~\ref{subsec:astrometry}, and aperture photometry
results (see Sec.~\ref{subsec:photometry} and Sec.~\ref{sec:prog:fiphot})
were subjected to External Parameter Decorrelation (EPD, 
Sec.~\ref{subsec:trendfiltering}), and also to the Trend Filtering
Algorithm (\citep[TFA; see Sec.~\ref{subsec:trendfiltering} or][]{kovacs2005}. 
We searched the \lcs{} of field
G154 for box-shaped transit signals using the BLS algorithm of
\citet{kovacs2002}. A very significant periodic dip in brightness was
detected in the $I\approx\hatcurACCmag$ magnitude star \hatcurACCgsc{}
(also known as \hatcurACCtwomass{}; $\alpha = \hatcurACCra$, $\delta =
\hatcurACCdec$; J2000), with a depth of $\sim\hatcurALCdip$\,mmag, a
period of $P=\hatcurALCPshort$\,days and a relative duration (first to
last contact) of $q\approx0.078$, equivalent to a duration of
$Pq\approx4.1$~hours.

\begin{figure*}
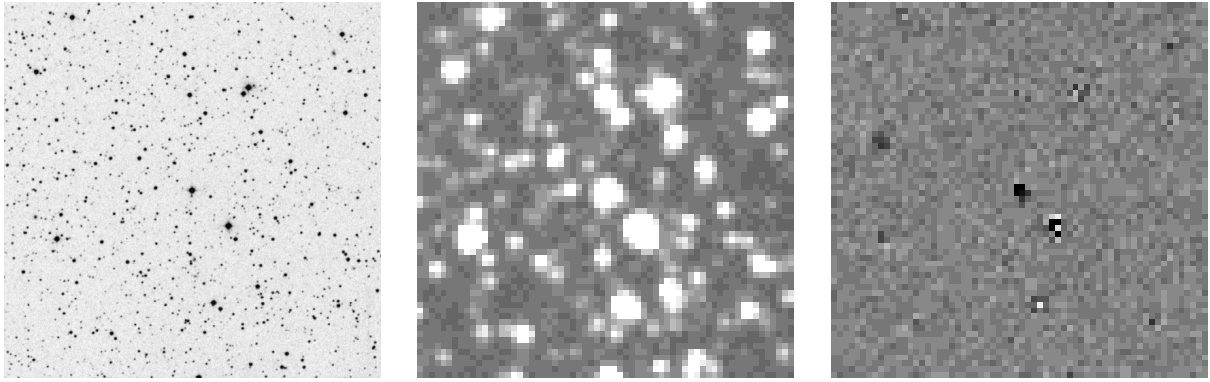

\begin{center}
\resizebox{50mm}{!}{\includegraphics{img/hatp7/hatp7-dss.eps}}\hspace*{5mm}%
\resizebox{50mm}{!}{\includegraphics{img/hatp7/hatp7-ref.eps}}\hspace*{5mm}%
\resizebox{50mm}{!}{\includegraphics{img/hatp7/t1.eps}}
\end{center}
\caption{Stamps showing the vicinity of the star HAT-P-7. 
All of the stamps have the same size, covering an area of $15.7'\times 15.7'$
on the sky and centered on HAT-P-7. 
The left panel is taken from the POSS-1 survey 
(available, e.g. from the STScI Digitized Sky Survey web page).
The middle panel shows the same area, as 
the HATNet telescopes see it. This stamp was cut from the photometric 
reference image (as it was used for the image subtraction process),
that was derived from the $\sim 20$ sharpest and cleanest images 
of the HAT-8 telescope. The right panel shows the convolution 
residual images averaged on the $\sim 160$ frames acquired by 
the HAT-8 telescope during the transit. 
The small dip at the center of the image can be seen well. Some
residual structures at the positions of brighter stars also present.}
\label{fig:hatp7stamps}
\end{figure*}

In addition, the star happened to fall in the overlapping area between
fields G154 and G155. Field G155, centered at $\alpha = 19^{\rm h}
48^{\rm m}$, $\delta = +45^\circ00'$, was also observed over an
extended time in between 2004 July 27 and 2005 September 20 by the
\mbox{HAT-6} (Arizona) and \mbox{HAT-9} (Hawaii) telescopes. We
gathered 1220 and 10260 data-points, respectively (which independently
confirmed the transit), yielding a total number of 16620 data-points.

After the announcement and the publication of the planet HAT-P-7b
\citep{pal2008hatp7}, all of the images for the fields G154 and G155
were re-analyzed by the method of image subtraction photometry.
Based on the astrometric solution\footnote{The astrometric
solutions have been already obtained at this point since the source 
identification and the centroid coordinates were
already required earlier by aperture photometry.}, the images
were registered to the coordinate system of one of the images that was found
to be a proper reference image (Sec.~\ref{subsec:imageregistering}).
From the set of registered frames approximately a dozen of them
have been chosen to create a good signal-to-noise ratio
master reference image for the image subtraction procedure.
These frames were selected to be the sharpest ones, i.e. where the 
overall profile sharpness parameter, $S$ (see 
Sec.~\ref{sec:stardetection:coordshapemodel}) were the largest
among the images (note that large $S$ corresponds to small FWHM,
i.e. to sharp stars). Moreover, such images were chosen from the ones
where the Moon was below the horizon (see also Fig.~\ref{fig:astromres}
and the related discussion). The procedure was repeated for
both fields G154 and G155. The intensity levels of these
individual sharp frames were then transformed to the same level
involving the program \texttt{ficonv}, with a formal kernel 
size of $1\times 1$ pixels ($B_{\rm K}=0$, $N_{\rm kernel}=1$, 
$K^{(1)}=\delta^{(00)}$). Such an intensity level transformation
corrects for the changes in the instrumental stellar brightnesses
due to the varying airmass, transparency and background level.
These images were then combined (Sec.~\ref{sec:prog:ficombine})
in order to have a single master convolution reference image. This step
was performed for both of the fields. The reference
images were then used to derive the optimal convolution transformation,
and simultaneously the residual (``subtracted'') images were also
obtained by \texttt{ficonv}. 
For each individual object image, both the result of 
the convolution kernel fit and the residual image were saved
to files for further processing.
For the fit, we have employed a discrete kernel basis with the size
of $7\times 7$ pixels and we let a spatial variation of 4th polynomial order 
for both the kernel parameters and the background level. Due to
the sharp profiles (the profile FWHMs were between $2.0\dots 2.4$), 
this relatively small kernel size were sufficient for our purposes. The residuals
on the subtracted images were subjected to aperture photometry,
based on the considerations discussed in Sec.~\ref{subsec:subtractedphotometry}.
For the proper image subtraction-based photometry, one needs 
to derive and to use the fluxes on the reference image as well. These
fluxes were derived using aperture photometry, and the instrumental
raw magnitudes were transformed to the catalogue magnitudes with
a fourth order polynomial transformation. The residual of this fit
was nearly $0.05$\,mags for both fields, thus the fluxes of the
individual stars have been well determined, and this transformation
yielded proper reference fluxes even for the faint and the 
blended stars. The results of the image subtraction photometry
were then processed similarly to the normal aperture photometry
results (see also Fig.~\ref{fig:photometrypipeline}), and the respective
light curves were de-trended involving both the EPD and TFA algorithms.

For a comparison, the light curve residuals for the normal aperture photometry
and the image subtraction photometry are plotted on the left
and middle panel of Fig.~\ref{fig:g154lcstat}. In general,
the image subtraction photometry yielded light curve residuals smaller
by a factor of $\sim 1.2 - 1.5$. The gain achieved by the image subtraction
photometry is larger for the fainter stars. It is important to note that
in the case of the star HAT-P-7, the image subtraction photometry improved 
the photometric quality\footnote{In the case of a star having
periodic dips in its light curve, the scatter is derived only
from the out-of-transit sections.} by a factor of $\sim 1.8$: 
the rms of the out-of-transit section in the aperture photometry
light curve were $6.75$\,mmag while the image subtraction method
yielded an rms of $3.72$\,mmag. The lower limit of the intrinsic
noise of this particular star is $2.8$\,mmag (see also 
the right panel of Fig.~\ref{fig:g154lcstat}).
In Fig.~\ref{fig:hatp7stamps}, we display some image stamps from the 
star HAT-P-7 and its neighborhood. Since the dip of $\sim 7$\,mmag 
during the transits of HAT-P-7b is only $\sim 2$ times larger than 
the overall rms of the light curve, individual subtracted frames
does not significantly show the ``hole'' at the centroid position of
the star, especially because this weak signal is distributed among
several pixels. Therefore, on the right panel of Fig.~\ref{fig:hatp7stamps},
all of the frames acquired by the telescope HAT-8 during the
transit have been averaged in order to show a clear visual
detection of the transit. Albeit the star HAT-P-7 is a well isolated one,
such visual analysis of image residuals can be relevant when
the signal is detected for stars whose profiles are significantly merged.
In such cases, either the visual analysis or a more precise quantification
of this ``negative residual'' (e.g. by employing the star detection 
and characterization algorithms of Sec.~\ref{subsec:stardetection})
can help to distinguish which star is the variable.

\begin{figure*}
\begin{center}
\resizebox{160mm}{!}{\includegraphics{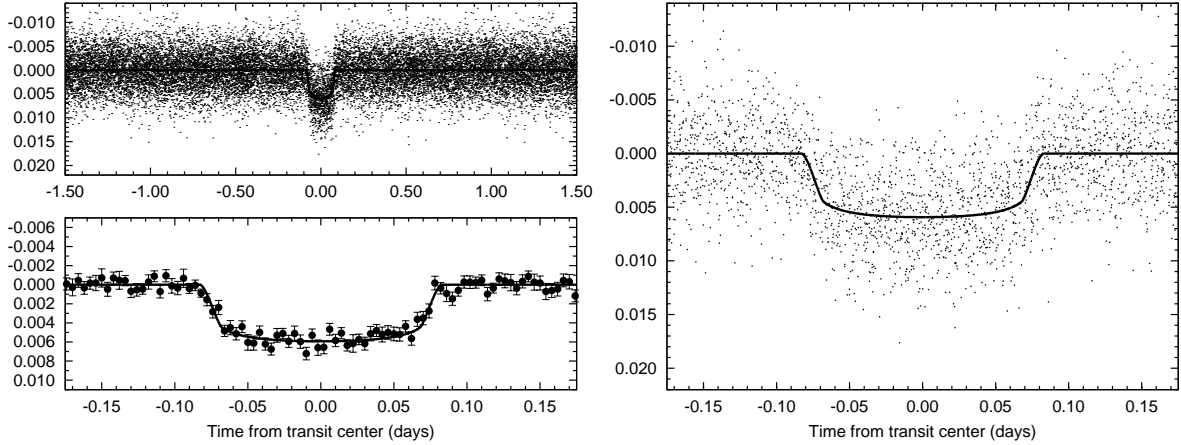}}
\end{center}
\caption%
{
	Upper left panel: the complete \lc{} of \hatcurA{} with all of the 
	16620 points, unbinned instrumental \band{I}
	photometry obtained with four telescopes of HATNet (see text for
	details), and folded with the period of $P = \hatcurALCPprec$~days
	(the result of a joint fit to all available data,
	Sec.~\ref{sec:hatp7:jointfit}). The superimposed curve shows the 
	best model fit using quadratic limb darkening. 
	Right panel: The transit zoomed-in (3150 data points are shown).
	Lower left panel: same as the right panel, with the points binned with 
	a bin size of $0.004$ in days. 
}
\label{fig:hatp7:hatnetlc}
\end{figure*}

The combined HATNet \lc{}, yielded by the image
subtraction photometry and de-trended by the EPD and TFA is plotted
on Fig.~\ref{fig:hatp7:hatnetlc}. Superimposed on
these plots is our best fit model (see Sec.~\ref{sec:hatp7:analysis}).
We note that TFA was run in signal reconstruction mode, i.e. systematics
were iteratively filtered out from the observed time series assuming that
the underlying signal is a trapeze-shaped transit 
\citep[see Sec.~\ref{subsec:trendfiltering} and][for additional details]{kovacs2005}.
We note that fields G154 and G155 both intersect the field of view of the
Kepler mission \citep{borucki2007}, and more importantly, HAT-P-7 lies
in the Kepler field.


\phdsection{Follow-up observations}
\label{sec:hatp7:followup}

\phdsubsection{Reconnaissance spectroscopy}

Following the HATNet photometric detection, \hatcurA{} (then a transit
{\em candidate}) was observed spectroscopically with the CfA Digital
Speedometer \citep[DS, see][]{latham1992} at the \flwos{} Tillinghast reflector, in
order to rule out a number of blend scenarios that mimic planetary
transits \citep[e.g.][]{brown2003,odonovan2007}, as well as to characterize
the stellar parameters, such as surface gravity, effective temperature,
and rotation. Four spectra were obtained over an interval of 29 days. 
These observations cover $45$\,\AA{} in a single echelle order centered
at $5187$\,~\AA{}, and have a resolving power of $\lambda/\Delta\lambda
\approx 35,\!000$.  Radial velocities were derived by
cross-correlation, and have a typical precision of $1$\,\kms. Using
these measurements, together with collaborators, we have ruled out an unblended companion of stellar
mass (e.g.~an M dwarf orbiting an F dwarf), since the radial velocities
did not show any variation within the uncertainties. The mean
heliocentric radial velocity of \hatcurA{} was measured to be
$-11$\,~\kms.  Based on an analysis similar to that described in
\cite{torres2002}, the DS spectra indicated that the host star is a
slightly evolved dwarf with $\logg = 3.5$ (cgs), $\teffstar = 6250$\,K
and $\vsini \approx 6\,\kms$.

\begin{figure*}
\begin{center}
\resizebox{160mm}{!}{\includegraphics{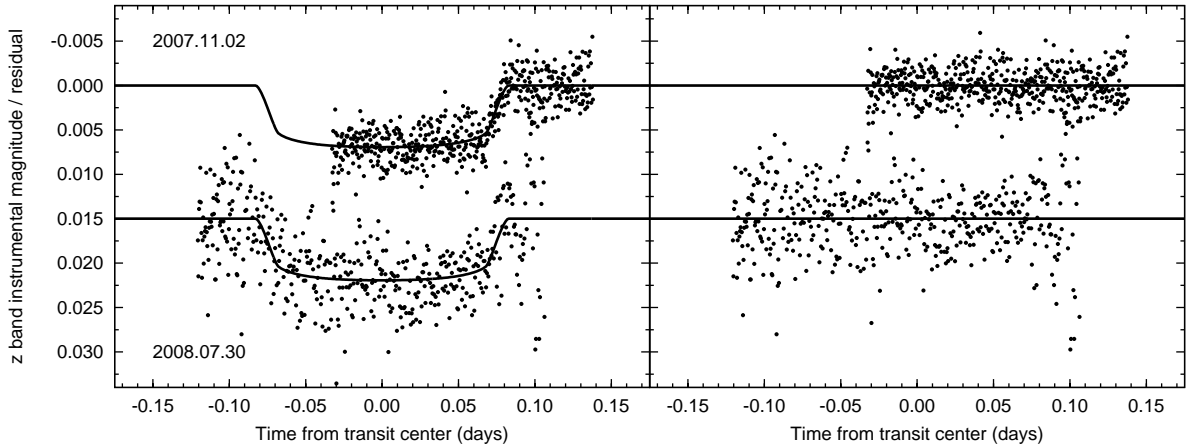}}
\end{center}
\caption%
{
	Left panel: unbinned instrumental Sloan \band{z} partial 
	transit photometry acquired by the KeplerCam at the \flwof{} 
	telescope on 2007 November 2 and 2008 July 30; superimposed is 
	the best-fit transit model \lc{}.
	Right panel: the difference between the KeplerCam observation 
	and model (on the same vertical scale).
}
\label{fig:hatp7:flwo48lc}
\end{figure*}

\phdsubsection{High resolution spectroscopy}

\begin{figure}
\begin{center}
\resizebox{80mm}{!}{\includegraphics{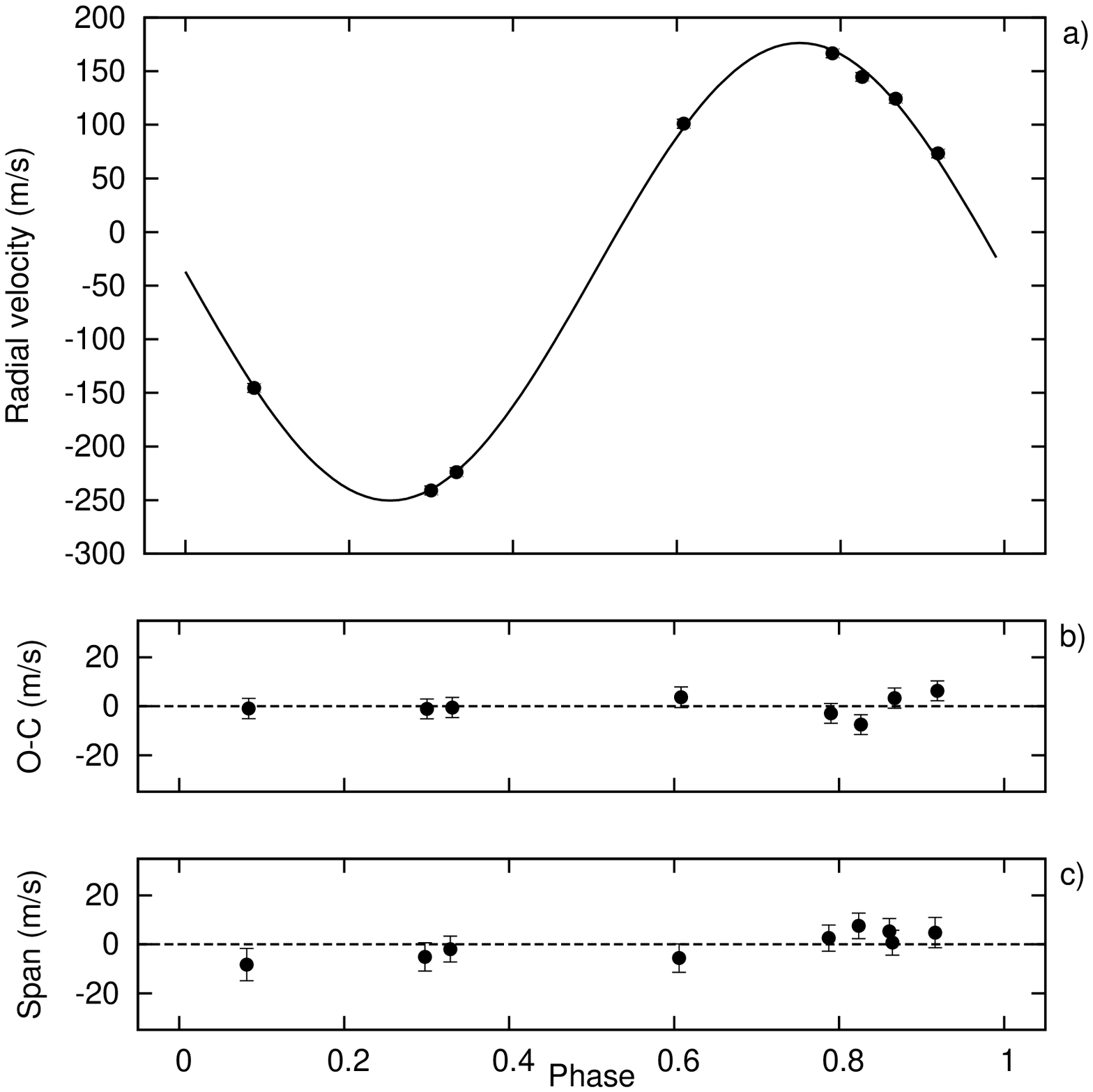}}
\end{center}
\caption{
(a) Radial-velocity measurements from Keck for \hatcurA{}, along with an
orbital fit, shown as a function of orbital phase, using our best fit
as period (see Sec.~\ref{sec:hatp7:jointfit}). The center-of-mass velocity has
been subtracted.
(b) Phased residuals after subtracting the orbital fit 
(also see Sec.~\ref{sec:hatp7:jointfit}). The rms 
variation of the residuals is about $3.8$\,\ms. 
(c) Bisector spans (BS) for the 8 Keck spectra plus the single template
spectrum, computed as described in the text.  The mean value has been
subtracted. Due to the relatively small errors comparing to the RV
amplitude, the vertical scale on the (b) and (c) panels differ from the
scale used on the top panel. 
\label{fig:rvbis}}
\end{figure}

For the characterization of the radial velocity variations and for the
more precise determination of the stellar parameters, we obtained 8
exposures with an iodine cell, plus one iodine-free template, using the
HIRES instrument \citep{vogt1994} on the Keck~I telescope, Hawaii,
between 2007 August 24 and 2007 September 1. The width of the
spectrometer slit was $0\farcs86$ resulting a resolving power of
$\lambda/\Delta\lambda \approx 55,\!000$, while the wavelength coverage
was $\sim3800-8000$\,\AA\@. The iodine gas absorption cell was used to
superimpose a dense forest of $\mathrm{I}_2$ lines on the stellar
spectrum and establish an accurate wavelength fiducial
\citep[see][]{marcy1992}. Relative radial velocities in the Solar System
barycentric frame were derived as described by \cite{butler1996},
incorporating full modeling of the spatial and temporal variations of
the instrumental profile. The final radial velocity data and their
errors are listed in Table~\ref{tab:rvs}. The folded data, with our
best fit (see Sec.~\ref{sec:hatp7:jointfit}) superimposed, are plotted in
Fig.~\ref{fig:rvbis}a.

\phdsubsection{Photometric follow-up observations}

Partial photometric coverage of a transit event of \hatcurA{} was
carried out in the Sloan \band{z} with the KeplerCam CCD on the
\mbox{1.2 m} telescope at FLWO, on 2007 November 2. The total number of
frames taken from \hatcurA{} was 514 with cadence of 28 seconds. During
the reduction of the KeplerCam data, we used the following method.
After bias and flat calibration of the images, an astrometric
transformation (in the form of first order polynomials)  between the
$\sim450$ brightest stars and the 2MASS catalog was derived, as
described in Sec.~\ref{subsec:astrometry}, yielding a residual 
of $\sim 0.2 - 0.3$ pixel.
Aperture photometry was then performed
using a series of apertures with the radius of 4, 6 and 8 pixels
in fixed positions calculated from this
solution and the actual 2MASS positions. The instrumental magnitude
transformation was obtained using $\sim350$ stars on a frame taken near
culmination of the field. The transformation fit was initially weighted
by the estimated photon- and background-noise error of each star, then
the procedure was repeated by weighting with the inverse variance of the
\lcs{}. From the set of apertures we have chosen the aperture for which
the out-of-transit (OOT) rms of \hatcurA{} was the smallest; the radius
of this aperture is 6 pixels. The resulted light curve has been 
presented in the discovery paper of \cite{pal2008hatp7}. More
recently, in 2008 July 30, we have obtained an additional 
complete light curve for the transit of HAT-P-7b, also
in Sloan \band{z} with the KeplerCam CCD.

The two follow-up light curves from 2007 November 2 and 2008 July 30 
were then de-correlated against trends 
using the complete data, involving a simultaneous fit
for the light curve model function parameters and the EPD parameters
(see also Sec.~\ref{sec:hatp7:analysis}). These fits
yielded a light curve with an overall rms of $1.83$\,mmag 
and $4.23$\,mmag for these two nights, respectively.
In both cases, the cadence of the individual photometric measurements
were 28 seconds. For the first night the residual scatter of $1.83$\,mmag
is a bit larger than the expected rms of
1.5mmag, derived from the photon noise (1.2mmag) and scintillation noise
-- that has an expected amplitude of 0.8mmag, based on the observational
conditions and the calculations of \citet{young1967} -- 
possibly due to unresolved trends and other noise sources. For
the second night, the photometric quality was significantly worse,
due to the high variations in the transparency\footnote{For 2007 November 2,
the scatter of the \emph{raw} magnitudes were $\sim 14$\,mmag while
on the night of 2008 July 30, the raw magnitude rms were more than 
$15$ times higher, nearly $0.24$\,mag.}.
The resulting light curves are shown in Fig.~\ref{fig:hatp7:flwo48lc},
superimposed with our best fit model (Sec.~\ref{sec:hatp7:analysis}).

\phdsubsection{Excluding blend scenarios}

Following \cite{torres2007}, we explored the possibility that the
measured radial velocities are not real, but instead caused by
distortions in the spectral line profiles due to contamination from a
nearby unresolved eclipsing binary.  In that case the ``bisector span''
of the average spectral line should vary periodically with amplitude
and phase similar to the measured velocities themselves
\citep{queloz2001,mandushev2005}. We cross-correlated each Keck spectrum
against a synthetic template matching the properties of the star
(i.e.~based on the SME results, see Sec.~\ref{sec:hatp7:stellarparameters}),
and averaged the correlation functions over all orders blueward of the
region affected by the iodine lines. From this representation of the
average spectral line profile we computed the mean bisectors, and as a
measure of the line asymmetry we computed the ``bisector spans'' as the
velocity difference between points selected near the top and bottom of
the mean bisectors \citep{torres2005}. If the velocities were the result
of a blend with an eclipsing binary, we would expect the line bisectors
to vary in phase with the photometric period with an amplitude similar
to that of the velocities. Instead, we detect no variation in excess of
the measurement uncertainties (see Fig.~\ref{fig:rvbis}c). 
We have also tested the significance of the correlation between the
radial velocity and the bisector variations. 
Therefore, we conclude
that the velocity variations are real and that the star is orbited by a
Jovian planet. We note here that the mean bisector span ratio relative
to the radial velocity amplitude is the smallest ($\sim 0.026$) among
all the HATNet planets, indicating an exceptionally high confidence
that the RV signal is not due to a blend with an eclipsing binary
companion.


\phdsection{Analysis}
\label{sec:hatp7:analysis}

The analysis of the available data was done in four steps. First, an
independent analysis was performed on the HATNet, the radial velocity
(RV) and the high precision photometric follow-up (FU) data,
respectively. Analysis of the HATNet data yielded an initial value for
the orbital period and transit epoch. The initial period and epoch were
used to fold the RV's, and phase them with respect to the predicted
transit time for a circular orbit. The HATNet and the RV epochs
together yield a more accurate period, since the time difference
between the discovery \lc{} and the RV follow-up is fairly long; more
than 3 years.  Using this refined period, we can extrapolate to the
expected center of the KeplerCam partial transit, and therefore obtain
a fit for the two remaining key parameters describing the light curve:
$a/R_\star$ where $a$ is the semi-major axis for a circular orbit, and
the impact parameter $b\equiv(a/R_\star)\cos i$, where $i$ is the
inclination of the orbit.

Second, using as starting points the initial values as derived above,
we performed a joint fit of the HATNet, RV and FU data, i.e.~fitting
\emph{all} of the parameters simultaneously. The reason for such a
joint fit is that the three separate data-sets and the fitted
parameters are intertwined. For example, the epoch (depending partly on
the RV fit) has a relatively large error, affecting the extrapolation
of the transit center to the KeplerCam follow-up.

In the discovery report, in all of the above procedures, we used the 
downhill simplex method (DHSX, Sec.~\ref{sec:prog:lfit}) 
to search for the best fit values and the method of refitting to synthetic
data sets (called EMCE, see also Sec.~\ref{sec:prog:lfit})
to find out the error of the adjusted parameters. The refined
analysis based on the HATNet light curves reduced by the
method of image subtraction photometry and an additional
photometric measurement from the night of 2008 July 30 was also
involved. In this analysis the extended Markov Chain Monte-Carlo 
algorithm (XMMC) was employed, also in the form of an implementation
found in the program \texttt{lfit}. As it was mentioned in 
Sec.~\ref{sec:prog:lfit}, the XMMC method used in this
particular analysis has also been aided by the DHSX
minimization (as a first iteration) 
and used as a sanity check of the chain convergence
(see also Sec.~\ref{sec:hatp7:analysis}). Both of these 
error estimation methods (EMCE and XMMC) yield a Monte-Carlo set 
of the \emph{a posteriori} distribution of the fit parameters,
that were subsequently used in the derivation of the final
planetary, orbital and stellar characteristics.

The third step of the analysis was the derivation of the stellar
parameters, based on the spectroscopic analysis of the host star (high
resolution spectroscopy using Keck/HIRES), and the physical modeling of
the stellar evolution, based on existing isochrone models. As the
fourth step, we then combined the results of the joint fit and stellar
parameter determination to determine the planetary and orbital
parameters of the \hatcurAb{} system.  In the following 
we summarize these steps.

\begin{table}
\caption%
{	Relative radial velocity (RV) and bisector span (BS) measurements 
	of \hatcurA{}. The RV and BS data points, as well as their formal 
	errors are given in units of m/s.
}
\label{tab:rvs}
\begin{center}
\footnotesize
\begin{tabular}{rrrrr}
\hline
$\mathrm{BJD}$& ${\rm RV}$ & $\sigma_{\rm RV}$ & ${\rm BS}$ & $\sigma_{\rm BS}$  \\
\hline
2454336.73121		&	-		&	-		&$ 5.30 $ & $  5.36$ \\
2454336.73958		&  $+124.40$		&   $		1.63  $ &$ 0.68 $ & $  5.10$ \\
2454336.85366		&  $+ 73.33$		&   $		1.48  $ &$ 4.82 $ & $  6.17$ \\
2454337.76211		&  $-223.89$		&   $		1.60  $ &$-1.94 $ & $  5.30$ \\
2454338.77439		&  $+166.71$		&   $		1.39  $ &$ 2.58 $ & $  5.35$ \\
2454338.85455		&  $+144.67$		&   $		1.42  $ &$ 7.60 $ & $  5.22$ \\
2454339.89886		&  $-241.02$		&   $		1.46  $ &$-5.13 $ & $  5.77$ \\
2454343.83180		&  $-145.42$		&   $		1.66  $ &$-8.30 $ & $  6.58$ \\
2454344.98804		&  $+101.05$		&   $		1.91  $ &$-5.62 $ & $  5.80$ \\	
\hline
\end{tabular}
\end{center}
\end{table}

\phdsubsection{Independent fits}
\label{sec:hatp7:independentfit}

For the independent fit procedure, we first analyzed the HATNet \lcs{},
as observed by the \mbox{HAT-6}, \mbox{HAT-7}, \mbox{HAT-8} and
\mbox{HAT-9} telescopes. Using the initial period and transit length
from the BLS analysis, we fitted a model to the 214 cycles of
observations spanned by all the HATNet data.  Although at this stage we
were interested only in the epoch and period, we have used the transit
\lc{} model with the assumption of quadratic limb darkening, where the
flux decrease was calculated using the models provided by
\citet{mandel2002}. In principle, fitting the epoch and period as two
independent variables is equivalent to fitting the time instant of the
centers of the first and last observed individual transits,
$T_{\mathrm{c,first}}$ and $T_{\mathrm{c,last}}$, with a constraint
that all intermediate transits are regularly spaced with period $P$.
Note that this fit takes into account {\em all}\/ transits that
occurred during the HATNet observations, even though it is described
only by $T_{\mathrm{c,first}}$ and $T_{\mathrm{c,last}}$.
The fit yielded 
$T_{\mathrm{c,first}}=2453153.0924\pm0.0021$ (BJD) and
$T_{\mathrm{c,last}} =2453624.9044\pm0.0023$ (BJD). 
the correlation between these two epochs turned out to be:
$C(T_{\mathrm{c,first}},T_{\mathrm{c,last}})=-0.53$. 
The  period 
derived from the 
$T_{\mathrm{c,first}}$
and $T_{\mathrm{c,last}}$ epochs  was
$P^{(1)}=2.20480\pm0.00049$~days.
Using these values, we found that there were 326 cycles between 
$T_{\mathrm{c,last}}$ 
and the end of the RV campaign. The epoch
extrapolated to the approximate time of RV measurements was
$T_{\mathrm{c,RV}}=2454343.646\pm0.008$ (BJD). Note that the error in
$T_{\mathrm{c,RV}}$ is 
much smaller than the period itself
($\sim2.2$\,days), 
so there is no ambiguity in the number of elapsed
cycles when folding the periodic signal. 

We then analyzed the radial velocity data in the following way. We
defined the $N_{\rm tr}\equiv0$ transit as that being closest to the
end of the radial velocity measurements. This means that the first
transit observed by HATNet (at $T_{\mathrm{c,first}}$) was the $N_{\rm
tr,first}=-569$ event. Given the short period, we assumed that the
orbit has been circularized \citep{hut1981} (later verified; see below). 
The orbital fit is linear if we choose the radial velocity zero-point
$\gamma$ and the amplitudes $A$ and $B$ as adjusted values, namely:
\begin{equation}
v(t) = \gamma + A\cos\left[\frac{2\pi}{P}(t-t_0)\right] + 
	B\sin\left[\frac{2\pi}{P}(t-t_0)\right],
\end{equation}
where $t_0$ is an arbitrary time instant (chosen to be $t_0=2454342.6$
BJD), $K\equiv\sqrt{A^2+B^2}$ is the semi-amplitude of the RV
variations, and $P$ is the initial period $P^{(1)}$ taken from the
previous independent HATNet fit. The actual epoch can be derived from
the above equation, since for circular orbits the transit center occurs when
the RV curve has the most negative slope.
For circular orbits, the transit occurs at
the time instant when the RV curve has the smallest time-derivative,
the actual epoch of the transit must be:
\begin{equation}
T_c 
 =t_0+\frac{P}{2\pi}\arg(-B,A)
 =t_0+\frac{P}{2\pi}\mathop{\mathrm{arc~tan}}\nolimits\left(-\frac{A}{B}\right).
\label{eq:rvtransitepoch}
\end{equation}
Using the equations above, we derived the initial epoch of the $N_{\rm
tr} = 0$ transit center to be $T_c=2454343.6462\pm0.0042\equiv
T^{(1)}_{\mathrm{c},-29}$ (BJD). We also performed a more general
(non-linear) fit to the RV in which we let the eccentricity float. 
This fit yielded an eccentricity consistent with zero, namely
$e\cos\omega=-0.003\pm0.007$ and $e\sin\omega= 0.000\pm0.010$.
Therefore, we adopt a circular orbit in the further analysis.

\begin{figure*}
\begin{lcmd}
\small
\# Downhill simplex best fit value: \\
 2453153.09286 2454678.76582 ~ 213.35  ...  0.07619  0.2061 13.4529 ... 2759.01432 \\
\# XMMC values: \\
 2453153.09356 2454678.76621 ~ 211.99  ...  0.07478  0.1864 13.4551 ... 2766.28256 \\
 2453153.09446 2454678.76562 ~ 213.75  ...  0.07625  0.3036 13.6665 ... 2768.77404 \\
 2453153.09472 2454678.76638 ~ 213.41  ...  0.07616  0.0714 13.6635 ... 2769.49600 \\
 2453153.09473 2454678.76449 ~ 213.61  ...  0.07489  0.0190 13.2588 ... 2766.48611 \\
 2453153.09468 2454678.76509 ~ 215.68  ...  0.07541  0.0542 13.2887 ... 2769.26264 \\
 2453153.09496 2454678.76499 ~ 214.70  ...  0.07685  0.1223 13.3642 ... 2767.24865 \\
 2453153.09465 2454678.76477 ~ 214.68  ...  0.07736  0.1542 13.2951 ... 2767.86983 \\
 2453153.09474 2454678.76425 ~ 214.25  ...  0.07708  0.3420 13.3437 ... 2768.80460 \\
 2453153.09371 2454678.76438 ~ 213.48  ...  0.07622  0.3270 13.3973 ... 2766.88065 \\
 2453153.09358 2454678.76455 ~ 216.24  ...  0.07376  0.0295 13.3136 ... 2767.79127 \\
 2453153.09317 2454678.76618 ~ 211.07  ...  0.07474  0.2070 13.6488 ... 2765.08235 \\
 ............................................................................. \\
\# \\
\# Accepted transitions / total iterations: 4000/30299 \\
\# Total acceptance ratio : 0.13202 +/- 0.00209 \\
\# Theoretical probability: 0.14493 [independent:10=23-13-0 (total-constrained-linear)] \\
\# \\
\# Correlation lengths:  \\
\#       2.64    2.38    1.80    ....  1.47    0.83    2.20    ...  \\
\# \\
\# chi\^{ }2 values: \\
\#	minimal: 0.928961 \\
\# Appropriate values for this chi\^{ }2: \\
\# 2453153.09286 2454678.76582   213.35  ...  0.07619  0.2061 13.4529 ... \\
\# \\
\# Errors and correlations (projected Fisher matrix): \\
\#      0.00085      0.00110     1.92  ...  0.00132  0.1272  0.1906   \\
\# \\
\#  ~1.000 -0.195 -0.002~ ... -0.013 -0.023 -0.002 \\
\#  -0.195 ~1.000 -0.064~ ... ~0.043 ~0.047 ~0.683 \\
\#  -0.002 -0.064 ~1.000~ ... -0.003 ~0.003 -0.046 \\
\#  .............................................. \\
\#  -0.013 ~0.043 -0.003~ ... ~1.000 ~0.728 ~0.256 \\  
\#  -0.023 ~0.047 -0.003~ ... ~0.728 ~1.000 ~0.389 \\  
\#  -0.002 ~0.683 -0.046~ ... ~0.256 ~0.389 ~1.000 \\  
\# \\
\# Errors and correlations (statistical, around the best fit): \\
\#      0.00109      0.00134     2.04  ...  0.00118  0.1050  0.2221  \\
\# \\
\#  ~1.000 -0.149 -0.015  ... ~0.017 ~0.018 ~0.018 \\
\#  -0.149 ~1.000 -0.126  ... ~0.039 ~0.102 ~0.796 \\
\#  -0.015 -0.126 ~1.000  ... -0.025 -0.037 -0.086 \\
\#  .............................................. \\ 
\#  ~0.017 ~0.039 -0.025  ... ~1.000 ~0.554 ~0.124 \\  
\#  ~0.018 ~0.102 -0.037  ... ~0.554 ~1.000 ~0.318 \\  
\#  ~0.018 ~0.796 -0.086  ... ~0.124 ~0.318 ~1.000 
\end{lcmd}
\caption{The output of the program
\texttt{lfit} showing the results of the extended 
Markov Chain Monte-Carlo (XMMC) analysis related to
the HAT-P-7(b) planetary system. The parameters in the output
are $T_{-569}$, $T_{+123}$, $K$, 
$R_{\rm p}/R_\star$, $b^2$ and $\zeta/R_\star$, respectively. For clarity,
the other parameters were cut from the output list.
}\label{fig:hatp7lfitxmmcout}
\end{figure*}

\begin{figure}
\begin{center}
\resizebox{80mm}{!}{\includegraphics{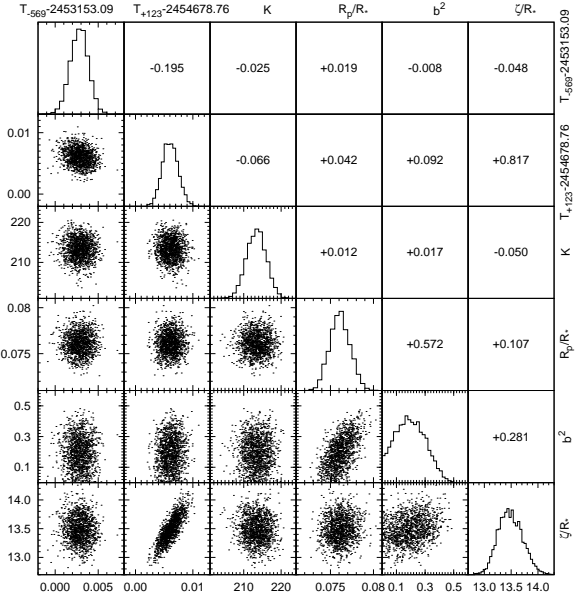}}
\end{center}
\caption{Probability distributions and mutual correlations 
of the adjusted parameters $T_{-569}$, $T_{+123}$, $K$, 
$R_{\rm p}/R_\star$, $b^2$ and $\zeta/R_\star$ for the planet HAT-P-7b.
These are the only adjusted parameters of the analysis that are 
explicitly related to the physical properties of the planet and its orbit.
The derivation of these distributions were performed exploiting the
extended Markov Chain Monte-Carlo (XMMC) algorithm as 
it is implemented in the program \texttt{lfit} (the related output
is shown partially in Fig.~\ref{fig:hatp7lfitxmmcout}). 
See text for further details.
}\label{fig:hatp7xmmc}
\end{figure}

Combining the RV epoch $T^{(1)}_{\mathrm{c},-29}$ with the first epoch
observed by HATNet ($T_{\mathrm{c,first}}$), we obtained a somewhat
refined period, $P^{(2)}=2.204732\pm0.000016$~days. This was fed back
into phasing the RV data, and we performed the RV fit again to the
parameters $\gamma$, $A$ and $B$. The fit yielded
$\gamma=-37.0\pm1.5$\,\ms, $K\equiv\sqrt{A^2+B^2}= 213.4\pm2.0$\,\ms{}
and $T^{(2)}_{\mathrm{c},-29}=2454343.6470\pm0.0042$ (BJD). This epoch
was used to further refine the period to get
$P^{(3)}=2.204731\pm0.000016$\,d, where the error calculation assumes
that $T_{\mathrm{c},-29}$ and $T_{\mathrm{c},-569}$ are uncorrelated. At
this point we stopped the above iterative procedure of refining the
epoch and period; instead a final refinement of epoch and period was
obtained through performing a joint fit, (as described later in
Sec.~\ref{sec:hatp7:jointfit}). We note that in order to get a reduced
chi-square value near unity for the radial velocity fit, it was
necessary to quadratically increase the noise component with an
amplitude of $3.8$~\ms, which is well within the range of stellar
jitter observed for late F stars; see \cite{butler2006}.

Using the improved period $P^{(3)}$ and the epoch $T_{\mathrm{c},-29}$,
we extrapolated to the center of KeplerCam follow-up transit ($N_{\rm
tr}=29$).  Since the follow-up observation only recorded a partial
event (see Fig.~\ref{fig:hatp7:flwo48lc}), this extrapolation was necessary to
improve the \lc{} modeling. For this, we have used a quadratic
limb-darkening approximation, based on the formalism of
\citet{mandel2002}. The limb-darkening coefficients were based on the
results of the SME analysis (notably, \teffstar; see
Sec.~\ref{sec:hatp7:stellarparameters} for further details), which yielded
$\gamma_1^{(z)}=0.1329$ and $\gamma_2^{(z)}=0.3738$.  Using these
values and the extrapolated time of the transit center, we adjusted the
\lc{} parameters: the relative radius of the planet $p=R_p/R_\star$,
the square of the impact parameter $b^2$ and the quantity
$\zeta/R_\star=(a/R_\star)(2\pi/P)(1-b^2)^{-1/2}$ as independent
parameters \citep[see][for the choice of parameters]{bakos2007hatp5}. The
result of the fit was $p=0.0762\pm0.0012$, $b^2=0.205\pm0.144$ and
$\zeta/R_\star=13.60\pm0.83$\,day$^{-1}$, where the uncertainty of the
transit center time due to the relatively high error in the transit
epoch $T_{\mathrm{c},-29}$ was also taken into account in the error
estimates.

\phdsubsection{Joint fit based on the aperture photometry data and the
single partial follow-up light curve}
\label{sec:hatp7:jointfit}

The results of the individual fits described above provide the starting
values for a joint fit, i.e.~a simultaneous fit to all of the available
HATNet, radial velocity and the partial follow-up \lc{} data.  The
adjusted parameters were $T_{\mathrm{c},-569}$, the time of first
transit center in the HATNet campaign, $m$, the out-of-transit
magnitude of the HATNet \lc{} in \band{I} and the previously defined
parameters of $\gamma$, $A$, $B$, $p$, $b^2$ and $\zeta/R_\star$.  We
note that in this joint fit {\em all}\/ of the transits in the HATNet
\lc{} have been adjusted simultaneously, tied together by the
constraint of assuming a strictly periodic signal; the shape of all
these transits were characterized by $p$, $b^2$ and $\zeta/R_\star$
(and the limb-darkening coefficients) while the distinct transit center
time instants were interpolated using $T_{\mathrm{c},-569}
=T_{\mathrm{c,first}}$ and $A$, $B$ via the RV fit. For initial values
we used the results of the independent fits (Sec.~\ref{sec:hatp7:independentfit}).
The error estimation based on method refitting to synthetic data sets
gives the distribution of the adjusted values, and moreover, this
distribution can be used directly as an input for a Monte-Carlo
parameter determination for stellar evolution modeling, as described
later (Sec.~\ref{sec:hatp7:stellarparameters}).

Final results of the joint fit were:
$T_{\mathrm{c},-569}=2453153.0924\pm0.0015$~(BJD), 
$m=9.85053\pm0.00015$\,mag,
$\gamma=\hatcurARVgamma$\,\ms,
$A=33.8\pm0.9$\,\ms,
$B=210.7\pm1.9$\,\ms,
$p=0.0763\pm0.0010$, 
$b^2=0.135_{-0.116}^{+0.149}$ and 
$\zeta/R_\star=13.34\pm0.23$~$\mathrm{day^{-1}}$. 
Using the distribution of these parameters, it is
straightforward to obtain the values and the errors of the additional 
parameters derived from the joint derived fit, namely 
$T_{\mathrm{c},-29}$, $a/R_\star$, $K$ and $P$. 
All final fit parameters are listed in Table~\ref{tab:parameters}. 

\phdsubsection{Joint fit based on the image subtraction photometry 
data and both of the follow-up light curves}
\label{sec:hatp7:jointfitrefined}
Involving the additional recent follow-up photometry data from 2008 July 30 and 
the HATNet light curve obtained by the method based on image subtraction,
we repeated the analysis of the available data. In this new analysis, 
the method of extended Markov Chain Monte-Carlo (XMMC) has been employed
to derive the best fit parameters and their \emph{a posteriori} distributions.
Due to the presence of a complete photometric follow-up light curve,
we have used a slightly different set of parameters. Moreover, the 
trend filtering based on the EPD algorithm has been performed simultaneously
with the fit. Thus, the set of adjusted parameters that are related
to the physical properties of the planetary system were the following:
the center of the first transit measured by the HATNet telescopes, 
$T_{\mathrm{c},-569}$; the transit center of the last follow-up photometry
$T_{\mathrm{c},+123}$, the radial velocity semi-amplitude $K$, 
the light curve parameters $R_{\rm p}/R_\star$, $b^2$ and $\zeta/R_\star$.
Additionally, the out-of-transit magnitudes (both for the HATNet photometry
and the two follow-up photometry), the zero-point of the radial velocity
$\gamma$, and the EPD coefficients for the two follow-up photometry
were also included in the fit. The EPD was performed up to the
first order against the 
profile sharpness parameters ($S$, $D$, $K$), the hour angle and
the airmass. In the case of the HATNet photometry, we incorporated an additional 
parameter, an instrumental blend factor whose inclusion was based
on the experience that HATNet light curves tend to slightly underestimate 
the depth of the transits. To have a general purpose analysis,
we extended the parameter set with the Lagrangian orbital 
elements $k=e\cos\omega$ and $h=e\sin\omega$, but based our assumption
for circular orbits, these were fixed to be zero in the case of HAT-P-7b.

The XMMC analysis was performed in three ways. First, a full XMMC
run was accomplished, involving all of the $23$ parameters discussed 
below ($6$ physical parameters, $3$ out-of-transit magnitudes, the 
radial velocity zero-point, the $2\times 5$ EPD coefficients,
the instrumental blend factor and the fixed Lagrangian orbital elements).
Second, we have separated the $2\times 5$ linear EPD coefficients
from the merit function and run the Markov chains while minimizing the
$\chi^2$ accordingly in each step of the chain. Third, we derived the 
best fit parameters using the downhill simplex algorithm and during
the XMMC run we kept the EPD coefficients to be fixed to their best
fit values. All of these fits yielded a successful convergence and 
all of the sanity checks mentioned in Sec.~\ref{sec:prog:lfit}
were adequate, namely a) the \emph{a posteriori} distribution centers 
of the adjusted parameters (median values)
agreed well with the downhill simplex best fit values, b) the 
chain acceptance ratio was in agreement with the theoretical expectations,
c) the correlation lengths for the parameter chains were sufficiently
small, all of them were smaller than $\sim 2.6$, and d) the covariance
estimations from the Fisher information matrix agreed well, within 
a factor of $\sim 1.2$, with
the statistical covariances derived from the \emph{a posteriori} distributions.
See also Fig.~\ref{fig:hatp7lfitxmmcout}, that shows the (slightly 
clarified and simplified) output of the \texttt{lfit} program related
to this particular analysis.
In all of the cases, we have used a Gaussian \emph{a priori} distribution
for the transitions, where the covariance matrix of this Gaussian were
derived from the Fisher matrix evaluated at the downhill simplex best
fit value. In Fig.~\ref{fig:hatp7xmmc} the distributions and some
statistics for the $6$ parameters related to the physical planetary
(and orbital) parameters are displayed. The plots in 
Fig.~\ref{fig:hatp7xmmc} clearly show how the proper selection
of the adjusted parameters can help to reduce the mutual correlations.
The only significant correlation is between $\zeta/R_\star$
$T_{\mathrm{c},+123}$. This correlation is resulted from 
the lack of a good quality complete follow-up photometry (due to 
its large scatter, the contribution of the 
second follow-up light curve is relatively smaller).

For the final set of the
parameters we accepted the distribution that was derived using 
the third method mentioned above (i.e. when in the XMMC runs the 
$2\times 5$ EPD parameters were fixed to their best fit values). The
derived best fit parameters that are related to physical quantities
were the following:
$T_{\mathrm{c},-569}=2453153.09286\pm0.00105$~(BJD), 
$T_{\mathrm{c},+123}=2454678.76582\pm0.00137$~(BJD), 
$K=213.4\pm1.9 $\,\ms,
$p=R_{\rm p}/R_\star=0.7619\pm0.0009$, 
$b^2=0.206\pm0.103$ and 
$\zeta/R_\star=13.45\pm0.22$~$\mathrm{day^{-1}}$. Comparing to
these values with the ones presented in Sec.~\ref{sec:hatp7:jointfit},
the improvements in the parameter uncertainties are quite conspicuous. 
Especially,
the new, image subtraction based HATNet light curve has decreased
the uncertainty in the first transit epoch of $T_{\mathrm{c},-569}$
with its significantly better quality.
In the further analysis, we incorporated these distributions in order
to derive the final stellar, planetary and orbital parameters.

\begin{table}
\caption{Stellar parameters for 
\hatcurA{}. The values of effecitve temperature, metallicity and projected 
rotational velocity are based on purely spectroscopic data while the 
other ones are derived from the both the spectroscopy and the joint light
curve and stellar evolution modelling.\label{tab:stellar}}
\begin{center}
\begin{tabular}{lrl}
\hline
Parameter		&  Value			& Source	\\
\hline
$\teffstar$ (K)		&  \hatcurASMEteff		& SME 		\\
$[\mathrm{Fe/H}]$	&  \hatcurASMEzfeh		& SME 		\\
$v \sin i$ (\kms)	&  \hatcurASMEvsin		& SME 		\\
$M_\star$ ($M_{\sun}$)  &  \hatcurAYYm			& Y$^2$+LC+SME	\\
$R_\star$ ($R_{\sun}$)  &  \hatcurAYYr			& Y$^2$+LC+SME	\\
$\log g_\star$ (cgs)    &  \hatcurAYYlogg		& Y$^2$+LC+SME	\\
$L_\star$ ($L_{\sun}$)  &  \hatcurAYYlum			& Y$^2$+LC+SME	\\
$M_V$ (mag)		&  \hatcurAYYmv   		& Y$^2$+LC+SME	\\
Age (Gyr)		&  \hatcurAYYage			& Y$^2$+LC+SME	\\
Distance (pc)		&  \hatcurAXdist			& Y$^2$+LC+SME	\\
\hline
\end{tabular}
\end{center}
\end{table}

\phdsubsection{Stellar parameters}
\label{sec:hatp7:stellarparameters}

The results of the joint fit enable us to refine the parameters of the
star. First, the iodine-free template spectrum from Keck was used for
an initial determination of the atmospheric parameters. Spectral
synthesis modeling was carried out using the SME software
\citep{valenti1996}, with wavelength ranges and atomic line data as
described by \citet{valenti2005}. We obtained the following initial
values: effective temperature $\hatcurASMEteff$\,K, surface gravity
$\log g_\star = \hatcurASMElogg$ (cgs), iron abundance
$\mathrm{[Fe/H]}=\hatcurASMEzfeh$, and projected rotational velocity
$v\sin i=\hatcurASMEvsin$\,\kms.  The rotational velocity is slightly
smaller than the value given by the DS measurements. The temperature
and surface gravity correspond to a slightly evolved \hatcurAYYspec{}
star. The uncertainties quoted here and in the remaining of this
discussion are twice the statistical uncertainties for the values given
by the SME analysis. This reflects our attempt, based on prior
experience, to incorporate systematic errors (e.g. \cite{noyes2008};
see also \cite{valenti2005}). 
Note that the previously
discussed limb darkening coefficients, $\gamma_1^{(z)}$,
$\gamma_2^{(z)}$, $\gamma_1^{(I)}$ and $\gamma_2^{(I)}$ have been taken
from the tables of \cite{claret2004} by interpolation to the
above-mentioned SME values for $\teffstar$, $\log g_\star$, and
$\mathrm{[Fe/H]}$.

As described by \cite{sozzetti2007}, $a/R_\star$ is a better luminosity
indicator than the spectroscopic value of $\log g_\star$ since the
variation of stellar surface gravity has a subtle effect on the line
profiles. Therefore, we used the values of $\teffstar$ and
$\mathrm{[Fe/H]}$ from the initial SME analysis, together with the
distribution of $a/R_\star$ to estimate the stellar properties from
comparison with the Yonsei-Yale (Y$^2$) stellar evolution models by
\cite{yi2001}.  Since a Monte-Carlo set for $a/R_\star$ values has been
derived during the joint fit, we performed the stellar parameter
determination as follows. For a selected value of $a/R_\star$, two
Gaussian random values were drawn for $\teffstar$ and $\mathrm{[Fe/H]}$
with the mean and standard deviation as given by SME (with formal SME
uncertainties doubled as indicated above).Using these
three values, we searched the nearest isochrone and the corresponding
mass by using the interpolator provided by \citet{demarque2004}. 
Repeating this procedure for values of $a/R_\star$, $\teffstar$,
$\mathrm{[Fe/H]}$, the set of the \emph{a posteriori} distribution of
the stellar parameters was obtained, including the mass, radius, age,
luminosity and color (in multiple bands).  The age determined in this
way is $2.2$~Gy with a statistical uncertainty of $\pm 0.3$~Gy;
however, the uncertainty in the theoretical isochrone ages is about
1.0~Gy. Since the corresponding value for the surface gravity of the
star, $\log g_\star=\hatcurAYYlogg$ (cgs), is well within 1-$\sigma$ of
the value determined by the SME analysis, we accept the values from the
joint fit as the final stellar parameters. These parameters are
summarized in Table~\ref{tab:stellar}.

We note that the Yonsei-Yale isochrones contain the absolute magnitudes and
colors for
different photometric bands from $U$ up to $M$, providing an easy
comparison of the estimated and the observed colors. Using these data,
we determined the $V-I$ and $J-K$ colors of the best fitted stellar
model:
$(V-I)_{\rm YY}=0.54\pm0.02$ and 
$(J-K)_{\rm YY}=0.27\pm0.02$.
Since the colors for the infrared bands provided by \citet{yi2001} and
\citet{demarque2004} are given in the ESO photometric standard system,
for the comparison with catalog data, we converted the infrared color
$(J-K)_{\rm YY}$ to the 2MASS system $(J-K_S)$ using the
transformations given by \citet{carpenter2001}. The color of the best
fit stellar model was $(J-K_S)_{\rm YY}=0.25\pm0.03$, which is in
fairly good agreement with the actual 2MASS color of \hatcurA{}:
$(J-K_S)=0.22\pm0.04$. We have also compared the $(V-I)_{\rm YY}$ color
of the best fit model to the catalog data, and found that although
\hatcurA{} has a low galactic latitude, $b_{\rm II}=13\fdg8$, the model
color agrees well with the observed TASS color of $(V-I)_{\rm
TASS}=\hatcurACCtassvi$ \citep[see][]{droege2006}. Hence, the star is not
affected by the interstellar reddening within the errors, since
$E(V-I)\equiv(V-I)_{\rm TASS}-(V-I)_{\rm YY}=0.06\pm0.07$.  For
estimating the distance of \hatcurA, we used the absolute magnitude
$M_V=\hatcurAYYmv$ (resulting from the isochrone analysis, see also 
Table~\ref{tab:stellar}) and the $V_{\rm TASS}=\hatcurACCtassmv\pm0.06$
observed magnitude. These two yield a distance modulus of $V_{\rm
TASS}-M_V=7.51\pm0.28$, i.e.~distance of $d=\hatcurAXdist$\,pc.

\phdsubsection{Planetary and orbital parameters}

The determination of the stellar properties was followed by the
characterization of the planet itself. Since Monte-Carlo distributions
were derived for both the \lc{} and the stellar parameters, the final
planetary and orbital data were also obtained by the statistical
analysis of the \emph{a posteriori} distribution of the appropriate
combination of these two Monte-Carlo data sets.  We found that the mass
of the planet is $M_p=\hatcurAPPmlong$\,\mjup, the radius is
$R_p=\hatcurAPPrlong$\,\rjup{} and its density is
$\rho_p=\hatcurAPPrho$\,\gcmc.  We note that in the case of binary
systems with large mass and radius ratios (such as the one here) there
is a strong correlation between $M_p$ and $R_p$ \citep[see
e.g.][]{beatty2007}. This correlation is also exhibited here with
$C(M_p,R_p)=\hatcurAPPmrcorr$.  The final planetary parameters are also
summarized at the bottom of Table~\ref{tab:parameters}.

Due to the way we derived the period, i.e.
$P=(T_{\mathrm{c},-29}-T_{\mathrm{c},-569})/540$, one can expect a large
correlation between the epochs $T_{\mathrm{c},-29}$,
$T_{\mathrm{c},-569}$ and the period itself. Indeed,
$C(T_{\mathrm{c},-569},P)=-0.783$ and $C(T_{\mathrm{c},-29},P)=0.704$,
while the correlation between the two epochs is relatively small;
$C(T_{\mathrm{c},-569},T_{\mathrm{c},-29})=-0.111$. It is easy to show
that if the signs of the correlations between two epochs $T_{\rm A}$
and $T_{\rm B}$ (in our case $T_{\mathrm{c},-29}$ and
$T_{\mathrm{c},-569}$) and the period are different, respectively, then
there exists an optimal epoch $E$, which has the smallest error among
all of the interpolated epochs. We note that $E$ is such that it
also exhibits the smallest correlation with the period. If
$\sigma(T_{\rm A})$ and $\sigma(T_{\rm B})$ are the respective
uncorrelated errors of the two epochs, then
\begin{equation}
E=\left[\frac{T_{\rm A}\sigma(T_{\rm B})^2+T_{\rm B}\sigma(T_{\rm A})^2}
	{\sigma(T_{\rm B})^2+\sigma(T_{\rm A})^2}\right]
\end{equation}
where square brackets denote the time of the transit event nearest to
the time instance $t$. In the case of \hatcurAb{}, $T_{\rm A}\equiv
T_{\mathrm{c},-569}$ and $T_{\rm B}\equiv T_{\mathrm{c},-29}$, the
corresponding epoch is the event $N_{\rm tr}=-280$ at $E\equiv
T_{\mathrm{c},-280}=\hatcurALCT$ (BJD).  The final ephemeris and
planetary parameters are summarized in Table~\ref{tab:parameters}.

\begin{table}
\caption{Orbital and planetary parameters
for \hatcurA{}. The parameters are derived from the joint modelling
of the photometric, radial velocity and spectroscopic data.
\label{tab:parameters}}
\begin{center}
\begin{tabular}{lr}
\hline
Parameter			&  Value		\\
\hline	
$P$ (days)			& $\hatcurALCP$ 	\\
$E$ (${\rm BJD}-2,\!400,\!000$)	& $\hatcurALCMT$	\\
$T_{14}$ (days)$^{\rm a}$	& $\hatcurALCdur$	\\
$T_{12}=T_{34}$ (days)$^{\rm a}$& $\hatcurALCingdur$	\\
\hline	
$a/R_\star$			& $\hatcurAPPar$	\\
$R_p/R_\star$			& $\hatcurALCrprstar$	\\
$b \equiv a \cos i/R_\star$	& $\hatcurALCimp$	\\
$i$ (deg)		       	& $\hatcurAPPi$  	\\
Transit duration (days)\dotfill & $0.1461\pm 0.0016$ 	\\
$(\gamma_1,\gamma_2)$ $^{\rm b}$& $(0.1195,0.3595)$	\\
\hline
$K$ (\ms)		       	& $\hatcurARVK$		\\
$\gamma$ (\kms)		     	& $\hatcurARVgamma$	\\
$e$				& $0$ (adopted)		\\
\hline	
$M_p$ ($\mjup$)			& $\hatcurAPPmlong$	\\
$R_p$ ($\rjup$)			& $\hatcurAPPrlong$	\\
$C(M_p,R_p)$			& $\hatcurAPPmrcorr$	\\
$\rho_p$ (\gcmc)		& $\hatcurAPPrho$	\\
$a$ (AU)		       	& $\hatcurAPParel$	\\
$\log g_p$ (cgs)		& $\hatcurAPPlogg$	\\
$T_{\rm eq}$ (K)		& $\hatcurAPPteff$	\\
\hline
\end{tabular}
\end{center}
\footnotesize
${ }^{\rm a}$ \ensuremath{T_{14}}: total transit duration, time
between first to last contact; \ensuremath{T_{12}=T_{34}}:
ingress/egress time, time between first and second, or third and fourth
contact. 
\end{table}


\phdchapter{Follow-up observations}
\label{chapter:followup}

\newcommand{\hatcurB}{HAT-P-2}
\newcommand{\hatcurBb}{HAT-P-2\lowercase{b}}

\newcommand{\hatcurBCCra}{\ensuremath{16^{\mathrm{h}}20^{\mathrm{m}}36^{\mathrm{s}}.36}}
\newcommand{\hatcurBCCdec}{\ensuremath{+41^{\circ}02'53''.1}}
\newcommand{\hatcurBCCmag}{\ensuremath{8.71}}
\newcommand{\hatcurBCCtwomass}{2MASS~16203635+4102531}
\newcommand{\hatcurBCCgsc}{GSC~03065-01195}
\newcommand{\hatcurBCCtassmv}{8.71}
\newcommand{\hatcurBCCtassmvshort}{8.7}
\newcommand{\hatcurBCCtassvi}{\ensuremath{0.55\pm0.06}}

\newcommand{\hatcurBLCdip}{\ensuremath{5.2}}				
\newcommand{\hatcurBLCrprstar}{\ensuremath{0.0724\pm0.0010}}		%
\newcommand{\hatcurBLCimp}{\ensuremath{0.354^{+0.087}_{-0.156}}}	%
\newcommand{\hatcurBLCdur}{\ensuremath{0.1790\pm0.0013}}		%
\newcommand{\hatcurBLCingdur}{\ensuremath{0.0136\pm 0.0012}}		%
\newcommand{\hatcurBLCP}{\ensuremath{5.6334697\pm0.0000074}}		%
\newcommand{\hatcurBLCPprec}{\ensuremath{5.6334697}}			%
\newcommand{\hatcurBLCPshort}{5.6335}					%
\newcommand{\hatcurBLCT}{\ensuremath{2,454,342.42616\pm0.00064}}	
\newcommand{\hatcurBLCMT}{\ensuremath{54,342.42616\pm0.00064}}		%

\newcommand{\hatcurBSMEteff}{\ensuremath{6290\pm60}}			%
\newcommand{\hatcurBSMEzfeh}{\ensuremath{+0.14\pm0.08}}			%
\newcommand{\hatcurBSMElogg}{\ensuremath{4.16\pm0.03}}			%
\newcommand{\hatcurBSMEvsin}{\ensuremath{20.8\pm0.3}}			%

\newcommand{\hatcurBYYm}{\ensuremath{1.34\pm0.04}}			%
\newcommand{\hatcurBYYmshort}{\ensuremath{1.34}}			%
\newcommand{\hatcurBYYmlong}{\ensuremath{1.343\pm0.040}}		%
\newcommand{\hatcurBYYr}{\ensuremath{1.60^{+0.09}_{-0.07}}}		%
\newcommand{\hatcurBYYrshort}{\ensuremath{1.60}}			%
\newcommand{\hatcurBYYrlong}{\ensuremath{1.599^{+0.088}_{-0.069}}}	%
\newcommand{\hatcurBYYrho}{\ensuremath{0.462\pm0.058}}			%
\newcommand{\hatcurBYYlogg}{\ensuremath{4.158\pm0.031}}			%
\newcommand{\hatcurBYYlum}{\ensuremath{3.6^{+0.5}_{-0.3}}}		%
\newcommand{\hatcurBYYmv}{\ensuremath{3.36\pm0.12}}			%
\newcommand{\hatcurBYYage}{\ensuremath{2.7\pm0.5}}			%
\newcommand{\hatcurBYYspec}{F8}
\newcommand{\hatcurBYYsigma}{\ensuremath{0.0046\pm0.0004}}

\newcommand{\hatcurBRVK}{\ensuremath{958.9\pm13.9}}			%
\newcommand{\hatcurBRVgammaK}{\ensuremath{318.4\pm6.6}}
\newcommand{\hatcurBRVgammaL}{\ensuremath{77.0\pm10.4}}
\newcommand{\hatcurBRVgammaS}{\ensuremath{-19868.9\pm9.9}}

\newcommand{\hatcurBPPi}{\ensuremath{87\fdg2^{+1.2}_{-0.9}}}		%
\newcommand{\hatcurBPPg}{\ensuremath{172\pm16}}				%
\newcommand{\hatcurBPPlogg}{\ensuremath{4.23\pm0.04}}			%
\newcommand{\hatcurBPPar}{\ensuremath{9.21^{+0.37}_{-0.40}}}		%
\newcommand{\hatcurBPParel}{\ensuremath{0.0686\pm0.0007}}		%
\newcommand{\hatcurBPPrho}{\ensuremath{7.63^{+1.14}_{-1.09}}}		%

\newcommand{\hatcurBPPm}{\ensuremath{8.8^{+0.2}_{-0.3}}}		%
\newcommand{\hatcurBPPmshort}{\ensuremath{8.84}}			%
\newcommand{\hatcurBPPmlong}{\ensuremath{8.84^{+0.22}_{-0.29}}}		%
\newcommand{\hatcurBPPr}{\ensuremath{1.12^{+0.07}_{-0.05}}}		%
\newcommand{\hatcurBPPrshort}{\ensuremath{1.12}}			%
\newcommand{\hatcurBPPrlong}{\ensuremath{1.123^{+0.071}_{-0.054}}}	%
\newcommand{\hatcurBPPmrcorr}{\ensuremath{0.68}}			%
\newcommand{\hatcurBPPteff}{\ensuremath{1525_{-30}^{+40}}}		%
\newcommand{\hatcurBPPtheta}{\ensuremath{0.79\pm0.04}}

\newcommand{\hatcurBXdist}{\ensuremath{118\pm8}}


Now we shift our attention to another system, one of the 
eccentric transiting planetary systems of \hatcurBb{}. 
At the time of its discovery, \hatcurBb{} was the longest period 
and most massive transiting extrasolar planet (TEP), and the only one 
on an eccentric orbit \citep{bakos2007hatp2}. In the following,
other TEPs have also been discovered with significant
orbital eccentricity, and long period: GJ~436b \citep{gillon2007},
HD~17156b \citep{barbieri2007} and XO-3b \citep{johnskrull2008}.
(See, e.g. \texttt{http://exoplanet.eu} for an up-to-date 
database for transiting extrasolar planets.)

Planet \hatcurBb{} was detected as a transiting object 
during the campaign of the HATNet telescopes \citep[]{bakos2002,bakos2004}, 
and Wise HAT telescope \citep[WHAT][]{shporer2006}. The HATNet 
telescopes and the WHAT telescope gathered $\sim26,000$ 
individual photometric measurements. The planetary transit 
was followed up by the FLWO 1.2m telescope, utilizing the KeplerCam 
detector. The planetary properties have been confirmed by radial velocity
measurements and bisector analysis of the spectral line profiles. The latter
has shown no bisector variations, excluding the possibilities of 
a hierarchical triplet or a blended eclipsing binary.

Recently, the spin-orbit alignment of the \hatcurB{}(b) system was
measured by \cite{winn2007} and \cite{loeillet2008}. Both studies
reported an alignment consistent with zero within an uncertainty of 
$\sim10^\circ$. These results are exceptionally interesting since
short period planets are thought to be formed at much larger distances
from their parent star and migrated inward while the orbital eccentricity
is damped yielding an almost circular orbit \citep{dangelo2006}. 
Physical mechanisms such as Kozai interaction between the transiting planet
and an unknown massive companion on an inclined orbit
could result tight eccentric orbits \citep{fabrycky2007,takeda2008}. 
However, in such a scenario, the spin-orbit alignment can be expected to
be significantly larger than the measured. For instance, in the case of
XO-3b \citep{hebrard2008}, the reported alignment is 
$\lambda=70^\circ\pm15^\circ$.
In multiple planetary systems, planet-planet scattering can also yield
eccentric orbits \citep[see e.g.][]{ford2007}.

The physical properties of the host star \hatcurB{} have been controversial
since different methods for stellar characterization resulted
stellar radii between $\sim1.4\,R_{\odot}$ and $\sim1.8\,R_{\odot}$. Moreover,
the actual distance of the system also had large systematic
errors, since the reported Hipparcos distance seemed to be significantly
larger than what could be expected from the absolute luminosity
(coming from the stellar evolution modelling).

In this chapter new photometric and spectroscopic observations
of the planetary system \hatcurB{}(b) are presented, and
I demonstrate how the photometry package can be used 
in the case of a follow-up observation.
The new photometric measurements
significantly improve the light curve parameters, therefore some
of the stellar parameters are more accurately constrained. In addition,
radial velocity measurements based on spectroscopic observations have 
resulted significantly smaller
uncertainties, which, due to the orbital eccentricity, also affect
the results of the stellar evolution modelling. In Sec.~\ref{sec:hatp2:photometric},
we summarize our photometric observations of this system, while in 
Sec.~\ref{sec:hatp2:rv} we describe briefly the issues related to the 
radial velocity data points. The details of a new formalism used in 
the characterization of the radial velocities is discussed in 
Sec.~\ref{sec:hatp2:eof} and the steps of the complete analysis are 
described in Sec.~\ref{sec:hatp2:analysis}. 
We summarize our results in Sec.~\ref{sec:hatp2:discussion}.




\phdsection{Photometric observations and reductions}
\label{sec:hatp2:photometric}

\begin{figure}
\begin{center}
\resizebox{80mm}{!}{\includegraphics{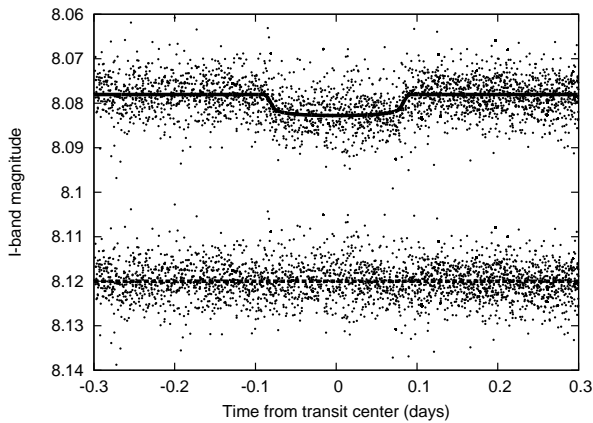}}
\end{center}
\caption{The folded HATNet light curve of 
\hatcurB{} \citep[published in][]{bakos2007hatp2}, showing the 
points only nearby the transit. The upper panel is superimposed with our
best-fit model and the lower panel shows the fit residual.
See text for further details.}\label{fig:hatp2:hatnetlc}
\end{figure}

\begin{figure}
\begin{center}
\resizebox{80mm}{!}{\includegraphics{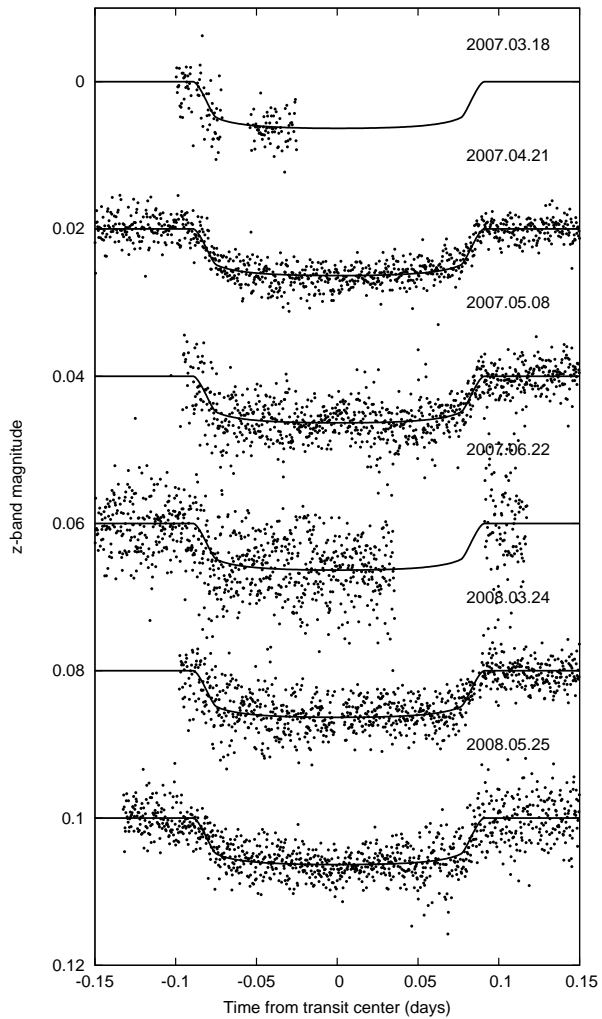}}
\end{center}
\caption{Follow-up light curves of \hatcurB{}. The light curves
were acquired on 2007 March 18, 2007 April 21, 2007 May 08, 2007 June 22,
2008 March 24 and 2008 May 25, while the respective transit 
sequence numbers were $N_{\rm tr}=-6$, $0$, $+3$, $+11$, $+60$ and $+71$.
All of these light curves are superimposed with our best-fit model.
See text for further details.}\label{fig:hatp2:lc}
\end{figure}

In the present analysis we utilize photometric data obtained by
the HATNet telescopes \citep[published in][]{bakos2007hatp2} and by the 
KeplerCam detector
mounted on the FLWO~1.2m telescope. The photometry of HATNet have already
been presented in \cite{bakos2007hatp2}. These HATNet data are plotted on 
Fig.~\ref{fig:hatp2:hatnetlc}, superimposed with our new best-fit model 
(see Sec.~\ref{sec:hatp2:analysis} for details on light curve modelling).
We observed the planetary transit
six times, on 2007 March 18, 2007 April 21, 2007 May 08, 2007 June 22,
2008 March 24 and 2008 May 25, yielding 4 nearly complete and 2 partial
transit light curves. One of these follow-up light curves (2007 April 21)
has already been published in the discovery paper. 
All of our high precision
follow-up photometry data are plotted on Fig.~\ref{fig:hatp2:lc}, along with
our best-fit transit light curve model (see also Sec.~\ref{sec:hatp2:analysis}).

The frames taken by the KeplerCam detector have been calibrated and 
reduced in the following similar fashion for all of the observations for
the six nights. Prior to the real calibration, all 
pixels which are saturated (or blooming) have been marked 
(\texttt{fiign}, see Sec.~\ref{sec:prog:fiign}), forcing
them to be omitted from the upcoming photometry. During the calibration
of the frames we have used standard bias, dark and sky-flat 
corrections. 

Following the calibration, the detection of stars
and the derivation of the astrometrical solution was done
in two steps. 
First, an initial astrometrical transformation 
was derived using the $\sim 50$ brightest and non-saturated
stars (whose parameters were derived by the program \texttt{fistar},
see Sec.~\ref{sec:prog:fistar}) from each frame, and 
by using the 2MASS catalogue \citep{skrutskie2006} 
as a reference. The transformation itself has been
obtained by the program \texttt{grmatch} (Sec.~\ref{sec:prog:grmatch}),
with a second-order polynomial fit. Since the astrometrical
data found in the 2MASS catalogue was obtained by the same kind 
of telescope, 
one could expect significantly better astrometrical data from the 
FLWO~1.2\,m telescope due to the numerous individual frames taken
at better spatial resolution. Indeed, an internal catalog
which was derived from the detected stellar centroids by registering them
to the same reference system has shown an internal precision
$\sim 0.005$\,arc~sec for the brighter stars while the 2MASS catalog
reports an uncertainty that is larger by an order of magnitude: 
nearly $\sim 0.06$\,arc~sec.
Therefore, in the second step of the astrometry, we used this 
new catalog to derive the individual astrometrical solutions for
each frame, still using a second-order polynomial fit. We note here
that this method also corrects for the systematic errors in the
photometry yielded by the proper motion of the stars.

Using the above astrometrical solutions, we performed aperture photometry
(with the program \texttt{fiphot}, Sec.~\ref{sec:prog:fiphot})
on fix centroids, employing a set of five apertures between 7.5 and 17.5 pixels in radius. 
The results of the aperture photometry were then transformed to the same 
instrumental magnitude system using a correction to the spatial variations and
the differential extinction (the former depends on the celestial 
coordinates while the latter depends on the intrinsic colors 
of the stars). Both corrections were  
linear in the pixel coordinates and linear in the colors. Experience
shows that significant correlations can occur between the
instrumental magnitudes and some of the external parameters
of the light curves (such as the FWHM of the stars, subpixel positions).
Although one should de-trend against these correlations using purely 
out-of-transit data (both before ingress and after egress), 
we have carried out such an external parameter decorrelation (EPD)
simultaneously with the light curve modelling (Sec.~\ref{sec:hatp2:analysis})
due to the lack of 
out-of-transit data in several cases. After the simultaneous light curve
modelling and de-trending, we chose the aperture for each night that
yielded the smallest residual. 
In all of the cases this ``best aperture'' was neither the smallest
nor the largest one from the set, confirming our assumptions for
selecting a good aperture series. We note here that since all
of the stars on the frames were well isolated, such choice
of different radii of the apertures does not result in any systematics,
because stars are not blended by any of these apertures.
In addition, due to the high flux of \hatcurB{} and the comparison 
stars, the frames were slightly extrafocal (in order to avoid
saturation). This resulted different FWHM per night for the stars
and therefore the optimal apertures yielding the highest signal-to-noise
ratio also have different radii for each night.



\phdsection{Radial velocity observations}
\label{sec:hatp2:rv}

In the discovery paper of \hatcurBb{} \citep{bakos2007hatp2}, $13$ individual radial
velocity measurements were reported that were utilizing the 
HIRES instrument \citep{vogt1994} on the Keck~I telescope,
on Mauna Kea, Hawaii, plus $10$ measurements from the Hamilton echelle 
spectrograph at the Lick Observatory \citep{vogt1987}. In the last year,
we have acquired $14$ additional radial velocity measurements
using the HIRES instrument on Keck. In the analysis, 
we have also used the online radial velocity data obtained by 
the OHP/SOPHIE spectrograph at out-of-transit (i.e.~omitting the
measurements for the Rossiter-McLaughlin effect), published 
by \cite{loeillet2008}. 
With these additional $8$ observations, we have $27+10+8=45$ 
high precision RV data points at hand for a refined analysis.

\begin{figure}
\begin{center}
\resizebox{80mm}{!}{\includegraphics{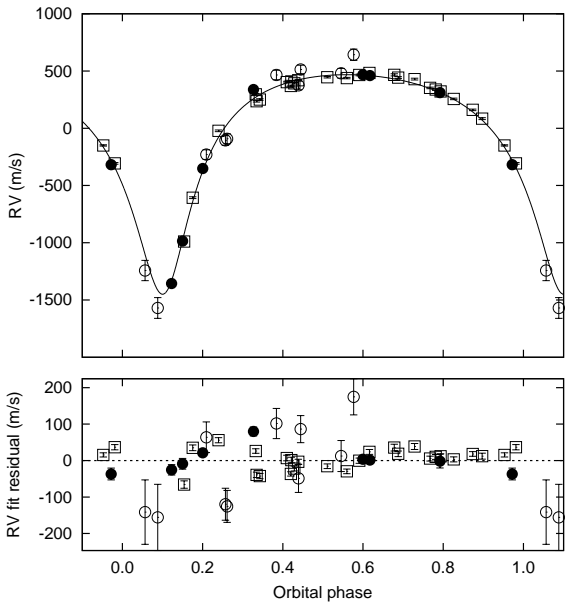}}
\end{center}
\caption{Radial velocity measurements 
for \hatcurB{} folded with
the best-fit orbital period. Filled dots represent
the OHP/SOPHIE data, open circles show the Lick/Hamilton, while
the open boxes mark the Keck/HIRES observations. In the upper panel, 
all of these three
RV data sets are shifted to zero mean barycentric velocity.
The RV data are superimposed with our best-fit model. The lower panel
shows the residuals from the best-fit. Note the different vertical
scales on the two panels. The transit occurs at zero
orbital phase. See text for further details.}\label{fig:hatp2:rv}
\end{figure}

In Table~\ref{tab:hatp2:rvs} we collected all (previously published and our
newly obtained) radial velocity measurements. 
In Fig.~\ref{fig:hatp2:rv} we show the RV data, overplotted
with our best-fit model solution (for details of the fit,
see Sec.~\ref{sec:hatp2:analysis}).


\begin{table}
\caption{Comprehensive list of relative radial velocity measurements 
for \hatcurB{}. The Keck measurements marked with an asterix and the
Lick measurements are published in \citet{bakos2007hatp2}. The
OHP/SOPHIE data are taken from \citet{loeillet2008}.}\label{tab:hatp2:rvs}
\begin{center}
\begin{tabular}{lrrr}
\hline
${\rm BJD}-{2M4}$	& RV ($\rm m/s$)& $\sigma_{\rm RV} (\rm m/s)$	& Source	\\
\hline
53981.77748 \dotfill	&  $    12.0 $ 	&   $     7.3$	& Keck$^\star$ \\     
53982.87168 \dotfill	&  $  -288.3 $ 	&   $     7.9$	& Keck$^\star$ \\     
53983.81485 \dotfill	&  $   569.0 $ 	&   $     7.3$	& Keck$^\star$ \\     
54023.69150 \dotfill	&  $   727.3 $ 	&   $     7.8$	& Keck$^\star$ \\     
54186.99824 \dotfill	&  $   721.3 $ 	&   $     7.7$	& Keck$^\star$ \\     
54187.10415 \dotfill	&  $   711.0 $ 	&   $     6.7$	& Keck$^\star$ \\     
54187.15987 \dotfill	&  $   738.1 $ 	&   $     6.8$ 	& Keck$^\star$ \\     
54188.01687 \dotfill	&  $   783.6 $ 	&   $     7.1$ 	& Keck$^\star$ \\     
54188.15961 \dotfill	&  $   801.8 $ 	&   $     6.7$ 	& Keck$^\star$ \\     
54189.01037 \dotfill	&  $   671.0 $ 	&   $     6.7$ 	& Keck$^\star$ \\     
54189.08890 \dotfill	&  $   656.7 $ 	&   $     6.8$ 	& Keck$^\star$ \\     
54189.15771 \dotfill	&  $   640.2 $ 	&   $     6.9$ 	& Keck$^\star$ \\     
54216.95938 \dotfill 	&  $   747.7 $ 	&   $     8.1$ 	& Keck \\     
54279.87688 \dotfill 	&  $   402.0 $ 	&   $     8.3$ 	& Keck \\     
54285.82384 \dotfill 	&  $   168.3 $ 	&   $     5.7$ 	& Keck \\     
54294.87869 \dotfill 	&  $   756.8 $ 	&   $     6.5$ 	& Keck \\     
54304.86497 \dotfill 	&  $   615.5 $ 	&   $     6.2$ 	& Keck \\     
54305.87010 \dotfill 	&  $   764.2 $ 	&   $     6.3$ 	& Keck \\     
54306.86520 \dotfill 	&  $   761.4 $ 	&   $     7.6$ 	& Keck \\     
54307.91236 \dotfill 	&  $   479.1 $ 	&   $     6.5$ 	& Keck \\     
54335.81260 \dotfill 	&  $   574.7 $ 	&   $     6.8$ 	& Keck \\     
54546.09817 \dotfill 	&  $  -670.9 $ 	&   $    10.1$ 	& Keck \\     
54547.11569 \dotfill 	&  $   554.6 $ 	&   $     7.4$ 	& Keck \\     
54549.05046 \dotfill 	&  $   784.8 $ 	&   $     9.2$ 	& Keck \\     
54602.91654 \dotfill 	&  $   296.3 $ 	&   $     7.0$ 	& Keck \\     
54603.93210 \dotfill 	&  $   688.0 $ 	&   $     5.9$ 	& Keck \\     
54168.96790 \dotfill 	&  $  -152.7 $ 	&   $    42.1$ 	& Lick$^{\rm a}$ \\     
54169.95190 \dotfill 	&  $   542.4 $ 	&   $    41.3$ 	& Lick$^{\rm a}$ \\     
54170.86190 \dotfill 	&  $   556.8 $ 	&   $    42.6$ 	& Lick$^{\rm a}$ \\     
54171.03650 \dotfill 	&  $   719.1 $ 	&   $    49.6$ 	& Lick$^{\rm a}$ \\     
54218.80810 \dotfill 	&  $ -1165.2 $ 	&   $    88.3$ 	& Lick$^{\rm a}$ \\     
54218.98560 \dotfill 	&  $ -1492.6 $ 	&   $    90.8$ 	& Lick$^{\rm a}$ \\     
54219.93730 \dotfill 	&  $   -28.2 $ 	&   $    43.9$ 	& Lick$^{\rm a}$ \\     
54219.96000 \dotfill 	&  $   -14.8 $ 	&   $    43.9$ 	& Lick$^{\rm a}$ \\     
54220.96410 \dotfill 	&  $   451.6 $ 	&   $    38.4$ 	& Lick$^{\rm a}$ \\     
54220.99340 \dotfill 	&  $   590.7 $ 	&   $    37.1$ 	& Lick$^{\rm a}$ \\     
54227.50160 \dotfill 	&  $-19401.4 $ 	&   $     8.8$ 	& \sophie{}$^{\rm b}$ \\     
54227.60000 \dotfill 	&  $-19408.2 $ 	&   $     6.5$ 	& \sophie{}$^{\rm b}$ \\     
54228.58420 \dotfill 	&  $-19558.1 $	&   $    18.8$ 	& \sophie{}$^{\rm b}$ \\     
54229.59930 \dotfill 	&  $-20187.4 $ 	&   $    16.1$ 	& \sophie{}$^{\rm b}$ \\     
54230.44750 \dotfill 	&  $-21224.9 $ 	&   $    14.1$ 	& \sophie{}$^{\rm b}$ \\     
54230.60290 \dotfill 	&  $-20853.6 $ 	&   $    14.8$ 	& \sophie{}$^{\rm b}$ \\     
54231.59870 \dotfill 	&  $-19531.1 $ 	&   $    12.1$ 	& \sophie{}$^{\rm b}$ \\     
54236.51900 \dotfill 	&  $-20220.7 $ 	&   $     5.6$ 	& \sophie{}$^{\rm b}$ \\     
\hline
\end{tabular}
\end{center}
\end{table}



\phdsection{An analytical formalism for Kepler's problem}
\label{sec:hatp2:eof}

In this section we present a set of analytic relations (based 
on a few smooth functions defined in a closed form) 
that provides a straightforward solution of Kepler's 
problem, and consequently, time series
of RV data and RV model functions. Due to the analytic property, the
partial derivatives can also be obtained directly and therefore can be
utilized in various fitting and data analysis methods, including the
Fisher analysis of uncertainties and correlations.
The functions presented here are nearly as simple to manage 
as trigonometric functions. This section has three major parts.
In Sec.~\ref{sec:hatp2:eof:formalism}, the basics of the mathematical
formalism are presented, including the rules for calculating
partial derivatives. In Sec.~\ref{sec:hatp2:eof:miscformalism}, the
solution of the spatial problem is shown, supplemented with the inverse
problem, still using infinitely differentiable functions. This part
also discusses how transits constrain the phase of the radial velocity curve.
And finally, in Sec.~\ref{sec:hatp2:eof:lfit}, we show how the 
presented formalism can be implemented in practice, in the framework
of the \texttt{lfit} program and involving some of the built-in functions.

\phdsubsection{Mathematical formalism}
\label{sec:hatp2:eof:formalism}

The solution for the time evolution of Kepler's problem can be derived
in the standard way as given in various 
textbooks \citep[see, e.g.,][]{murray1999}. 
The restricted two body problem itself is an integrable ordinary differential 
equation. In the planar case, three independent integrals of motion exist
and one variable with uniform monotonicity (i.e. which is an affine 
function of time). The integrals are related to the well known orbital 
elements, that are used to characterize the orbit. These are
the semimajor axis $a$, the eccentricity $e$ and the longitude of 
pericenter\footnote{In two dimensions, the argument
of pericenter is always equal to the longitude of pericenter, i.e.
$\varpi\equiv\omega$} $\varpi$. 
The fourth quantity is the mean anomaly $M=nt$, where
$n=\sqrt{\mu/a^3}=2\pi/P$, the mean motion, which is zero at 
pericenter passage\footnote{The mass
parameter of Kepler's problem is denoted by $\mu\equiv \mathcal{G}(m_1+m_2)$,
where $m_1$ and $m_2$ are the masses of the two orbiting bodies
and $ \mathcal{G}$ is the Newtonian gravitational constant.}.
The solution to Kepler's problem can be given in terms of the
mean anomaly $M$ as defined as
\begin{equation}
E-e\sin E=M,
\end{equation}
where $E$ is the eccentric anomaly. The spatial coordinates are
\begin{eqnarray}
\xi  & = & \xi_0 \cos\varpi - \eta_0\sin\varpi, \label{ellipxi} \\
\eta & = & \xi_0 \sin\varpi + \eta_0\cos\varpi, \label{ellipeta}
\end{eqnarray}
where
\begin{eqnarray}
\xi_0  & = & a(\cos E-e), \label{ellip0xi} \\
\eta_0 & = & a\sqrt{1-e^2}\sin E \label{ellip0eta};
\end{eqnarray}
see also \citet{murray1999}, Sect.~2.4 for the derivation of these equations.
Since for circular orbits the longitude of pericenter and 
pericenter passage cannot be defined, and for nearly 
circular orbits, these can only be badly constrained; in these
cases it is useful to define a new variable, the mean 
longitude as $\lambda=M+\varpi$ to use instead of $M$. Since
$\varpi$ is an integral of the motion, $\dot\lambda=\dot M=n$. 
Therefore for circular orbits $\varpi\equiv0$ and 
\eqrefs{ellip0xi}{ellip0eta} should be replaced by
\begin{eqnarray}
\xi_0  & = & a\cos\lambda, \label{circ0xi} \\
\eta_0 & = & a\sin\lambda. \label{circ0eta}
\end{eqnarray}
To obtain an analytical solution to the problem, i.e. which is 
infinitely differentiable with respect to all of the orbital elements
and the mean longitude, first let us define the Lagrangian orbital
elements $k=e\cos\varpi$ and $h=e\sin\varpi$. 
Substituting \eqrefs{ellip0xi}{ellip0eta} into 
\eqrefs{ellipxi}{ellipeta} gives
\begin{equation}
\binom{\xi}{\eta}=a\left[\binom{c}{s}+\frac{e\sin E}{2-\ell}\binom{+h}{-k}-\binom{k}{h}\right], \label{kepler2dspatial}
\end{equation}
where $c=\cos(\lambda+e\sin E)$, $s=\sin(\lambda+e\sin E)$ and 
$\ell=1-\sqrt{1-e^2}$, the oblateness of the orbit. The derivation
of the above equation is straightforward, one should only keep in mind
that $E+\varpi=\lambda+e\sin E$. In the first part of this section
we prove that the quantities 
\begin{equation}
\eop(\lambda,k,h)=\left\{\begin{tabular}{ll} $0$ & if $k=0$ and $h=0$ \\ $e\sin E$  & otherwise\end{tabular}\right. \label{eopdef}
\end{equation}
and
\begin{equation}
\eoq(\lambda,k,h)=\left\{\begin{tabular}{ll} $0$ & if $k=0$ and $h=0$ \\ $e\cos E$  & otherwise\end{tabular}\right. \label{eoqdef}
\end{equation}
are analytic -- infinitely differentiable -- 
functions of $\lambda$, $k$ and $h$ for all real values
of $\lambda$ and for all $k^2+h^2=e^2<1$. In the following parts, we utilize
the partial derivatives of these analytic functions to obtain
the orbital velocities, and we also derive some other useful 
relations. In this section
we only deal with planar orbits, the three dimensional case is discussed
in the next section.

\phdsubsubsection{Partial derivatives and the analytic property}
A real function is analytic when all of its partial derivatives
exist, the partial derivatives are continuous functions and 
only depend on other analytic functions. It is proven in 
\cite{pal2009kepler} that the partial derivatives of 
$q=\eoq(\lambda,k,h)$ and $p=\eop(\lambda,k,h)$ are the following
for $(k,h)\ne(0,0)$:
\begin{eqnarray}
\frac{\partial q}{\partial\lambda} & = & \frac{-p}{1-q}, \label{dqdl}\\
\frac{\partial q}{\partial k} & = & \frac{c-k}{1-q} = \frac{\cos(\lambda+p)-k}{1-q}, \label{dqdk}\\
\frac{\partial q}{\partial h} & = & \frac{s-h}{1-q} = \frac{\sin(\lambda+p)-h}{1-q}  \label{dqdh}
\end{eqnarray}
and
\begin{eqnarray}
\frac{\partial p}{\partial\lambda} & = & \frac{q}{1-q}, \label{dpdl}\\
\frac{\partial p}{\partial k} & = & \frac{+s}{1-q} = \frac{+\sin(\lambda+p)}{1-q}, \label{dpdk}\\
\frac{\partial p}{\partial h} & = & \frac{-c}{1-q} = \frac{-\cos(\lambda+p)}{1-q}. \label{dpdh}
\end{eqnarray}
Since for all $k^2+h^2<1$, $q<1$ and therefore $1-q>0$, 
all of the above functions are continuous on their domains.
Since the $\sin(\cdot)$ and $\cos(\cdot)$ functions are 
analytic, therefore one can conclude that
the functions $\eoq(\cdot,\cdot,\cdot)$ and $\eop(\cdot,\cdot,\cdot)$
are also analytic. 

Substituting the definition of $p=\eop(\lambda,k,h)$ into 
\eqref{kepler2dspatial}, one can write
\begin{equation}
\binom{\xi}{\eta}=a\left[\binom{\cos(\lambda+p)}{\sin(\lambda+p)}+\frac{p}{2-\ell}\binom{+h}{-k}-\binom{k}{h}\right], \label{kepler2dspatial2}
\end{equation}
while the radial distance of the orbiting body from the center is 
$\sqrt{\xi^2+\eta^2}=r=a(1-q)$. For small eccentricities
in \eqref{kepler2dspatial2} the third term $(k,h)$ is negligible
compared to the first term $(\cos,\sin)$ while the second 
term $(h,-k)p/(2-\ell)$ is negligible compared to the third term.
Therefore for $e\ll1$, $p$ is proportional to the phase offset in the polar
angle of the orbiting particle (as defined from the geometric center
of the orbit) and $q$ is proportional to the distance 
offset relative to a circular orbit; both caused by the non-zero orbital
eccentricity. 

Since \eqref{kepler2dspatial2} is a combination of purely
analytic functions, the solution of Kepler's problem is analytic
with respect to the orbital elements $a$, $(k,h)$, and to
the mean longitude $\lambda$ in the domain $a>0$ and $k^2+h^2<1$.
We note here that this formalism omits the parabolic or hyperbolic solutions.
The formalism based on the Stumpff functions \citep[see][]{stiefel1971}
provides a continuous set of formulae for the elliptic, parabolic,
and hyperbolic orbits but this parametrization is still singular 
in the $e\to0$ limit.

\phdsubsubsection{Orbital velocities}
Assuming a non-perturbed orbit, i.e. when $(\dot k,\dot h)=0$, and $\dot a=0$
and when the mean motion $n=\dot\lambda$ is constant, the orbital velocities
can be directly obtained by calculating the partial derivative 
of \eqref{kepler2dspatial2} with respect to $\lambda$ and applying the 
chain rule since
\begin{equation}
\frac{\partial}{\partial t}\binom{\xi}{\eta}\equiv
\binom{\dot\xi}{\dot\eta}=
\left[\frac{\partial}{\partial \lambda}\binom{\xi}{\eta}\right]\frac{\partial\lambda}{\partial t}
=n\frac{\partial}{\partial \lambda}\binom{\xi}{\eta}. 
\end{equation}
Substituting the partial derivative \eqref{dpdl} into the
expansion of $\partial\xi/\partial\lambda$ and $\partial\eta/\partial\lambda$
one gets
\begin{equation}
\binom{\dot\xi}{\dot\eta}=\frac{an}{1-q}\left[\binom{-\sin(\lambda+p)}{+\cos(\lambda+p)}+\frac{q}{2-\ell}\binom{+h}{-k}\right]. \label{kepler2dvelocity}
\end{equation}
Note that \eqref{kepler2dvelocity} is also a combination of purely
analytic functions, the components of the orbital velocity are analytic
with respect to the orbital elements $a$, $(k,h)$, and to
the mean longitude $\lambda$.

It is also evident that the time derivative of \eqref{kepler2dvelocity} is
\begin{eqnarray}
\binom{\ddot\xi}{\ddot\eta} & = &\frac{-an^2}{(1-q)^3}\left[\binom{\cos(\lambda+p)}{\sin(\lambda+p)}+\right.  \label{kepler2daccel2} \\
& & \left.+\frac{p}{2-\ell}\binom{+h}{-k}-\binom{k}{h}\right]. \nonumber
\end{eqnarray}
Obviously, \eqref{kepler2daccel2} can be written as 
\begin{equation}
\binom{\ddot\xi}{\ddot\eta}=-\frac{n^2}{(1-q)^3}\binom{\xi}{\eta},
\end{equation}
which is equivalent to the equations of motion since 
$\mu=n^2a^3$ and $\sqrt{\xi^2+\eta^2}=r=a(1-q)$.

\phdsubsection{Additional constraints given by the transits}
\label{sec:hatp2:eof:miscformalism}

In the follow-up observations of planets discovered by transits in photometric
data series, the detection of variations in the RV signal is 
one of the most relevant steps, either to rule out transits of late-type dwarf 
stars, and/or blends, or to characterize the mass of the planet and the 
orbital parameters. Since transit timing constrains the epoch and orbital
period much more precisely than radial velocity alone, 
these two can be assumed  to be fixed in the analysis of the RV data.
However, this constraint also 
includes an additional feature. The mean longitude has to be 
shifted to the transits since it is $\pi/2$ only for circular orbits
at the time of the transit. It can be 
shown that the mean longitude at the time instance of the transit is
\begin{equation}
\lambda_{\rm tr}=\arg\left(k+\frac{kh}{2-\ell},1+h-\frac{k^2}{2-\ell}\right)-\frac{k(1-\ell)}{h},\label{eq:lamtran}
\end{equation}
therefore the mean longitude at the orbital phase $\varphi$ becomes
$\lambda=\lambda_{\rm tr}+2\pi\varphi$. Thus, the observed radial velocity
signal is proportional to the $\dot\eta$ component of the velocity vector,
namely 
\begin{eqnarray}
{\rm RV} & = & \gamma + K_0 v, \\
v & = & \dot\eta(\lambda_{\rm tr}+2\pi\varphi,k,h), \label{eq:rvbase}
\end{eqnarray}
where $\gamma$ is the mean barycentric velocity and $K_0$ is 
related to the semi-amplitude $K$ as $K_0=K\sqrt{1-e^2}$.
Consequently,
the partial derivatives of the $v=\dot\eta$ RV component,
$v=\dot\eta(\lambda_{\rm tr}+2\pi\varphi,k,h)$ with respect
to the orbital elements $k$ and $h$ are
\begin{eqnarray}
\frac{\partial v}{\partial k} & = & \frac{\partial \dot\eta}{\partial k}+
        \frac{\partial \dot\eta}{\partial\lambda}\frac{\partial \lambda_{\rm tr}}{\partial k}, \label{eq:drvdk}\\
\frac{\partial v}{\partial h} & = & \frac{\partial \dot\eta}{\partial h}+
        \frac{\partial \dot\eta}{\partial\lambda}\frac{\partial \lambda_{\rm tr}}{\partial h}. \label{eq:drvdh}
\end{eqnarray}

A radial velocity curve of a star, caused by the perturbation of
a single companion can be parametrized by six quantities: the semi-amplitude
of RV variations, $K$, the zero point, $G$, the Lagrangian orbital
elements, $(k,h)$, the epoch, $T_0$ (or equivalently the phase at an 
arbitrary fixed time instant) and the period $P$. In the cases of 
transiting planets, the later two are known since
the photometric observations of the transits constrain both quantities
with exceeding 
precision (relative to the precision attainable purely by the RV data).
Therefore, one has to fit only four quantities, i.e. 
$\mathbf{a}=(K,G,k,h)$.

\begin{figure*}
\begin{lcmd}
\small
lfit~~~ -x "eoc(l,k,h)=cos(l+eop(l,k,h))" $\backslash$ \\
\ptab	-x "eos(l,k,h)=sin(l+eop(l,k,h))" $\backslash$ \\
\ptab	-x "J(k,h)=sqrt(1-k*k-h*h)" $\backslash$ \\
\ptab	-x "lamtranxy(x,y,k,h)=arg(k+x+h*(k*y-h*x)/(1+J(k,h)),h+y-k*(k*y-h*x)/(1+J(k,h)))- \\
\ptab	\ptab	(k*y-h*x)*J(k,h)/(1+k*x+h*y)" $\backslash$ \\
\ptab	-x "lamtran(l0,k,h)=lamtranxy(cos(l0),sin(l0),k,h)" $\backslash$ \\
\ptab	-x "prx0(l,k,h)=(+eoc(l,k,h)+h*eop(l,k,h)/(1+J(k,h))-k)" $\backslash$ \\
\ptab	-x "pry0(l,k,h)=(+eos(l,k,h)-k*eop(l,k,h)/(1+J(k,h))-h)" $\backslash$ \\
\ptab	-x "rvx0(l,k,h)=(-eos(l,k,h)+h*eoq(l,k,h)/(1+J(k,h)))/(1-eoq(l,k,h))" $\backslash$ \\
\ptab	-x "rvy0(l,k,h)=(+eoc(l,k,h)-k*eoq(l,k,h)/(1+J(k,h)))/(1-eoq(l,k,h))" $\backslash$ \\
\ptab	-x "prx1(l,k,h)=prx0(l+lamtranxy(0,1,k,h),k,h)" $\backslash$ \\
\ptab	-x "pry1(l,k,h)=pry0(l+lamtranxy(0,1,k,h),k,h)" $\backslash$ \\
\ptab	-x "rvx1(l,k,h)=rvx0(l+lamtranxy(0,1,k,h),k,h)" $\backslash$ \\
\ptab	-x "rvy1(l,k,h)=rvy0(l+lamtranxy(0,1,k,h),k,h)" $\backslash$ \\
\ptab	-x "rvbase(l,k,h)=rvy1(l,k,h)" $\backslash$ \\
\ptab	\dots
\end{lcmd}%
\caption{Macro definitions for 
\texttt{lfit}, implementing some functions related to radial velocity analysis.
All of the above functions are based on the eccentric offset functions
\texttt{eop(.,.,.)} and \texttt{eoq(.,.,.)} as defined by equations
(\ref{eopdef}) and (\ref{eoqdef}).}
\label{fig:lfitrvmacros}
\end{figure*}

\phdsubsection{Practical implementation}
\label{sec:hatp2:eof:lfit}

The eccentric offset functions $\eop(\lambda,k,h)$ and $\eoq(\lambda,k,h)$
are implemented in the program \texttt{lfit} (see also 
Sec.~\ref{sec:prog:lfit}). This program does not provide further functionality
related to the radial velocity analysis, however, the macro definition
capabilities of the program can be involved in order to define 
some more useful functions which then can be directly applied in
real problems. The shell script pieces shown in Fig.~\ref{fig:lfitrvmacros}
demonstrate how equations~(\ref{eq:lamtran}) and (\ref{eq:rvbase}) 
are implemented in practice. The parametric derivatives of these
functions, such as equation~(\ref{eq:drvdk}) or (\ref{eq:drvdh}) are
then derived automatically by \texttt{lfit}, using the partial derivatives
of the base functions $\eop(\lambda,k,h)$ and $\eoq(\lambda,k,h)$ as
well as the chain rule.



\phdsection{Analysis of the HAT-P-2 planetary system}
\label{sec:hatp2:analysis}

In this section we briefly describe the analysis of the available
photometric and radial velocity data of HAT-P-2 in order to determine the
planetary parameters as accurately as possible. The modelling was
done in three major steps in an iterative way. The first step 
was the modelling of the light curve and the radial velocity data 
series. Second, this was followed by the determination of the stellar 
parameters. In the last step, by combining the light curve parameters 
with the stellar properties, we obtained the physical 
parameters (mass, radius) of the planet. 

To model transit light curves taken in optical or near-infrared
photometric passbands, we include the effect of the stellar
limb darkening. We have used the formalism of \cite{mandel2002} 
to model the flux decrease during transits under the 
assumption of quadratic limb darkening law.
Since the limb darkening coefficients are the function
of the stellar atmospheric parameters (such as effective 
temperature $\teffstar$, 
surface gravity $\log g_\star$ and metallicity), 
the whole light curve analysis should be preceded
by the initial derivation of these parameters. These parameters
were obtained by collaborators, 
using the iodine-free template spectrum obtained by the 
HIRES instrument on Keck~I and employing the Spectroscopy Made Easy
software package \citep{valenti1996}, supported by the atomic line
database of \cite{valenti2005}. This analysis yields the 
$\teffstar$, $\log g_\star$, $\mathrm{[Fe/H]}$ and the projected
rotational velocity $v\sin i$. The result of the SME analysis when all
of these values have been adjusted simultaneously were 
$\log g_\star=4.22\pm0.14$\,(CGS), $\teffstar=6290\pm110$\,K, 
$\mathrm{[Fe/H]}=0.12\pm0.08$ and $v\sin i=20.8\pm0.2\,\rm{km\,s^{-1}}$.

The limb darkening coefficients
are then derived for $z'$ and $I$ photometric bands by interpolation, 
using the tables provided by \cite{claret2000} and \cite{claret2004}. 
The initial values for the coefficients were
$\gamma_1^{(z)}=0.1430$,
$\gamma_2^{(z)}=0.3615$,
$\gamma_1^{(I)}=0.1765$, and
$\gamma_2^{(I)}=0.3688$.
After the first iteration, with the knowledge of
the stellar parameters, the SME analysis is repeated
by fixing the surface gravity to the value yielded by the stellar evolution
modelling. This can be done in a straightforward way: 
the normalized semimajor axis $a/R_\star$ can be obtained from the transit light
curve model parameters, the orbital eccentricity and
the argument of pericenter. As it was pointed out by 
\cite{sozzetti2007}, the ratio $a/R_\star$ is a more effective luminosity
indicator than the stellar surface gravity, since the stellar density 
is related to  
\begin{equation}
\rho_\star \propto (a/R_\star)^3. \label{rhovsar}
\end{equation}
Since \hatcurBb{} is a quite massive planet, i.e. 
$M_{\rm p}/M_\star \sim 0.01$, relation~(\ref{rhovsar}) requires
a significant correction, which also depends on observable quantities 
\citep[see][for more details]{pal2008param}. In our case, this
correction is not negligible since $M_{\rm p}/M_\star$ is comparable
to the typical relative uncertainties in the light curve parameters.

\phdsubsection{Light curve and radial velocity parameters}

The first major step of the analysis is the determination of the 
light curve and radial velocity parameters. 
We performed a joint fit
by adjusting the light curve and radial velocity parameters simultaneously
as described below.

The parameters can be classified into three major groups. 
The light curve parameters that are related to
the \emph{physical} properties of the planetary system are the
transit epoch $E$, the period $P$, the fractional planetary 
radius $p\equiv R_{\rm p}/R_\star$, the impact parameter $b$, and the 
normalized semimajor axis $a/R_\star$. The physical radial velocity parameters
are the RV semi-amplitude $K$, the orbital eccentricity $e$ and
the argument of pericenter $\omega$. In the third group there are
parameters that are not related to the physical properties of the
system, but are rather instrumentation specific ones. 
These are the out-of-transit instrumental magnitudes of the 
follow-up (and HATNet) light curves, and the RV zero-points $\gamma_{\rm Keck}$,
$\gamma_{\rm Lick}$ and $\gamma_{\rm OHP}$ of the three individual data 
sets\footnote{Since in the reduction of the 
\cite{loeillet2008} data a synthetic
stellar spectrum was used as a reference, $\gamma_{\rm OHP}$ is the
physical barycentric radial velocity of the system. 
In the reductions of the Keck and Lick data, we used one of the 
spectra as a template, therefore the zero-points of these two are arbitrary,
lack any real physical interpretation.}.

To minimize the correlation between the adjusted parameters, we use 
a slightly different parameter set. Instead of adjusting the epoch and
period, we fitted the first and last available transit center time, 
$T_{-148}$ and $T_{+71}$.
Here indices note the transit event number:
the $N_{\rm tr}\equiv0$ event was defined as
the first complete follow-up light curve taken on 2007 April 21,
the first available transit observation from the HATNet data was
the event $N_{\rm tr}\equiv-148$ and the last follow-up was
observed on 2008 May 25, was event $N_{\rm tr}\equiv+71$. Note
that assuming equidistant transit cadences, all of the transit centers
available in the HATNet and follow-up photometry are constrained
by these two transit instances \citep[see][]{bakos2007hatp5,pal2008hatp7}.
Similarly, instead of the eccentricity $e$ and argument of pericenter $\omega$,
we have adjusted the Lagrangian orbital elements $k\equiv e\cos\omega$
and $h\equiv e\sin\omega$. These elements show no correlation
in practice, moreover, the radial velocity curve is an analytic
function of these even for $e\to 0$ cases (although in the case
of \hatcurBb{} this is irrelevant because $e$ is non-zero).
As it is known in the literature \citep{winn2007b,pal2008lcdiff}, 
the impact parameter $b$ and
$a/R_\star$ are also strongly correlated, especially for 
small $p\equiv R_{\rm p}/R_\star$ values. Therefore, 
as it was suggested by \cite{bakos2007hatp5},
we chose the parameters $\zeta/R_\star$ and $b^2$ for fitting instead
of $a/R_\star$ and $b$, where for eccentric orbits $\zeta/R_\star$ is 
related to $a/R_\star$ as
\begin{equation}
\frac{\zeta}{R_\star} = \left(\frac{a}{R_\star}\right)\frac{2\pi}{P}\frac{1}{\sqrt{1-b^2}}\frac{\sqrt{1-e^2}}{1+h}.
\end{equation}
The quantity $\zeta/R_\star$ is related to the transit duration as
$T_{\rm dur}=2 (\zeta/R_\star)^{-1}$, if the duration is defined 
between the time instants when the center of the planet crosses the limb
of the star inwards and outwards. 

\phdsubsection{Effects of the orbital eccentricity}

Let us denote the projected radial distance between the center of the 
planet and the center of the star (normalized by $R_\star$) by $d$.
As it was shown in \cite{pal2008lcdiff}, $d$ can 
be parametrized in a second order approximation as 
\begin{equation}
d^2=(1-b^2)\left(\frac{\zeta}{R_\star}\right)^2(\Delta t)^2+b^2, \label{zlinear}
\end{equation}
where $\Delta t$ is the time between the actual observation time and the
intrinsic transit center. The \emph{intrinsic transit} center is defined
when the planet reaches its maximal tangential velocity 
during the transit. Although the tangential velocity cannot be measured
directly, the intrinsic transit center is determined by purely 
the radial velocity data, without any knowledge of the transit 
geometry\footnote{In other words, predictions can only be made 
for the intrinsic transit center in cases where the planet 
was discovered by a radial velocity survey and initially we have no
further constraint for the geometry of the system.}.
For eccentric orbits the impact parameter $b$ is 
related to the orbital inclination $i$ as
\begin{equation}
b=\left(\frac{a}{R_\star}\right)\cos i\frac{1-e^2}{1+h}.
\end{equation}
In order to have a better description of the transit light curve,
we used a higher order expansion in the $d(\Delta t)$ function 
(Eq.~\ref{zlinear}).
For circular orbits, such an expansion is straightforward. To derive
the expansion for elliptic orbits, we employed the method of Lie-integration
which gives the solution of any ordinary differential 
equation (here, the equations for the two-body problem)
in a recursive series for the Taylor expansion with respect to 
the independent variable (here, the time). It can be shown
involving the Taylor expansion of the orbital motion that 
the normalized projected distance $d$ up to fourth order is:
\begin{eqnarray}
d^2 & = & b^2\left[1-2R\varphi-(Q-R^2)\varphi^2-\frac13QR\varphi^3\right] + \nonumber \\
 & & \left(\frac{\zeta}{R_\star}\right)^2(1-b^2)\Delta t^2\left[1-\frac13Q\varphi^2+\frac12QR\varphi^3\right], \label{d2expand}
\end{eqnarray}
where 
\begin{equation}
Q  =  \left(\frac{1+h}{1-e^2}\right)^3, \label{corrQ}
\end{equation}
and 
\begin{equation}
R  = \frac{1+h}{(1-e^2)^{3/2}}k. \label{corrR}
\end{equation}
Here $n=2\pi/P$ is the mean motion, and $\varphi$ is 
defined as $\varphi=n\Delta t$. For circular orbits, 
$Q=1$ and $R=0$, and for small eccentricities ($e\ll1$), 
$Q\approx 1+3h$ and $R\approx k$.
The leading order correction term in $\varphi$, $-2b^2R\varphi$, is related
to the time lag between the photometric and intrinsic transit centers.
The \emph{photometric transit center} is defined halfway between
the instants when the center of the planet crosses the limb of the
star inward and outward. 
It is easy to show by solving the equation
$d(\varphi)=1$, yielding two solutions ($\varphi_{\rm I}$
and $\varphi_{\rm E}$), that this phase lag is:
\begin{eqnarray}
\Delta\varphi & = & \frac{\varphi_{\rm I}+\varphi_{\rm E}}{2} = \\
& = & -\frac{b^2 R}{\left(\frac{\zeta}{R_\star}\frac{1}{n}\right)^2(1-b^2)-(Q-R^
2)b^2} \approx \\
& \approx & -\left(\frac{a}{R_\star}\right)^{-2}\frac{b^2 k}{(1+h)\sqrt{1-e^2}},
\end{eqnarray}
which can result in a time lag of several minutes.

In \eqref{d2expand}, the third order terms 
in $\varphi$ describe the asymmetry between the slopes of 
the ingress and egress parts of the light curve. 
For some other aspects of 
light curve asymmetries see \cite{loeb2005} and \cite{barnes2007}.
In the cases when no
assumptions are known for the orbital eccentricity, 
we cannot treat the parameters $R$ and $Q$ as independent
since the intrinsic transit center 
and $R$ have an exceptionally high correlation. 
However, if we assume a simpler model function, with only third order terms
in $\varphi$ with fitted coefficients present, i.e.
\begin{eqnarray}
d^2 & = & b^2\left[1-\varphi^2-\frac13C\varphi^3\right] + \nonumber \\
 & & \left(\frac{\zeta}{R_\star}\right)^2(1-b^2)\Delta t^2\left[1-\frac13\varphi^2+\frac12C\varphi^3\right], \label{d2expand2}
\end{eqnarray}
yields a non-zero value for the $C$ coefficient for asymmetric light 
curves. In the case of \hatcurBb{}, the derived values for $Q$ and $R$ are
$Q=2.23\pm0.10$ and $R=-0.789\pm0.021$ (derived from the values of $k$ and $h$,
see Sec.~\ref{sec:hatp2:jointfit}), thus the coefficient for the third 
order term in $\varphi$ is $QR=-1.75\pm0.13$. Using \eqref{d2expand2},
for an ``ideal'' light 
curve (with similar parameters of $k$, $h$, $\zeta/R_\star$ and $b^2$ as for
\hatcurBb{}), the best fit value for $C$ is $C=-2.23$, which is close 
to the value of $QR\approx-1.75$. The difference between the best fit 
value of $C$ and the fiducial value of $QR$ is because in \eqref{d2expand2}
the coefficients for the first and second order terms were fixed to be 
$0$ and $1$, respectively. Although this asymmetry can be 
measured directly (without leading to any degeneracy 
between the fit parameters), in practice we need extreme photometric precision to
obtain a significant detection for a non-zero $C$ parameter: assuming
a photometric time series for a single transit of \hatcurBb{} 
with $5\,{\rm sec}$ cadence where each 
individual measurement has a photometric error of $0.01$\,mmag(!), 
the uncertainty in $C$ is $\pm0.47$, equivalent to a 5-$\sigma$
detection of the light curve asymmetry. This detection would be hard
for ground-based instrumentation (i.e. for a 1-$\sigma$ detection one
should achieve a photometric precision of $0.05$\,mmag at the same cadence).
Space missions like {\it Kepler} \citep{borucki2007} will
be able to detect orbital eccentricity relying only on photometry
of primary transits.

\phdsubsection{Joint fit}
\label{sec:hatp2:jointfit}

As it was discussed before, in order to achieve a self-consistent 
fit, we performed a simultaneous fit on all of the light curve
and radial velocity data. We have involved \eqref{d2expand}, 
to model the light curves, where the parameters $Q$ and $R$ were
derived from the actual values of $k$ and $h$, using equations \eqref{corrQ}
and \eqref{corrR}.
To find the best-fit values for the parameters we used the downhill
simplex algorithm \citep[see][]{press1992} and we used the method
of refitting to synthetic data sets to get an \emph{a posteriori}
distribution for the adjusted values. The final results of the fit
were 
$T_{-148}=2453379.10281\pm0.00141$,
$T_{+71} =2454612.83271\pm0.00075$,
$K=958.9\pm13.9\,\mathrm{m\,s^{-1}}$,
$k=-0.5119\pm0.0040$,
$h=-0.0543\pm0.0098$,
$R_{\rm p}/R_\star\equiv p=0.0724\pm0.0010$,
$b^2=0.125\pm0.073$,
$\zeta/R_\star=12.090\pm0.046\,\mathrm{day}^{-1}$,
$\gamma_{\rm Keck}=318.4\pm6.6\,\mathrm{m\,s^{-1}}$,
$\gamma_{\rm Lick}=77.0\pm30.4\,\mathrm{m\,s^{-1}}$,
$\gamma_{\rm OHP}=-19868.9\pm9.8\,\mathrm{m\,s^{-1}}$. 
The uncertainties
of the out-of-transit magnitudes were between $(6\dots21)\times10^{-5}\,{\rm mag}$
for the follow-up light curves and $16\times10^{-5}\,{\rm mag}$ for the 
HATNet data. The fit resulted a normalized $\chi^2$ value of $0.995$. 
As it is described in the following 
subsection, the resulted distribution has been used then as an input
for the stellar evolution modelling.

\begin{table}
\caption{Stellar parameters for 
\hatcurB{}. The values of effecitve temperature, metallicity and projected 
rotational velocity are based on purely spectroscopic data (SME) while the 
other ones are derived from the both the spectroscopy (SME) and the 
joint modelling (LC+Y$^2$).}
\label{tab:hatp2:stellar}
\begin{center}
\begin{tabular}{lcl}
\hline
Parameter		& Value			& Source \\
\hline
$T_{\rm eff}$ (K)	&  \hatcurBSMEteff	& SME$^{\rm a}$ \\
$[\mathrm{Fe/H}]$	&  \hatcurBSMEzfeh	& SME \\
$v \sin i$ (\kms)	&  \hatcurBSMEvsin	& SME \\
$M_\star$ ($M_{\sun}$)  &  \hatcurBYYm		& Y$^2$+LC+SME$^{\rm a}$ \\
$R_\star$ ($R_{\sun}$)  &  \hatcurBYYr		& Y$^2$+LC+SME \\
$\log g_\star$ (cgs)    &  \hatcurBYYlogg	& Y$^2$+LC+SME \\
$L_\star$ ($L_{\sun}$)  &  \hatcurBYYlum	& Y$^2$+LC+SME \\
$M_V$ (mag)		&  \hatcurBYYmv   	& Y$^2$+LC+SME \\
Age (Gyr) 		&  \hatcurBYYage	& Y$^2$+LC+SME \\
Distance (pc)		&  \hatcurBXdist	& Y$^2$+LC+SME \\
\hline
\end{tabular}
\end{center}
\end{table}

\phdsubsection{Stellar parameters}
\label{sec:hatp2:stellarparams}

The second step of the analysis was the derivation of the physical
stellar parameters. Following the complete Monte-Carlo way of 
parameter estimation, as it was described by \cite{pal2008hatp7},
we calculated the distribution of the stellar density, derived from
the $a/R_\star$ values. To be more precise, the density of the
star is 
\begin{equation}
\rho_\star  =  \rho_0 - \frac{\Sigma_0}{R_\star}, \label{rhostar}
\end{equation}
where both $\rho_0$ and $\Sigma_0$ are directly related to observable 
quantities, namely
\begin{eqnarray}
\rho_0   & = & \frac{3\pi}{GP^2}\left(\frac{a}{R_\star}\right)^3, \label{rho0star}\\
\Sigma_0 & = & \frac{3K\sqrt{1-e^2}}{2PG\sin i}\left(\frac{a}{R_\star}\right)^2.
\end{eqnarray}
In \eqref{rhostar}, the only unknown quantity is the radius of the star,
which can be derived using a stellar evolution model, and 
it depends on a luminosity indicator (that is, in practice, the surface
gravity or the density of the star), a color indicator 
(which is the $\teffstar$
effective surface temperature, given by the SME analysis) and
the stellar composition (here $\mathrm{[Fe/H]}$). Therefore, one
can write
\begin{equation}
R_\star=R_\star(\rho_\star,\teffstar,\mathrm{[Fe/H]}). \label{rstar}
\end{equation}
Since both $\teffstar$ and $\mathrm{[Fe/H]}$ are known
from stellar atmospheric analysis, \eqref{rhostar}
and \eqref{rstar} have two unknowns, and thus this set of equations can be solved 
iteratively. Note that in order to solve \eqref{rstar}, supposing its
parameters are known in advance, one has to use a certain stellar
evolutionary model. Such models are available in tabulated form, therefore
the solution of the equation requires the inversion of the interpolating
function on the tabulated data. Thus, \eqref{rstar} is only a symbolical
notation for the algorithm which provides the solution. Moreover, if
the star is evolved, the isochrones and/or evolutionary tracks for the
stellar models intersect themselves, resulting an ambiguous solution
(i.e. it is not a ``function'' any more).
For \hatcurB{}, however, the solution of \eqref{rstar} is definite since
the host star is a main sequence star.
To obtain the physical parameters (e.g. the stellar radius),
we used the stellar evolutionary
models of \cite{yi2001}, by interpolating the values of $\rho_\star$,
$\teffstar$ and $\mathrm{[Fe/H]}$ using the interpolator provided by 
\cite{demarque2004}. 

The procedure described above has been applied to all of the parameters
in the input set, where the values of $\rho_0$ have been derived 
from the values of $a/R_\star$ and the orbital period $P$ using
\eqref{rho0star}, while the values for $\teffstar$ and $\mathrm{[Fe/H]}$
have been drawn from Gaussian random variables with the mean and
standard deviation of the first SME results 
($\teffstar=6290\pm110$\,K and $\mathrm{[Fe/H]}=0.12\pm0.08$).
This step resulted the \emph{a posteriori} distribution of the 
physical stellar parameters, including the surface gravity. The 
value and uncertainty for the latter was $\log g_\star=4.16\pm0.04$\,(CGS),
which is slightly smaller than the value provided by the SME analysis.
To reduce the uncertainties in $\teffstar$ and $\mathrm{[Fe/H]}$, 
we repeated the SME modelling by fixing the value of $\log g_\star$
to the above. This second SME run resulted $\teffstar=6290\pm60\,{\rm K}$
and $\mathrm{[Fe/H]}=0.14\pm0.08$. Following, we updated the values for
the limb darkening parameters 
($\gamma_1^{(z)}=0.1419$,
$\gamma_2^{(z)}=0.3634$,
$\gamma_1^{(I)}=0.1752$, and
$\gamma_2^{(I)}=0.3707$),
and repeated the simultaneous light curve
and radial velocity fit. The results of this fit were then used to
repeat the stellar evolution modelling, which yielded among other
parameters $\log g_\star=4.158\pm0.031$\,(CGS). 
Since the value of $\log g_\star$ did not change significantly, 
we accepted these stellar parameter values as final ones. 
The stellar parameters are 
summarized in Table~\ref{tab:hatp2:stellar} and the light curve
and radial velocity parameters are listed in the top two blocks
of Table~\ref{tab:hatp2:parameters}.

\phdsubsection{Planetary parameters}

In the previous two steps of the analysis, we determined the light 
curve, radial velocity and stellar parameters. In order to get the
planetary parameters, we combined the two Monte-Carlo data sets 
that yield their \emph{a posteriori} distribution in a consistent way.
For example, the mass of the planet is calculated using
\begin{equation}
M_{\rm p}=\frac{2\pi}{P}\frac{K\sqrt{1-e^2}}{G\sin i}
\left(\frac{a}{R_\star}\right)^2 R_\star^2,
\end{equation}
where the values for the period $P$, RV semi-amplitude $K$, eccentricity
$e$, inclination $i$, and normalized semimajor axis $a/R_\star$
were taken from the results of the light curve and RV fit while
the values for $R_\star$ were taken from the respective points of 
the stellar parameter distribution.
From the distribution of the planetary parameters, 
we obtained the mean values and uncertainties. We derived
$M_{\rm p}=8.84^{+0.22}_{-0.29}\,M_{\rm Jup}$ for
the planetary mass, $R_{\rm p}=1.123^{+0.071}_{-0.054}\,R_{\rm Jup}$
for the radius while the correlation between these parameters
were $C(M_{\rm p},R_{\rm p})=0.68$. 
The planetary parameters 
are summarized in the lower block of Table~\ref{tab:hatp2:parameters}.

Due to the eccentric orbit and
the lack of the knowledge of the heat redistribution of the incoming
stellar flux, the surface temperature of the planet can be constrained
with difficulties. Assuming complete heat redistribution, the 
surface temperature can be estimated by time averaging the 
incoming flux which varies as $1/r^2=a^{-2}(1-e\cos E)^{-2}$ due to the 
orbital eccentricity. The time average of $1/r^2$ is
\begin{equation}
\left<\frac{1}{r^2}\right>=\frac{1}{T}\int\limits_{0}^{T}\frac{\mathrm{d}t}{r^2(
t)}=
\frac{1}{2\pi}\int\limits_{0}^{2\pi}\frac{\mathrm{d}M}{r^2(M)},
\end{equation}
where $M$ is the mean anomaly of the planet. Since $r=a(1-e\cos E)$
and $\mathrm{d}M=(1-e\cos E)\mathrm{d}E$, where $E$ is the eccentric anomaly,
the above integral can be calculated analytically and the result is 
\begin{equation}
\left<\frac{1}{r^2}\right>=\frac{1}{a^2\sqrt{1-e^2}}. \label{semicorr}
\end{equation}
Using this time averaged weight for the incoming flux, we 
derived $T_{\rm p}=\hatcurBPPteff\,{\rm K}$. However, the planet surface 
temperature would be $\sim 2975\,{\rm K}$ on the dayside during periastron
and assuming no heat redistribution, while the equilibrium temperature
would be only $\sim 1190\,{\rm K}$ if the planet was always 
at that of apastron. Thus, we conclude that the surface
temperature can vary by a factor of $\sim 3$, depending on the
actual atmospheric dynamics.

\begin{table}
\caption{Spectroscopic and light curve solutions for 
\hatcurB{}, and inferred planet parameters, derived from
the joint modelling of photometric, spectroscopic and radial 
velocity data.}\label{tab:hatp2:parameters}
\begin{center}
\begin{tabular}{lr}
\hline
Parameter					& Value 		\\
\hline
$P$ (days)			        	& $\hatcurBLCP$ 	\\
$E$ (HJD$-2,\!400,\!000$)			& $\hatcurBLCMT$	\\
$T_{14}$ (days)$^{\rm a}$ 			& $\hatcurBLCdur$	\\
$T_{12} = T_{34}$ (days)$^{\rm a}$ 		& $\hatcurBLCingdur$	\\
$R_{\rm p}/R_\star$		              	& $\hatcurBLCrprstar$	\\
\hline
$K$ (\ms)			               	& $\hatcurBRVK$		\\
$k\equiv e\cos\omega$				& $-0.5119\pm0.0040$	\\
$h\equiv e\sin\omega$				& $-0.0543\pm0.0098$ 	\\
$e$						& $0.5148\pm0.0038$	\\
$\omega$					& $186.1^\circ\pm1.1^\circ$	\\
\hline
$a/R_\star$			               	& $\hatcurBPPar$	\\
$b$						& $\hatcurBLCimp$	\\
$i$ (deg)			               	& $\hatcurBPPi{}^{\circ}$	\\
\hline
$M_{\rm p}$ ($M_{\rm Jup}$)			& $\hatcurBPPmlong$	\\
$R_{\rm p}$ ($R_{\rm Jup}$)			& $\hatcurBPPrlong$	\\
$C(M_{\rm p},R_{\rm p})$			& $\hatcurBPPmrcorr$	\\
$\rho_{\rm p}$ (g~cm$^{-3}$)			& $\hatcurBPPrho$	\\
$a$ (AU)			                & $\hatcurBPParel$	\\
$\log g_{\rm p}$ (cgs)				& $\hatcurBPPlogg$	\\
$T_{\rm eff}$ (K)				& $\hatcurBPPteff$ (see $^{\rm b}$) \\
\hline
\end{tabular}
\end{center}
\scriptsize
\noindent \hspace*{1ex} $^{\rm a}$
	\ensuremath{T_{14}}: total transit duration,
	time between first to 
	last contact; \ensuremath{T_{12}=T_{34}}: ingress/egress time, 
	time between first and second, or third and fourth contact.

\noindent \hspace*{1ex} $^{\rm b}$
	This effective temperature assumes uniform
	heat redistribution while the irradiance is averaged on
	the orbital revolution. See text for further details about the issue
	of the planetary surface temperature.
\end{table}

\begin{figure}
\begin{center}
\resizebox{80mm}{!}{\includegraphics{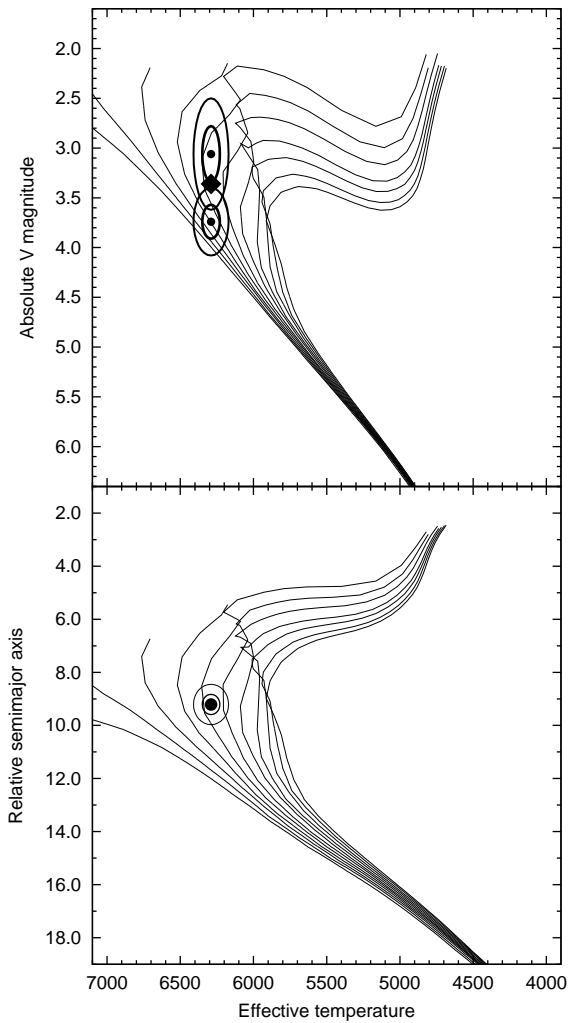}}
\end{center}\vspace*{-5mm}
\caption{Stellar evolutionary isochrones from the Yonsei-Yale models,
showing the isochrones for $[\rm{Fe/H}]=0.14$ stars, between $0.5$ and 
$5.5$\,Gyrs (with a cadence of $0.5$\,Gyrs). The stellar color is indicated by 
the effective temperature, while the left panel shows the luminosity
using the absolute V magnitude $M_{\rm V}$ and the
right panel uses the ratio $a/R_{\star}$ as a luminosity indicator.
In the left panel, the isochrones are overplotted by the 1-$\sigma$ and 
2-$\sigma$ confidence ellipsoids, defined by the effective temperature,
and the absolute magnitude estimations from the TASS catalogue and 
the two Hipparcos reductions (older: upper ellipse, recent: lower ellipse).
The diamond indicates the $M_{\rm V}$ magnitude derived from our best fit
stellar evolution models. On the right plot, the confidence ellipsoid
for the effective temperature and $a/R_{\star}$ is
shown.\label{fig:hatp2:isochrones}}
\end{figure}

\phdsubsection{Photometric parameters and the distance of the system}

The stellar evolution modelling (see Sec.~\ref{sec:hatp2:stellarparams}) also 
yields the absolute magnitudes and colors for the models for various
photometric passbands. We compared the obtained colors and absolute
magnitudes with other observations. First, the $V-I$ color of the 
modelled star was compared with the observations. The TASS catalogue
\citep{droege2006} has magnitudes for this star, 
$V_{\rm TASS}=8.71\pm0.04$ and $I_{\rm TASS}=8.16\pm0.05$, i.e.
the observed color of the star is $(V-I)_{\rm TASS}=0.55\pm0.06$.
The stellar evolution modelling resulted a color of 
$(V-I)_{\rm YY}=0.552\pm0.016$,  which is in perfect agreement with
the observations. The absolute magnitude of the star in $V$ band
is $M_V=3.36\pm0.12$, also given by the stellar evolution models.
This therefore yields a distance modulus of $V_{\rm TASS}-M_V=5.35\pm0.13$,
which is equivalent to a distance of $117\pm7$\,pc, assuming no 
interstellar reddening. This distance value 
for the star is placed right between the distance values found in the two
different available Hipparcos reductions of \cite{perryman1997}
and \cite{vanleeuwen2007a,vanleeuwen2007b}: \cite{perryman1997}
reports a parallax of $7.39\pm0.88$\,mas, equivalent to
a distance of $135\pm18$\,pc while \cite{vanleeuwen2007a,vanleeuwen2007b}
states a parallax of $10.14\pm0.73$\,mas, equivalent to a distance
of $99\pm7$\,pc. In the two panels of Fig.~\ref{fig:hatp2:isochrones}, 
stellar evolutionary isochrones are shown for the metallicity of \hatcurB{},
superimposed by the effective temperature and various 
luminosity estimations based on both the above discussion (relying only
on various Hipparcos distances and TASS apparent magnitudes) 
and the constraints yielded by the stellar evolution modelling.
The 2MASS magnitude of the star
in $J$ band is $J_{\rm 2MASS}=7.796\pm0.027$ while the stellar evolution
models yielded an absolute magnitude of $M_J=2.465\pm0.110$. Thus,
the distance modulus here is $J_{\rm 2MASS}-M_J=5.33\pm0.11$, equivalent
to a distance of $116\pm6$\,pc, confirming the distance derived from
the photometry taken from the TASS catalogue.




\phdsection{Discussion}  
\label{sec:hatp2:discussion}

We presented refined planetary, stellar and orbital parameters 
for the \hatcurB{}(b) transiting extrasolar planetary system. Our improved
analysis was based on numerous radial velocity data points, 
including both new measurements and data
taken from the literature. We have also carried out high precision
follow-up photometry. The refined parameters have uncertainties that
are smaller by a factor of $\sim2$ in the planetary parameters and a factor
of $\sim 3-4$ in the orbital parameters than the previously reported 
values of \cite{bakos2007hatp2}. We note that the density
of the planet turned out to be significantly smaller that the 
value by \cite{bakos2007hatp2}, namely $\rho_{\rm p}=7.6\pm1.1\,{\rm g\,cm^{-3}}$,
moreover, the uncertainty reported by \cite{bakos2007hatp2} was significantly
larger. In our analysis we did not rely on the distance of the system, i.e. 
we did not use the absolute magnitude as a luminosity indicator. Instead,
our stellar evolution modelling was based on the density of the star,
an other luminosity indicator related to precise
light curve and RV parameters. We have compared the estimated distance of the
system (which was derived from the absolute magnitudes, known 
from the stellar modelling) with the Hipparcos distances. We found that our 
newly estimated distance falls between the two values available from
the different reductions of Hipparcos raw data. 

The improved orbital eccentricity and argument of pericenter allow us
to estimate the time of the possible secondary transits. We found 
that secondary transits occur at the orbital phase of 
$\phi_{\rm sec}=0.1886\pm0.0020$, i.e. 1\,day 1 hour and 30 minutes
($\pm$ 16 minutes) after primary transit events. 

The zero insolation planetary isochrones of \cite{baraffe2003} give
an expected radius of $R_{\rm p,Baraffe03}=1.02\pm0.02\,R_{\rm Jup}$,
that is slightly smaller than the measured radius 
of $\hatcurBPPr\,R_{\rm Jup}$. 
The work of \cite{fortney2007} takes into account not 
only the evolutionary age and the total mass of the planet but the 
incident stellar flux and the mass of the planet's core. By scaling
the semimajor axis of \hatcurBb{} to one that yields the same incident
flux from a solar-type star on a circular orbit, taking into account
both the luminosity of the star and the correction for the orbital
eccentricity given by \eqref{semicorr}, we 
obtained $a'=0.033\pm0.003\,{\rm AU}$. 
Using
this scaled semimajor axis, the interpolation based on the tables 
provided by \cite{fortney2007} yields radii between 
$R_{\rm p,Fortney,0}=1.142\pm0.003\,R_{\rm Jup}$ (core-less planets) and
$R_{\rm p,Fortney,100}=1.111\pm0.003\,R_{\rm Jup}$ (core-dominated planets, with a
core of $M_{\rm p,core}=100\,M_{\oplus}$). Although these values agree nicely with
our value of $R_{\rm p}=\hatcurBPPrlong\,R_{\rm Jup}$, the 
relatively large uncertainty of $R_{\rm p}$ excludes any 
further conclusion for the size of the planet's core. Recent models of
\cite{baraffe2008} also give the radius of the planet as the function of
evolutionary age, metal enrichment and an optional insolation for 
equivalent to scaled semimajor axis of $a'=0.045\,{\rm AU}$. Using this 
latter insolation, their models yield 
$R_{\rm p,Baraffe08,0.02}=1.055\pm0.006\,R_{\rm Jup}$ (for metal poor, $Z=0.02$ planets)
and
$R_{\rm p,Baraffe08,0.10}=1.008\pm0.006\,R_{\rm Jup}$ (for more metal rich, $Z=0.10$ planets).
These values are slightly smaller than the actual radius of \hatcurBb{},
however, the actual insolation of \hatcurBb{} is roughly two times larger than
the insolation implied by $a'=0.045\,{\rm AU}$. Since the 
respective planetary radii of \cite{baraffe2008} for zero insolation
give $R^{(0)}_{\rm p,Baraffe08,0.02}=1.009\pm0.006\,R_{\rm Jup}$
and $R^{(0)}_{\rm p,Baraffe08,0.10}=0.975\pm0.006\,R_{\rm Jup}$ for the
respective cases of $Z=0.02$ and $Z=0.10$ metal enrichment, an
extrapolation for a two times larger insolation would put the expected planetary
radius in the range of $\sim 1.10\,R_{\rm Jup}$. This is consistent with
the models of \cite{fortney2007} as well as with the measurements. However, 
as discussed earlier in the case of \cite{fortney2007} models,
the uncertainty in $R_{\rm p}$ does not let us properly constrain the metal 
enrichment.

\hatcurBb{} will remain an interesting target, as a member of an
emerging heavy-mass population. Further photometric measurements will
refine the light curve parameters and therefore more precise stellar
parameters can also be obtained. This will yield smaller uncertainties
in the physical planetary radius, thus some parameters of the 
planetary evolution models, such as the metal enrichment can be obtained
more explicitly. Moreover, observations of secondary eclipses will reveal
the planetary atmosphere temperature which now is poorly constrained. Since 
the secondary eclipse occurs shortly after periastron passage, the 
temperature and therefore the contrast might be high enough to detect
the occultation with a good signal-to-noise ratio.


\phdchapter{Summary}
\label{chapter:summary}

Transiting extrasolar planets are the only group among the extrasolar
planets whose basic physical parameters, such as mass and radius
can be determined without any ambiguity. Therefore, these planets
provide a great opportunity to determine other properties, such 
as the characteristics of the planetary interior or their
atmosphere. 
Recently, wide-field photometric surveys became the most prominent 
observation techniques for detecting transiting planets and
these surveys yielded several dozens of discoveries. Since such wide-field
surveys yield massive amount of data which cannot be efficiently
and consistently processed by the available existing software solutions, 
I started developing a new package in order to overcome the related
problems. The development of this package has been related to the 
Hungarian-made Automated Telescope Network (HATNet)
project, one of the most successful initiatives searching for
transiting extrasolar planets. 

The aims of my work were both implementing the algorithms related 
to the photometric reduction in a form of a standalone software
package, as well as applying these programs in the analysis of the 
HATNet data. Additionally, the photometric reduction is intended to work
on data obtained by other facilities, typically 1m-class
telescopes (such as the 
48" telescope at Fred Lawrence Whipple Observatory 
or the Schmidt telescope at the Piszk\'estet\H{o} Mountain Station). 

Of course, both the confirmation of planetary candidates
and the characterization of known objects require other types of technologies
such as spectroscopy, radial velocity measurements and stellar evolution
modelling. In order to perform a consistent determination of the 
planetary, orbital and stellar parameters of transiting exoplanetary
systems, my work also focused on to include these additional types of 
measurements and methods in the data analysis.

In this PhD thesis I presented a new software package intended to perform
photometric data reduction on massive amount of astronomical images. 
Existing software solutions do not provide a consistent 
framework for the reduction of images acquired by wide-field and undersampled 
instrumentation. During the development of the related algorithms and the 
implementation, I focused on the issues related to these problems
in order to have a homogeneous reduction environment, ranging from the 
calibration of frames to the final light curve generation and analysis. 
This new package has been successfully applied in processing
the images of the HATNet
and led to the discovery and confirmation of almost a dozen of 
transiting extrasolar planets.


\section*{Acknowledgments}

First of all, I would like to thank my parents and family for
their immutable support during the years of my studies and in my whole life. 

I would like to thank G\'asp\'ar Bakos for inviting to the 
project and for the possibility to be a member of people
working in the field of transiting extrasolar planets. I also thank
my supervisor, B\'alint \'Erdi for the opportunity to be a PhD 
student at the E\"otv\"os University and for his help in the 
proofreading. I am grateful to the hospitality of the Harvard-Smithsonian
Center for Astrophysics, where this work has been partially carried
out. I would like to thank G\'asp\'ar, his wife Krisztina Meiszel and
and other friends, Istv\'an Cziegler, G\'abor F\H{u}r\'esz, Bence 
Kocsis, M\'aria Pet\H{o} and D\'avid V\'egh for their 
continuous and great help related to the life in Cambridge and around Boston.

I thank Brigitta Sip\H{o}cz for her comments, ideas for improvements 
and bug reports related to the data analysis programs. I also thank 
collaborators at the Harvard-Smithsonian CfA and the Konkoly Observatory, 
Joel Hartman, G\'abor Kov\'acs, G\'eza Kov\'acs, Robert Noyes
and Guillermo Torres for their help.

I would like to say thanks to my friends Bal\'azs Dianiska, \'Agnes 
K\'osp\'al, Andr\'as L\'aszl\'o and Judit Szul\'agyi for their 
encouragement and for their
valuable comments on this dissertation. Like so, I thank Eric Agol, Daniel 
Fabrycky, Bence Kocsis and Joshua Winn for their help during the
preparation of various articles related to this field of science. 
I also thank Edward Miller, one of my former roommates for his help
on the earlier versions of the draft. 

Last but not least, I would like to thank P\'eter \'Abrah\'am, Ferenc Horvai
and Csaba Kiss for their present support and the opportunity to 
continue the related research in the Konkoly Observatory. 


{}


\onecolumn
\appendix

\end{document}